\documentclass[usenatbib,useAMS]{mnras}
\pdfoutput=1
\usepackage{epic,eepic}
\usepackage{graphicx}
\usepackage{color}
\usepackage{wasysym}

\usepackage[english]{babel}
\usepackage[latin1, applemac]{inputenc}
\usepackage{deluxetable}

\newcommand{\mgii}{Mg\,{\sc ii}{\rm \,}}
\newcommand{\caii}{Ca\,{\sc ii}{\rm \,}}
\newcommand{\civ}{C\,{\sc iv}\,}
\newcommand{\feii}{Fe\,{\sc ii}\,}
\newcommand{\feiii}{Fe\,{\sc iii}\,}
\newcommand{\oiii}{[O\,{\sc iii}]\,}
\newcommand{\contrast}{\ensuremath{m_{\mathrm{nuc}}-m_{\mathrm{host}}}}
\newcommand{\mnuc}{\ensuremath{m_{\mathrm{nuc}}}}
\newcommand{\mhost}{\ensuremath{m_{\mathrm{host}}}}

\def\ttim{\emph{\sc Tiny Tim}\hspace{2pt}}
\def\md{\emph{\sc MultiDrizzle}\hspace{2pt}}

\def\mean#1{\left< #1 \right>}

\title{A \emph{Hubble Space Telescope} Imaging Study of Four FeLoBAL Quasar Host Galaxies}

\author[Lawther, D., Vestergaard, M., Fan, X.]{Lawther, D.,$^1$ Vestergaard, M.,$^1$ $^2$ Fan, X.$^2$\\
	$^1$ Dark Cosmology Centre, Niels Bohr Institute, University of Copenhagen.\\
	$^2$ Steward Observatory, University of Arizona, 933 N. Cherry Avenue, 85721 Tucson, AZ, USA}

\begin{document}
%\selectlanguage{english}

\maketitle

\begin{abstract}
We study the host galaxies of four Iron Low-Ionization Broad Absorption-line Quasars (FeLoBALs) using \emph{Hubble Space Telescope} imaging data, investigating the possibility that they represent a transition between an obscured AGN and an ordinary optical quasar. In this scenario, the FeLoBALs represent the early stage of merger-triggered accretion, in which case their host galaxies are expected to show signs of an ongoing or recent merger. Using PSF subtraction techniques, we decompose the images into host galaxy and AGN components at rest-frame ultraviolet and optical wavelengths. The ultraviolet is sensitive to young stars, while the optical probes stellar mass. In the ultraviolet we image at the BAL absorption trough wavelengths so as to decrease the contrast between the quasar and host galaxy emission. We securely detect an extended source for two of the four FeLoBALs in the rest-frame optical; a third host galaxy is marginally detected. In the rest-frame UV we detect no host emission; this constrains the level of unobscured star formation. Thus, the host galaxies have observed properties that are consistent with those of non-BAL quasars with the same nuclear luminosity, i.e., quiescent or moderately starforming elliptical galaxies. However, we cannot exclude starbursting hosts that have the stellar UV emission obscured by modest amounts of dust reddening. Thus, our findings also allow the merger-induced young quasar scenario. For three objects, we identify possible close companion galaxies that may be gravitationally interacting with the quasar hosts.
\end{abstract}


\section{Introduction}\label{sec:intro}

Considering the central gravitational potentials required to power quasar activity, galaxies that have harbored quasars should host inactive `relic' black holes with masses of order $10^8M_{\astrosun}$ today \citep{Soltan1982}. Indeed, studies of stellar dynamics in galaxy bulges reveal that most or all massive galaxies at low redshift contain a central supermassive black hole (SMBH) with masses of approximately $10^{6}M_{\astrosun}$--$10^{9}M_{\astrosun}$ \citep[e.g.,][]{Kormendy2001}.  Unless there are alternative ways of growing black holes to such high masses, the ubiquity of inactive SMBHs suggests that most galaxies underwent an active phase at some point in their lifetimes. Given the large energy output involved, the onset of quasar activity may affect the evolution of the ambient gas (and therefore, the future star formation). The observed correlations between black hole mass and host galaxy bulge luminosities and stellar velocity dispersions \citep[e.g.,][]{Magorrian1998,Ferrarese2000,Tremaine2002,Marconi2003,Gultekin2009} support this picture. Therefore, to understand the evolution of massive galaxies, we need to understand the quasar phase.

However, the triggering mechanisms for quasar activity are as yet poorly understood. Several authors have proposed that galaxy mergers may trigger quasar activity by forcing large amounts of gas to sink towards the SMBH at the centre of the galaxy, and that the subsequent quasar activity may subsequently expel some of this gas through radiation pressure \citep[e.g.,][]{Sanders1988b,Fabian1999}. This scenario has been explored in numerical simulations of galaxy mergers including supermassive black hole components capable of accreting gas and exerting feedback on their surroundings \citep[e.g.,][]{Barnes1991,Hopkins2005,DiMatteo2005}; the feedback prescriptions applied may represent direct radiation pressure due to the central source or kinetic feedback from outflowing gas launched at small radii. According to the simulations performed by \citet{Hopkins2005}, merger-triggered quasars are intrinsically brightest around the time at which the two galaxies coalesce. However, this phase also displays the highest gas column densities ($N_{\mathrm{H}}\apprge10^{24}$ cm$^{-2}$) along the line of sight to the black hole. Thus, the intrinsically most luminous quasar phase is heavily obscured to a distant observer. As the quasar exerts radiative feedback on the obscuring gas and dust at the centre of the host galaxy, it becomes detectable as a reddened quasar, before entering an unobscured phase, and eventually becomes quiescent as the lack of gas in the nucleus starves the SMBH of fuel.

How would this young evolutionary phase reveal itself to observations? \citet{Glikman2012} find that the reddest quasars at a given redshift tend to be intrinsically the most luminous, as expected for the scenario described above. \citet{Urrutia2008} and \citet{Glikman2015} show that the reddest quasars (at $z<1$ and $z\approx2$, respectively) tend to show disturbed morphologies, indicative of recent or ongoing merger activity. However, \citet{Mechtley2016} demonstrate that inactive galaxies at $z\approx2$ show evidence of merger activity at a similar rate to quasar hosts. This raises the possibility that the `red quasars' found by \citet{Glikman2015} tend to be merger-hosted simply due to their selecting\footnote{The FIRST-2MASS sample are detected in 20 cm and in 2 micron surveys, while being optically faint; details of their selection criteria are presented by \citet{Glikman2012}.} heavily reddened quasars, where the excess reddening is due to merger-fueled starburst activity, with no causal relationship between the AGN activity and the merger. As noted by \citet{Cisternas2011}, the interpretation of the observed merger fractions depends critically on our knowledge of the timescales involved for the (observable) merger and quasar activity. 

An alternative approach is to search for observational evidence of the purported `blowout' phase. Broad Absorption Line (BAL) quasars, which display outflow velocities ranging from $\sim2000$ km s$^{-1}$ up to $\sim0.1c$, have been considered as candidate transition objects \citep[e.g.,][]{Weymann1991}. They are classified as high-ionization BAL quasars (HiBALs), which display broad absorption only in high-ionization lines such as \civ $\lambda1549$, or as low-ionization BAL quasars (Lo\-BALs), which additionally display broad absorption in low-ionization transitions such as \mgii $\lambda$ 2798. Fe\-Lo\-BALs are a subclass of Lo\-BAL quasars that display absorption in excited states of \feii and \feiii in addition to the Lo\-BAL absorption lines \citep[e.g.,][]{Becker1997,Hall2002}. 

In the context of AGN unification scenarios, some authors have proposed that BAL outflows may be present in all quasars, but are only observed in absorption at certain orientation angles \citep[e.g.,][]{Elvis2000}. This orientation-based model is supported by the work of \citet{Gallagher2007}, who find the mid-infrared emission in HiBAL and non-BAL quasars to be statistically indistinguishable; \citet{Schulze2017} also find this for LoBAL quasars in the redshift range $0.6<z<2.5$. This argues against a large covering fraction for the BAL-absorbing material. Conversely, based on the lower levels of $[$O \textsc{iii}$]$ observed for LoBALs, \citet{Voit1993} suggest that the BAL absorbing material has a high covering fraction, and speculate that Lo\-BALs may be young quasars in the process of expelling an optically thick cocoon of gas and dust; we note, however, that \citet{Schulze2017} do not find significantly reduced $[$O \textsc{iii}$]$ emission in stacked $z\sim1.5$ LoBAL spectra. \citet{Canalizo2001} find a connection between Lo\-BALs and merger-triggered Ultraluminous Infrared Galaxies at low redshift: three out of six of their objects for which the BAL status could be unambiguously tested are indeed Lo\-BALs. Indeed, FeLoBALs are overrepresented amongst the reddest quasars \citep{Urrutia2009,Glikman2012}, suggesting that Fe\-Lo\-BAL activity may play a role in the proposed transition scenario. Interestingly, \citet{Farrah2012} report an anticorrelation between absorption strength and star formation activity for a sample of 31 FeLo\-BALs observed with the \emph{Spitzer Space Telescope}. This finding adds observational support to suggestions \citep[e.g.,][]{Granato2004} that the kinetic feedback due to Lo\-BAL outflows may represent the sought-after quenching mechanism for star formation in massive galaxies. Such quenching is otherwise often modeled using a semi-empirical AGN feedback prescription \citep[e.g.,][]{Croton2006,Bower2006}. On the other hand, recent detections of FeLo\-BAL variability are best explained by a low covering fraction for BAL absorption, with the BAL-absorbing gas situated at $\sim$pc radii from the central black hole \citep{Vivek2012,McGraw2015}. Such findings are difficult to reconcile with a scenario where BAL outflows provide galaxy-wide quenching of star formation. 

One way to test the evolutionary scenario for FeLoBAL quasars is to compare their host galaxy properties to those of non-BAL quasars. If the FeLoBAL absorption is intrinsically present in all quasars but with a small covering factor, e.g., in the disk-wind scenario of \citet{Elvis2000}, their host galaxies are expected to be massive elliptical galaxies with little recent star formation, as seen for non-BAL quasar hosts at $z\apprle0.2$ \citep[e.g.,][]{Nolan2001,Dunlop2003}, or perhaps massive ellipticals that are still actively star-forming at rates of $\sim100M_{\astrosun}$ yr$^{-1}$ \citep{Floyd2012}\footnote{For $2<z<3$ AGN, star formation rates appear to increase with the luminosity of the central source, reaching $\sim600M_{\astrosun}$ yr$^{-1}$ for the brightest quasars \citep{Harris2016}.}. On the other hand, if FeLoBALs are young quasars triggered by mergers, they may display merger-triggered starburst activity, and/or display interacting companion galaxies or highly disturbed morphologies due to recent interactions. Here we present \emph{Hubble Space Telescope} (\emph{HST}) imaging data for four overlapping-trough FeLoBAL quasars, as defined by \citet{Hall2002}. These objects are so strongly absorbed by \feii and \feiii that their ultraviolet (UV) spectra are barely recognizable as quasar spectra (Figure \ref{fig:0300_spec}). The quasar continuum emission is reduced by a factor $\sim10$ for these objects. Assuming that the BAL obscuration is centrally concentrated and does not absorb the host galaxy emission, this will reduce the nucleus-to-host brightness contrast when using a filter that covers the deepest BAL absorption. Thus, the heavy attenuation of the UV nuclear emission aids our study of the host galaxy stellar populations.

This paper is structured as follows. In \S \ref{sec:data} we outline our sample selection criteria and describe the \emph{HST} observations. We describe the image processing in \S \ref{sec:method_calibration}. In \S \ref{sec:fitting} we describe our image decomposition strategy, and present the host galaxy modeling results. We discuss our findings with regard to the nature of the FeLoBAL host galaxies in \S \ref{sec:discussion}. We use a cosmology with Hubble constant $H_0=67.8\pm0.9$ km s$^{-1}$ Mpc$^{-1}$ and matter density parameter $\Omega_\mathrm{m}=0.308\pm0.012$ throughout.

\section{Sample selection and \emph{HST} imaging observations}\label{sec:data}

\subsection{FeLoBAL quasar sample selection}\label{sec:sample}

One important objective of this study is to measure the rest-frame UV and optical brightnesses of FeLo\-BAL host galaxies, in an effort to characterize the stellar population. We expect the quasar continuum emission to outshine the host galaxy in the rest-frame UV, especially if the stellar population is quiescent and therefore faint in the UV \citep[e.g.,][]{Kinney1996,Bruzual2003}. Host galaxy detection is therefore challenging for these high-redshift quasars. The very broad absorption troughs of FeLoBAL quasars may provide a `natural coronagraph' effect aiding host galaxy detection. If the absorbing material is concentrated near the nucleus, as generally expected due to the high velocities of the BAL-absorbing gas being attributed to the central engine \citep[although \emph{cf.}][]{Faucher-Giguere2012}, the nuclear emission will be absorbed more heavily than that of the host galaxy. In this study, we therefore select rest-frame UV filter bandpasses that coincide with the BAL absorption, with the aim of reducing the nuclear-to-host brightness contrast. 

Due to their red colours, heavily-absorbed FeLoBAL quasars are often selected as serendipitous targets (as opposed to quasar candidates) by the Sloane Digital Sky Survey \citep[SDSS][]{York2000} photometric selection pipeline \citep{Stoughton2002}. To identify unusual BAL quasars, \citet{Hall2002} perform a visual inspection of all spectra which could not be identified in an automated fashion by the SDSS Early Data Release (EDR) spectroscopic pipeline, along with a re-inspection of all objects identified as quasars. These authors find 18 unusual BAL quasars in the EDR, along with two unidentified objects that may or may not be BALs. We select our FeLoBALs from this sample of unusual objects. The authors note that this sample is not statistically complete, as the EDR selection criteria varied during the spectroscopic observing campaign. The total number of quasar spectra in the EDR is $\sim8000$. However, the observed rarity of these `unusual BALs' relative to ordinary quasars is likely to be extremely sensitive to selection effects.

\begin{figure*}
	\advance\leftskip-3cm
	\centering
	\includegraphics[scale=0.29,clip=true,trim=2cm 1cm 1cm 0]{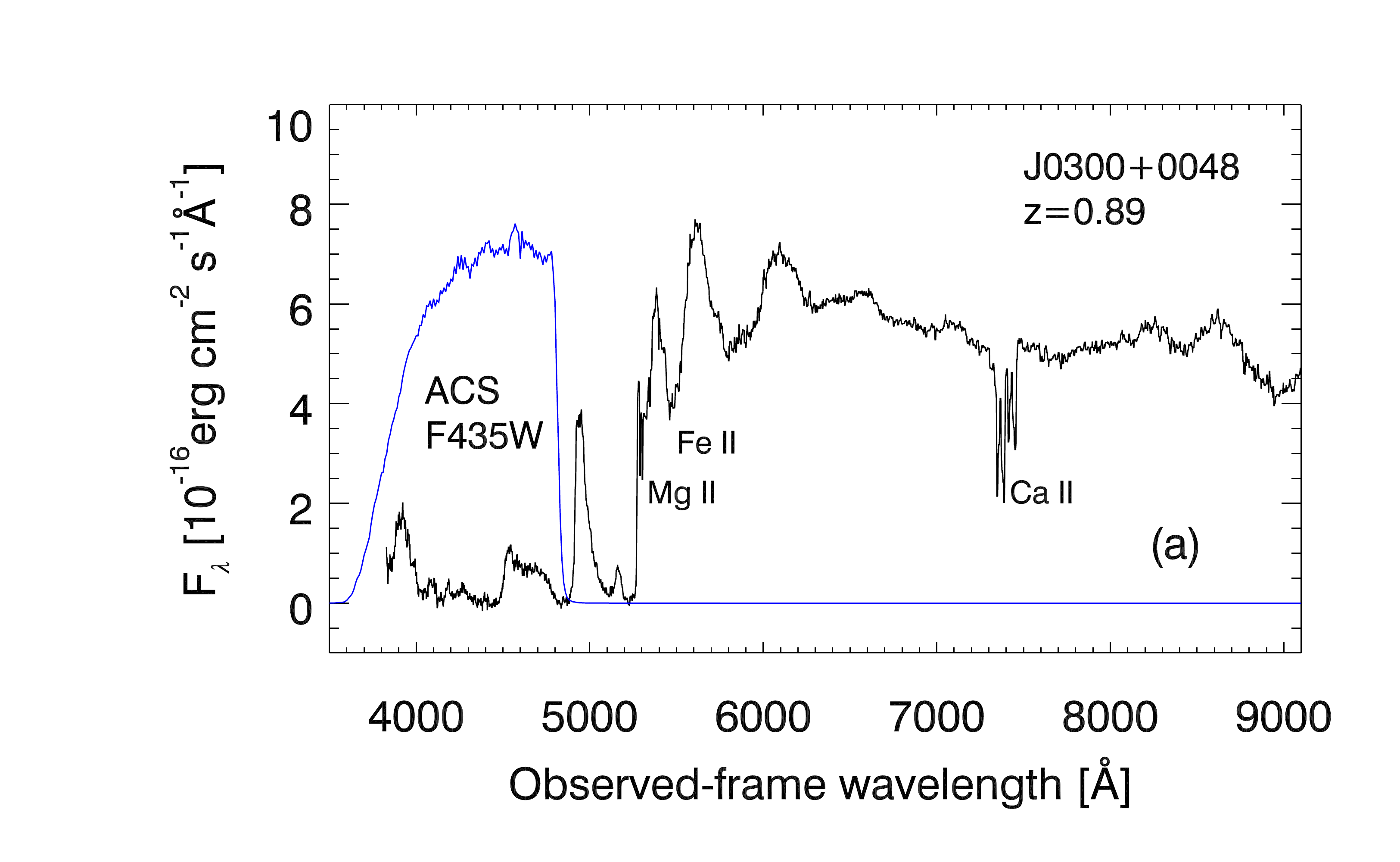}
	\includegraphics[scale=0.29,clip=true,trim=2cm 1cm 1cm 0]{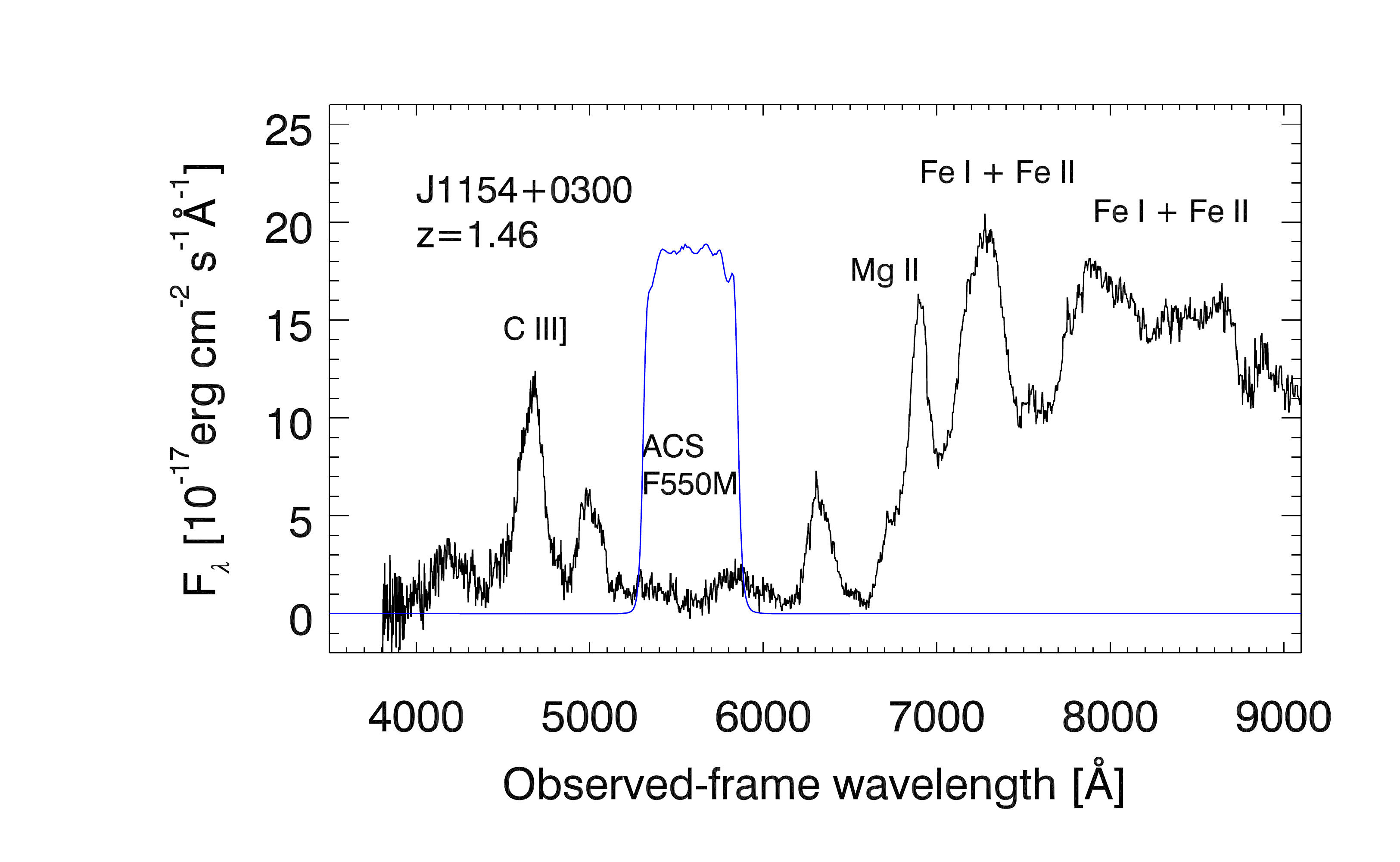}
	\includegraphics[scale=0.29,clip=true,trim=2cm 1cm 1cm 0]{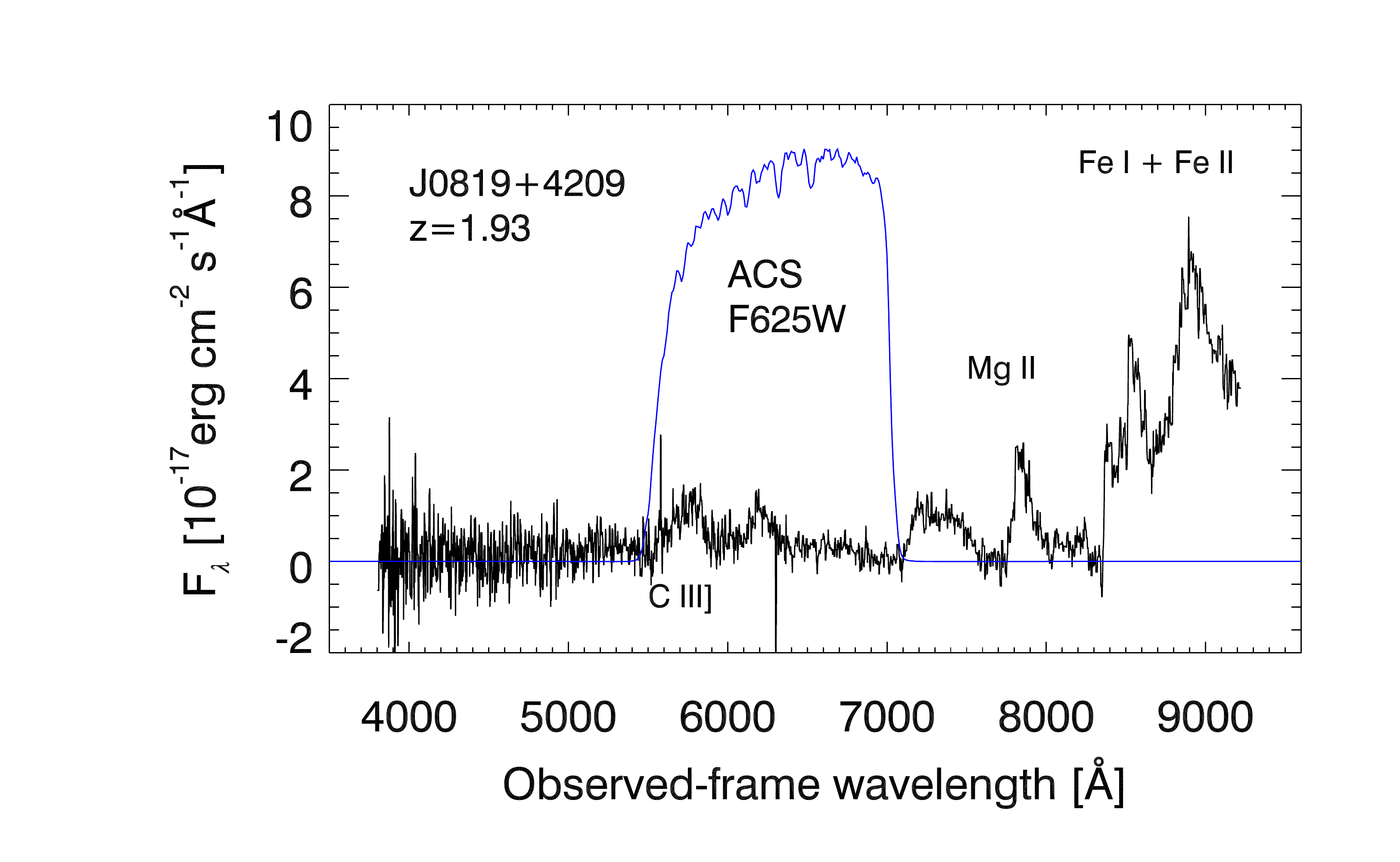}
	\includegraphics[scale=0.29,clip=true,trim=2cm 1cm 1cm 0]{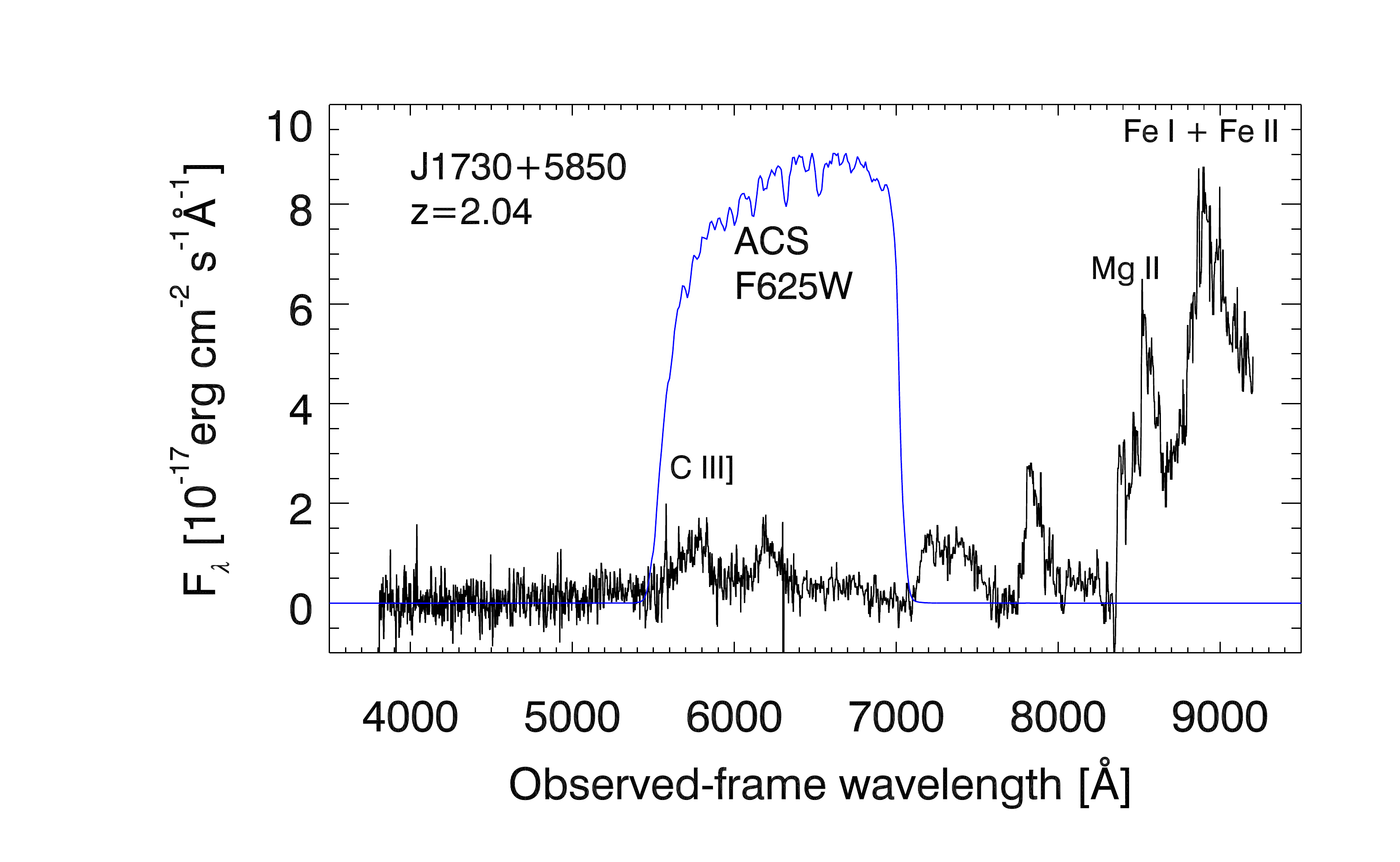}
	\centering
	\caption{SDSS Data Release 8 spectra of the four Fe\-LoBAL quasars in our sample, demonstrating our utilization of the 'natural coronagraph' offered by the BAL absorption (\S \ref{sec:sample}). The flux densities are median-smoothed using a 5-pixel smoothing window. The (arbitrarily scaled) throughput curve of the ACS/WFC broadband filter chosen for each quasar is shown as a blue curve. While we show selected emission line identifications here, we direct the reader to \citet{Hall2002} for detailed redshift determinations.}\label{fig:0300_spec}
\end{figure*}

Our main selection criterion is that the BAL trough should be sufficiently broad and deep that we can utilize the `natural coronagraph' effect for our UV broad-band imaging observations. The `overlapping-trough' class of Fe\-LoBAL quasars, five of which are presented by \citet{Hall2002}, fulfill this requirement: J0300+0048, J1154+0300, J0819+4209, J1730+5850 and J0437-0045. One of these objects, J0437-0045 located at $z=2.82$, was not selected as it would require more than 4 orbits to be detected with NICMOS. The remaining four Fe\-LoBAL objects, residing at $0.89\le z\le2.04$, were observed under the \emph{HST} GO-10237 program (PI: X. Fan). Figure \ref{fig:0300_spec} shows the rest-frame optical-UV spectra of the four Fe\-LoBAL quasars, extracted from the SDSS Data Release 8. The imaging filters selected for the ACS/WFC observations are overlaid.

We use the redshifts provided by \citet{Hall2002}, who present detailed discussion of this issue. The redshift determination is secure (if perhaps imprecise) for J0300+0048, as it is based on intrinsic \caii\,absorption, and confirmed by broad Hydrogen emission lines. For J1154+0300, the redshift is based on various optical iron emission lines redwards of \mgii. For the remaining objects, the redshift is less secure, as it is based on various emission and absorption features in the BAL-absorbed part of the spectrum, where the continuum level is unknown.

\subsection{\emph{Hubble Space Telescope} observations}\label{sec:hst_observations}

Each Fe\-LoBAL quasar was observed in Cycle 13 in two \emph{HST} imaging bands, covering the rest-frame optical (with NICMOS) and UV (with ACS) wavelengths. The rest-frame optical regime is sensitive to the stellar mass, while the rest-frame UV is sensitive to ongoing star formation. For both instruments, the observations were performed using a four-point dither pattern utilizing non-integer pixel shifts. This allows us to identify cosmic ray hits and hot/cold pixels during image combination, and provides an improved spatial resolution in the combined images. The dither patterns `ACS-WFC-DITHER-BOX' and `NIC-SPIRAL-DITH' were used for the ACS/WFC and NICMOS observations, respectively\footnote{To allow observations to be distributed over multiple orbits, the dither pattern was in some cases implemented using the POS-TARG mode. However, the resulting pixel offsets and exposure time distributions are the same as for the specified standard dither patterns.}. Table \ref{tab:datasummary} summarizes the observations, including the imaging filters and exposure times adopted for each source.

\paragraph*{ACS/WFC:} Imaging of the rest-frame UV regime was obtained using the Advanced Camera for Surveys (ACS) Wide Field Camera (WFC).  The ACS/WFC has a pixel scale of 0.05 arcsec$^2$ and a $202\times202$-arcsec field of view. When selecting a filter to utilize the `natural coronagraph' effect, we considered both the absorption strength and the observing efficiency. Although use of narrow-band filters would maximize the integrated absorption strength in the bandpass, we elect to use broad- or medium-band filters, so as to avoid unreasonably long exposure times. We estimate that the quasar continuum emission is reduced by a factor of 7 to 15 in the selected ACS/WFC filters for objects J0300+0048 (Figure \ref{fig:0300_spec}). The intrinsic continuum levels in the relevant bandpasses are difficult to determine for the other quasars, as the SDSS spectra lack regions of unabsorbed continuum emission.

\paragraph*{NICMOS:} The \emph{HST} Near Infrared Camera and Multi-Object Spectrometer (NICMOS) was used to observe rest-frame optical wavelengths. The NICMOS 2  detector (hereafter NIC2) has a pixel size of $0.076$ arcsec $\times$ $0.075$ arcsec, and a $19.2$ arcsec $\times$ $19.2$ arcsec field of view. The PSF core FWHM is around $0.14$ arcsec for wavelengths below $\sim1.6\mu$m, i.e., the images are somewhat undersampled for the bluer NIC2 filters\footnote{For two quasars we are able to achieve Nyquist sampling in our \md-combined images, see \S \ref{sec:method_calib_multidrizzle}.}. Due to lack of spectral coverage in the rest-frame optical, we do not have the opportunity to utilize a `natural coronagraph' effect here. On the other hand, we expect a smaller intrinsic (i.e., unabsorbed) nucleus-to-host contrast in our NICMOS imaging, as we expect the host galaxies to be brighter, and the quasar power-law continuum fainter, in the rest-frame optical relative to the UV. While the host galaxy is brightest relative to the active nucleus at near-infrared wavelengths $\sim1\mu$m, we chose somewhat bluer bandpasses so as to allow a robust comparison with quasar host samples available at the time of our \emph{HST} proposal (\S \ref{sec:discussion_comparison}). We select broad-band filters, as a high signal-to-noise ratio is required for our image decomposition analysis. These bandpasses also cover the \oiii narrow emission line. As the narrow line-emitting region can extend to kpc scales for some AGN \citep[e.g.,][]{Pogge1989}, there is a risk of contamination of the host galaxy signal. However, Lo\-BAL quasars generally display weak \oiii emission \citep{Weymann1991}, mitigating this issue.

\paragraph*{Limiting apparent magnitudes:}

The limiting apparent magnitudes reached for a $10\sigma$ detection of a point source for each observation is listed in Table \ref{tab:datasummary}.  Our observations have a depth equal to or greater than that of the NICMOS imaging of the quasar samples of \citet{Kukula2001} and \citet{Hutchings2002} at similar redshifts. This allows a robust comparison to these studies, which were state-of-the-art at the time of our \emph{HST} proposal; we also reach a similar depth to later studies that are better matched to our sample in terms of AGN luminosity (\S \ref{sec:discussion}).

\subsection{Point Spread Function Star Observations}\label{sec:maintext_psf}

No separate Point Spread Function (PSF) star observations were performed for this \emph{HST} observing program; our original intention was to use analytical PSF templates for this purpose. However, the current consensus on such models is that they do not reproduce the PSF core accurately, and therefore perform poorly in PSF-host decomposition studies of bright quasars \citep[e.g.,][]{Kim2008,Mechtley2012}. We confirm these findings in a series of preliminary tests (Appendix \ref{sec:appendix_psf}). We therefore use stellar observations as PSF models in this study. For the ACS imaging we construct a stacked PSF template using stars observed in the ACS field during the science observations. For the NICMOS imaging we stack archival stellar observations performed within three months of the science observation.  We discuss our PSF modeling strategies and considerations, including the selection criteria for PSF star observations in each bandpass, in Appendix \ref{sec:appendix_psf}.

% ; use of PSF star observations is therefore preferable when performing PSF-galaxy decompositions for bright quasars

\section{Processing and calibration of the data}\label{sec:method_calibration}

We calibrate the individual exposures using tools from the \textsc{PyRAF/STSDAS}\footnote{\textsc{STSDAS} and \textsc{PyRAF} are products of the Space Telescope Science Institute, which is operated by AURA for NASA.} package \emph{hst\_calib}; we outline this calibration for ACS/WFC and NICMOS/NIC2 in \S \ref{sec:method_calibration_acs} and \S \ref{sec:method_calibration_nicmos}, respectively. After calibrating the individual images, we combine the individual exposures for a given quasar using the \md package (\S \ref{sec:method_calib_multidrizzle}), thereby obtaining a single, combined ACS/WFC image and a single, combined NICMOS/NIC2 image for each quasar.

\subsection{ACS individual exposure image calibration}\label{sec:method_calibration_acs}

We calibrate the ACS/WFC imaging data following the method of the ACS pipeline calibration, as outlined in the ACS Data Handbook \citep{ACSdata}. We process the raw imaging data using the \emph{`calacs'} script, version 2012.2; this version includes a correction for charge transfer efficiency (CTE) degradation. Cosmic ray hits and hot/cold pixels are identified in an automated fashion during a preliminary processing using the \md algorithm (\S \ref{sec:method_calib_multidrizzle}). Upon visual inspection of the single-exposure images we identified and flagged a few additional bad pixels not identified as such by the \md processing. These included several `hot pixels' that were not included in the static bad pixel mask for the detector, and that are not bright enough to be flagged by \md. 

\subsection{NICMOS individual exposure image calibration}\label{sec:method_calibration_nicmos}

The NICMOS images display signatures of detector anomalies that require treatment in addition to the pipeline processing. These anomalies are common in NICMOS data, and are described in detail in the NICMOS Data Handbook \citep{NICMOSdata}. We process the raw images using the \textsc{PyRAF/STSDAS} script \emph{`calnica'}. Here we briefly describe the additional processing steps applied.

\paragraph*{Pedestal offset and amplifier glow:} The single-exposure images display a residual flatfield pattern due to the so-called pedestal offset. As advised by \citet{NICMOSdata}, we apply the \emph{`biaseq'} script partway through the \emph{`calnica'} processing, and apply the task \emph{`pedsub'} after \emph{`calnica'} processing. The \emph{`biaseq'} task is designed to remove the non-linear component of the NICMOS DC bias offset; \emph{`pedsub'} addresses the linear component. However, we find that the application of these two tasks does not remove the anomalous background structure completely. We measure a residual amplifier glow at a count-rate of between 0.03 counts s$^{-1}$ and 0.07 counts s$^{-1}$ above the median background level. As it is vital for our host galaxy analysis that we can determine the sky background level precisely, we perform an additional correction to the large-scale background structure, based on a method presented by \citet{Hsaio2010}. These authors find that the residual background in NIC2 F110W and F160W imaging, after flatfielding and \emph{`pedsub'} processing, can be modeled by two components. The first component scales with exposure time and is dominated by the amplifier glow contribution, while the second component scales with the background level. Scaling these components as prescribed by Hsiao et al. (their Equation 4) overestimates the amplifier glow signal in our data. We instead scale and subtract the two components using least-squares minimization, using the \textsc{mpfit} software\footnote{The \textsc{IDL} package \textsc{mpfit} is written by Craig Markwardt and is available from \url{http://cow.physics.wisc.edu/\~craigm/idl/fitting.html}.}, and masking bright sources for the fitting procedure. Compared to the pipeline-processed images, the standard deviation of the mean pixel intensity, measured in background-dominated regions, is reduced by $\approx47$\%\,for three of our quasars. The improvement is more modest ($\approx13$\%) for J1730+5850, as this observation was not impacted by SAA persistence (see below). While the subtraction of an empirical template increases the photon shot noise in the resulting images, this increase is in all cases less than 2\%.

\paragraph*{Photon persistence:} The NICMOS detector suffers from photon persistence, i.e., pixels amassing a large amount of charge during an exposure generate a spurious signal in subsequent exposures.  Photon persistence gives rise to two issues identified in our data. Firstly, the passage of the \emph{HST} through the South Atlantic Anomaly (SAA) between exposures results in an increased flux of cosmic rays. As the persistence behavior varies across the detector, this bombardment causes a persistent signal in a characteristic pattern. We see this pattern for all observations apart from that of  J1730+5850. We use the \emph{`SAAclean'} algorithm \citep{Barker2007} to remove the SAA persistence signal; we find that this algorithm removes the SAA signal more effectively when combined with the Hsaio et al. treatment described above. Secondly, the quasar itself leaves a point-like persistent signal at its previous dither position; we mask these regions before image combination.

\paragraph*{Nonlinearity correction:} The final calibration step for our single-exposure images addresses the count rate dependent nonlinearity identified by \citet{Bohlin2006}. We process each image with the \textsc{PyRAF/STSDAS} script \emph{rnlincor}, which models the non-linearity as a power-law with an empirically determined wavelength-dependent exponent, and corrects the measured count rate in each pixel using this model.

\subsection{Combination of dithered images}\label{sec:method_calib_multidrizzle}

We combine the dithered single-exposure images using the \textsc{PyRAF/STSDAS} package `\md'\footnote{The `\md' routine is based on the drizzle algorithm developed by \citet{Fruchter2002}}. This results in a single combined NICMOS image, and a single combined ACS image, for each quasar. `\md' identifies cosmic rays and other bad pixels in the single-exposure images by comparison with a median image shifted into a common reference frame. We use the resulting bad pixel maps as the sole method of cosmic ray identification for the ACS images, and as a supplementary method for the NICMOS data. For the ACS data, we weight the input pixels using the inverse of the error image generated by the \emph{`calacs'} pipeline. For the NICMOS data, several observations suffer cosmic ray hits near the PSF core, and the affected pixels are not suitable for inclusion in the combined image. As the inverse-error weighting scheme biases output images towards low fluxes when combining a small amount of input frames \citep{Cracraft2007}, we instead use exposure-time weighting for the NICMOS data. 

Table \ref{tab:multidrizzle} lists the `\md' settings that we select for the final image combination for each quasar. For each image we select the settings that yield the narrowest PSF FWHM in terms of angular size, whilst avoiding image combination artifacts. We test for such artifacts via visual inspection, and by measuring the standard deviation of the combined weight image \citep{Fruchter2002}. Full Nyquist sampling is not achievable for J0300+0048(NIC2) and J0819+4209(NIC2) without introducing image artifacts. To test the sensitivity of our image decomposition to this issue, we repeat the modeling of \S \ref{sec:results_nic} for these two observations after broadening the data and PSF images to Nyquist sampling via convolution with a Gaussian smoothing kernel, as suggested by \citet{Kim2008}. This treatment makes no qualitative difference to our PSF-only and PSF-plus-Sersic modeling (\S \ref{sec:results_nic}) for these quasars.

\section{Imaging Analysis}\label{sec:fitting}

Here, we describe our 2D image decomposition strategy and define detection criteria (\S \ref{sec:method}), and present our modeling results for the ACS (\S \ref{sec:results_acs}) and NICMOS (\S \ref{sec:results_nic}) imaging of the Fe\-LoBAL quasars.

\subsection{Modeling Strategy for the Imaging Data}\label{sec:method}

We use \textsc{Galfit} \citep[version 3.0.5,][]{PengGALFIT3} to model the emission in the 2-dimensional \md-combined images of the Fe\-LoBAL quasars. This software models astronomical images using analytical surface brightness profiles that are convolved with the instrumental PSF. The best-fit model is determined via least-squares minimization, and the modeling therefore requires error maps for each astronomical image. For the ACS data we generate these using the inverse-error maps produced by \md, which include an estimate of the cumulative error due to photon noise, readout noise and all image calibration operations. For the NICMOS images, which are combined using exposure-time weighting (\S \ref{sec:method_calib_multidrizzle}), we calculate the per-pixel uncertainties as the standard deviation of the background level in the combined images, added in quadrature to the per-pixel photon-counting error. As the image combination causes the uncertainties of adjacent pixels to be correlated, these NICMOS error maps represent upper limits on the true uncertainty.

\paragraph*{PSF-only Models:}\label{sec:method_psffit}

To determine whether there is any significant extended emission present, we first model each combined image as a point source along with a sky-background component with linear gradients in the $x$ and $y$ directions (hereafter, \emph{PSF-only models}). The PSF and background components are fitted simultaneously. Any other bright stars or galaxies in the image are modeled using PSF or Sersic components, respectively. To minimize the influence of detector-edge anomalies, most of the PSF fits are performed using $201\times201$-pixel (around 66 kpc for $z=1$) cutouts of the data and of the PSF template, centred on the quasar PSF. We ensure that the best-fit PSF and sky background scalings are not significantly altered if larger image regions are adopted. For the J0300+0048(NIC2/F110W) observation we model the entire image, so as to determine the sky level more accurately; this is necessary because the field is somewhat crowded. The best-fit PSF scalings are insensitive to the input parameter guesses, and the modeling converges within 10-20 iterations of the fitting routine. Our NICMOS PSF templates tend to model the quasar PSF diffraction spikes poorly. We therefore perform all NICMOS modeling with the diffraction spikes masked, so that \textsc{Galfit} does not include them in its least-squares minimization. We do not include these regions in the radially-averaged brightness profiles for NICMOS (i.e., Figures \ref{fig:starfits_f110w} and \ref{fig:quasars_radialplots_nicmos}).

\paragraph*{PSF-plus-Sersic Models:}\label{sec:method_sersicfit}

To quantify the total brightness and flux distribution of any extended emission, we model each image a second time, now including a Sersic component to represent the extended emission (hereafter, \emph{PSF-plus-Sersic models}).  The Sersic profile is given by: 

\begin{equation}
\Sigma(r)=\Sigma_e\exp[ -\kappa((r/R_e)^{1/n}-1)]
\end{equation}

where $\Sigma(r)$ is the surface brightness at a given radius, $R_e$ is the half-light radius of the profile, and $n$ is the Sersic index; $\kappa$ is not a free parameter, being fully determined, for a given $n$, by the definition of $R_e$. Again, we perform the analysis using $201\times201$-pixel cutouts of the quasar image and of the PSF template, and fit all components simultaneously.  For J0300+0048(NIC2/F110W) we again find it necessary to model the entire image. We experiment with including an elliptical deformation of this profile as a free parameter in our FeLoBAL modeling, but find that the resulting host galaxy magnitudes are insensitive to this parameter, while the $\chi^2$ statistic of the fit only weakly depends on the ellipticity. We therefore impose radial symmetry for all Sersic profiles. We also find it necessary to impose constraints on the allowed values of $n$ and $R_e$. This is due to PSF mismatch, i.e., differences between the `true' instrumental PSF at the time of the quasar observations and the PSF template used to model the point source, as described in detail in Appendix \ref{sec:appendix_psf}. Given the significant PSF mismatch that our modeling suffers, and the faintness of the host galaxies relative to the PSF, the largest improvement in the $\chi^2$ statistic is achieved by minimizing $R_e$ and/or maximizing $n$ so as to create a narrow, unresolved feature that alleviates the mismatch in the central region. We therefore restrict $R_e$ to be larger than the FWHM of the quasar PSF in a given observation. Given our low sensitivity to the Sersic index, we fit models with constant values of $n$ and average over their host magnitudes, as described in detail below..

\paragraph*{Host Galaxy Detection Criteria:}

The sensitivity of a quasar host galaxy study depends strongly on the quality of the available PSF templates. For this work, we lack dedicated PSF star observations (\S \ref{sec:data}), and instead construct stacked stellar PSF templates using stacked observations of field stars (for ACS imaging) and of calibration stars (for NICMOS), as described in Appendix \ref{sec:appendix_psf}. To quantify the fidelity of these PSF templates, we perform an extensive series of star-star PSF subtraction tests (Appendix \ref{sec:appendix_psf}) and host galaxy decomposition simulations (Appendix \ref{sec:appendixB}).  Based thereon (as discussed in Appendix \ref{sec:appendix_detection}), we apply the following detection criteria:

\emph{1)} At least three contiguous 1-pixel bins in our azimuthally averaged intensity plots show a positive residual at a significance $>3\sigma$ after PSF-only modeling. This must occur outside the inner 0.2''  of the radial profile for NICMOS and ACS/F425W, outside the inner 0.3'' for ACS/F550M, or outside the inner 0.4'' for ACS/F625W. %These radial constraints are due to the severe mismatch in the PSF core, and a general tendency to oversubtract the inner regions when performing a PSF-only fit in the presence of extended flux.

\emph{2)} The PSF wings are not significantly ($>3\sigma$) oversubtracted in any azimuthally-averaged bin exterior to the radial constraint given in \emph{1)}. 

\subsection{ACS (Rest-Frame UV) Modeling Results}\label{sec:results_acs}

\paragraph*{PSF-only Modeling, ACS:}

\begin{figure*}
	\advance\leftskip-3cm
	\centering
	\includegraphics[scale=0.26,clip,trim={0 0 3.3cm 0}]{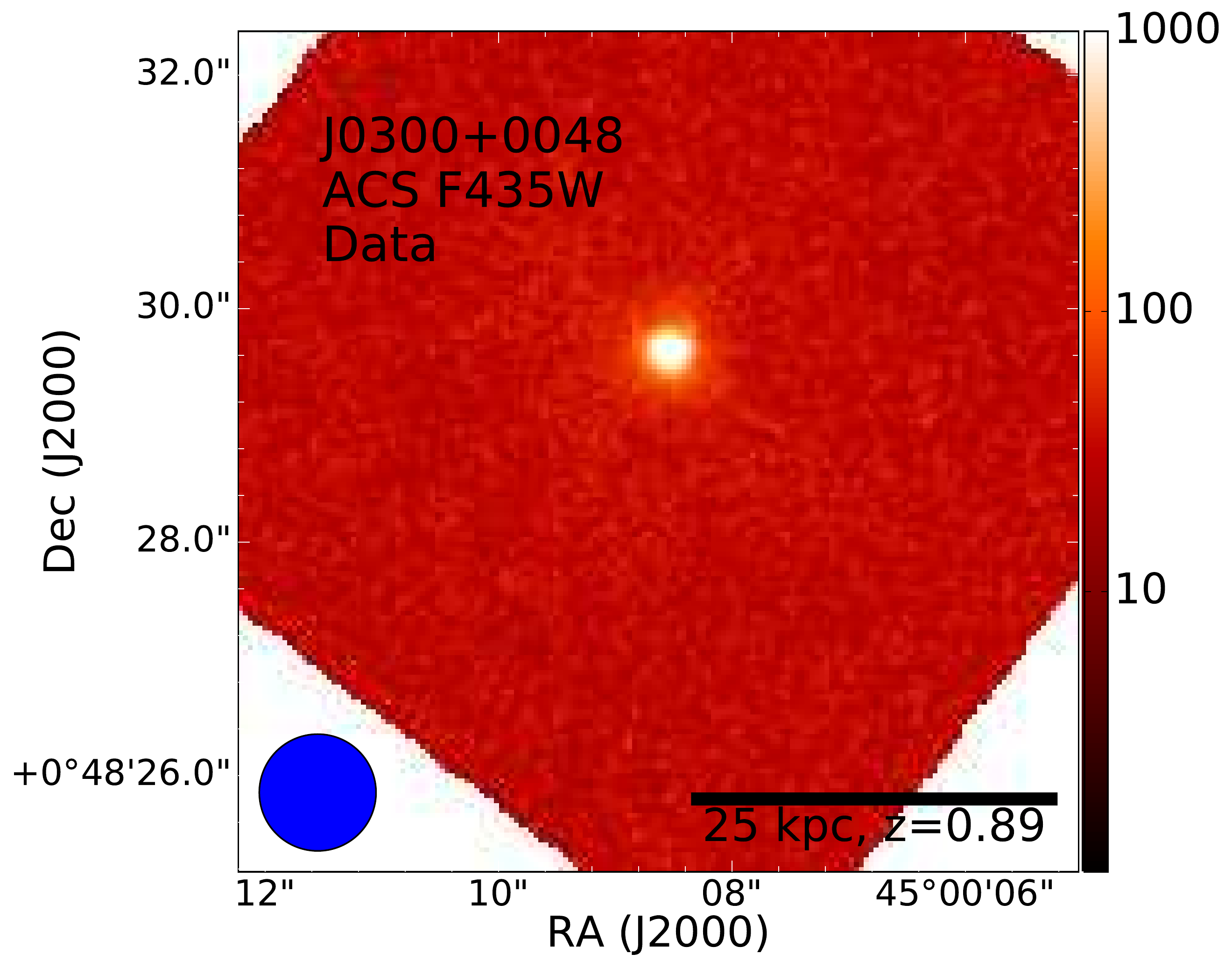}
	\includegraphics[scale=0.26,clip,trim={0 0 3.3cm 0}]{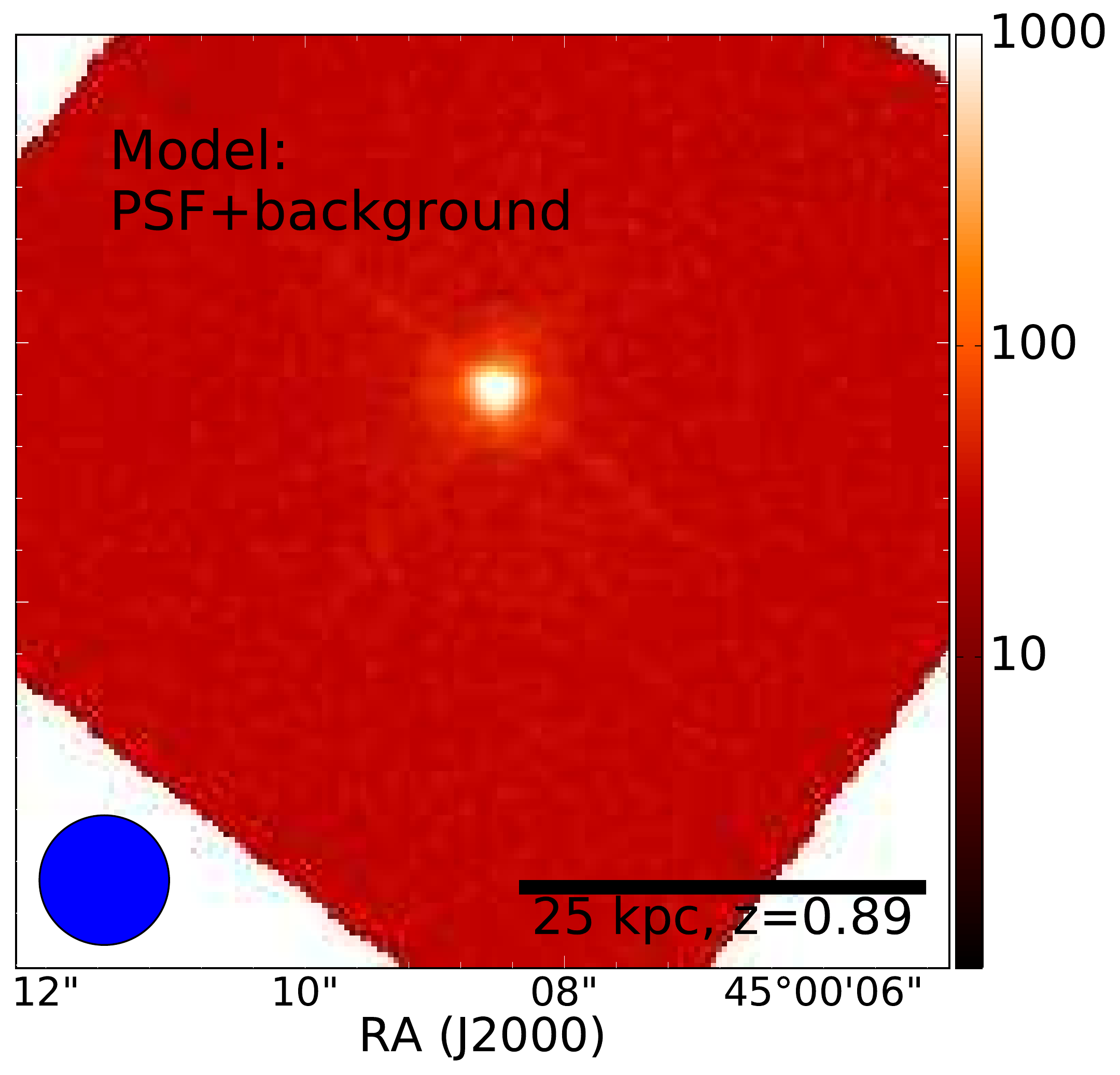}	 
	\includegraphics[scale=0.26]{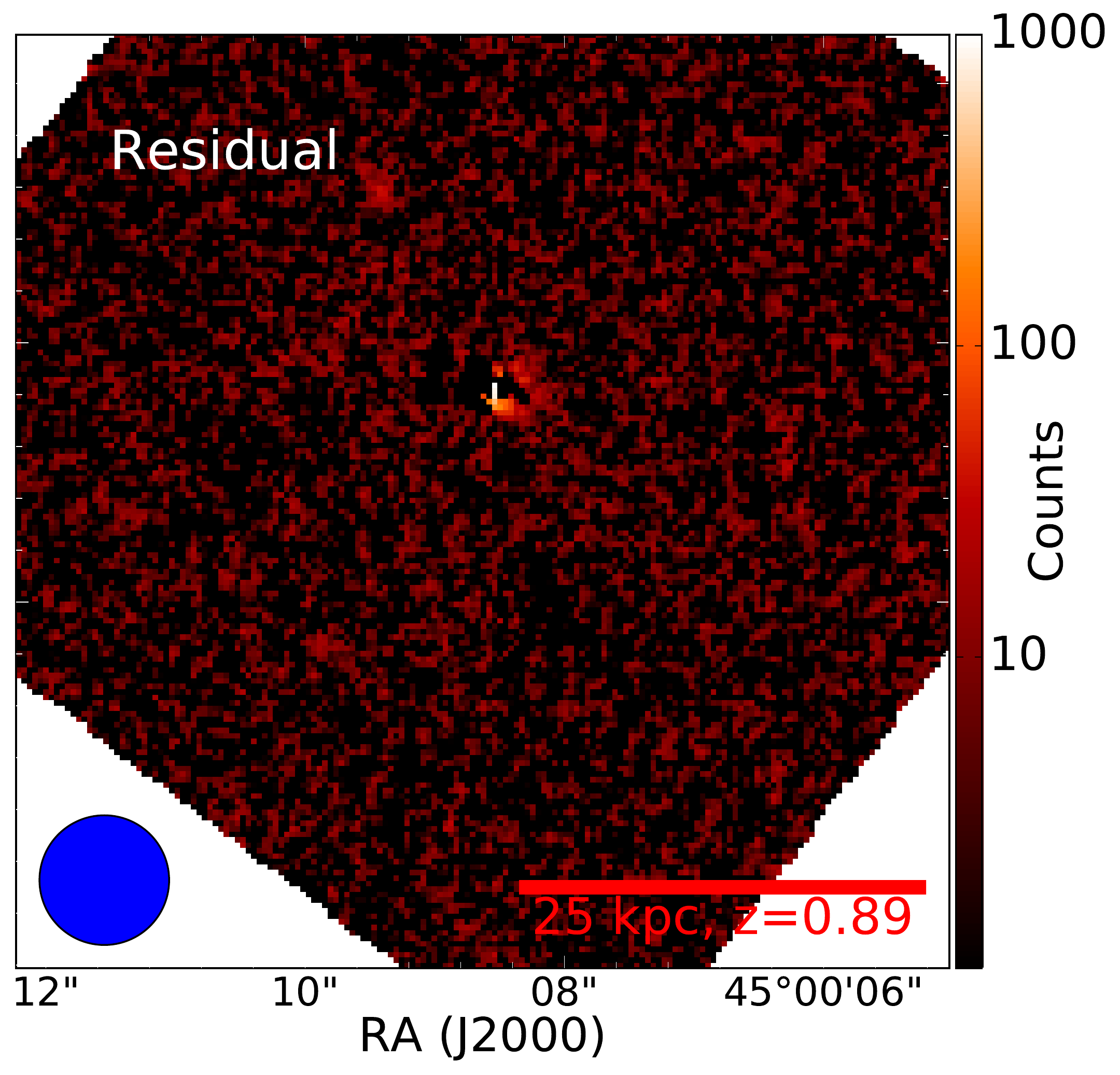}
	\includegraphics[scale=0.26,clip,trim={0 0 3.3cm 0}]{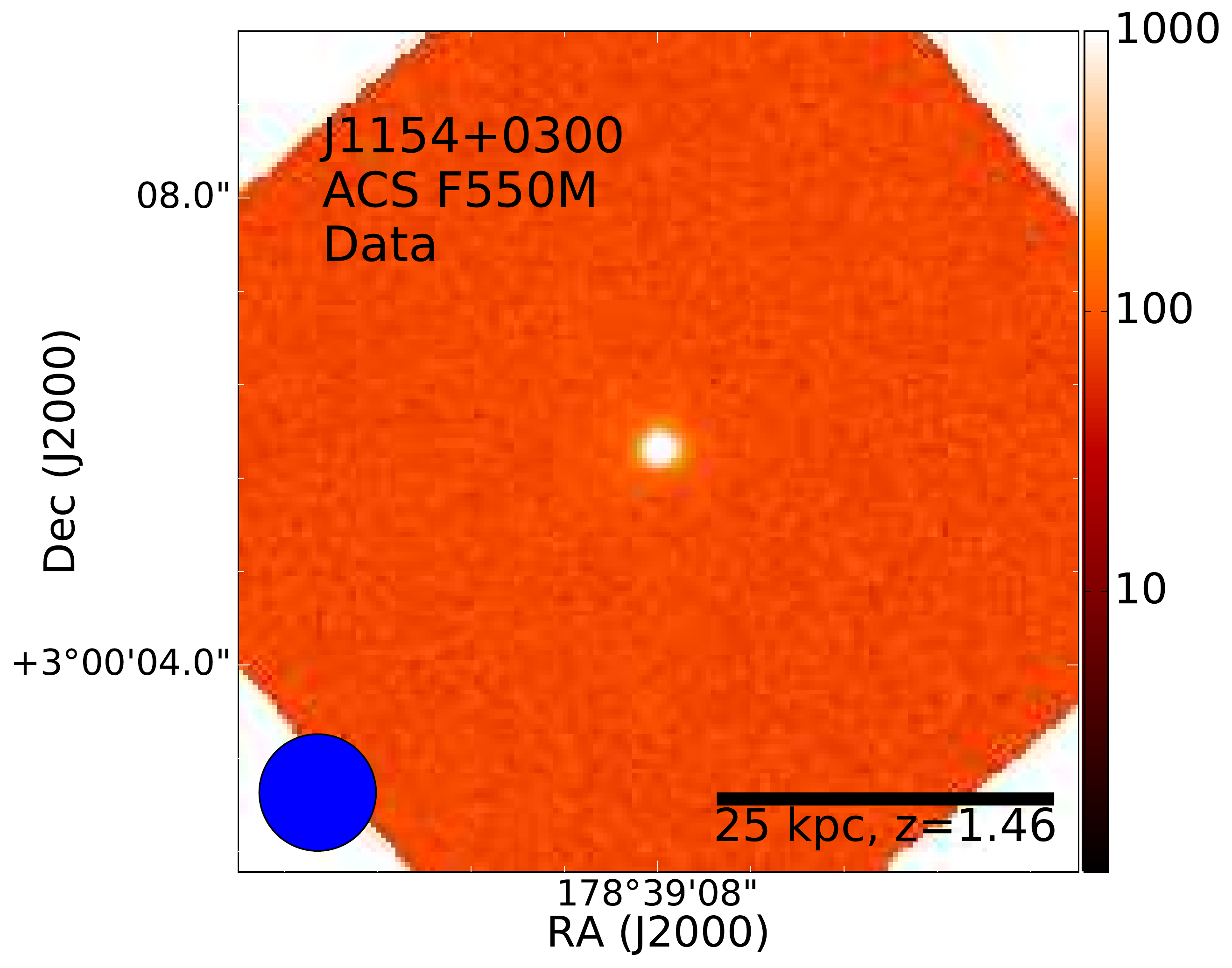}
	\includegraphics[scale=0.26,clip,trim={0 0 3.3cm 0}]{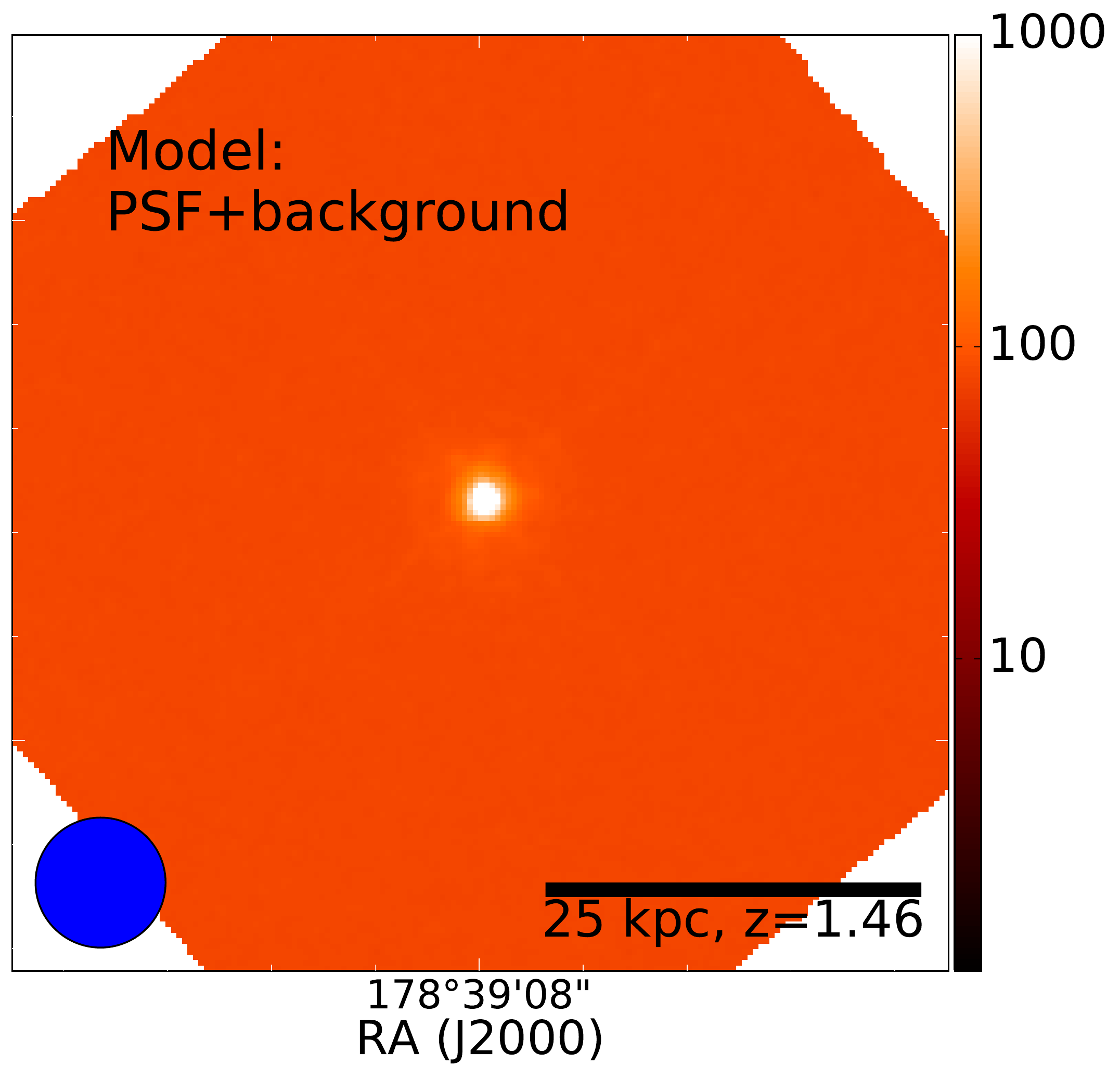}	 
	\includegraphics[scale=0.26]{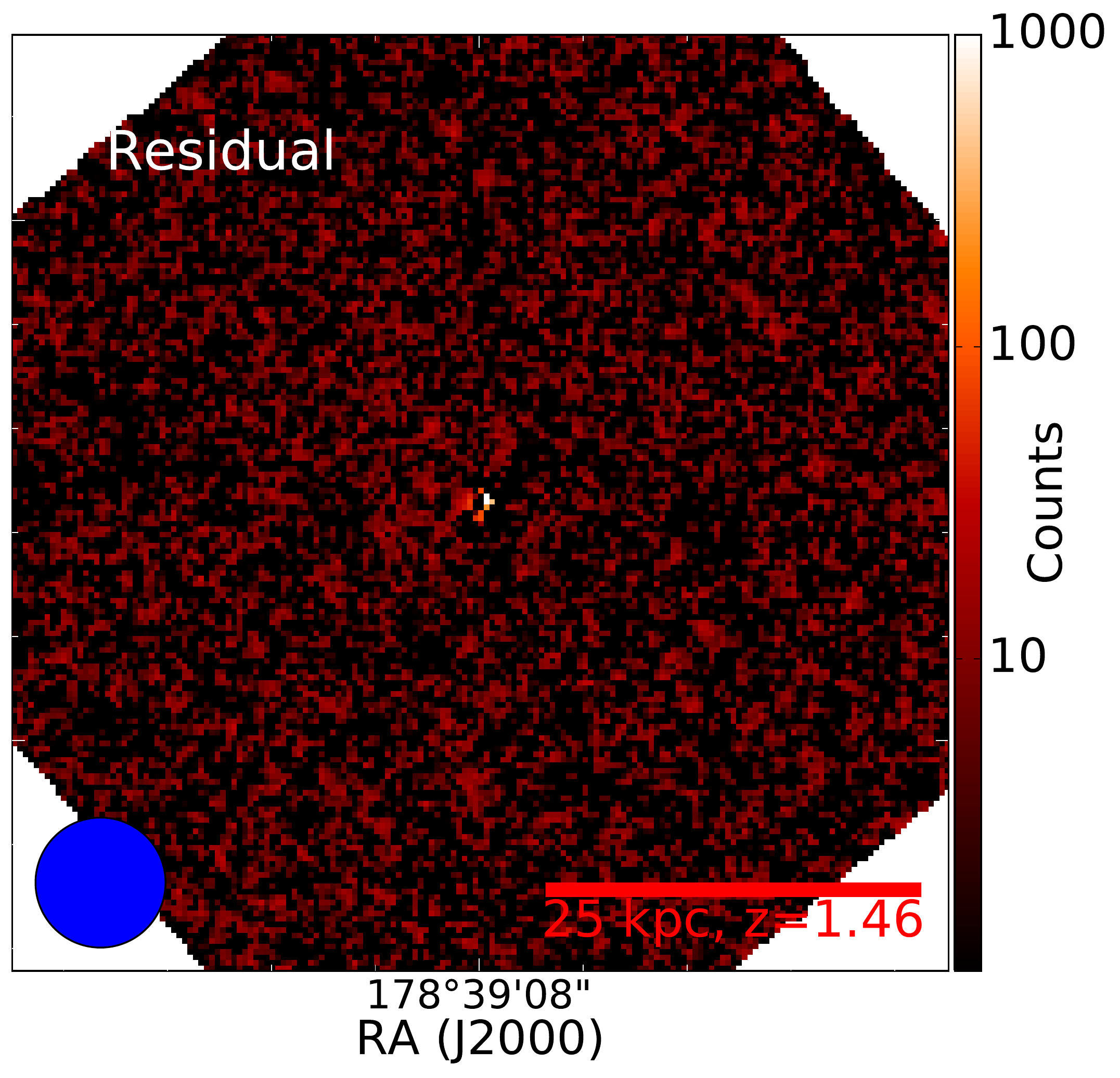}  	 
	\includegraphics[scale=0.26,clip,trim={0 0 3.3cm 0}]{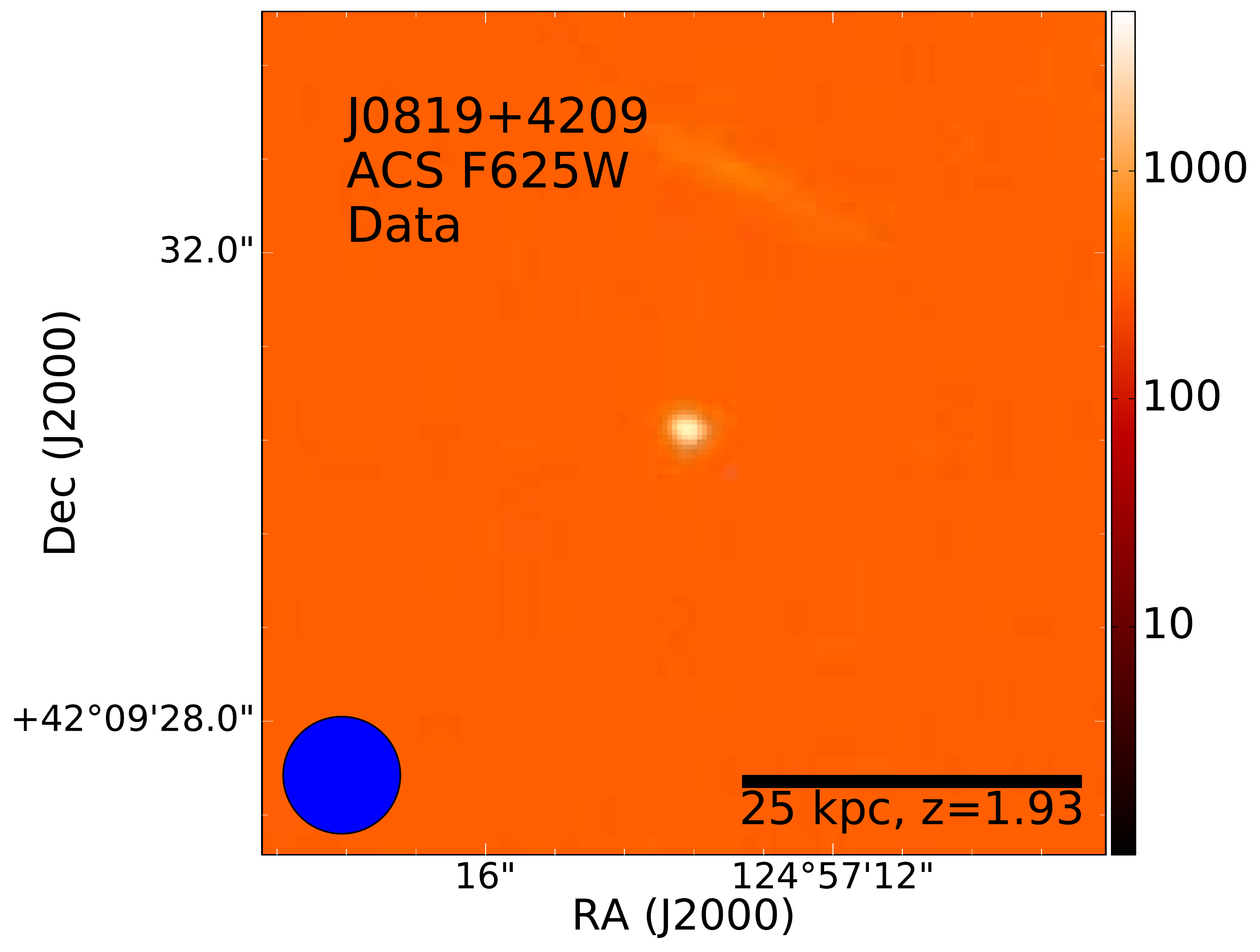}
	\includegraphics[scale=0.26,clip,trim={0 0 3.3cm 0}]{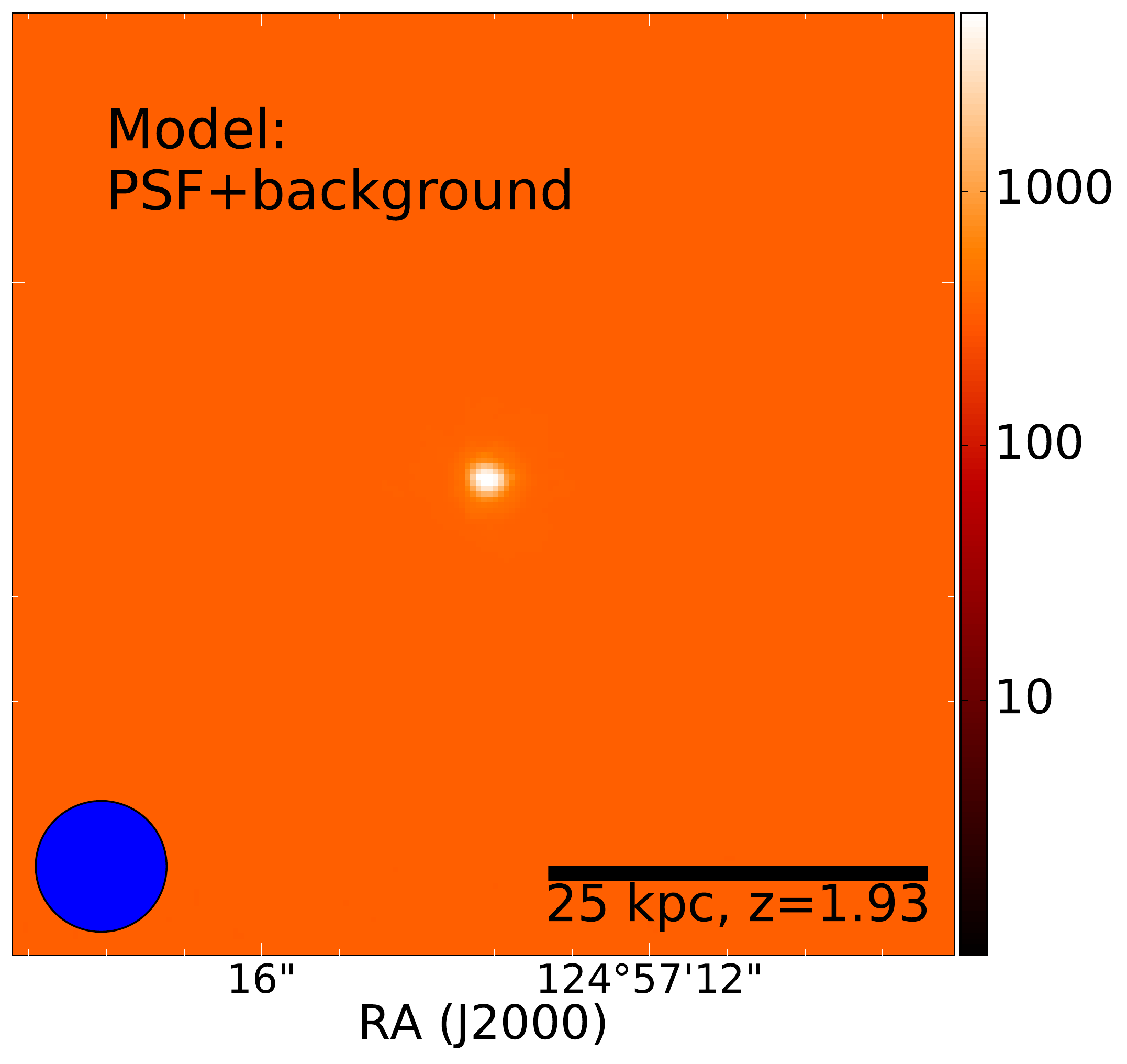}	 
	\includegraphics[scale=0.26]{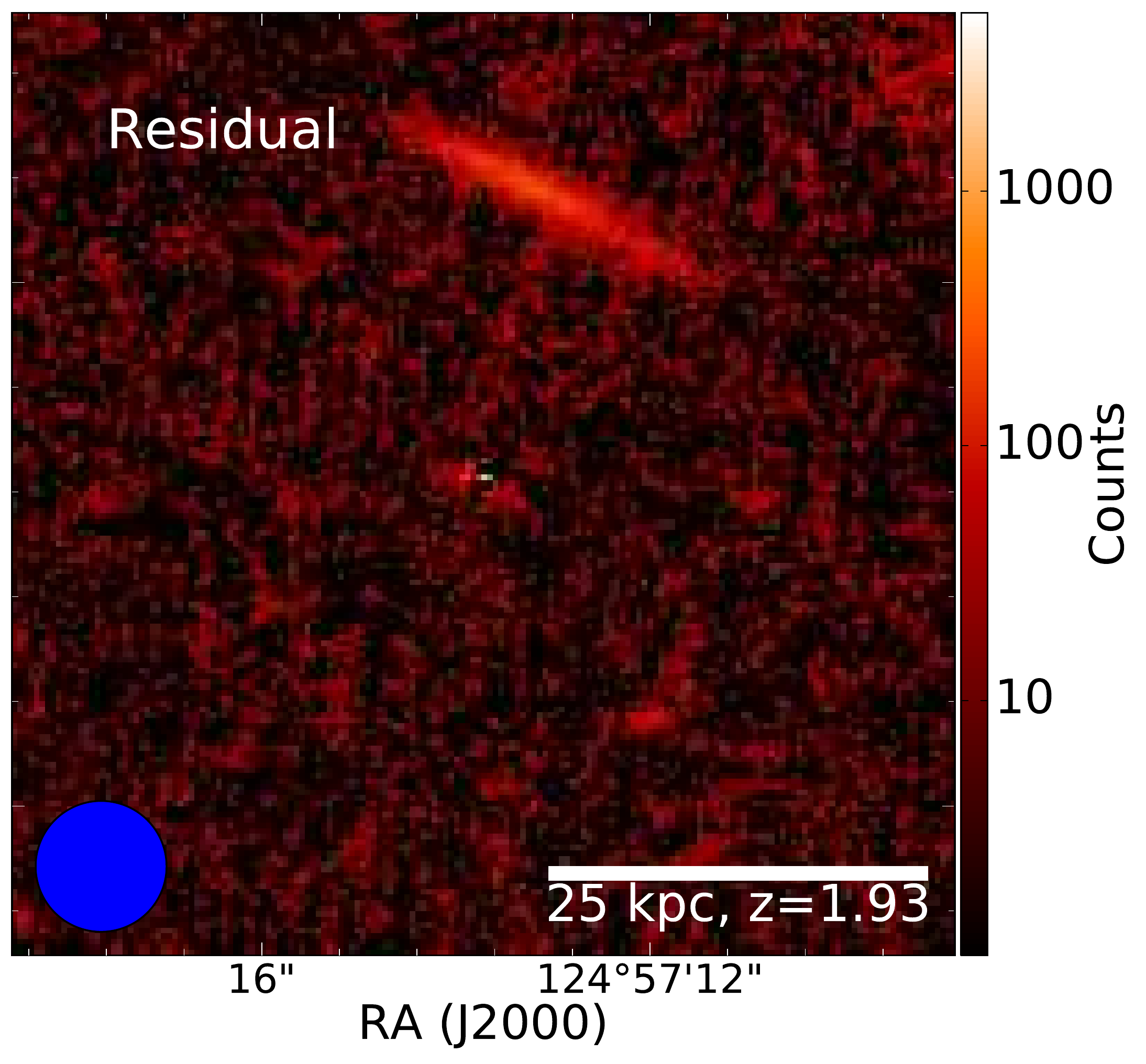}    
	\includegraphics[scale=0.26,clip,trim={0 0 3.3cm 0}]{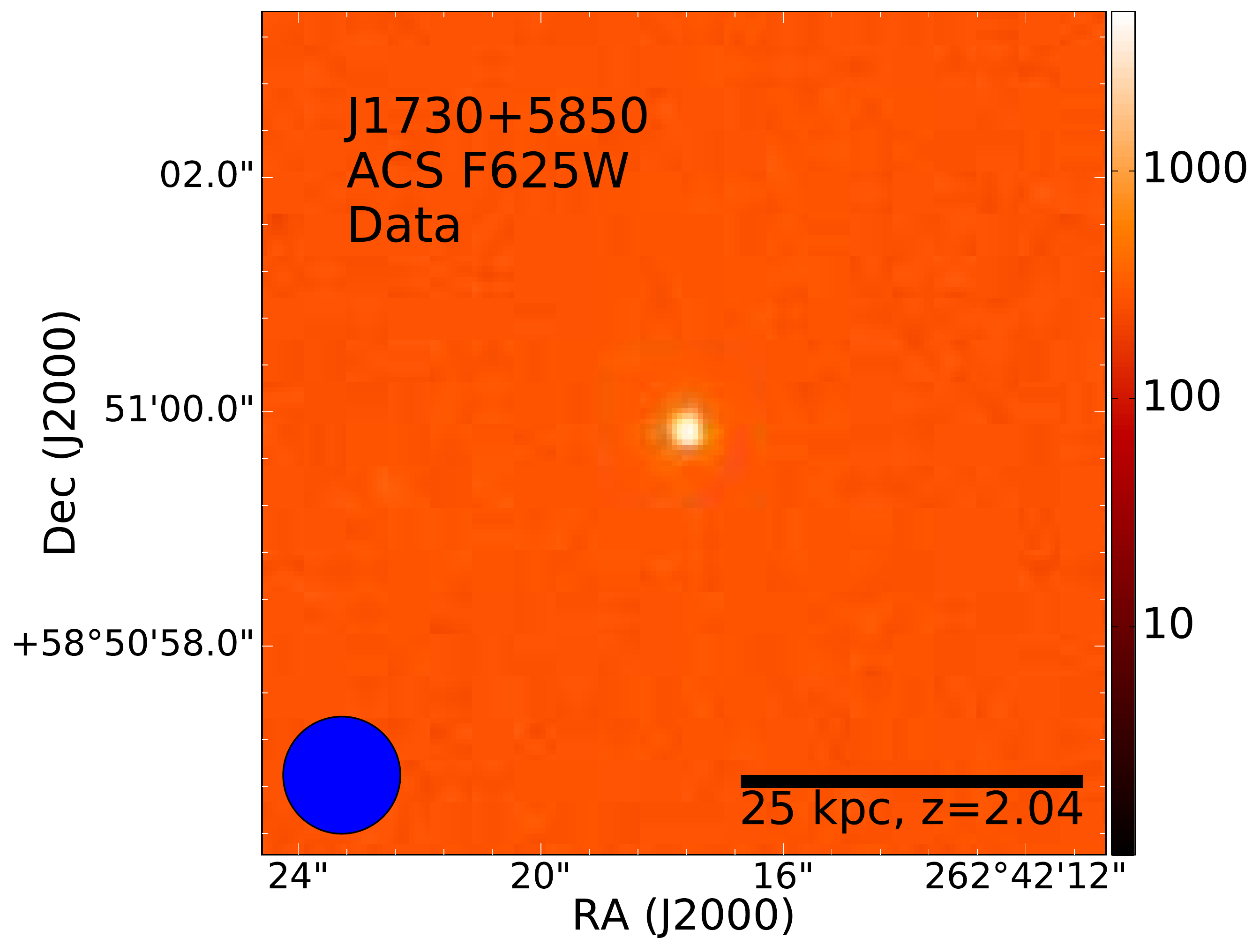}
	\includegraphics[scale=0.26,clip,trim={0 0 3.3cm 0}]{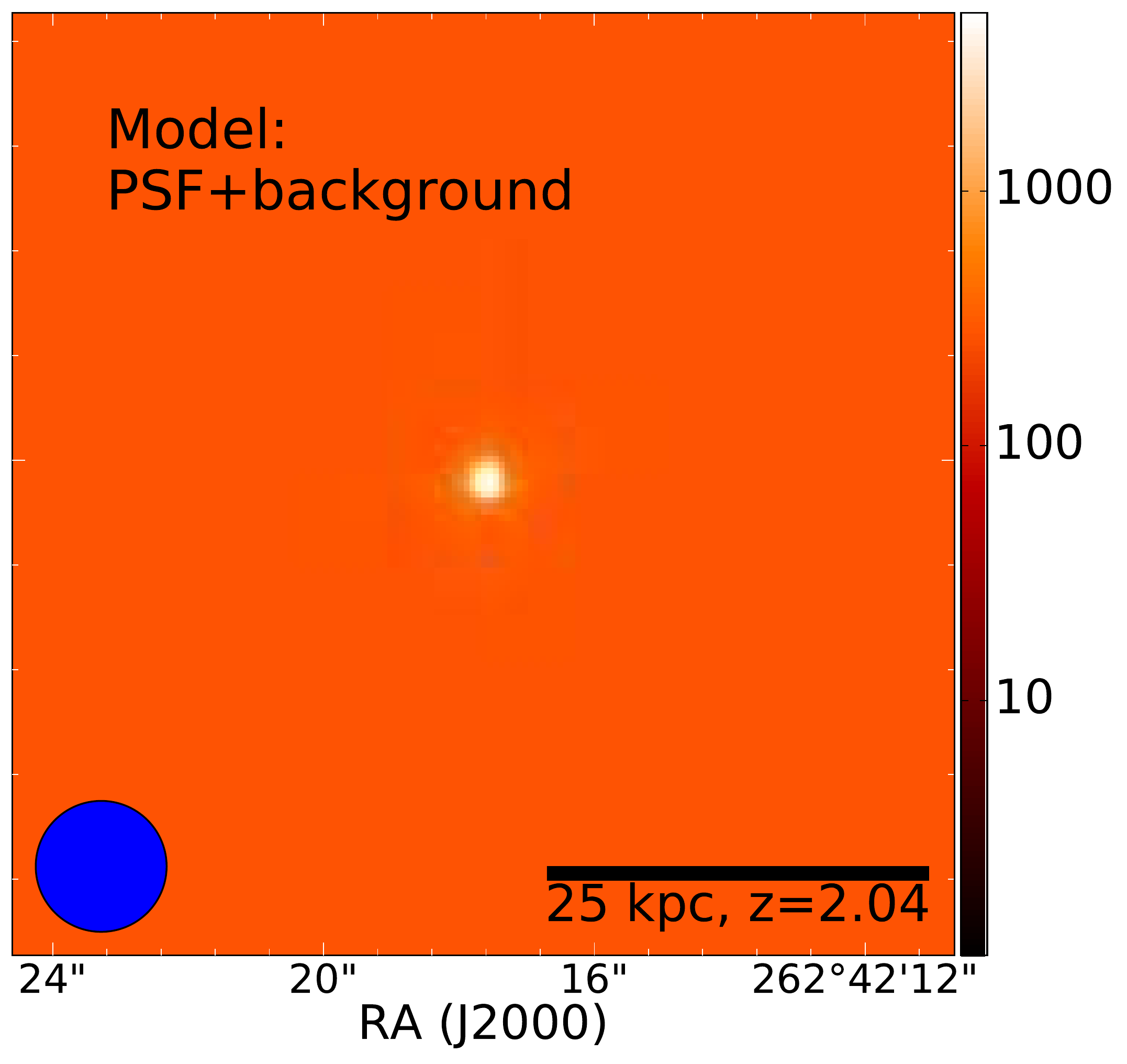}	 
	\includegraphics[scale=0.26]{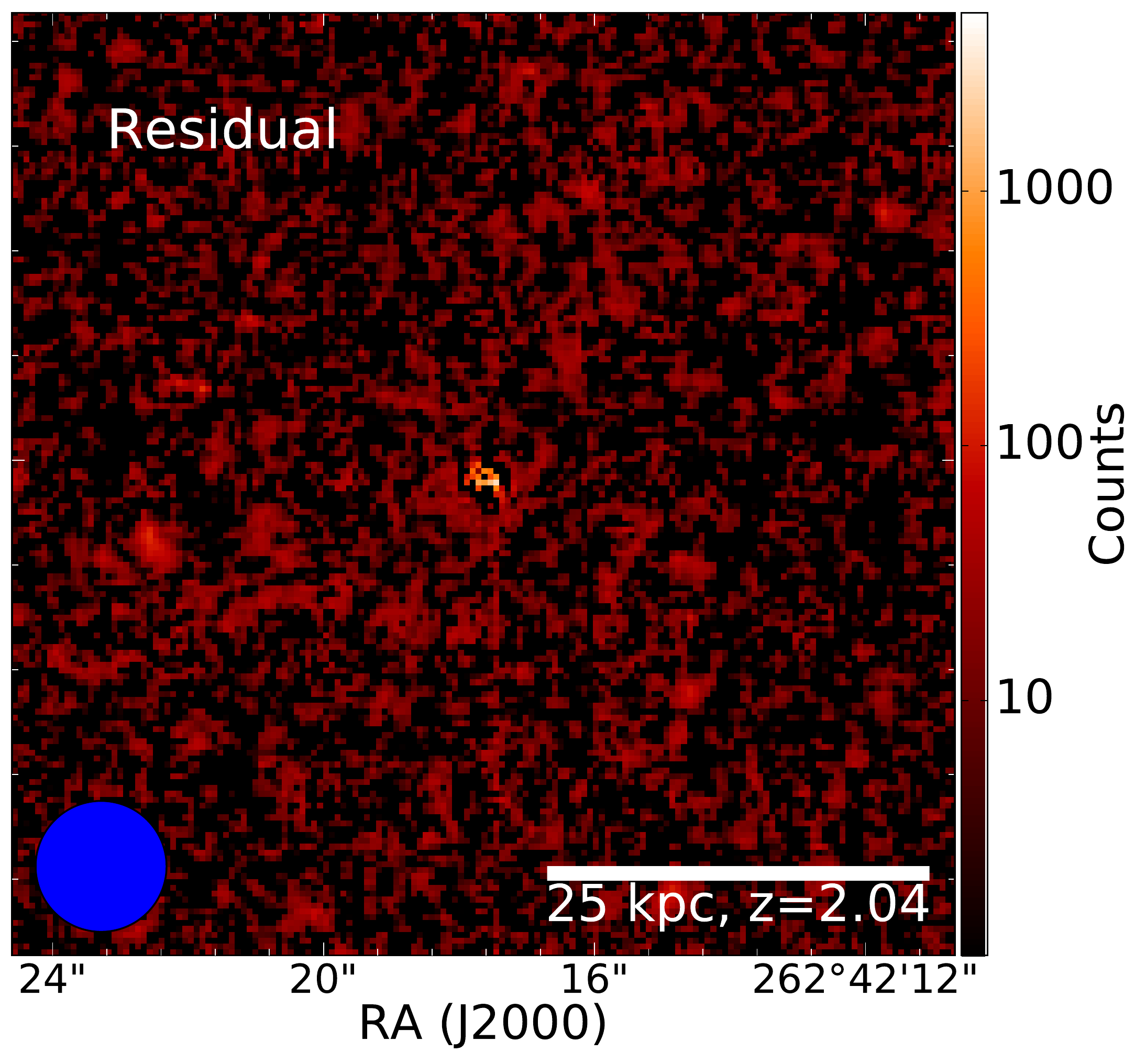}    
	\caption{PSF-only modeling of the ACS imaging of the Fe\-LoBAL quasars. The intensity is shown in units of total counts. The blue circle shows a diameter of 1 arcsecond; the image size is $7.2\times7.2$ arcseconds. The scale bars illustrate an angular size corresponding to 25 kpc at the quasar redshift. \emph{Left:} Original image. \emph{Centre:} Best-fit PSF model and background component. \emph{Right:} Residual after model subtraction. For J0819+4209, a close companion galaxy is visible in the residual image; however, we model and subtract this galaxy prior to our analysis of radial flux profiles (Figure \ref{fig:quasars_radialplots_acs}), to avoid attributing its flux to the quasar host galaxy. }\label{fig:quasars_2d_acs}
\end{figure*}

The quasar and template PSFs have FWHM of between 0.08-0.09 arcseconds (Table \ref{tab:psffits_acs}). All residual images appear consistent with point sources (Figure \ref{fig:quasars_2d_acs}). The quasars are relatively faint in ACS imaging, with best-fit AB magnitudes $m_{\mathrm{PSF}}>20.5$ mag for all objects; the ACS observations are, on average, 4.2 mag fainter than the NICMOS observations, corresponding to a factor $\sim$47 in flux. This confirms that our ACS observations sample the heavily absorbed BAL trough (Figure \ref{fig:0300_spec}). We present upper limits on \mhost\,in the rest-frame UV for each quasar in Table \ref{tab:psffits_acs_limits}.

\paragraph*{PSF-plus-Sersic modeling, ACS:} For J0300+0048, the Sersic component magnitude converges at \mhost$\sim24$ mag, and is unresolved, with $R_e\approx1$ kpc. Our simulation work indicates that this extended component may be a spurious result due to PSF mismatch, as it is fainter than our detection limit (Table \ref{tab:psffits_acs_limits}). For the three remaining quasars, the Sersic components diverge towards infinite faintness. In summary, we see no evidence of extended emission in the rest-frame UV for any of the Fe\-LoBAL quasars. As established by our simulation work (Appendix \ref{sec:appendixB}), the UV non-detections may be due to very centrally concentrated UV emission with scale size $R_e<1$ kpc (e.g., a nuclear starburst). A very extended UV source ($R_e\apprge10$ kpc) would also yield a non-detection, assuming \contrast$>2$ mag. Alternatively, the host galaxies may simply have quiescent or dust-obscured stellar populations, as discussed in \S \ref{sec:discussion_templates}. 

\newcommand{\radplotscale}{0.28}

\begin{figure*}
	\advance\leftskip-3cm
	\centering	 
	\includegraphics[trim={1cm 1cm 1cm 0.5cm},clip,scale=0.3]{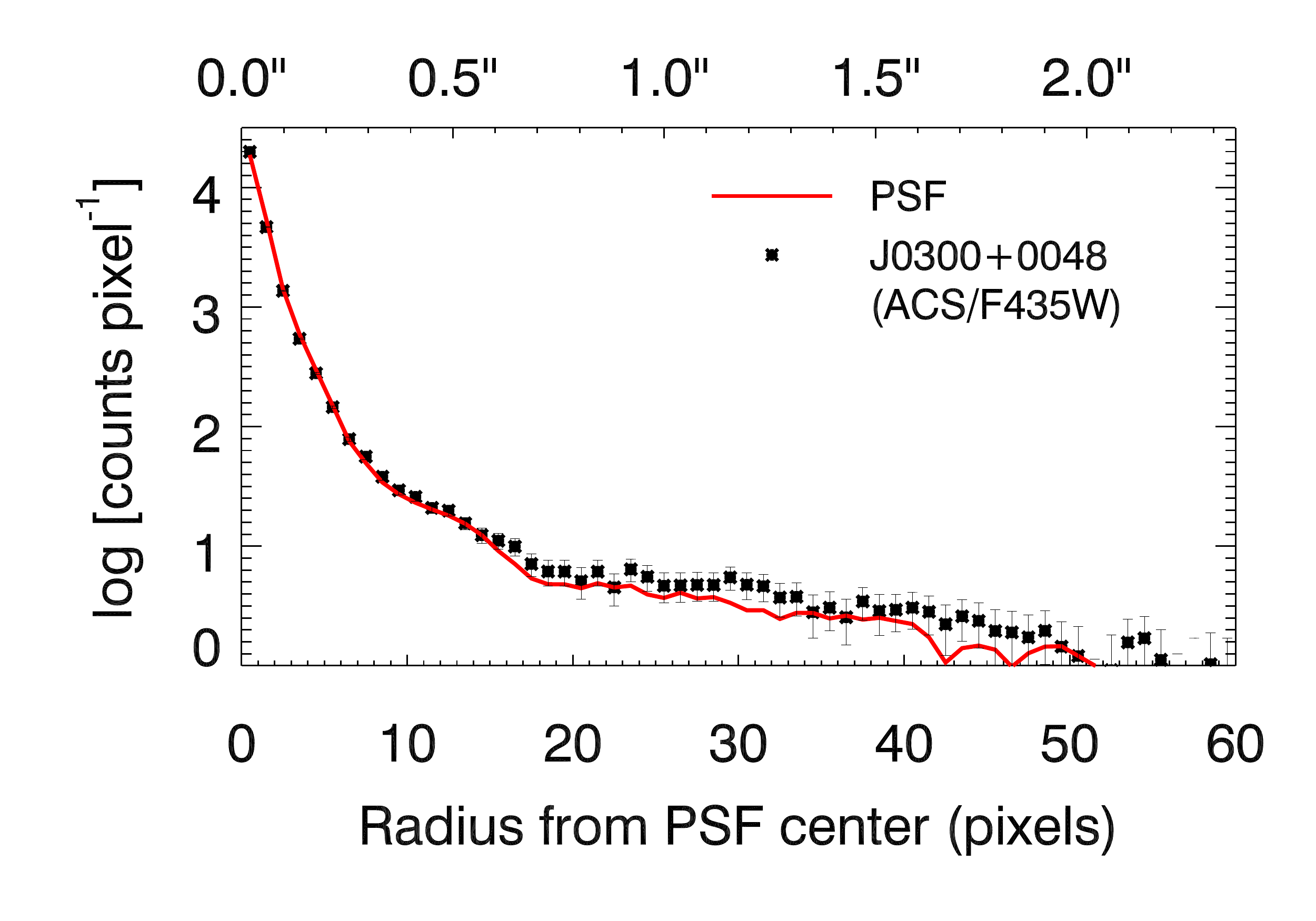}	 
	\includegraphics[trim={0 1cm 1cm 0.5cm},clip,scale=0.3]{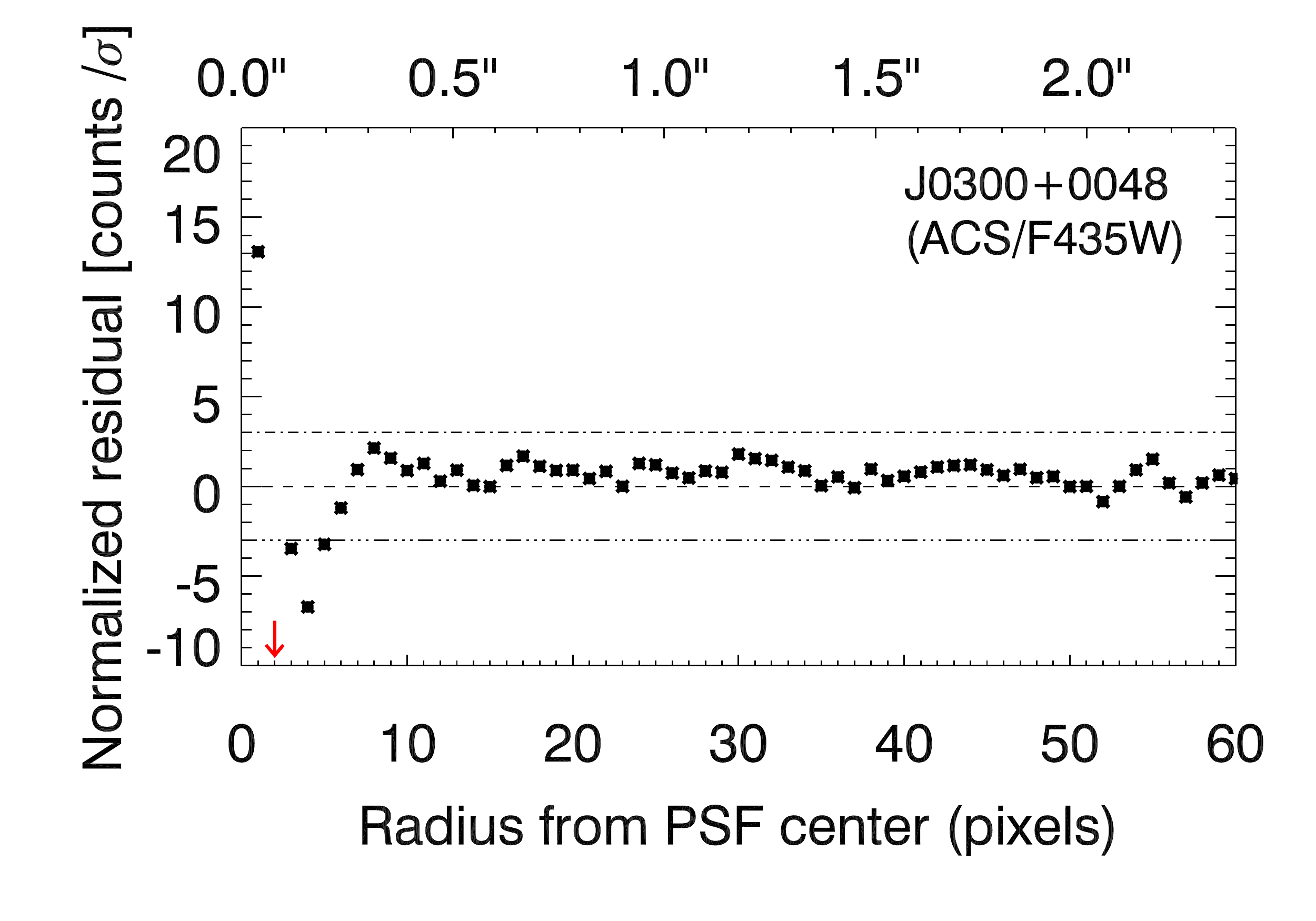}
	\includegraphics[trim={1cm 1cm 1cm 0.5cm},clip,scale=0.3]{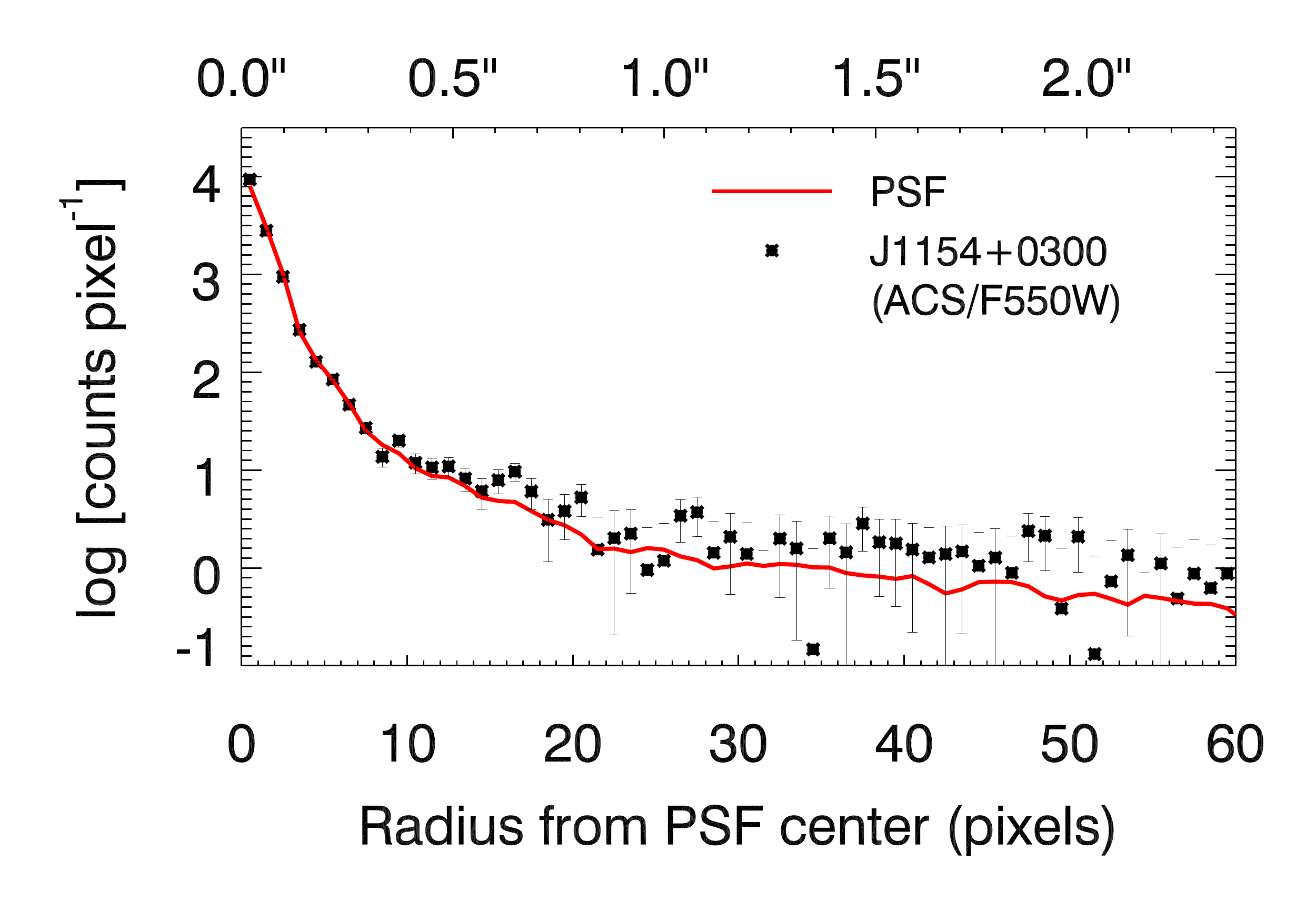}	 
	\includegraphics[trim={0 1cm 1cm 0.5cm},clip,scale=0.3]{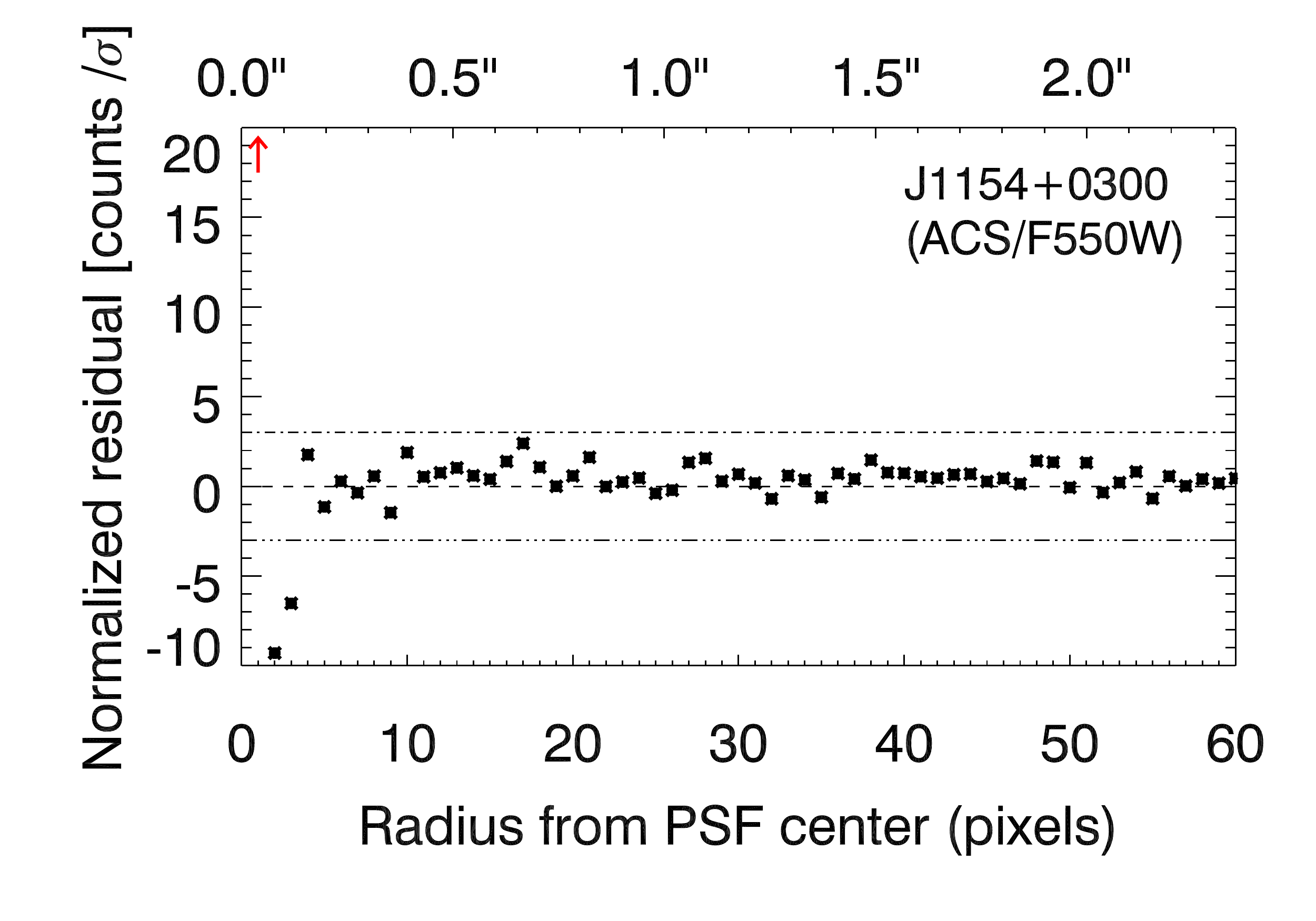}
	\includegraphics[trim={1cm 1cm 1cm 0.5cm},clip,scale=0.3]{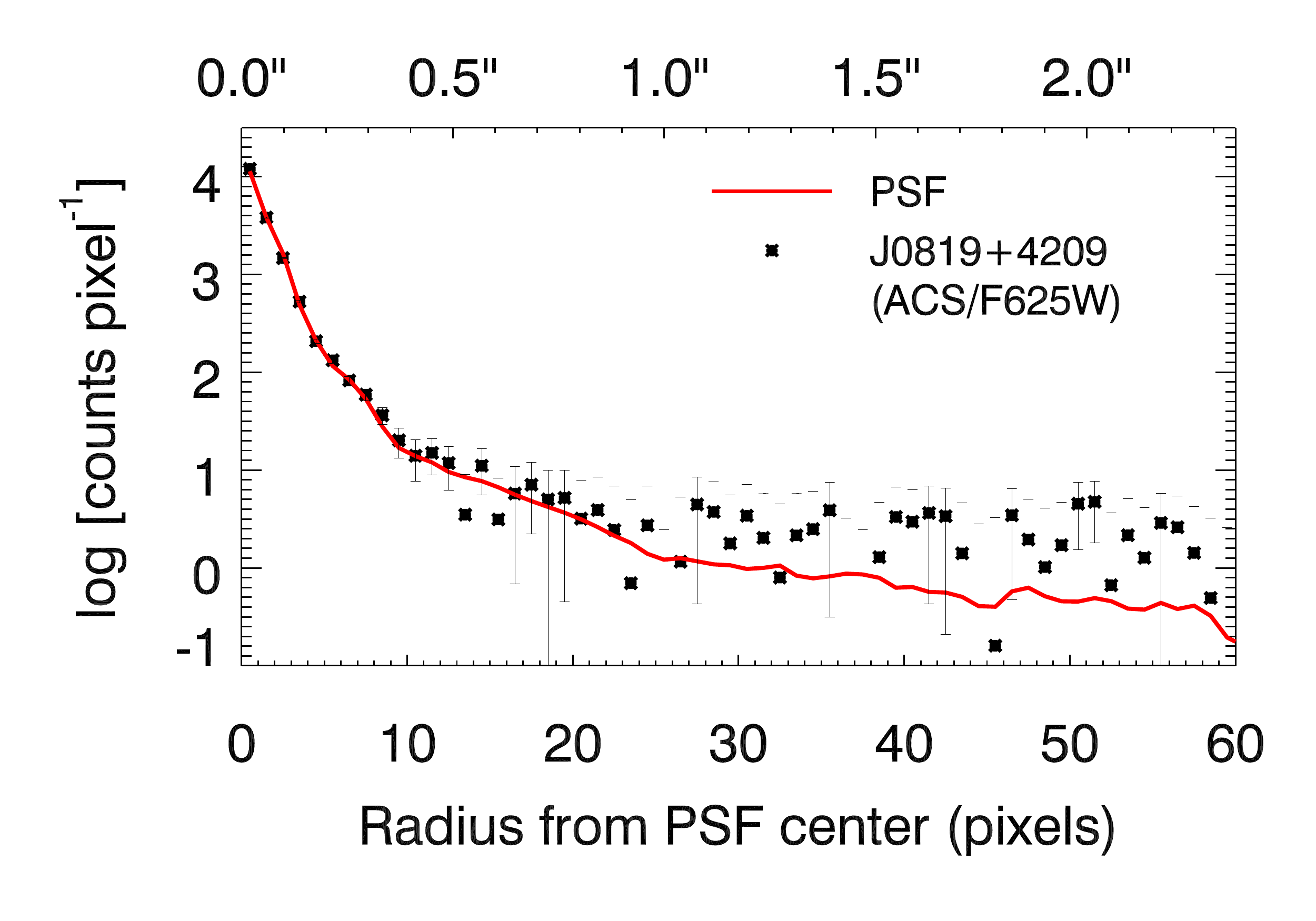}	 
	\includegraphics[trim={0 1cm 1cm 0.5cm},clip,scale=0.3]{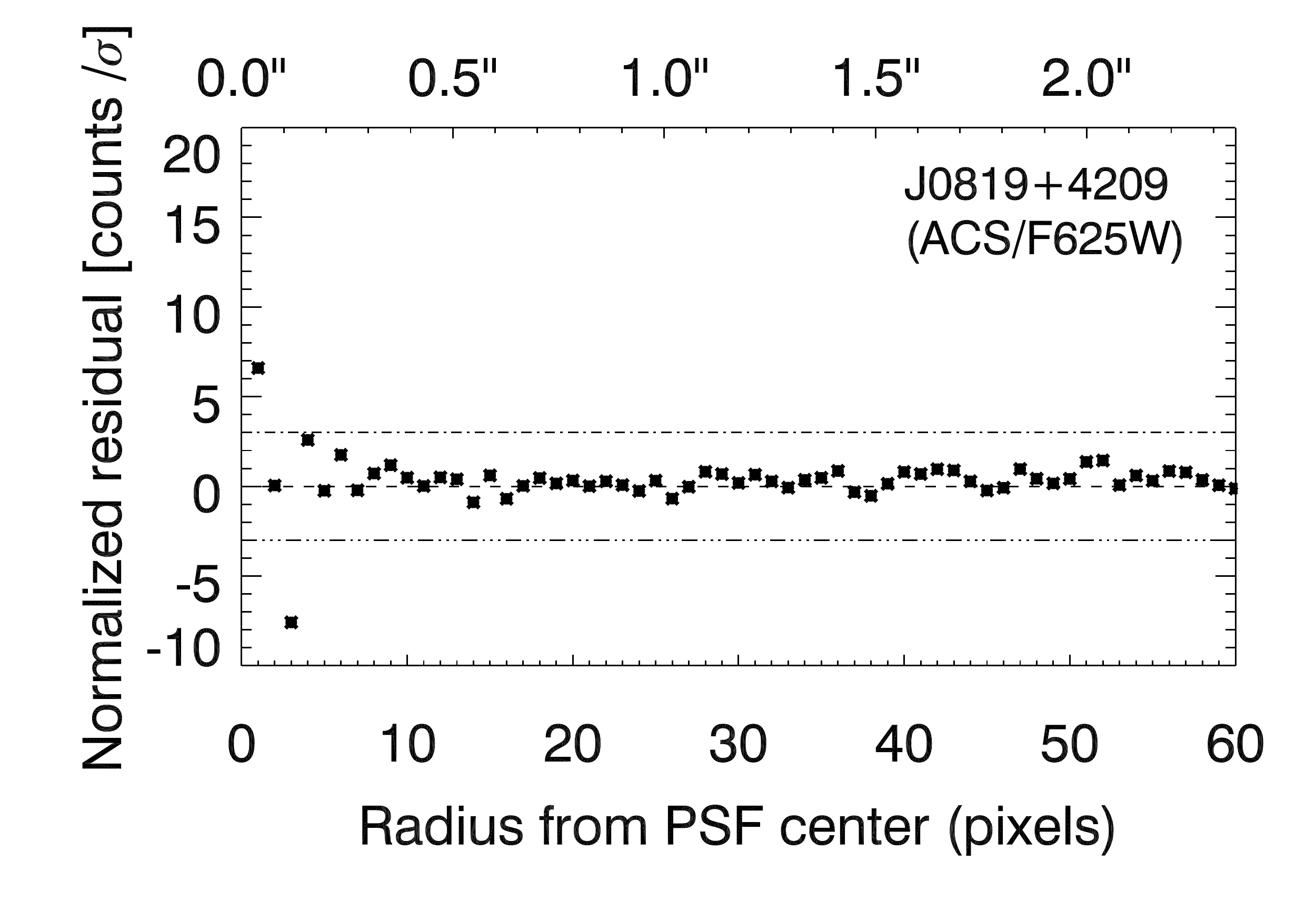}
	\includegraphics[trim={1cm 1cm 1cm 0.5cm},clip,scale=0.3]{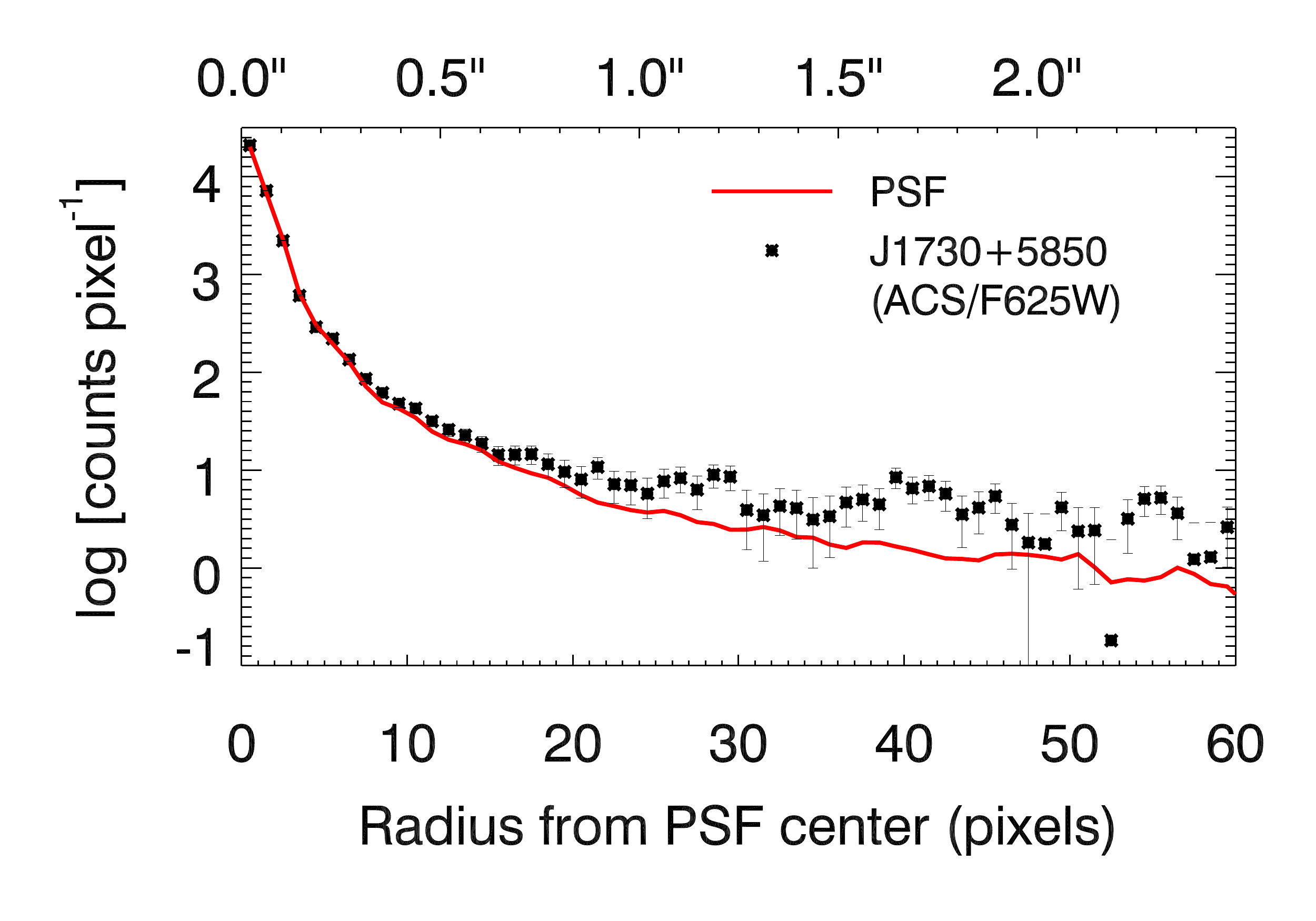}	 
	\includegraphics[trim={0 1cm 1cm 0.5cm},clip,scale=0.3]{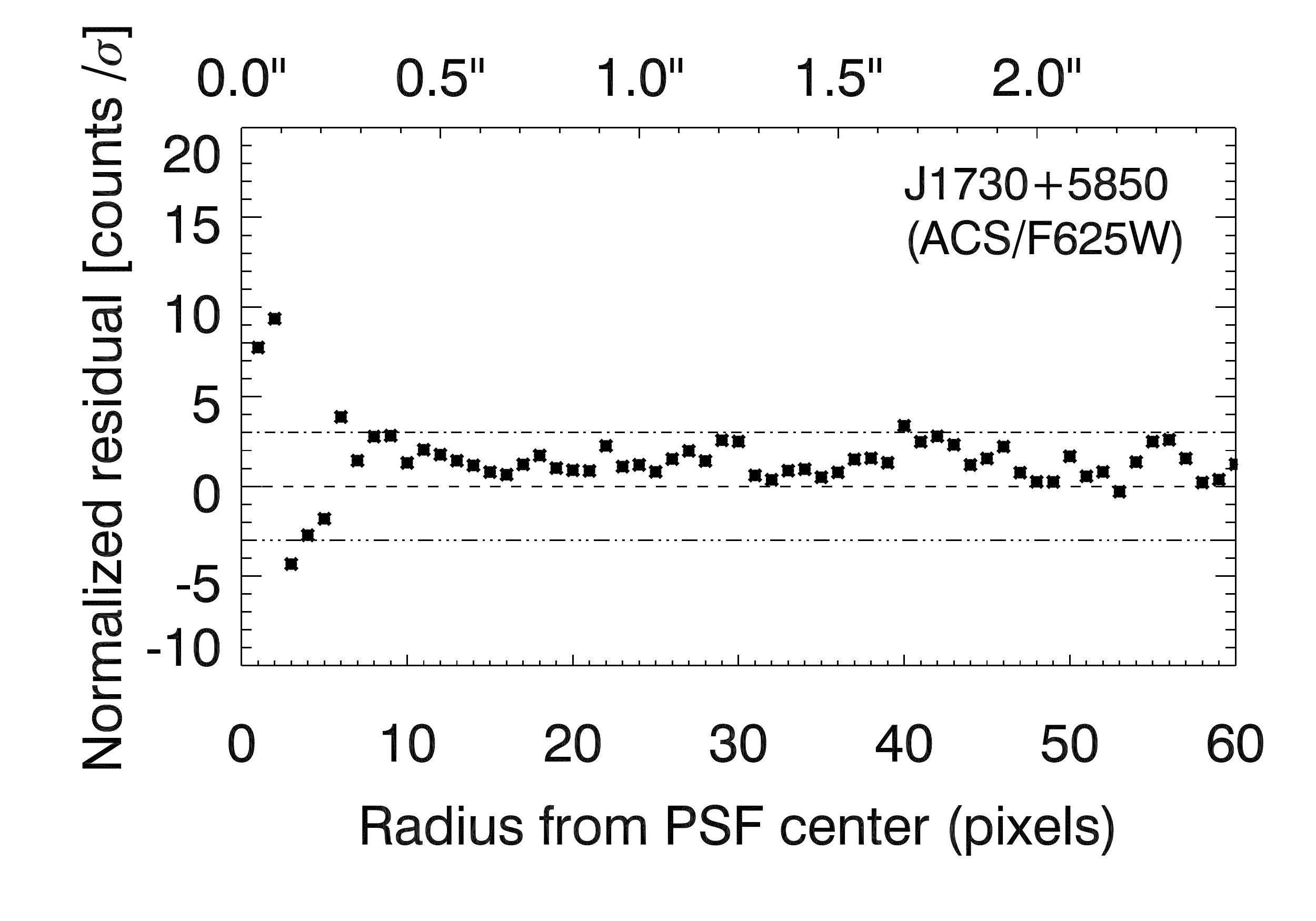}       	
	\caption{PSF-only modeling results for our ACS (rest-frame UV) imaging. \emph{Left:} radially-averaged intensity plot showing a scaled version of the stacked stellar PSF (red solid lines) and the sky-subtracted quasar emission profiles (black points). Emission attributed to companion galaxies is not included in these profiles. The error bars on the data points include a contribution due to photon noise in the PSF template. \emph{Right:} Offset of the sky-subtracted quasar profiles from the PSF templates, expressed in terms of the $1\sigma$ error bars on the data. The horizontal dashed, dash-dot, and dot-dot-dot-dash lines represent a zero offset, a $3\sigma$ positive residual, and a $3\sigma$ negative residual, respectively. Red arrows indicate annuli for which the offset exceeds the y-axis limits.}\label{fig:quasars_radialplots_acs}
\end{figure*}

\subsection{NICMOS (Rest-Frame Optical) Modeling Results}\label{sec:results_nic}

\begin{figure*}
	\advance\leftskip-2cm
	\centering	 
	\includegraphics[scale=0.26,clip,trim={0 0 3.3cm 0}]{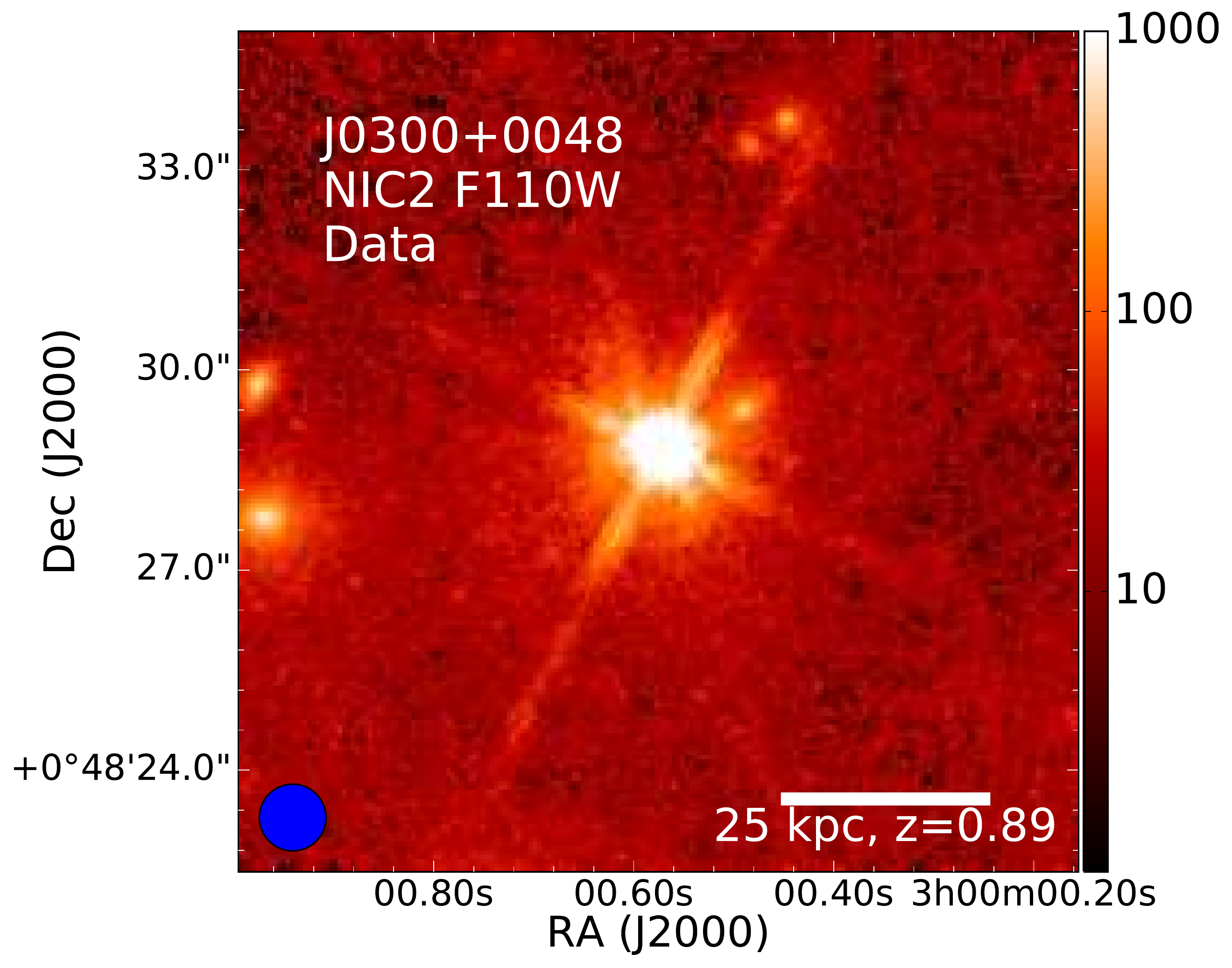}
	\includegraphics[scale=0.26,clip,trim={0 0 3.3cm 0}]{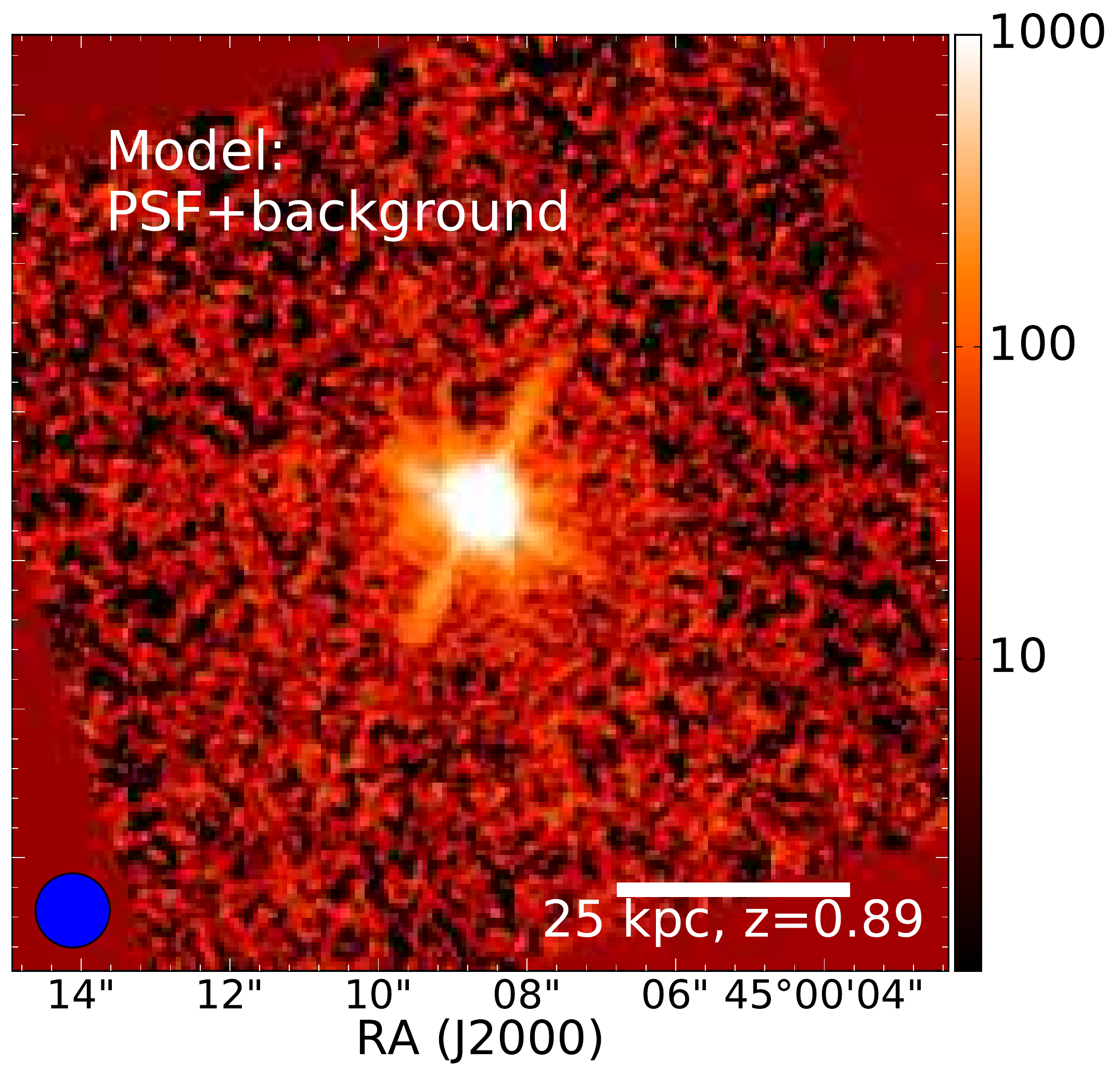}	 
	\includegraphics[scale=0.26]{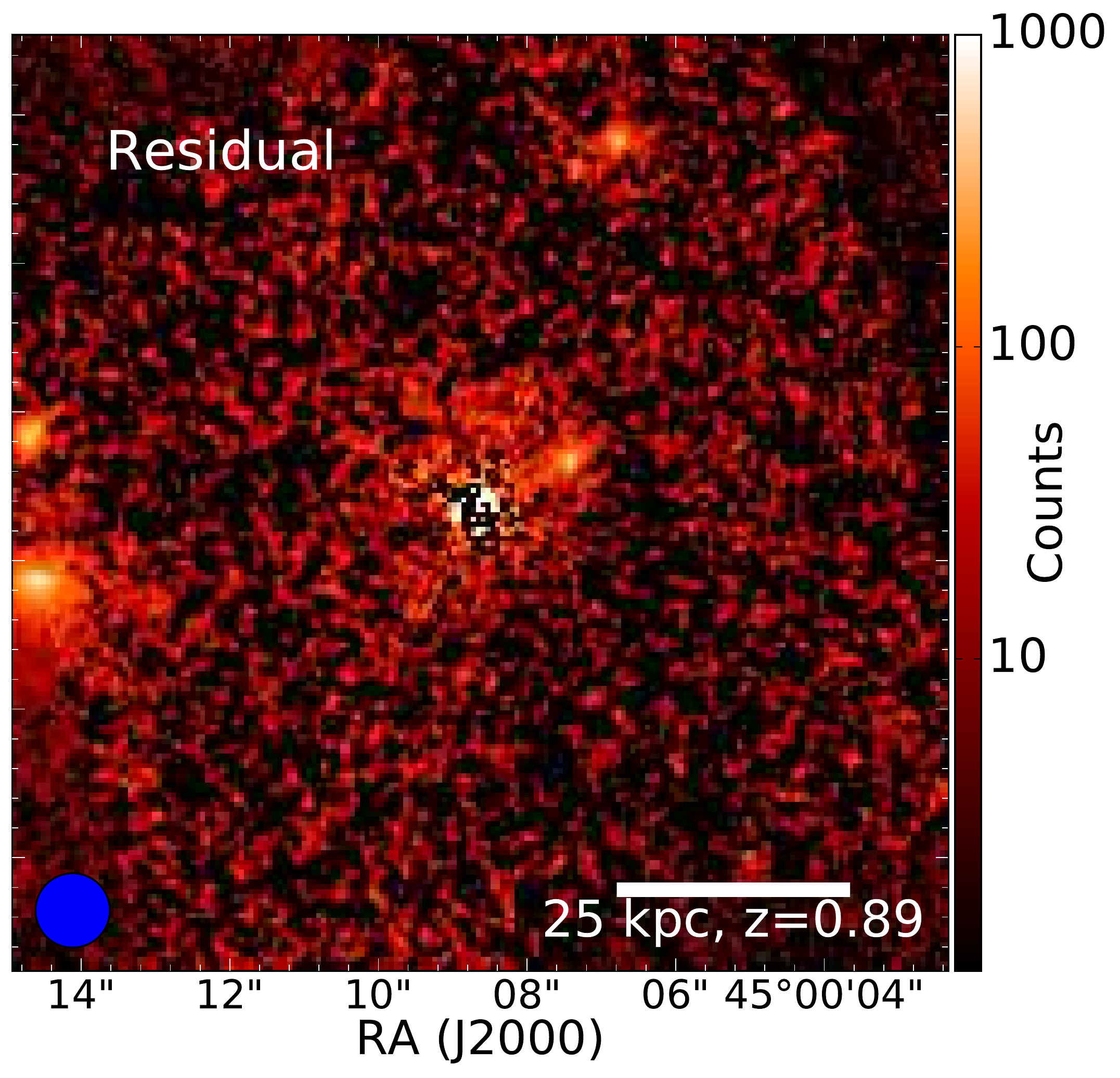}
	\includegraphics[scale=0.26,clip,trim={0 0 3.3cm 0}]{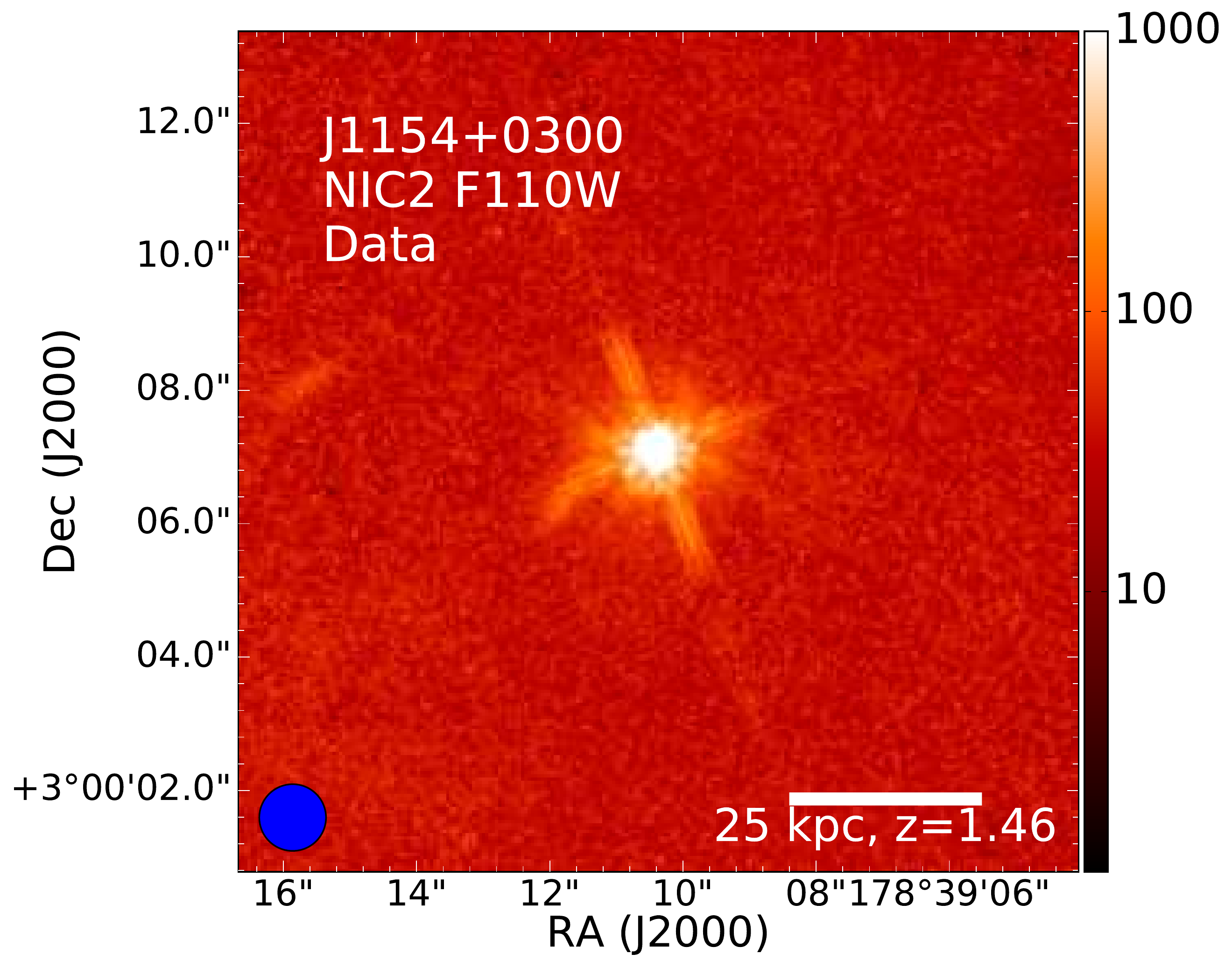}
	\includegraphics[scale=0.26,clip,trim={0 0 3.3cm 0}]{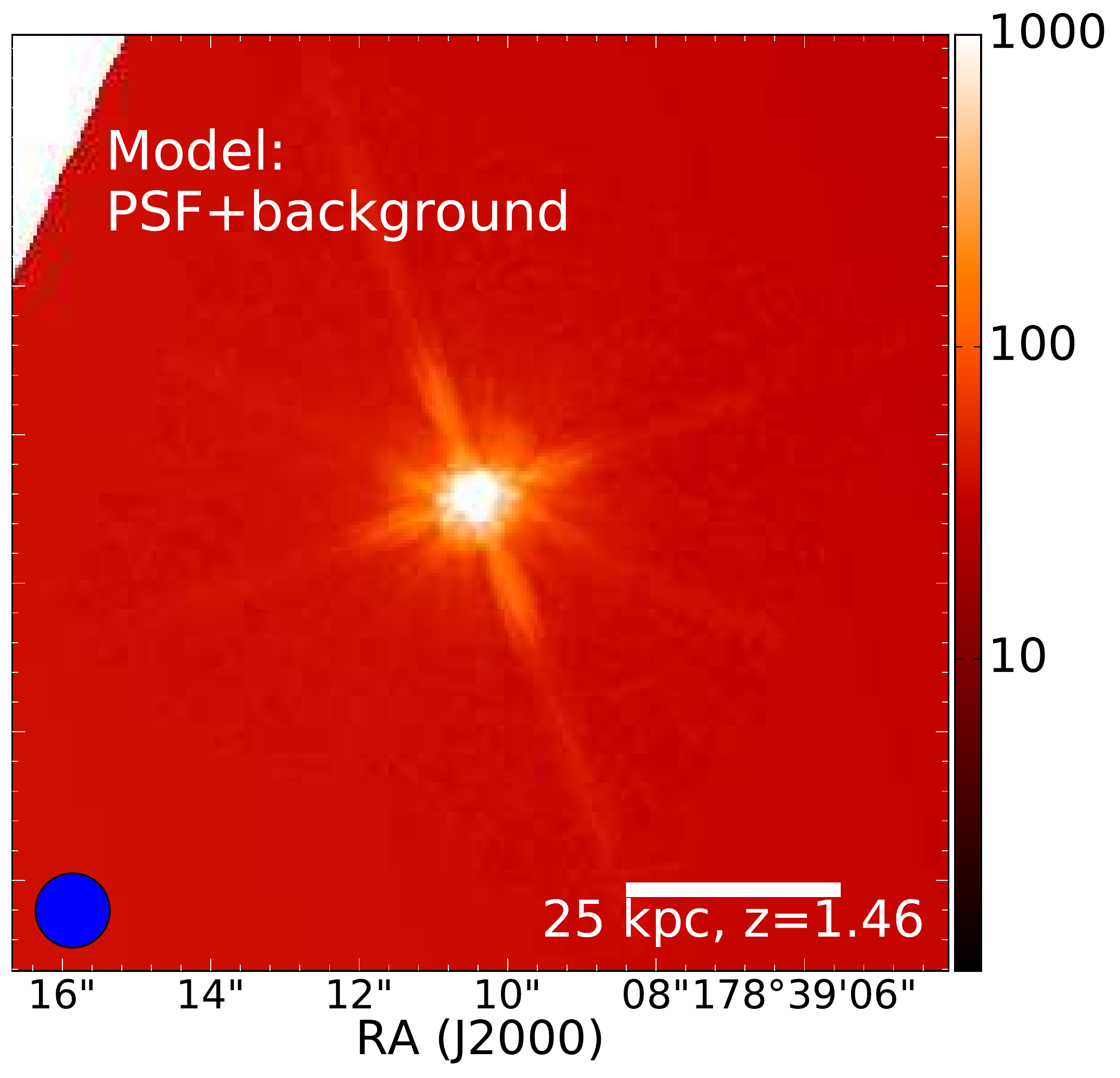}	 
	\includegraphics[scale=0.26]{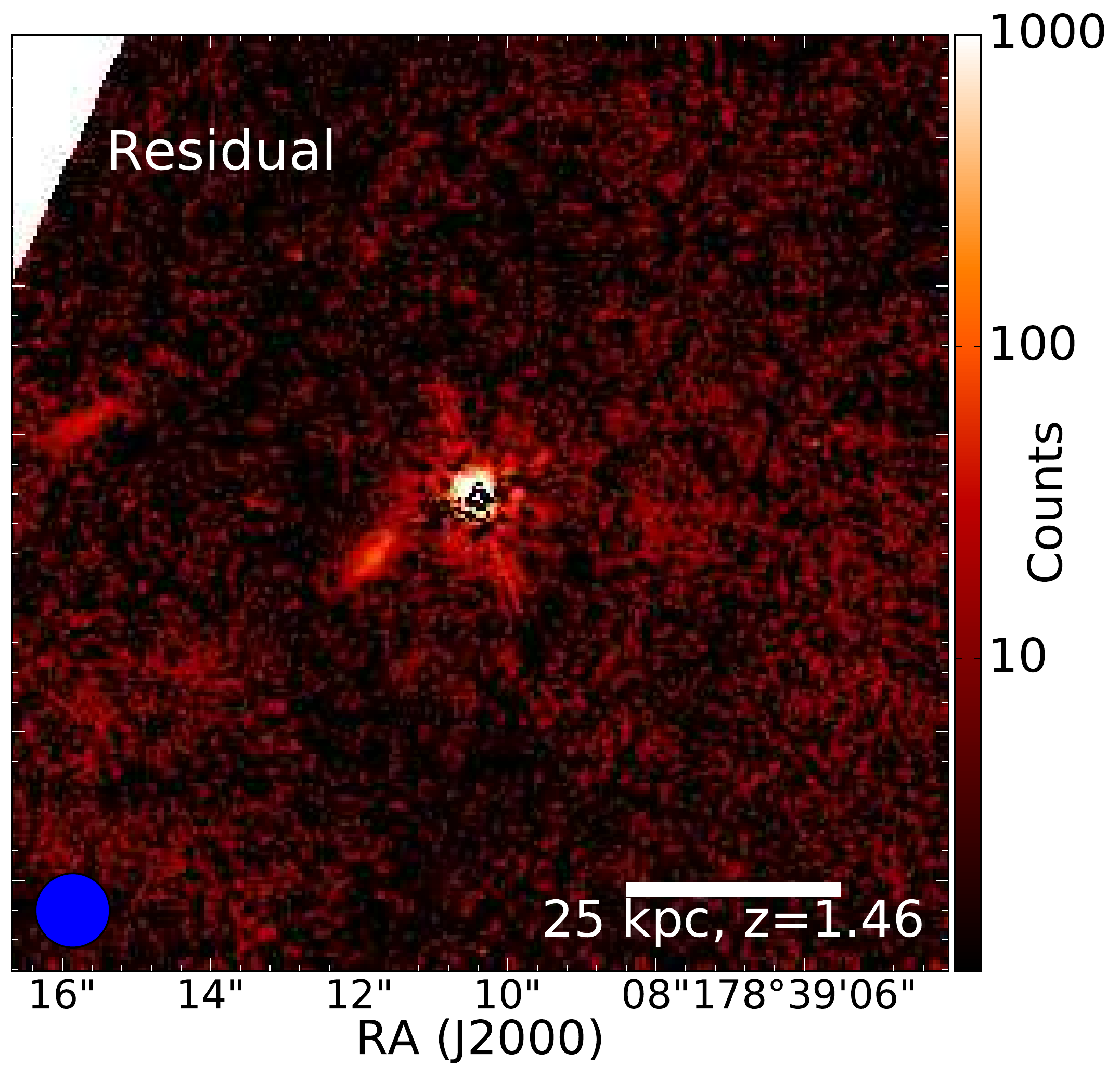}
	\includegraphics[scale=0.26,clip,trim={0 0 3.3cm 0}]{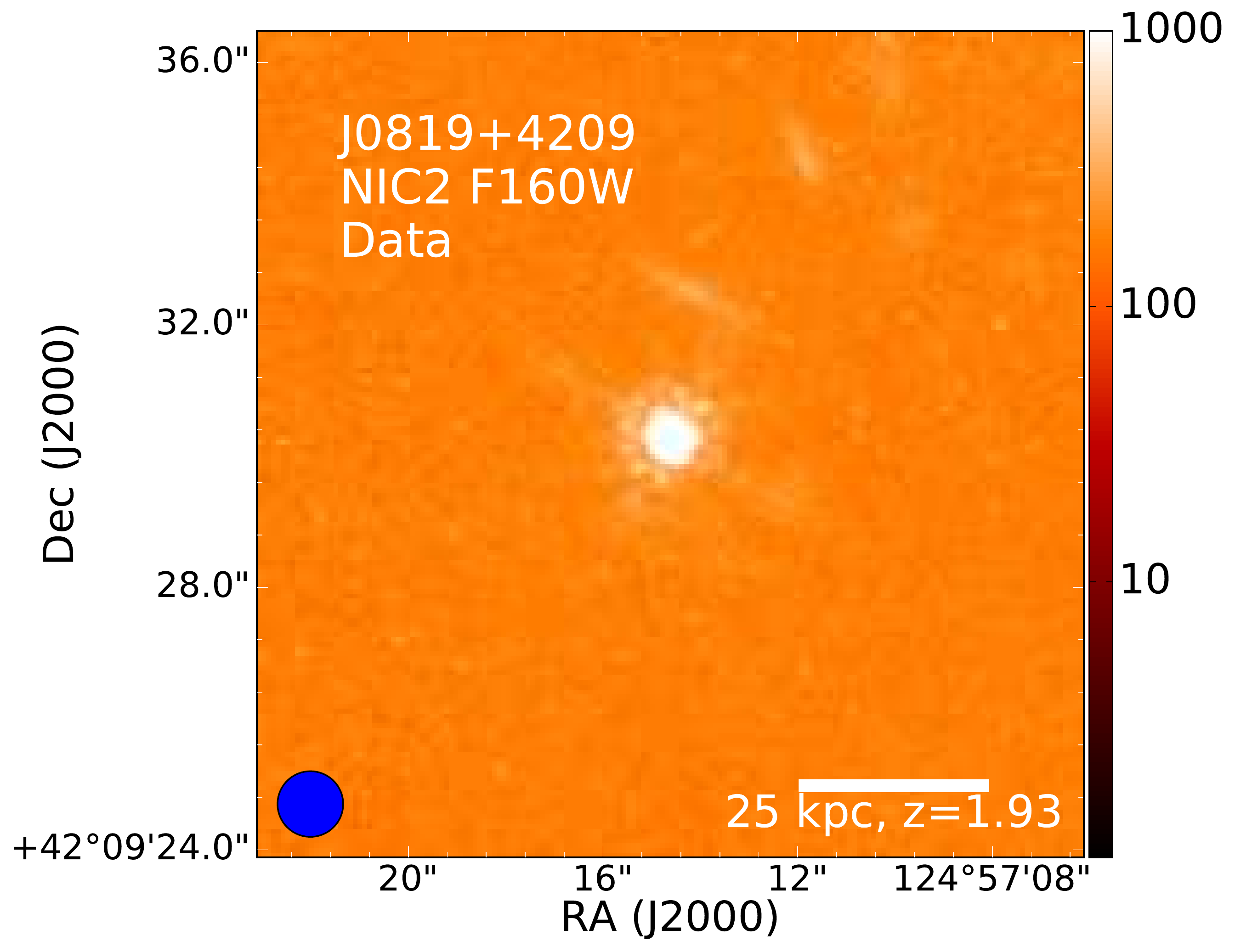}
	\includegraphics[scale=0.26,clip,trim={0 0 3.3cm 0}]{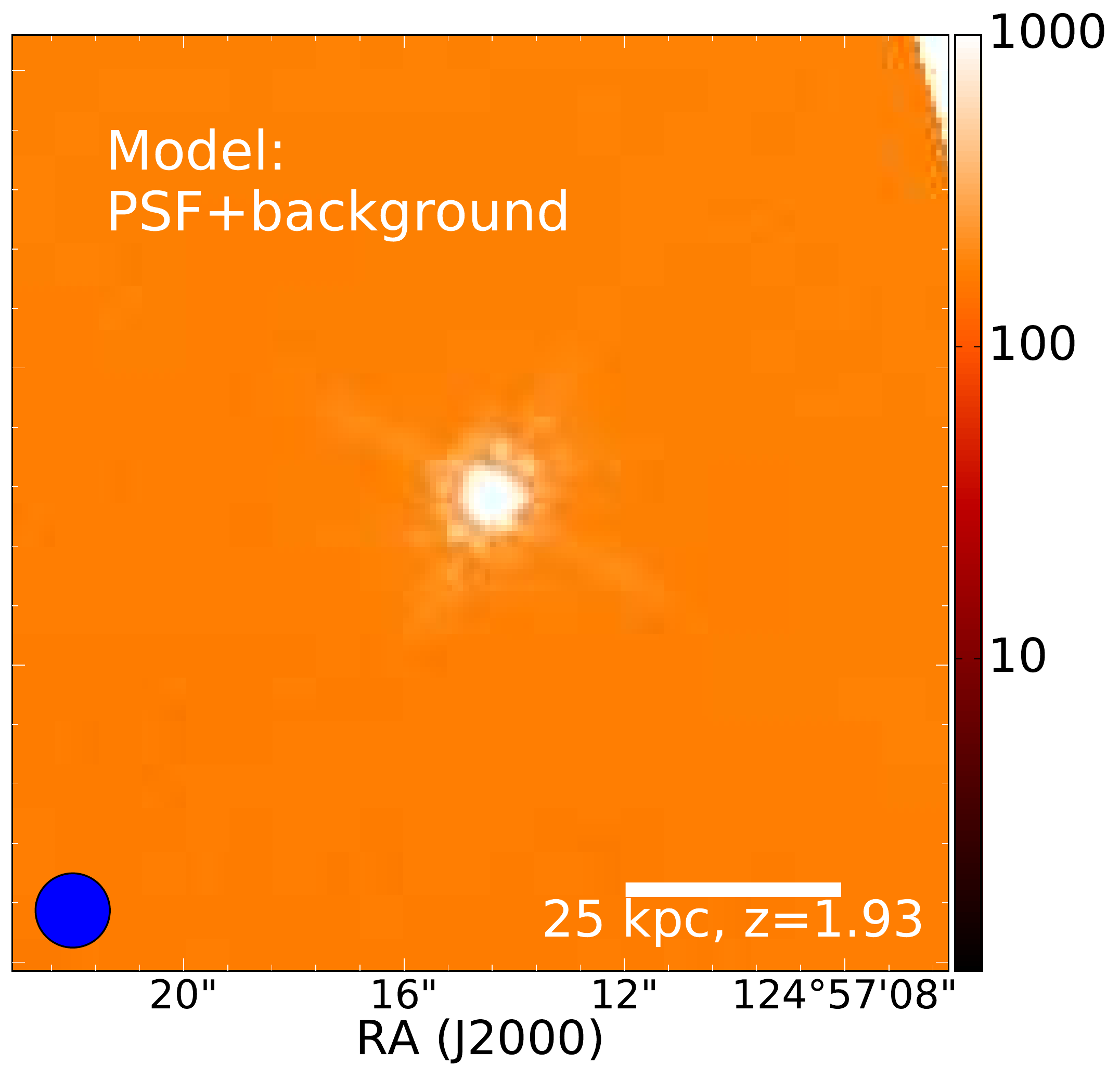}	 
	\includegraphics[scale=0.26]{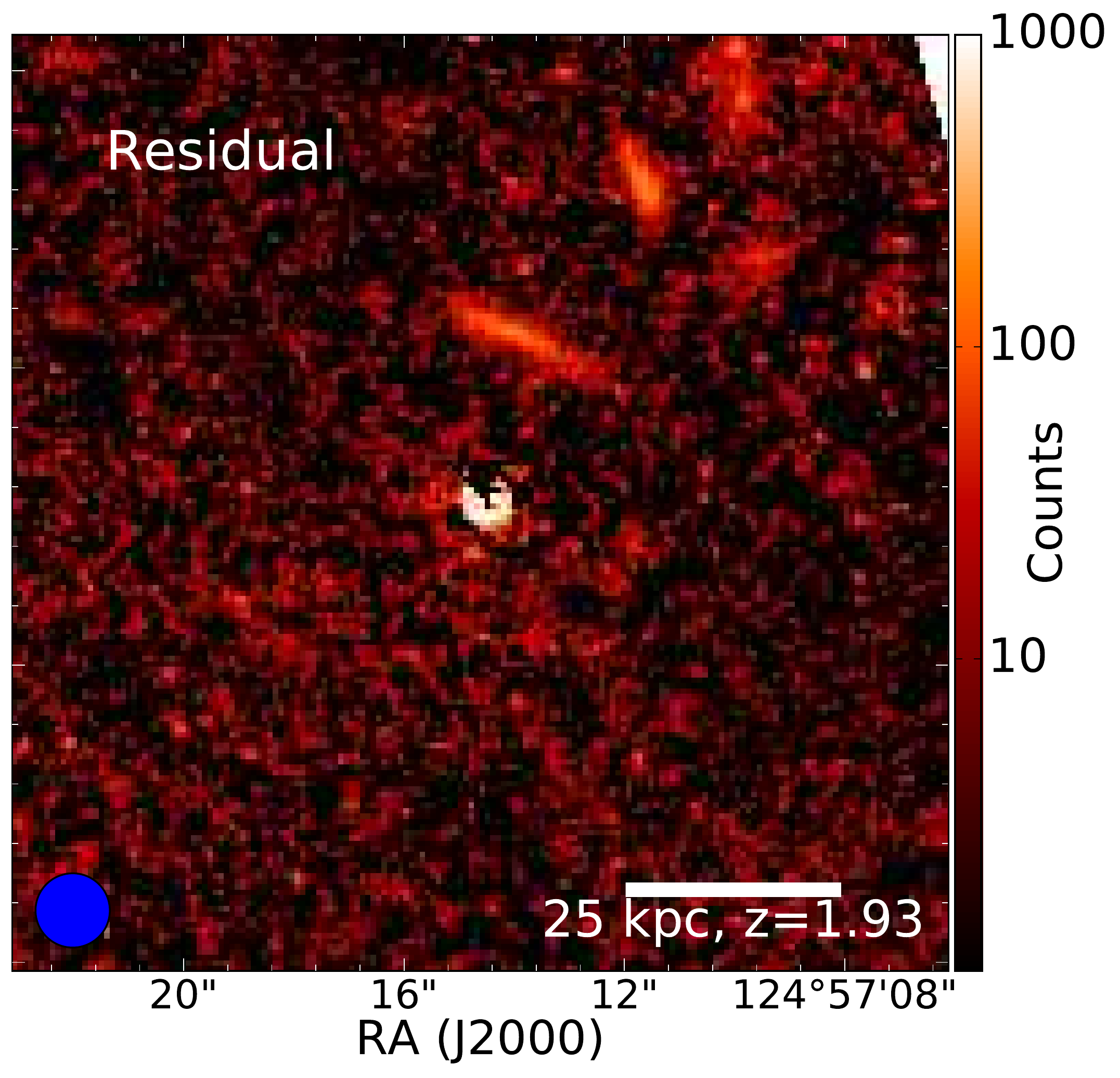}
	\includegraphics[scale=0.26,clip,trim={0 0 3.3cm 0}]{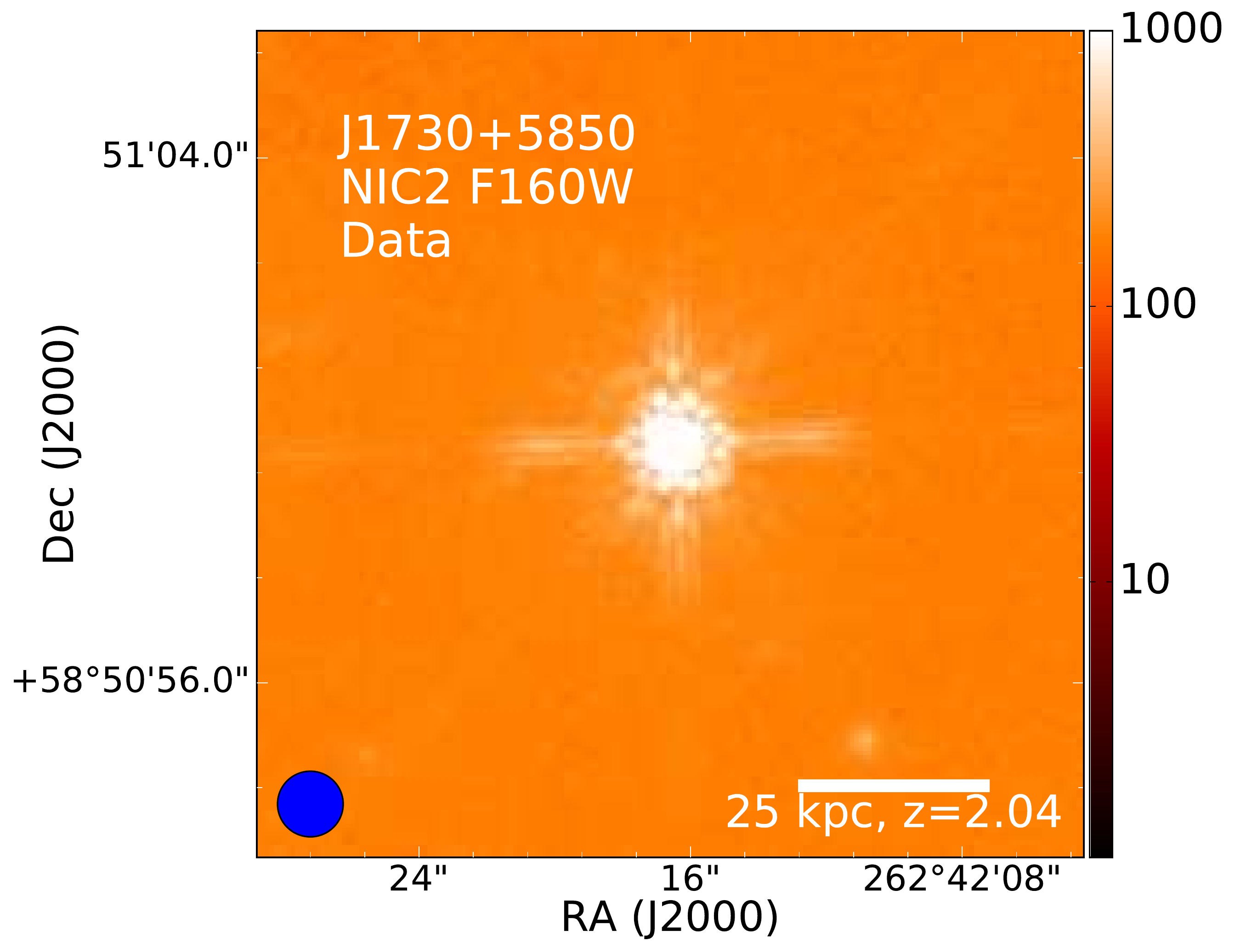}
	\includegraphics[scale=0.26,clip,trim={0 0 3.3cm 0}]{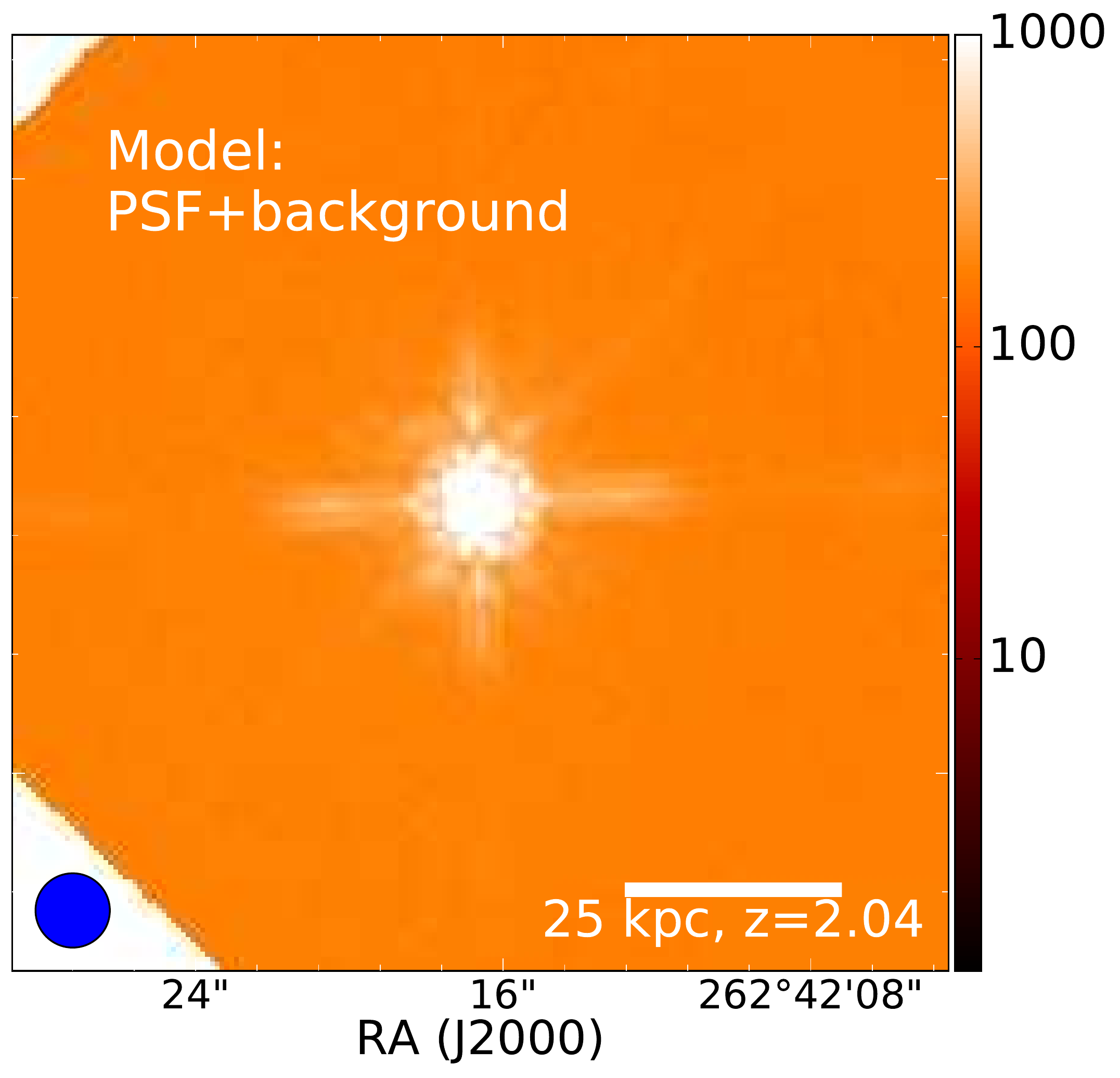}	 
	\includegraphics[scale=0.26]{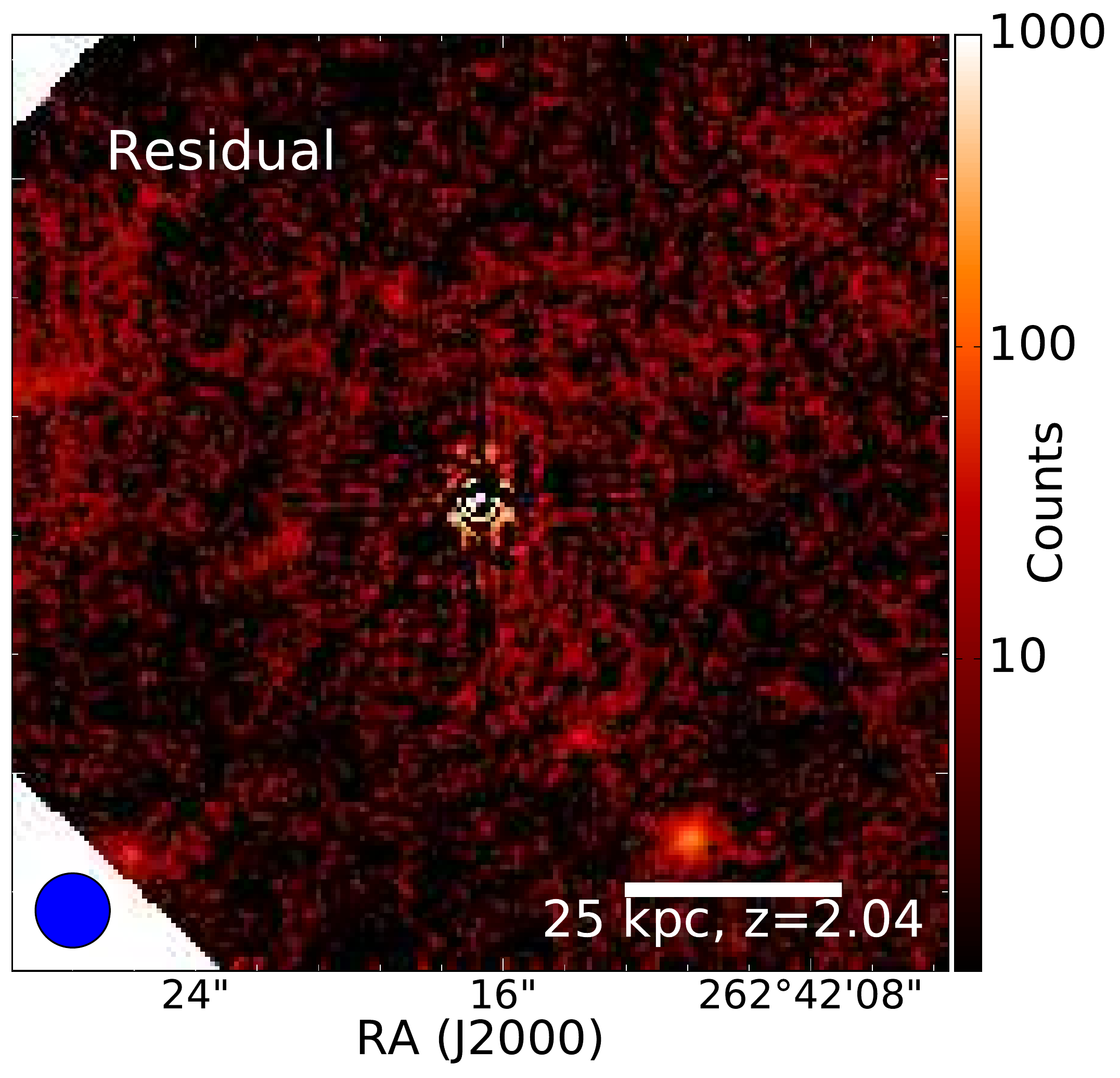} 	      	
	\caption{PSF-only modeling of NICMOS imaging of the Fe\-LoBAL quasars. The intensity is shown in units of total counts. The blue circle shows a diameter of 1 arcsecond; the image size is $12.6\times12.6$ arcseconds. The scale bars illustrate an angular size corresponding to 25 kpc at the quasar redshift. \emph{Left:} Original image. \emph{Centre:} Best-fit PSF model and background component. \emph{Right:} Residual after model subtraction. The PSF template for J0300+0048 has a low S/N (Appendix \ref{sec:appendix_psf}), resulting in a noisy residual image (top right). The distinct neighbouring galaxies visible for J0300+0048, J1154+0300 and J0819+4209 are modeled individually and subtracted before generating radial flux profiles (Figure \ref{fig:quasars_radialplots_nicmos}); their flux is thus not attributed to the quasar host galaxy.}\label{fig:quasars_2d_nicmos}
\end{figure*}

\begin{figure*}
	\advance\leftskip-3cm
	\centering	 
	\includegraphics[trim={1cm 1cm 1cm 0.5cm},clip,scale=0.3]{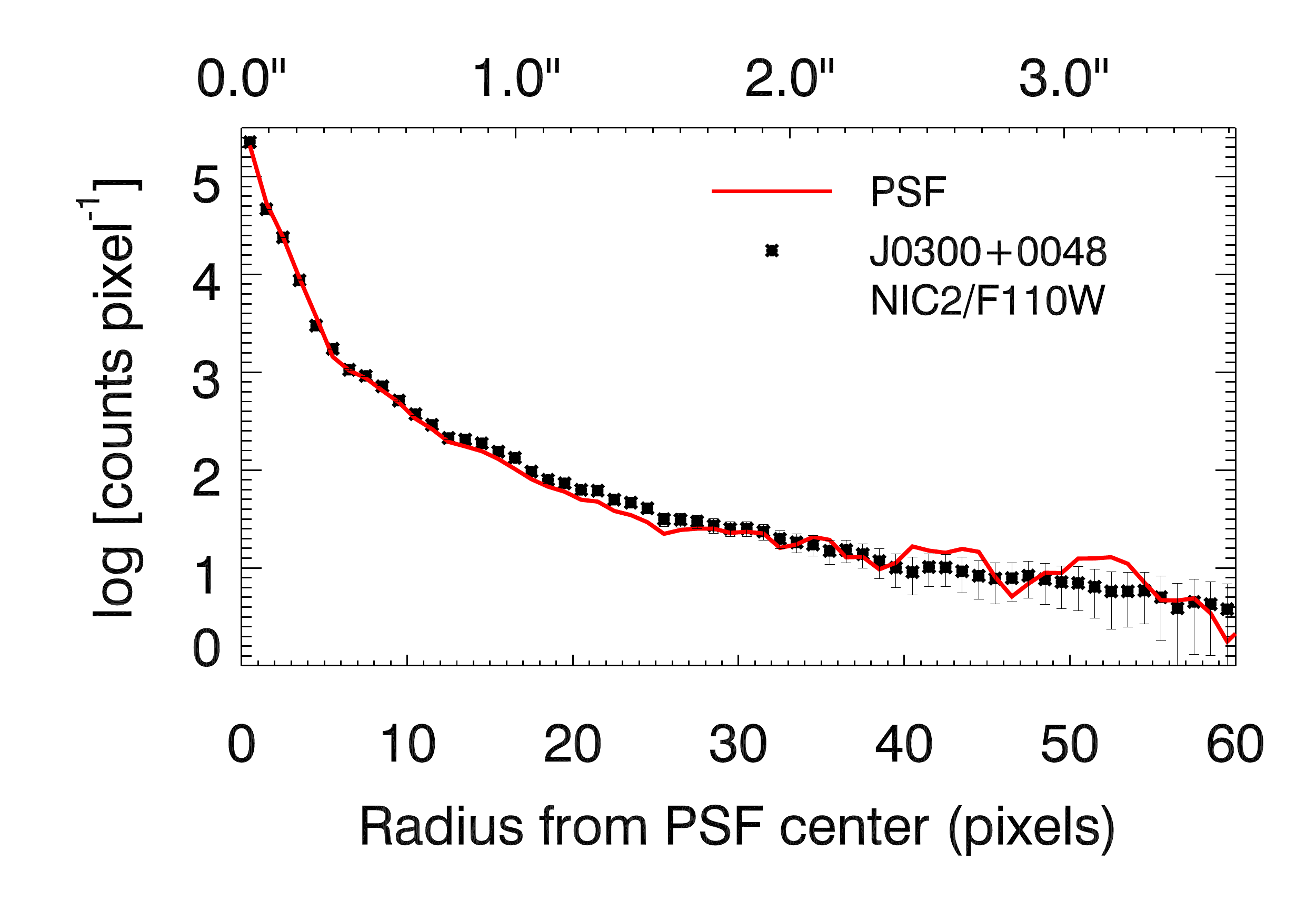}	 		 		
	\includegraphics[trim={0 1cm 1cm 0.5cm},clip,scale=0.3]{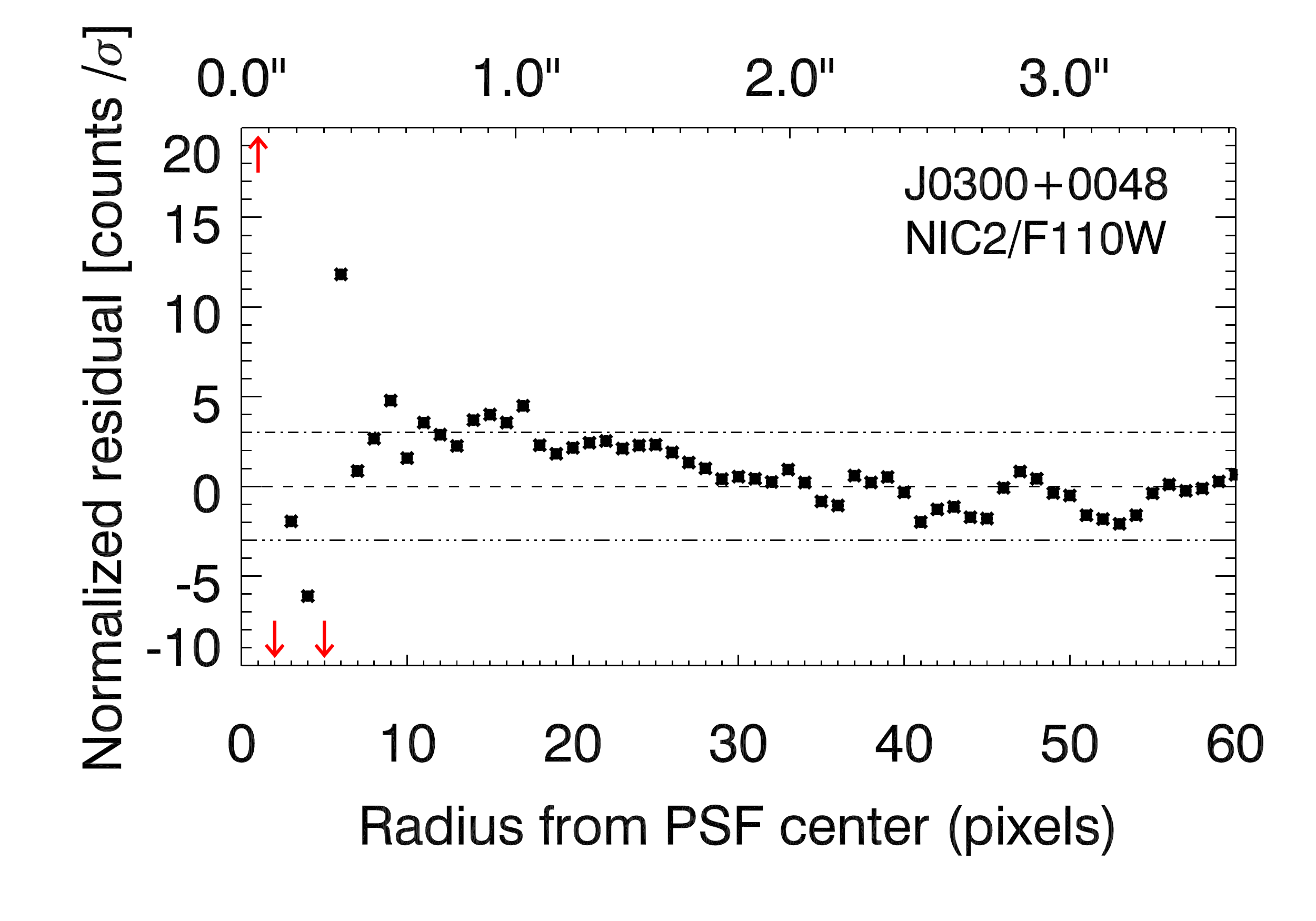}	
	\includegraphics[trim={1cm 1cm 1cm 0.5cm},clip,scale=0.3]{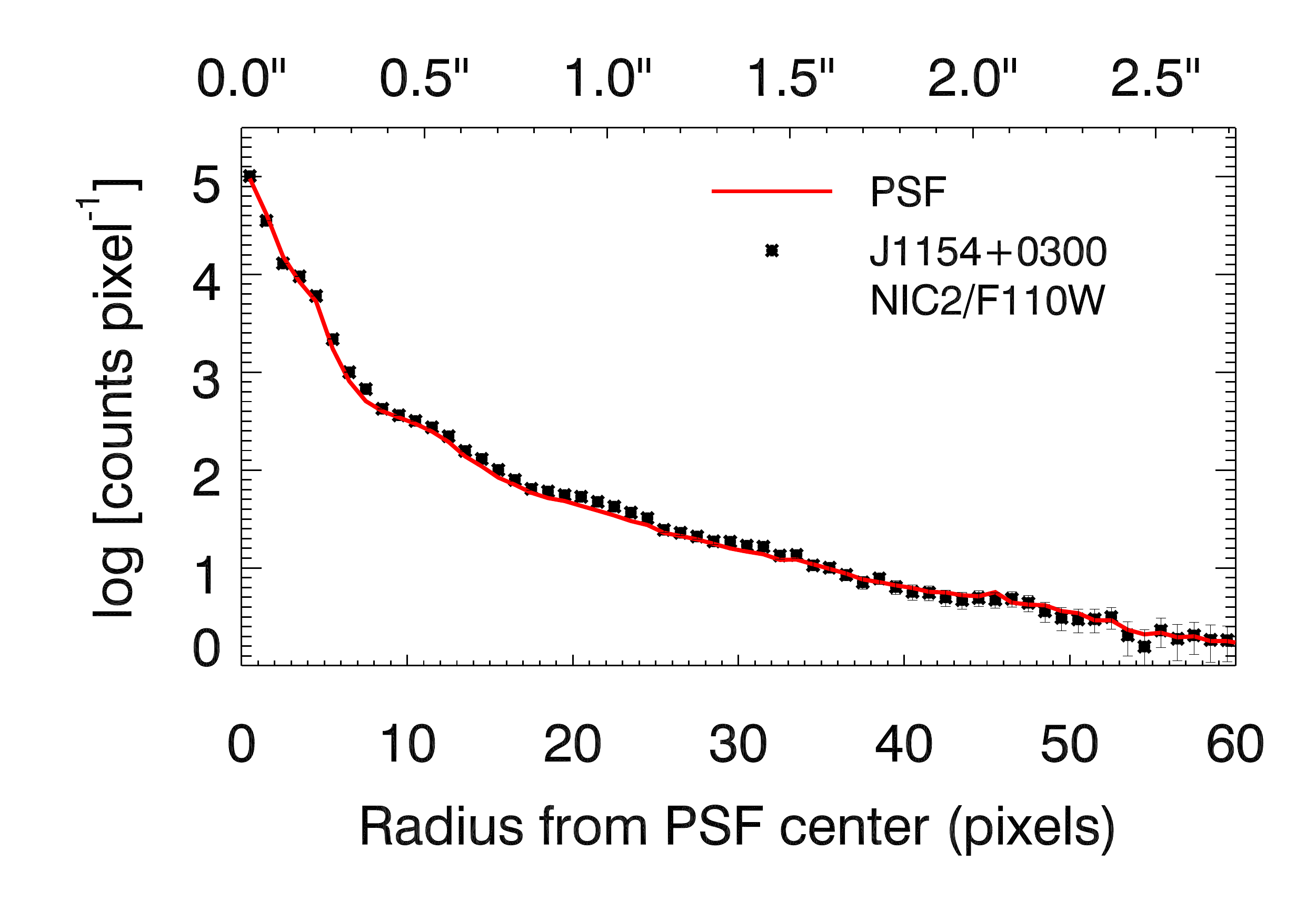}	 		 		
	\includegraphics[trim={0 1cm 1cm 0.5cm},clip,scale=0.3]{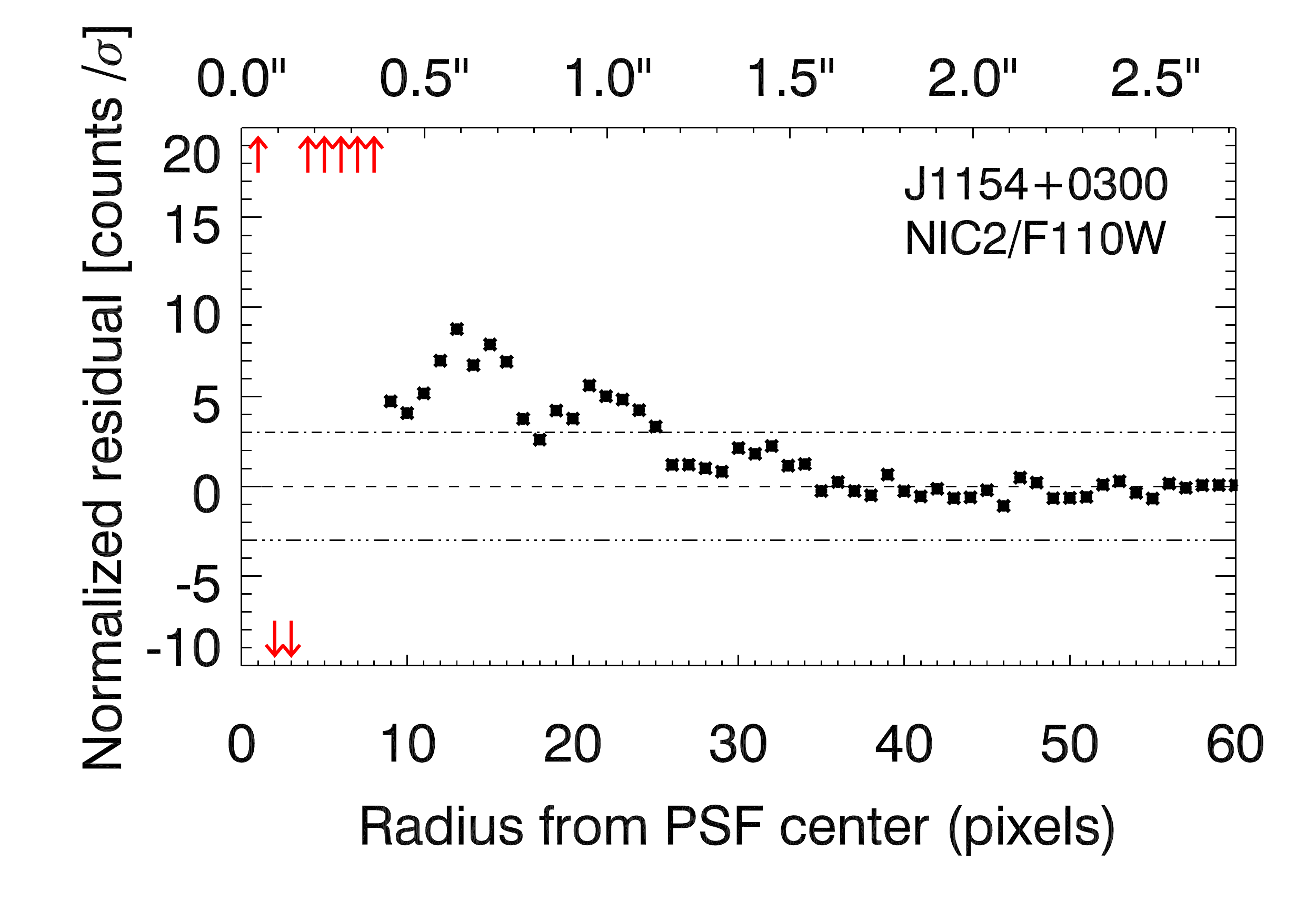}	
	\includegraphics[trim={1cm 1cm 1cm 0.5cm},clip,scale=0.3]{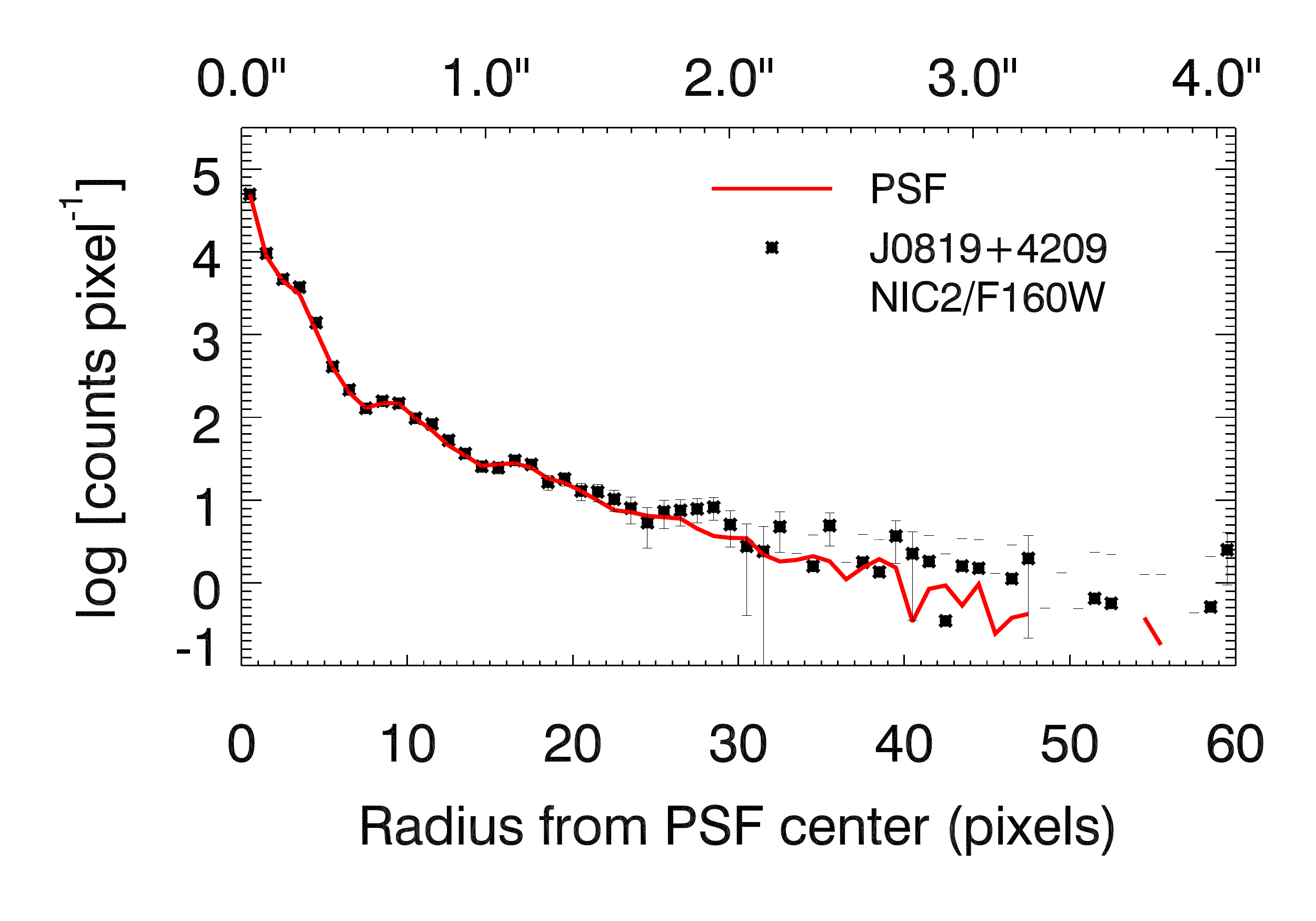}	 		 		
	\includegraphics[trim={0 1cm 1cm 0.5cm},clip,scale=0.3]{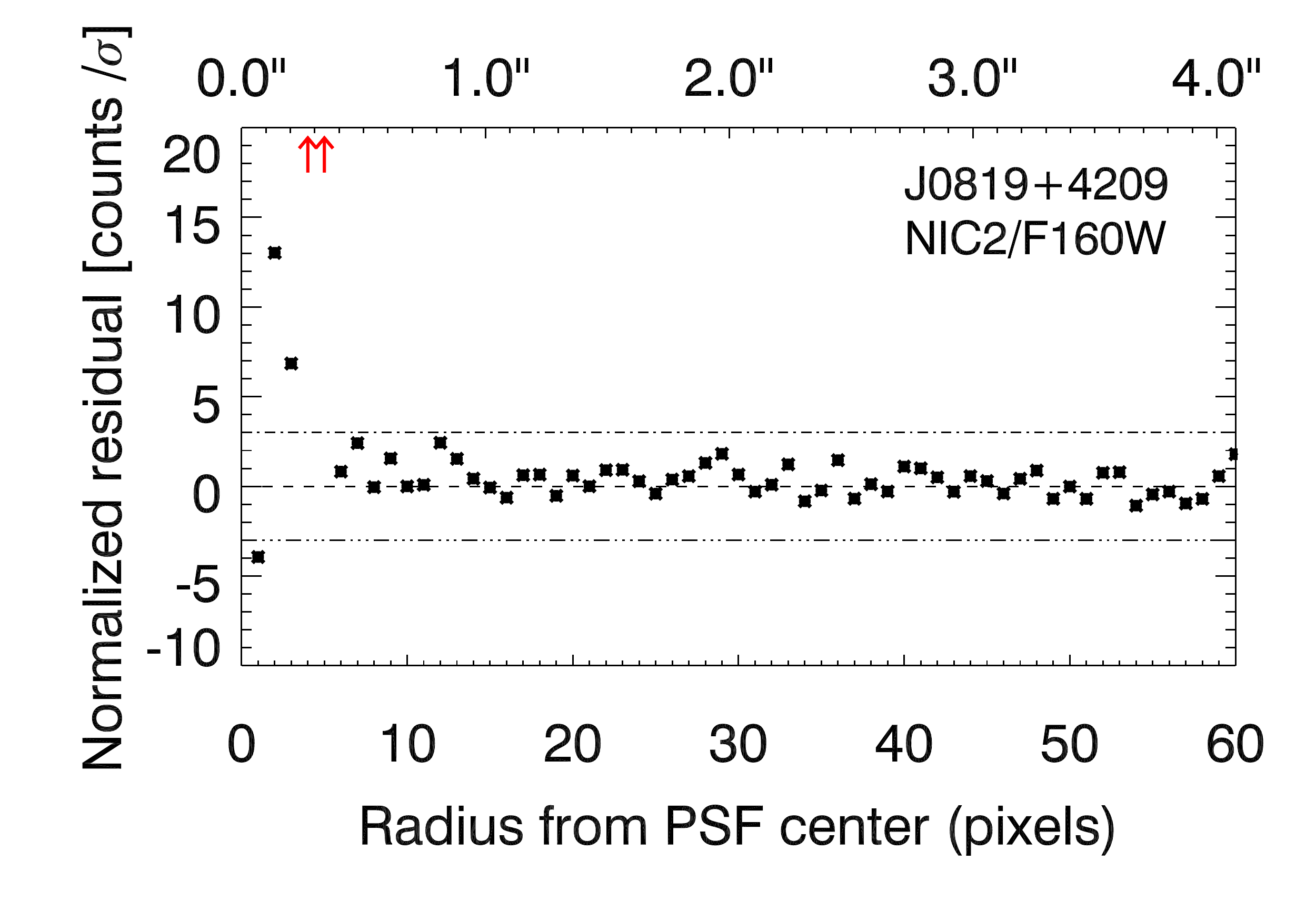}	
	\includegraphics[trim={1cm 1cm 1cm 0.5cm},clip,scale=0.3]{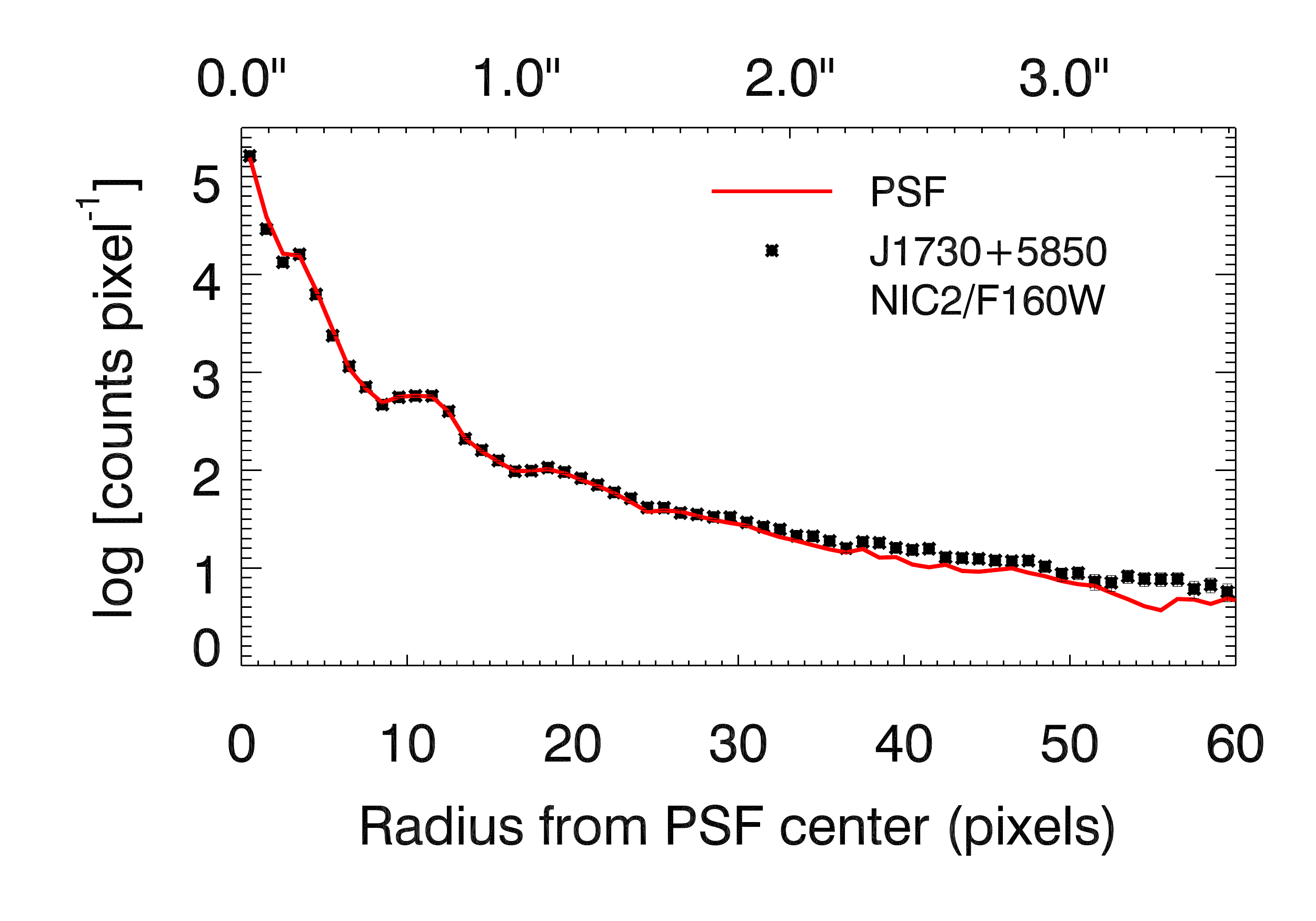}	 		 		
	\includegraphics[trim={0 1cm 1cm 0.5cm},clip,scale=0.3]{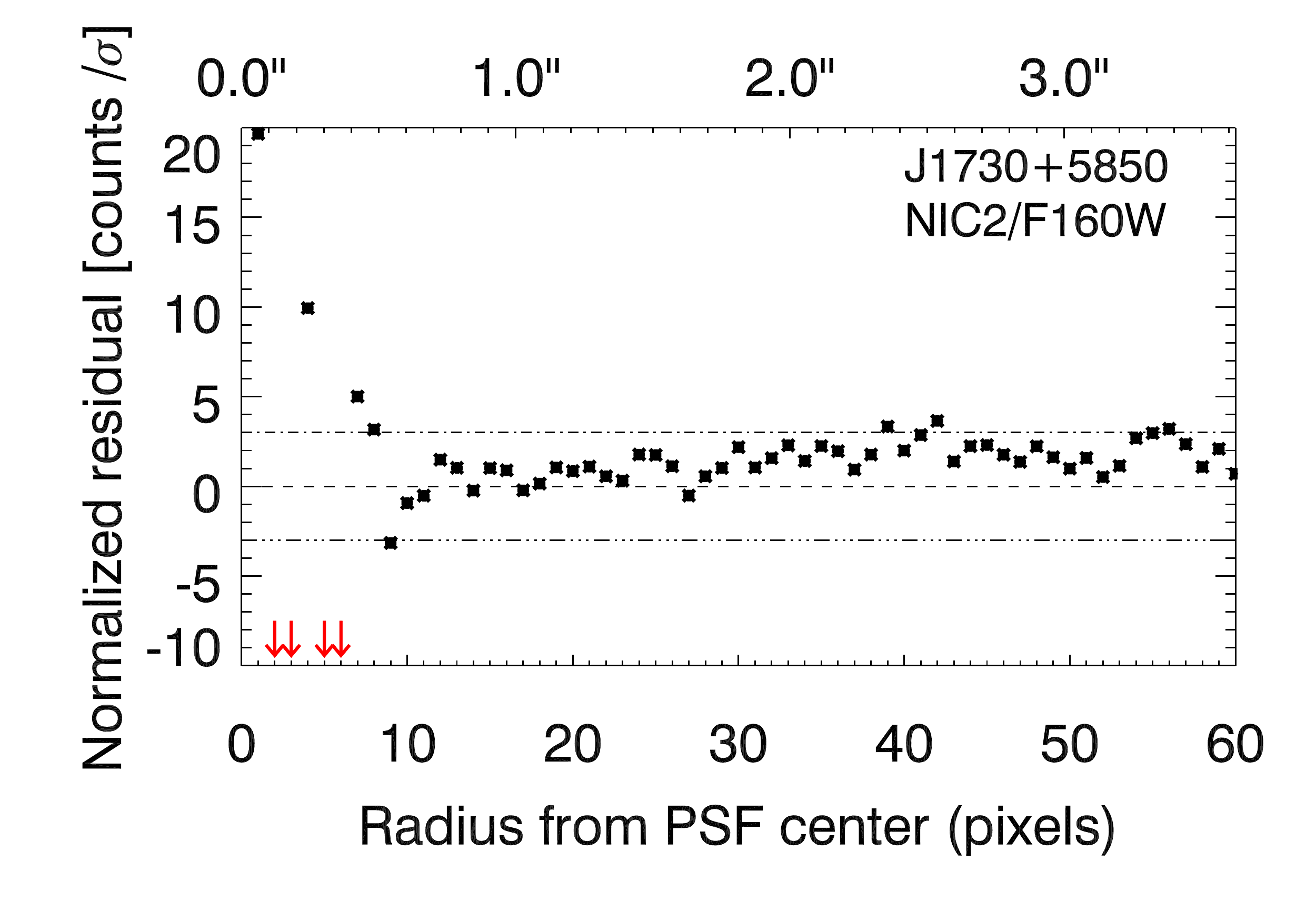}			       	
	\caption{PSF-only modeling results for our NICMOS imaging. See caption of Figure \ref{fig:quasars_radialplots_acs} for details.}\label{fig:quasars_radialplots_nicmos}
\end{figure*}

For J0300+0048, J1154+0300 and J0819+4209, the NICMOS imaging reveals neighbour galaxies at sufficiently small angular separations to potentially affect our image decompositions (Figure \ref{fig:quasars_2d_nicmos}). We model these galaxies using a single Sersic component; the fluxes attributed to them are not included in our analyses. We discuss these neighbouring galaxies further in \S \ref{sec:discussion_neighbors}.

\paragraph*{PSF-only Modeling, NICMOS:}

All the residual images show oversubtracted central pixels and residual PSF diffraction spikes due to mismatch between the PSF template and the quasar PSF (Figure \ref{fig:quasars_2d_nicmos}). As the PSF template for J0300+0048 has a low S/N relative to the quasar observation, the entire PSF-subtracted image is noisy, and the residual features have a lower significance than would be expected if using an ideal PSF. Nonetheless, J0300+0048 shows an extended residual that is detected at a significance of $\sim 3-5\sigma$ between radii of 0.7'' and 1.1'', extending to around 1.6'' at a lower significance (Figure \ref{fig:quasars_radialplots_nicmos}). J1154+0300 displays a bright, compact residual core at radii less than $\sim0.6$ arcseconds (corresponding to a physical transverse distance of around 4 kpc). Both these objects fulfill our detection criteria for extended emission. The very compact residual for J0819+4209 just barely fulfills our detection criteria, while the residuals for J1730+5850 are consistent with zero extended emission. 

\paragraph*{PSF-plus-Sersic Modeling, NICMOS:}

Our PSF-plus-Sersic modeling converges at a finite brightness for the Sersic component for all four quasars. In Table \ref{tab:sersicfit_nicmos_bestfit} we present the best-fitting models for each quasar. For three quasars, the best-fit solution has $n=0.75$, i.e., at the imposed lower limit for $n$. However, in terms of the $\chi^2$ statistic, the modeling does not strongly favor any particular values of the Sersic index, $n$. We therefore fit three different models with constant Sersic indices ($n=1$, $n=2.5$, and $n=4$) to each image, so as to explore a range of central concentrations. While non-BAL quasars in the local Universe tend to have giant elliptical hosts \citep{Dunlop2003}, suggesting $n\approx4$, we prefer not to assume an \emph{a priori} value of $n$ for these high-redshift host galaxies. For a morphologically diverse sample of template galaxies, our simulation work (Appendix \ref{sec:nicmos_sims}) suggests that the accuracy of the \mhost\,measurements are improved by averaging over the values obtained for $n=1$ and $n=4$ models, as opposed to adopting \mhost\,from the best-fit model. Note that the magnitudes themselves are averaged, i.e., the averaging is done in log-flux space. We present the average properties $\langle\mhost\rangle$ and $\langle R_e\rangle$ for a grid of model fits to each quasar (Table \ref{tab:sersicfit_nicmos}). The best-fit scale sizes $R_e$ are highly sensitive to the choice of $n$, and should therefore be regarded as order-of-magnitude estimates of the true $R_e$. For quasars J0300+0048 and J1154+0300, we find values of \contrast\,and $R_e$ that are consistent with those found for detected, resolved sources in our simulations. We therefore consider these objects to have detected host galaxies. For J0819+4209, the Sersic profile converges at the imposed lower limit $R_e=\mathrm{FWHM}_{\mathrm{PSF}}$, i.e., the extended flux is only marginally resolved. While we consider the detection of extended emission to be real, given the faintness of false-positive detections to \emph{bona fide} point sources in NICMOS data (Appendix \ref{sec:appendix_psf}), we consider our PSF-plus-Sersic modeling results to be highly uncertain for this object, as discussed below.

\paragraph*{Uncertainties on \boldmath{\mhost}, NICMOS:} For each quasar, we estimate ranges of uncertainty for \mhost\,by performing PSF-plus-Sersic modeling of real galaxies artificially redshifted to the quasar redshift, including a superimposed point source, and with appropriate noise properties for the NICMOS observations. Full details are presented in Appendix \ref{sec:appendixB}. These uncertainties are rather conservative, as they represent the largest error in \mhost\,found for any of our simulations for a given observation, covering a range of possible host galaxy morphologies. They are in rough agreement with the typical uncertainties found by \citet{Simmons2008} for \textsc{Galfit} image decomposition of simulated AGN with \contrast$\sim3$. We adopt the following uncertainties (in AB magnitudes) for the remainder of this work: \mhost$=19.6(\pm0.5)$ mag for J0300+0048, \mhost$=20.8(\pm0.5)$ mag for J1154+0300, and \mhost$=21.2_{-0.5}^{+1.3}$ mag for J0819+4209. The latter object has a relatively large uncertainty on \mhost\,due to the compactness of the detected host component. For J1730+5850, we see no evidence of extended emission. We therefore use our simulations to establish the limiting value of \contrast\,for which the host galaxy is no longer detected. This is done for a range of host morphologies, and we adopt the most conservative (i.e., brightest), upper limit \mhost. For J1730+5850 this yields \mhost$\ge20.0$. 

\paragraph*{Consistent \mhost\,When Using Alternative PSF Templates:}

We lack dedicated PSF star observations for this study. For our NICMOS analysis, we instead use stacked observations of the calibration star P330-E as a PSF template (Appendix \ref{sec:appendix_psf}). The stacking process unavoidably broadens the PSF slightly due to centring uncertainties. Also, the three-month time window that we allow for the PSF star observations relative to the quasar observation date may lead to additional mismatch due to long-term PSF variability \citep{Kim2008}. There is therefore a concern that the extended flux that we attribute to host galaxy emission may instead be caused by severe PSF mismatch. For J1154+0300, we perform additional tests using high-S/N PSF templates (Appendix \ref{sec:appendix_nicmos_highsnsample}). These were observed at time separations of years from the quasar observation, and are therefore expected to display significant PSF mismatch due to time evolution. We detect extended flux for J1154+0300 using any of these alternative PSF templates. While $R_e$ becomes unresolved for certain high-SN PSF templates, \mhost\,is in all cases consistent with our our original modeling to within the quoted uncertainties. We elect to use the results obtained using P330-E, as presented in Table \ref{tab:sersicfit_nicmos}, in the remainder of our analysis.

\section{Discussion}\label{sec:discussion}

\subsection{Similar Nucleus-to-Host Luminosity Ratios for FeLoBALs and Non-BAL Quasars}\label{sec:discussion_comparison}

\begin{figure*}
\centering
\includegraphics[width=0.7\linewidth]{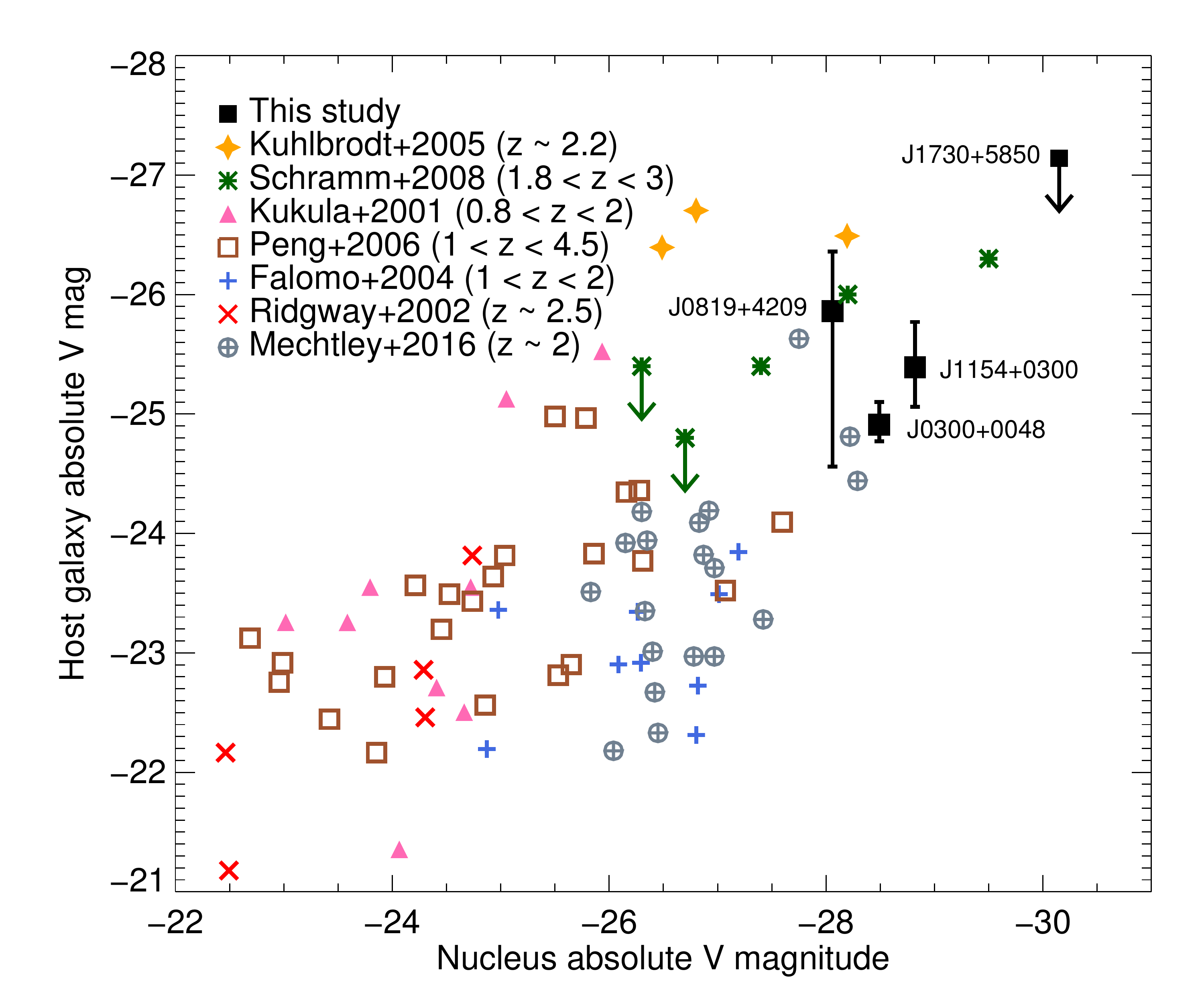}
\caption{Comparison of the rest-frame V band nuclear and host galaxy magnitudes of the FeLoBALs with those of non-BAL quasars taken from the literature. The black diamonds represent the four quasars studied in this work; the error bars are not statistical errors, but represent the most severe errors on \mhost\,found in our simulations at a given redshift and \contrast\,level. Downward-pointing arrows denote upper limits on the host galaxy magnitude (e.g., for J1730+5850). The pre-2008 results are compiled (and converted to a $\Omega_\lambda=0.7$ cosmology where necessary) by \citet{Schramm2008}.}
\label{fig:mvcompare}
\end{figure*}

In Figure \ref{fig:mvcompare} we compare the nuclear and host galaxy $V$ band absolute magnitudes for our FeLo\-BAL quasars with those of non-BAL quasars with host galaxy detections at similar redshifts. As the NIC2 filter bandpasses were chosen to sample the rest-frame $V$ band, and the spectral term in the K correction therefore should be minimal, we only apply the bandwidth-narrowing $(1+z)$ term to derive the absolute magnitudes. Given the $\sim2$ mag scatter in \contrast\,displayed by the non-BAL quasars, and the uncertainties on \mhost\,for our measurements, the four FeLo\-BALs are consistent with the existing $m_\mathrm{nuc}$ -- $m_\mathrm{host}$ relationship for non-BAL quasar host galaxies. Quasar J1730+5850 is highly luminous in the $V$ band; the upper limit host galaxy $V$ magnitude appears consistent with the existing relationship.

\subsection{No Direct Evidence for Ongoing Star Formation}\label{sec:discussion_templates}

In the local Universe, Lo\-BAL quasars are often found in ULIRG galaxies, which also display merger-induced star formation \citep[e.g.,][]{Canalizo2001}. Here, we use our measurements of \mhost\,in the rest-frame optical, along with the limiting values of \mhost\,in the UV, to investigate the presence of strong, unobscured star formation activity of the type seen in some local-Universe starburst galaxies. 

In Figure \ref{fig:kinneytemplates}, we show template spectra of a 1 Gyr single-age stellar population model \citep{Bruzual2003}, and of an unobscured starbursting galaxy \citep{Calzetti1994}, scaled to the F110W flux densities of the host galaxies of J0300+0048 and J1154+0300. The blue shaded region shows the error margins of the template scaling due to the uncertainties on \mhost. We also show the upper limits on the rest-frame UV host galaxy magnitude. For these two quasars, the UV non-detections exclude unobscured starbursting hosts, while being fully consistent with single-age stellar populations of age 1 Gyr or older, as expected for quiescent elliptical galaxies at $z\sim1$.  For J0819+4208 (not shown), \mhost\,is highly uncertain, and we cannot rule out an unobscured starburst. We apply the \citet{Calzetti2000} dust reddening law to the templates, and find that the template fluxes become marginally consistent with the UV non-detections for $E(B-V)\approx0.5$ (J0300+0048) or $E(B-V)\approx0.4$ (J1154+0300). Thus, our $HST$ observations cannot be used to discriminate a quiescent host from a vigorously star-forming, dust-obscured galaxy. We note that \citet{Glikman2012} find an overabundance of FeLo\-BALs in their sample of the reddest quasars. Many objects in their sample show a reddening of $E(B-V)\ge0.5$, which would yield UV non-detections for our observations, even for starbursting host galaxies.
 
\newcommand{\pskinney}{0.4972}
\newcommand{\trimkinney}{20pt 0pt 50pt 0pt}

\begin{figure*}
\centering
\includegraphics[trim=20pt 0pt 50pt 0pt,width=\pskinney\linewidth]{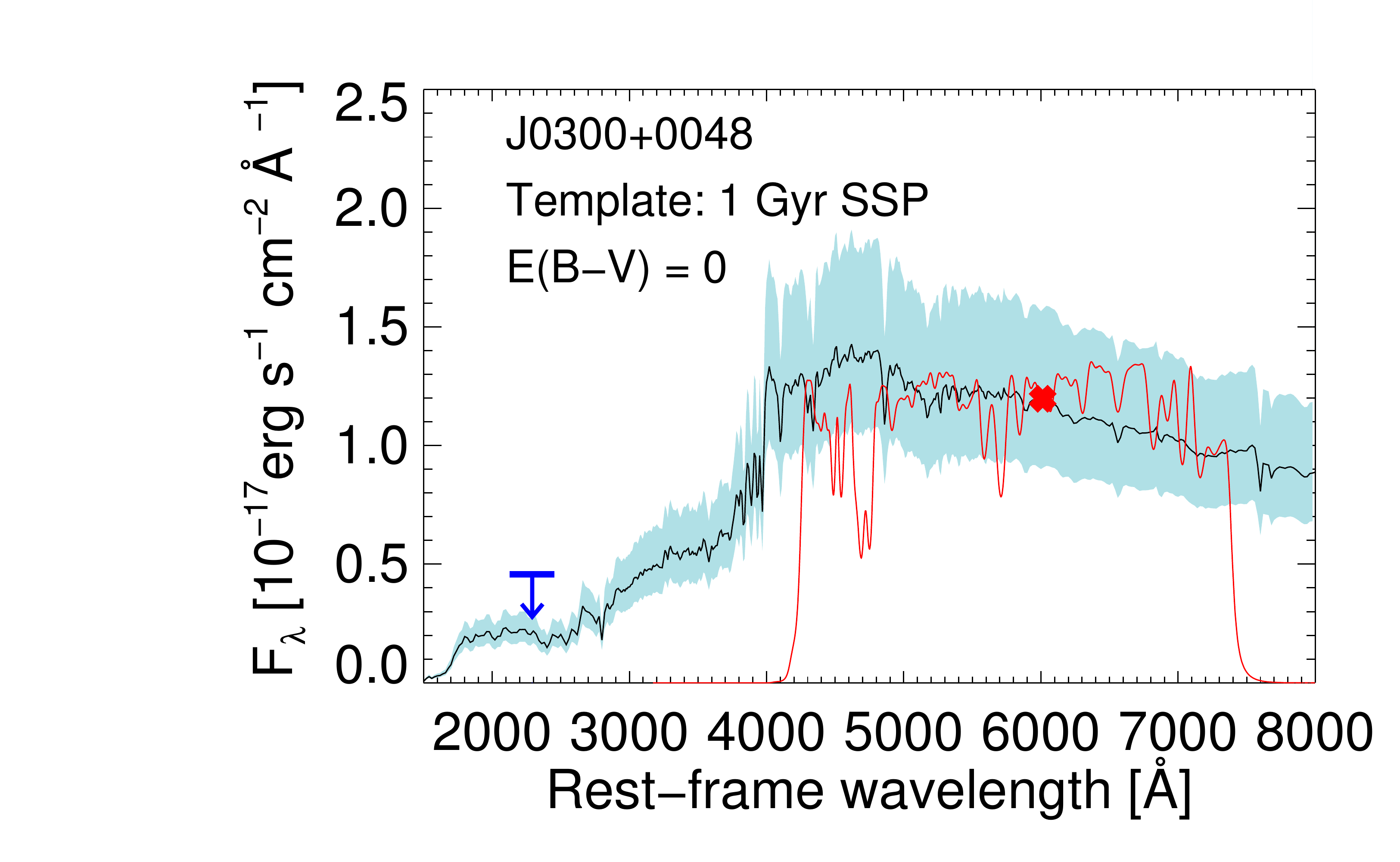}
\includegraphics[trim=20pt 0pt 50pt 0pt,width=\pskinney\linewidth]{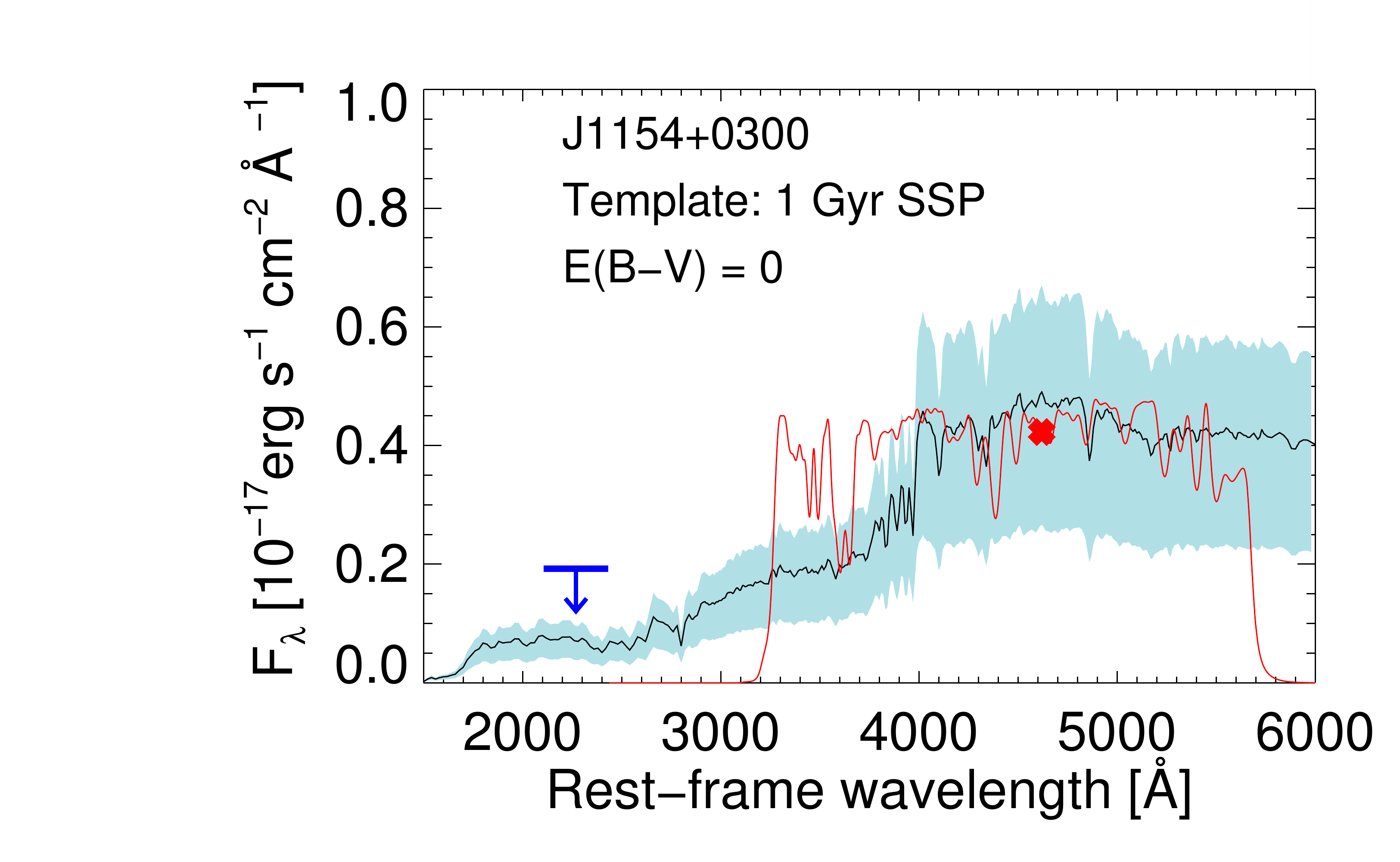}
\includegraphics[trim=20pt 0pt 50pt 0pt,width=\pskinney\linewidth]{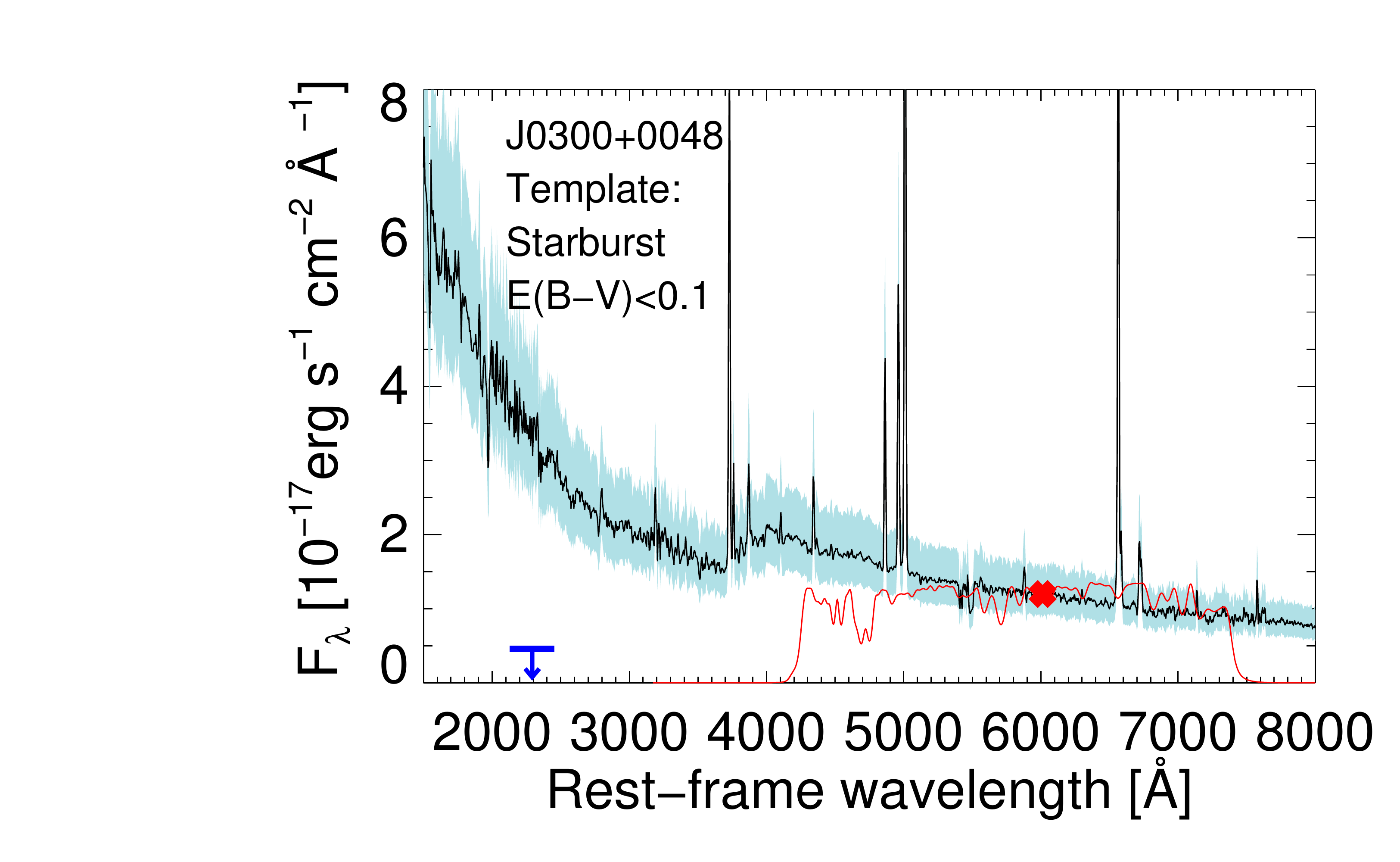}
\includegraphics[trim=20pt 0pt 50pt 0pt,width=\pskinney\linewidth]{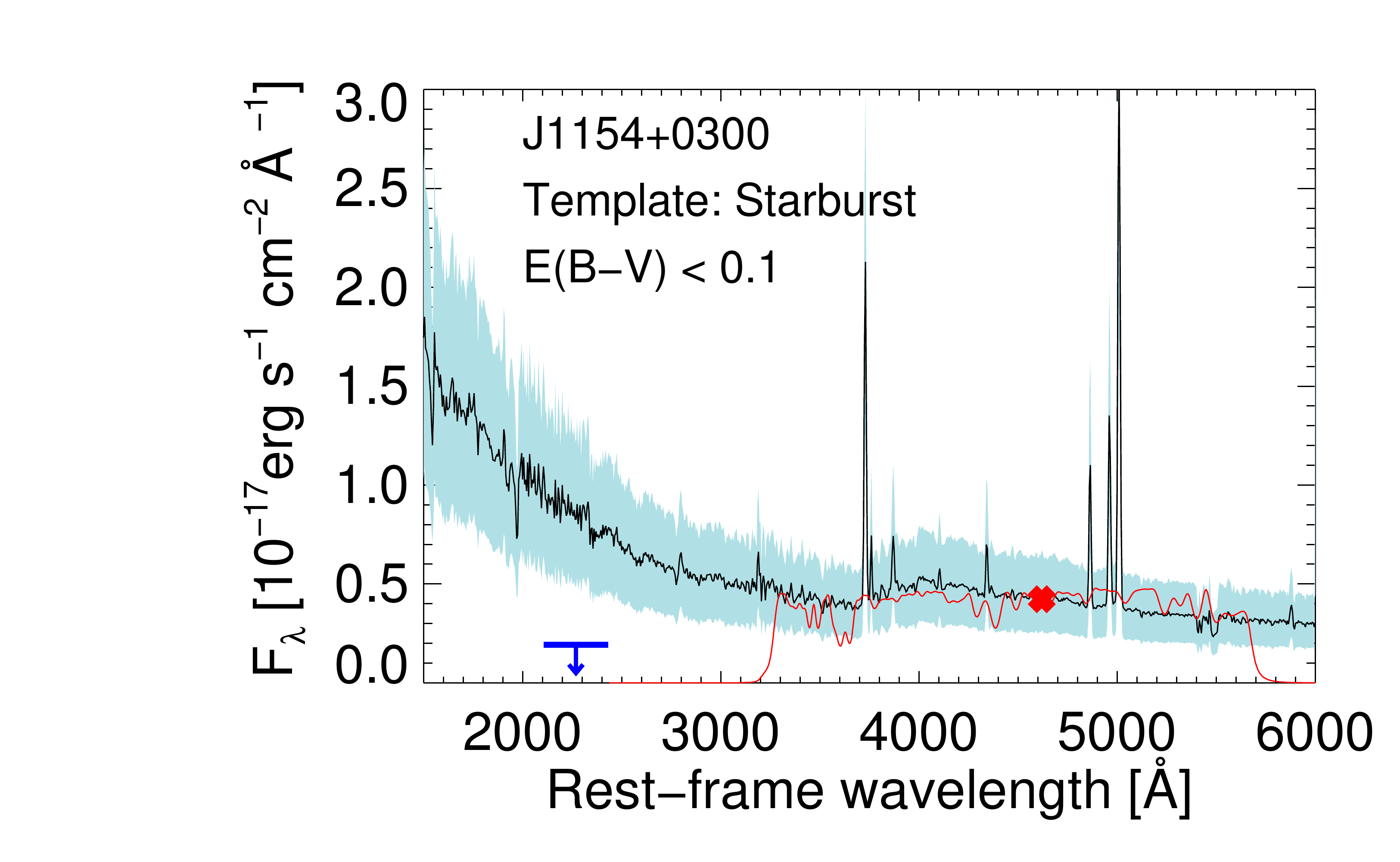}
\includegraphics[trim=20pt 0pt 50pt 0pt,width=\pskinney\linewidth]{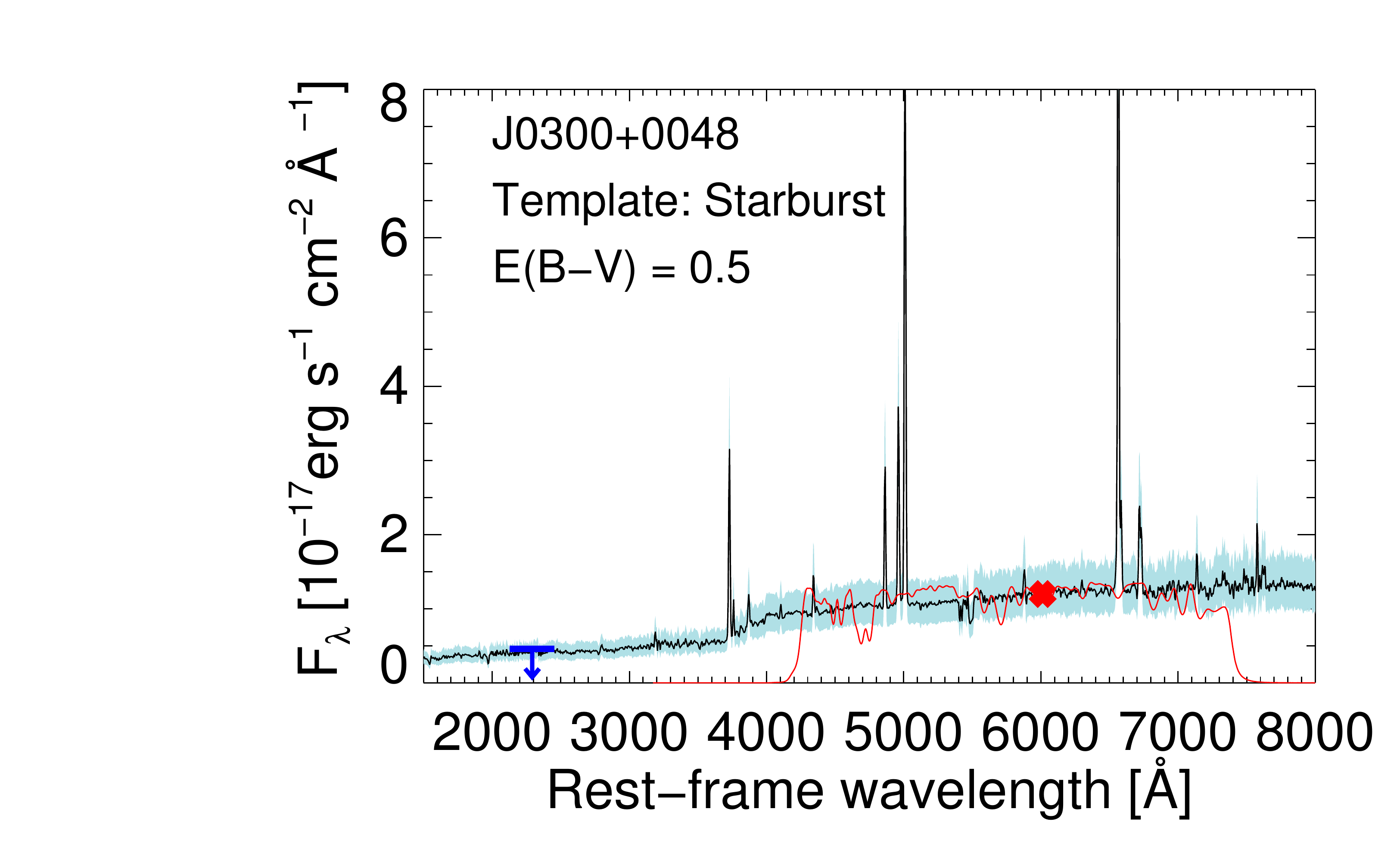}
\includegraphics[trim=20pt 0pt 50pt 0pt,width=\pskinney\linewidth]{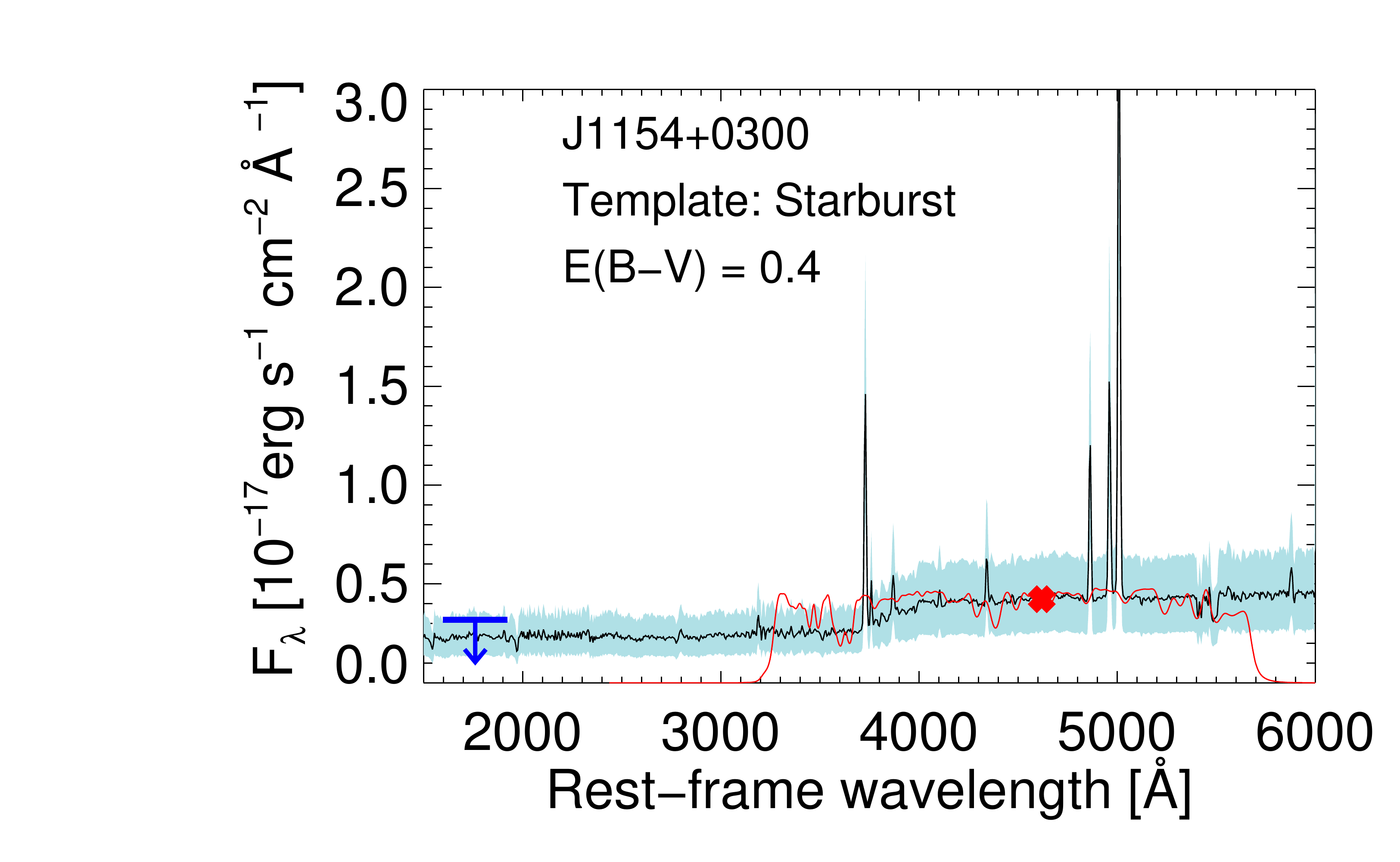}
\caption{UV-optical spectral galaxy templates (black curves), scaled to have observed fluxes, integrated over the NIC2 filter bandpass (thin red curves), corresponding to the $\langle\mhost\rangle$ that we find for J0300+0048 and J1154+0300 (\S \ref{sec:fitting}. The shaded area corresponds to the uncertainty on the template spectrum scaling, as derived from the simulations described in Appendix \ref{sec:appendixB}. The red cross shows the estimated host galaxy flux density at the bandpass pivot wavelength. The blue arrow shows the upper limit rest-frame UV host galaxy magnitude, as determined by the non-detection of extended flux in ACS imaging. The starburst template spectrum is provided by \citet{Calzetti1994}, while the 1 Gyr single stellar population template is based on the stellar synthesis model presented by \citet{Bruzual2003}. We apply the dust reddening law presented by \citet{Calzetti2000}.}
\label{fig:kinneytemplates}
\end{figure*}

\subsection{Comparison with Optical to Mid-Infrared SED Modeling}

\begin{figure*}
\centering
\includegraphics[width=0.495\linewidth]{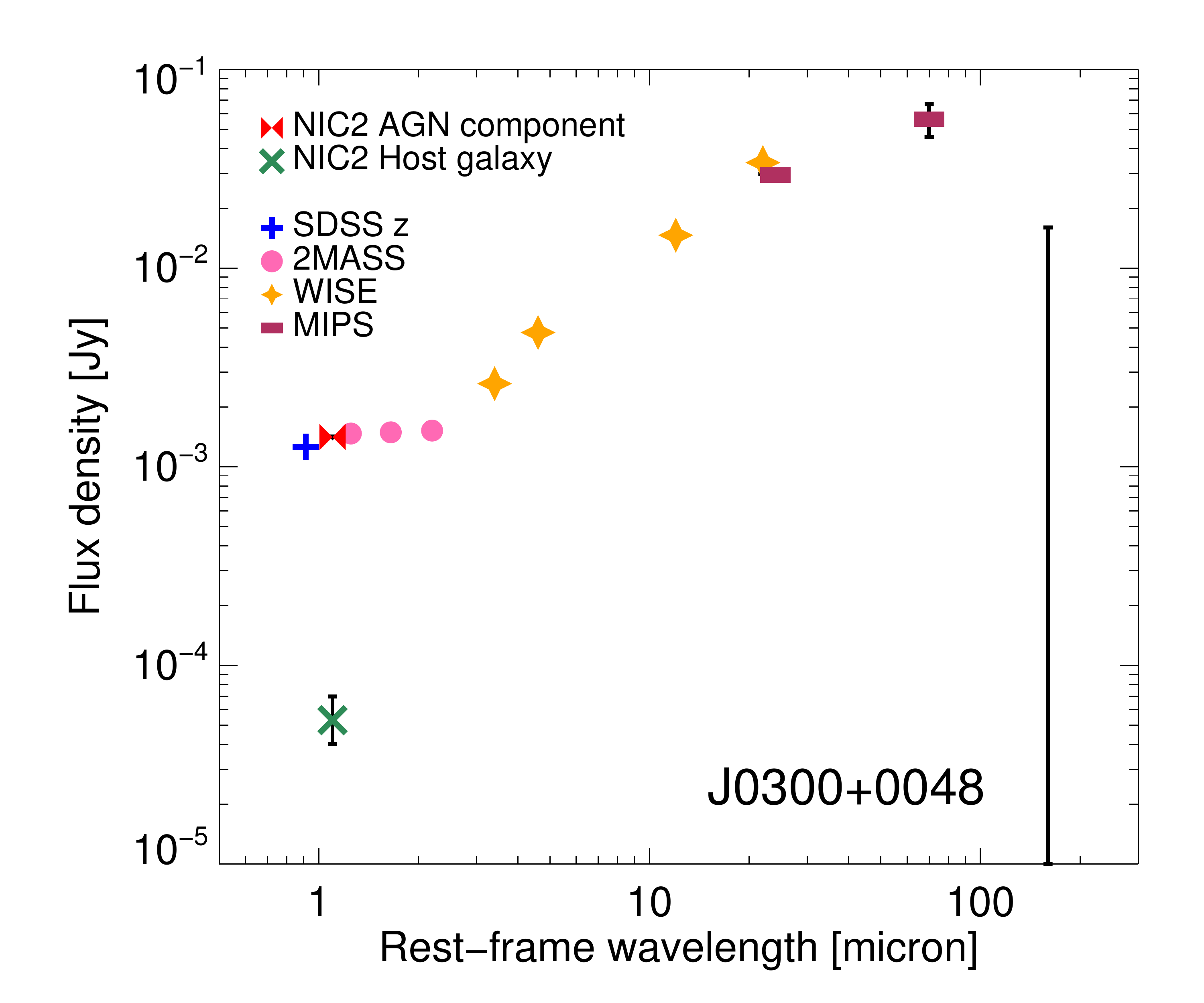}
\includegraphics[width=0.495\linewidth]{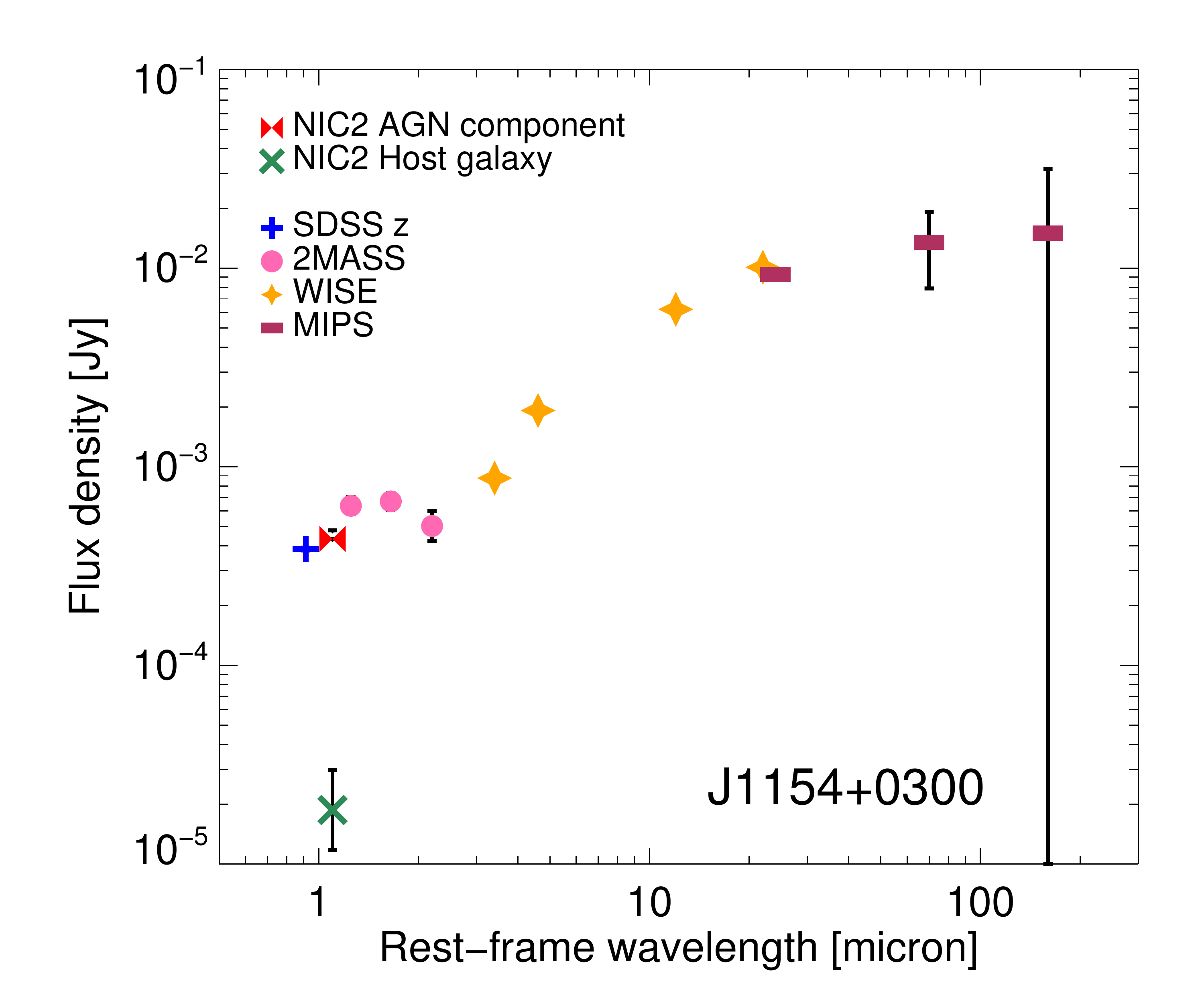}
\caption{Infrared photometry of the FeLoBAL quasars J0300+0048 and J1154+0300, taken from the SDSS, 2MASS, WISE and MIPS surveys, and originally compiled by \citet{Farrah2012}. We include the estimated host galaxy and AGN contribution (green and red diamonds, respectively) based on our image decomposition (\S \ref{sec:fitting}).}
\label{fig:1154_ir}
\end{figure*}

Two of the quasars in our sample are well-studied in the infrared by Farrah and coworkers. For J1154+0300, \citet{Farrah2010} estimate the star formation rate (SFR) in the host galaxy based on the luminosity of polycyclic aromatic hydrocarbon emission lines (PAH), finding SFR=900$\pm1100M_{\astrosun}$ yr$^{-1}$; they only detect one (of three) PAH lines at a significance $>1\sigma$, yielding a SFR consistent with zero. For J0300+0048 and J1154+0300, \citet{Farrah2012} estimate the starburst contribution to the total luminosities via modeling of the broad-band photometric SEDs using pure-AGN and starburst components. For J1154+0300, at 1.1 $\mu$m, they attribute a flux of around 0.07 mJy to the starburst-component (their Figure 4 panel 19), consistent with our estimate of the host galaxy flux (Figure \ref{fig:1154_ir},right panel). We note, however, that their results are consistent at the $3\sigma$ level with no starburst-component emission at 1.1 $\mu$m for this quasar. For J0300+0048, their best-fit starburst component \citep[][Figure 1, panel 3]{Farrah2012} is fainter than our host galaxy model (Figure \ref{fig:1154_ir},left panel). This discrepancy may be due to the lack of a quiescent stellar population component in their modeling. 

While these results suggest a lack of vigorous starburst activity in the FeLo\-BAL hosts, we note that the starburst component in the \citet{Farrah2012} modeling is primarily constrained by the MIPS 160 $\mu$m data-point, for which neither quasar is detected at the 3$\sigma$ level. For J1154+0300, especially, the starburst component scaling is not strongly constrained by the available upper limits.

\subsection{Neighbor Galaxies of the FeLoBALs}\label{sec:discussion_neighbors}

In NICMOS imaging, three of the four FeLo\-BALS in our sample have companion galaxies within an angular separation corresponding to 25 kpc at the quasar redshift; one of these companion galaxies (for J0819+4209) is also detected in ACS imaging. If these galaxies are physical neighbours to the FeLo\-BAL quasars, they may be involved in gravitational interactions with the quasar hosts, supporting the merger-triggering scenario for Fe\-LoBAL quasars. However, we lack redshifts of all three of these galaxies, and therefore cannot currently confirm that they are physically close neighbours. Obtaining spectroscopic redshifts for these galaxies would require prohibitively long exposure times ($\sim10$ hours to achieve S/N$\sim4$ in the rest-frame optical for the brightest neighbour galaxy, assuming a 2m-class telescope). 

The companion galaxy of J0819+4209 appears to have an edge-on disk morphology, and can be modeled satisfactorily with \textsc{Galfit} using a single Sersic component with $n=1$, yielding AB magnitudes of 23.0 mag in NICMOS, and 23.7 mag in the ACS imaging. We are unable to strongly constrain the redshift of this galaxy using our two-band photometry; the data are consistent with a starburst SED at $z\apprle2$ or a spiral (\emph{Sc}) galaxy SED at $z\apprle1.1$. If situated at the quasar redshift, this disk galaxy would have a scale size of 4.9 kpc, and an absolute magnitude of -22.9 (in roughly the rest-frame V band), comparable to the characteristic galaxy magnitude $M^\star_{\mathrm{AB}}=-22.67$ at $z\approx2$ \citep{Marchesini2007}. Thus, this object could plausibly be a massive spiral galaxy at $z\sim2$.

Likewise, we cannot place strong constraints on the redshifts of the companion galaxies of J0300+0048 and J1154+0300. These galaxies are modeled satisfactorily using a single Sersic profile with $n=4$, yielding AB magnitudes of 21.2 mag and 22.9 mag, respectively; these magnitudes are somewhat sensitive to the quasar PSF scaling, as they are partially hidden by the PSF wings (Figure \ref{fig:quasars_2d_nicmos}). The UV non-detections suggest quiescent or dust-obscured galaxies. Again, these galaxies would approximately correspond to the characteristic magnitude $M^\star_{\mathrm{R}}=-22.4$ at $z\sim1$ \citep{Dahlen2005}, and thus represent massive galaxies if they are local to the quasars.

\paragraph*{Probability of Chance Associations:}

As discussed above, we cannot determine the redshifts of the closest neighbour galaxies, and therefore cannot definitively establish whether these galaxies are physically close to the quasars; we note that the measured host galaxy properties are plausible for massive galaxies located at the quasar redshifts. We instead consider the probability that a field galaxy not associated with the quasar would be observed due to chance alignment.

The probability $P$ of a chance association in a small region of angular area $A$ is given by

\begin{equation}
P=1-\exp(-A\times n(m<m_{\mathrm{gal}}))
\end{equation}

\citep{Bloom2002,Perley2012}. Here, $n(m<m_{\mathrm{gal}})$ denotes the number density (per arcsec$^2$) of galaxies brighter than $m_{\mathrm{gal}}$, the apparent magnitude of the candidate associated galaxy. We use the J-band number densities presented by \citet{Cristobal-Hornillos2009}, supplemented with galaxy number counts from deep-field \emph{Subaru} observations \citep{Maihara2001}, to obtain $n(m<m_{\mathrm{gal}})$ in the F110W filter. For F160W, we use the number densities provided by \citet{Yan1998}. 

As the relative position uncertainty of our observations is negligible compared to the angular separation between the quasar and galaxy, we use for $A$ the area of a circle with a radius equal to this separation. This yields an estimate of the probability of a chance alignment of a galaxy at least as bright as observed, located at an angular separation at least as close to the quasar as observed. The probabilities of chance association within a circle defined by the quasar-galaxy separation are 2\% (J0300+0048), 9\% (J1154+0300), and 10\% (J0819+4209). The probability that all host galaxies are chance associations is less than 0.2\%. The probability that none of the host galaxies are chance alignments is given by

\begin{equation}
P(\mathrm{no\,chance\,alignments})=\prod_{k=1}^{3}(1-P_k)
\end{equation}

\citep[e.g.,][]{Bloom2002}, where $P_k$ denotes the probabilities of each individual galaxy being a chance association. Thus, the probability that all three galaxies are physically related to the FeLoBAL quasars is 80\%.

We note that there is no uniquely appropriate choice of $A$, given that we did not define criteria for candidate neighbour galaxies before examining the data. More conservatively, we can calculate the probability of a chance alignment in a set area around the quasar (e.g., a radius corresponding to a 25 kpc radius at the quasar redshift). The probabilities of one or more chance associations within a 25 kpc radius are 7\%, 20\%, and 17\% for J0300+0048, J1154+0300 and J0819+4209, respectively; in that case, the probability of all three galaxies being physically associated to the quasars is 62\%.

\paragraph*{Mass Ratios of Merger Candidates:}

Assuming that the neighbour galaxy is at the quasar redshift, and assuming a constant stellar mass-to-light ratio between the quasar host and the neighbour galaxy, we estimate their mass ratio. We find mass ratios of approximately 4:1 (J0300+0048), 7:1 (J1154+0300), and 5:1 (J0819+4209). All three putative mergers would be classified as minor mergers according to the standard cutoff ratio (3:1). Given the uncertainties on the quasar host galaxy magnitudes, we cannot exclude major mergers for J0300+0048 and J0819+4209.

\paragraph*{The merger fraction for FeLoBALs:}

Based upon the above discussion, we detect physically associated galaxies (i.e., members of the same galaxy group) for at least one, and at most three, FeLoBALs. We cannot determine whether the FeLoBALs are actually undergoing mergers given the available data: faint indicators of merger activity, such as tidal tails, bridges, or shells, cannot be detected in NICMOS imaging at $z\apprge1$ (Appendix \ref{sec:appendixB}). Should all three quasars be involved in gravitational interactions, the implied merger rate of 75\% for our FeLoBAL sample is consistent with the findings of \citet{Glikman2015} for luminous, dust-reddened quasars at $z\approx2$. While the measured luminosity ratios suggest minor mergers, we cannot exclude a major merger fraction of $\sim50\%$, given the uncertainties on \mhost. 

The minor merger fraction observed for inactive galaxies over the relevant redshift range is $\sim5$\%--15\%, while the major merger fraction is $\sim5$\%--20\%  \citep{Man2016}; a combined merger fraction of 75\% for quiescent galaxies is strongly excluded. We note that Man et al. base their merger classification exclusively on physical proximity, i.e., they would classify all three of our quasar-galaxy pairs as mergers if they fulfill their redshift-separation criterion. \citet{Treister2012} compile major merger fractions (based on visual merger classification) for AGN spanning a wide range of redshifts and luminosities, and find merger fractions exceeding $\sim25$\% for quasars with bolometric luminosities $L_{\mathrm{bol}}>10^{45}$ erg s$^{-1}$, with the merger fraction approaching 100\% for the most luminous quasars. Due to the extreme UV absorption, it is difficult to determine the intrinsic bolometric luminosities of our FeLoBAL sample. However, three of the FeLoBALs are detected in WISE W2 and W3 \citep{Wright2010}, allowing us to estimate the flux density at $10^{14}$ Hz ($3\mu$m) in the rest-frame, and calculate a guideline estimate of their luminosities using the bolometric correction BC$_{10^{14}\mathrm{Hz}}=9.12\pm2.62$ presented by \citet{Richards2006}. According to this approximation, these FeLoBALs are indeed intrinsically luminous quasars, with $L_{\mathrm{bol}}$ ranging from $1.5\times10^{47}$ erg s$^{-1}$ (J0300+0048) to $5.0\times10^{47}$ erg s$^{-1}$ (J1730+5850). We note that young stars in the host galaxy would contribute to the WISE photometry (especially in W3), so these luminosities will overestimate the brightness of the central source if the FeLoBALs indeed harbor obscured starbursts.

%However, they are likely intrinsically bright quasars, given their V-band luminosities ($M_V\le-28$ mag).

In conclusion, given the non-negligible probability of chance alignments and the small sample size, we cannot exclude that the FeLoBAL quasars have merger fractions comparable to quiescent galaxies at $z\apprge1$. On the other hand, the merger fraction may be as high as 75\%. Given that the FeLoBAL quasars are likely intrinsically highly luminous, this high merger rate would be consistent with the merger fraction observed for non-BAL quasars of similar luminosity.

\paragraph*{Environments of the FeLoBAL Quasars:} Via visual inspection of the ACS imaging, we identify between 9 and 25 galaxies per quasar at an angular separation corresponding to 200 kpc or less at the quasar redshift (Table \ref{tab:neighbor}). We exclude one face-on spiral galaxy that we classify as a foreground source due to its large angular size. The faintest of these candidate companion galaxies has an AB apparent magnitude of $m_V\sim23.8$ as modeled by \textsc{Galfit}. Using the \emph{HST} Exposure Time Calculator\footnote{\url{http://etc.stsci.edu/}}, we estimate that we are sensitive to extended sources as faint as $\mu_V\approx23.5$ mag arcsec$^{-2}$ at the 3$\sigma$ level in our ACS imaging. Assuming that all companion galaxies are located at the quasar redshift yields an average upper-limit galaxy number density of $1.1\times10^{-4}$ kpc$^{-2}$, within a radius of 200 kpc, for the four Fe\-LoBAL environments. This density is consistent with that found for (predominantly non-BAL) quasar environments at $z\sim1.5$ within a redshift interval of $\Delta z=0.15$ \citep{Karouzos2014}. For our NICMOS imaging, we find an average upper limit galaxy number density of $1.9\times10^{-4}$ kpc$^{-2}$ within an a 80 kpc radius. This result is consistent at the $2\sigma$ level with the number density found for ACS imaging, given the sampling uncertainty on the galaxy counts in the four NICMOS fields. In summary, we do not find evidence for the Fe\-LoBAL quasars residing in over-dense environments compared to other quasars; relative to quiescent galaxies, quasars have been reported to reside in over-dense regions at $\sim100$ kpc scales \citep[e.g.,][]{Serber2006}, although this finding may be an artifact due to control sample selection \citep{Karhunen2014}.

%For J0300+0048, we note the presence of a large-separation binary quasar \citep[$z=0.89$,][]{Hall2003}; as the other Fe\-LoBALs are not in binary systems, a causal link between the Fe\-LoBAL activity and the presence of another quasar is not likely.

\nocite{Kuhlbrodt2005} \nocite{Schramm2008} \nocite{Kukula2001} \nocite{Peng2006} \nocite{Falomo2004} \nocite{Ridgway2001} \nocite{Mechtley2016}

\section{Conclusion}\label{sec:conclusion}

We present host galaxy detections in the rest-frame optical for three FeLoBAL quasars, of a sample of four objects. The host galaxy luminosities of the FeLoBAL quasars are consistent with those found for non-BAL quasars with similar nuclear luminosities. None of the host galaxies are detected in the rest-frame UV. These results are consistent with the FeLoBAL hosts being either quiescent elliptical galaxies or dust-obscured starbursts. Three of the quasars have a companion galaxy at an angular separation corresponding to less than 25 kpc at the quasar redshift. Given the probability of chance alignments, at least one of these neighbour galaxies is very likely to be physically associated with the FeLoBAL; there is an 80\% chance that all three companion galaxies are physically associated. These companion galaxies may represent early-stage mergers, with intermediate stellar mass ratios of $\sim$5:1. While the FeLoBAL hosts do not appear to be late-stage major mergers, our NICMOS observations, while relatively deep, do not have the sensitivity required to exclude fainter indicators of merger activity such as morphological distortions and tidal tails.

\paragraph*{Acknowledgements: } We thank the anonymous referee for helpful comments and suggestions, which improved the presentation of the manuscript.

Based on observations made with the NASA/ESA Hubble Space Telescope, obtained at the Space Telescope Science Institute, which is operated by the Association of Universities for Research in Astronomy, Inc., under NASA contract NAS 5-26555. These observations are associated with program \# 10237. Some of the data presented in this paper were obtained from the Mikulski Archive for Space Telescopes (MAST). STScI is operated by the Association of Universities for Research in Astronomy, Inc., under NASA contract NAS5-26555.

DL and MV gratefully acknowledge support from the Danish Council for Independent Research via grant no. DFF 4002-00275. DL gratefully acknowledges additional support from the Instrument Center for Danish Astrophysics.

Funding for SDSS-III has been provided by the Alfred P. Sloan Foundation, the Participating Institutions, the National Science Foundation, and the U.S. Department of Energy Office of Science. The SDSS-III web site is \url{http://www.sdss3.org/}. SDSS-III is managed by the Astrophysical Research Consortium for the Participating Institutions of the SDSS-III Collaboration including the University of Arizona, the Brazilian Participation Group, Brookhaven National Laboratory, Carnegie Mellon University, University of Florida, the French Participation Group, the German Participation Group, Harvard University, the Instituto de Astrofisica de Canarias, the Michigan State/Notre Dame/JINA Participation Group, Johns Hopkins University, Lawrence Berkeley National Laboratory, Max Planck Institute for Astrophysics, Max Planck Institute for Extraterrestrial Physics, New Mexico State University, New York University, Ohio State University, Pennsylvania State University, University of Portsmouth, Princeton University, the Spanish Participation Group, University of Tokyo, University of Utah, Vanderbilt University, University of Virginia, University of Washington, and Yale University.

\bibliographystyle{mnras}
\bibliography{lawther_felobal_hosts}
\clearpage

\begin{deluxetable}{cccccccc}
	\tabletypesize{\small}
	\tablecaption{Observation log\label{tab:datasummary}}
	\tablewidth{0pt}
	\tablehead{
		\colhead{Object} & \colhead{Observation} & \colhead{Redshift} & \colhead{Filter} & \colhead{Number of} & \colhead{Number of} & \colhead{Total} & \colhead{10$\sigma$ detection} \\
		\colhead{} & \colhead{date} & \colhead{} & \colhead{} & \colhead{orbits} & \colhead{exposures\tablenotemark{a}} & \colhead{exp. time [s]} & \colhead{limit [mag]\tablenotemark{b}}\\
		\colhead{(1)} & \colhead{(2)} & \colhead{(3)} & \colhead{(4)} & \colhead{(5)} & \colhead{(6)} & \colhead{(7)} & \colhead{(8)} \\
	}
	\startdata
	\multicolumn{8}{c}{\textbf{ACS/WFC}} \\
	J0300+0048 & 2004-12-10 & 0.89 & F435W & 1 & 4 & 2185 & 27.0 \\
	J1154+0300 & 2005-05-16 & 1.46 & F550M & 2 & 4 & 5141 & 26.9 \\
	J0819+4209 & 2004-11-21 & 1.93 & F625W & 3 & 6 & 8150 & 27.9 \\
	J1730+5850 & 2006-12-09 & 2.04 & F625W & 4 & 5 & 6517 & 27.7 \\
	\\
	\hline
	\\
	\multicolumn{8}{c}{\textbf{NICMOS/NIC2}} \\
	J0300+0048 & 2004-12-30 & 0.89 & F110W & 1 & 5 & 2688 & 25.2 \\
	J1154+0300 & 2007-02-11 & 1.46 & F110W & 2 & 10 & 5887 & 25.9 \\
	J0819+4209 & 2005-02-19 & 1.93 & F160W & 3 & 12 & 8695 & 25.6 \\
	J1730+5850 & 2006-12-09 & 2.04 & F160W & 4 & 4 & 8191 & 25.9 \\
	\enddata
	\tablenotetext{a}{The total number of exposures across all dither points. All observations were performed using a four-point dither pattern, albeit distributed over several orbits. Due to an instrumental failure, the ACS observations of J1730+5850 lack the final exposures granted, and the dither pattern is unevenly sampled.}
	\tablenotetext{b}{The 10$\sigma$ detection limit for a point source in the combined image for this observation, calculated based on the standard deviation of the background level as measured in the combined images.}
\end{deluxetable}

\begin{center}
\begin{deluxetable}{ccccc}
	\tabletypesize{\small}
	\tablecaption{\md parameter settings\label{tab:multidrizzle}}
	\tablewidth{0pt}
	\tablehead{
		\colhead{Object} & \colhead{Kernel\tablenotemark{a}} & \colhead{Pixfrac\tablenotemark{b}} & \colhead{Scale\tablenotemark{c}} & \colhead{PSF FWHM\tablenotemark{d}}\\
		\colhead{} & \colhead{} & \colhead{} & \colhead{[pix ('')]} & \colhead{[pix ('')]}\\
		\colhead{(1)} & \colhead{(2)} & \colhead{(3)} & \colhead{(4)} & \colhead{(5)}\\
	}
	\startdata
	\multicolumn{5}{c}{\textbf{NICMOS/NIC2}}\\
	J0300+0048 & Square & 0.65 & 0.8 (0.0604'') & 1.89 (0.107'')\\
	J1154+0300 & Gaussian & 0.65 & 0.6 (0.0453'') & 2.16 (0.098'') \\
	J0819+4209 & Square & 0.6 & 0.9 (0.06795'') & 2.06 (0.140'') \\
	J1730+5850 & Gaussian & 0.7 & 0.8 (0.0604'') & 2.57 (0.155'') \\
	\multicolumn{5}{c}{\textbf{ACS/WFC}}\\
	J0300+0048 & Gaussian & 0.7 & 0.8 (0.0392'') & 2.04 (0.0800'') \\
	J1154+0300 & Square & 0.7 & 0.8 (0.0392'') & 2.06 (0.081'') \\ 
	J0819+4209 & Gaussian & 0.7 & 0.8 (0.0392'') & 2.21 (0.0866'') \\  
	J1730+5850 & Gaussian & 0.85 & 0.85 (0.04165'') & 2.40 (0.100'') \\ 
	\enddata
	\tablecomments{\md parameters adopted for the final combined image for each quasar. The PSF parameters are measured via a simple Gaussian fit.}
	\tablenotetext{a}{The convolution kernel used to distribute the flux of each input pixel in the output grid.} 
	\tablenotetext{b}{The factor by which each input pixel is resized before being transferred to the common WCS.}
	\tablenotetext{c}{The size of the output pixels, listed as a fraction of the input pixel size and in arcseconds (in parentheses).}
	\tablenotetext{d}{The FWHM of the FeLoBAL quasar PSF, expressed in pixels and (in parentheses) in arcseconds.}
\end{deluxetable}
\end{center}
\clearpage
\begin{deluxetable}{cccccc}
	\tabletypesize{\small}
	\tablecaption{PSF-only GALFIT Modeling\label{tab:psffits_acs}}
	\tablewidth{0pt}
	\tablehead{
		\colhead{Object} & \colhead{Filter} & \colhead{Quasar FWHM\tablenotemark{a}} & \colhead{PSF FWHM\tablenotemark{b}} & \colhead{$m_{\mathrm{PSF}}$\tablenotemark{c}}\\
		\colhead{} & \colhead{} & \colhead{[arcsec (pix)]} & \colhead{[arcsec (pix)]} & \colhead{[mag]} \\
		\colhead{(1)} & \colhead{(2)} & \colhead{(3)} & \colhead{(4)} & \colhead{(5)} \\
	}
	\startdata
	\textbf{ACS:}\\
	J0300+0048 & F435W & 0.08'' (2.03) & 0.09'' (2.29) & 20.51 \\
	J1154+0300 & F550M & 0.08'' (2.07) & 0.09'' (2.39) & 21.40 \\
	J0819+4209 & F625W & 0.09'' (2.21) & 0.08'' (2.07) & 22.62 \\
	J1730+5850 & F625W & 0.09'' (2.41) & 0.09'' (2.28) & 21.70 \\
	\textbf{NICMOS:} \\ 
	J0300+0048 & F110W & 0.11'' (1.81) & 0.13'' (2.10) & 16.05 \\
	J1154+0300 & F110W & 0.12'' (2.55) & 0.12'' (2.62) & 17.29 \\
	J0819+4209 & F160W & 0.15'' (2.15) & 0.14'' (2.09) & 18.96 \\
	J1730+5850 & F160W & 0.16'' (2.67) & 0.15'' (2.44) & 16.98 \\
	\enddata
	\tablenotetext{a}{Full Width at Half Maximum of the quasar emission, as measured using the \textsc{IRAF} task \emph{`imexamine'}.}
	\tablenotetext{b}{Full Width at Half Maximum of the stacked stellar PSF template used to model the point source emission.}
	\tablenotetext{c}{The apparent AB magnitude of the quasar in the listed bandpass, as determined by our PSF-only modeling.}
\end{deluxetable}

\begin{deluxetable}{cccc}
	\tabletypesize{\small}
	\tablecaption{Upper limits on FeLoBAL host galaxy flux in ACS observations\label{tab:psffits_acs_limits}}
	\tablewidth{0pt}
	\tablehead{
		\colhead{Object} & \colhead{Filter} & \colhead{\mnuc\tablenotemark{a}} & \colhead{Limiting \mhost\tablenotemark{b}} \\
		\colhead{} & \colhead{} & \colhead{[mag]} & \colhead{[mag]} \\
		\colhead{(1)} & \colhead{(2)} & \colhead{(3)} & \colhead{(4)} \\
	}
	\startdata
	J0300+0048 & F435W & 20.51 & 22.76 \\
	J1154+0300 & F550M & 21.40 & 23.15 \\
	J0819+4209 & F625W & 22.62 & 24.37 \\
	J1730+5850 & F625W & 21.70 & 23.70 \\
	\\
	\enddata
	\tablenotetext{a}{The nuclear (PSF) AB magnitude of the source, as determined by PSF-only modeling.}
	\tablenotetext{b}{The limiting AB magnitude of the host galaxy, calculated using the limiting nuclear-to-host contrast level determined in our simulations (Appendix \ref{sec:appendixB}). While we are ignorant of the scale size of non-detected host galaxies, the detection limits are in fact strongly sensitive to this parameter; the upper limits listed here assume $0.5$ kpc $\le R_e\le$ 10 kpc, for which range we quote the most conservative (i.e., brightest) upper limit.}
\end{deluxetable}

\clearpage
\begin{deluxetable}{ccccccc}
	\tabletypesize{\small}
	\tablecaption{Best-fit PSF-plus-Sersic GALFIT models of FeLoBAL quasars, NICMOS\label{tab:sersicfit_nicmos_bestfit}}
	\tablewidth{0pt}
	\tablehead{
		\colhead{Object} & \colhead{Filter} & \colhead{$m_\mathrm{nuc}$} & \colhead{$m_\mathrm{host}$} & \colhead{\contrast}& \colhead{$R_e$} & \colhead{$n$}\\
		\colhead{} & \colhead{} & \colhead{[mag]} & \colhead{[mag]} & [mag] & \colhead{[kpc]} & \colhead{}\\
		\colhead{(1)} & \colhead{(2)} & \colhead{(3)} & \colhead{(4)} & \colhead{(5)} & \colhead{(6)} & \colhead{(7)}\\
		%\\
	}
	\startdata
	J0300+0048 & F110W & $16.06\pm0.01$ & $19.64\pm0.32$ & $3.58$ & $10.4\pm6.0$ & $3.05\pm1.79$\\
	J1154+0300 & F110W & $17.34\pm0.00$ & $21.09\pm0.03$ & $3.75$ & $2.9\pm0.1$ & $[0.75]$\\
	J0819+4209 & F160W & $19.01\pm0.00$ & $21.48\pm0.03$ & $2.47$ & $[1.2]$ & $[0.75]$ \\
	J1730+5850 & F160W & $17.12\pm0.00$ & $20.26\pm0.15$ & $3.14$ & $[1.6]$ & $[0.75] $\\
	\\
	\enddata
	\tablecomments{Best-fit parameters for PSF-plus-Sersic modeling of the FeLoBALs in NICMOS imaging. Square brackets denote parameters that converge at the imposed limiting values ($0.75\le n\le5$, $R_e>\mathrm{FWHM}_{\mathrm{PSF}}$). The uncertainties listed are statistical errors as calculated by the \textsc{Galfit} software. As the $\chi^2$ minimization scheme does not explicitly account for PSF mismatch, or for mismatch between the surface flux distributions of the Sersic profile and the real galaxy, the listed uncertainties are likely to underestimate the true parameter uncertainties. Our simulations indicate that the average values of \mhost\,as determined using a range of Sersic indies (Table \ref{tab:sersicfit_nicmos}), are more accurate than the best-fit values presented here.} 
\end{deluxetable}

% old version! 'summary' with errors instead of explicit ranges!
%\begin{deluxetable}{cccccc}
%	\tabletypesize{\small}
%	\tablecaption{Summary of PSF-plus-Sersic modeling of FeLoBAL quasars, NICMOS\label{tab:sersicfit_nicmos}}
%	\tablewidth{0pt}
%	\tablehead{
%		\colhead{Object} & \colhead{Filter} & \colhead{$\mean{m_\mathrm{nuc}}$} & \colhead{$\mean{m_\mathrm{host}}$} & \colhead{$\mean{m_\mathrm{host}}-\mean{m_\mathrm{nuc}}$}& \colhead{$\mean{R_e}$}\\
%		\colhead{} & \colhead{} & \colhead{[mag]} & \colhead{[mag]} & [mag] & \colhead{[kpc]}\\
%		%\\
%	}
%	\startdata
%	J0300+0048 & F110W & 16.05 & $20.26^{+0.71}_{-0.65}$ & 4.21 & 6.99$^{+7.69}_{-4.71}$ \\
%	J1154+0300 & F110W & 17.32 & 20.11$^{+0.22}_{-0.18}$ & 2.79 & 1.29$^{+0.14}_{-0.13}$ \\
%	J0819+4209 & F160W & 19.02 & 21.19$^{+0.22}_{-0.20}$ & 2.17 & $<$1.25 \\
%	J1730+5850 & F160W & 16.92 & 22.75$^{+1.02}_{-1.78}$ & 5.83 & 9.36$^{+17.32}_{-6.19}$ \\
%	\\
%	\enddata
%	\tablecomments{Our data do not strongly constrain the Sersic index, $n$, or the Sersic component axis ratio, $b/a$. The values of \mnuc, \mhost, $n$ and $R_e$ presented here are average values calculated for a grid of model fits to each quasar, varying $n$ and $b/a$; our simulations suggest that this averaging minimizes the uncertainty due to our ignorance of these parameters. The tabulated `errors' are not statistical uncertainties - they represent minimum and maximum values obtained for $R_e$ and for \mhost\,across all models for that object.} 
%\end{deluxetable}

% new version! explicit ranges!
\begin{deluxetable}{lccccccc}
	\tabletypesize{\small}
	\tablecaption{Parameter ranges for PSF-plus-Sersic GALFIT modeling, NICMOS\label{tab:sersicfit_nicmos}}
	\tablewidth{0pt}
	\tablehead{
		\colhead{Object} & \colhead{$\mean{m_\mathrm{host}}$} &\colhead{Max. $m_\mathrm{host}$} & \colhead{Min. $m_\mathrm{host}$} & \colhead{$\mean{m_\mathrm{nuc}}-\mean{m_\mathrm{host}}$} &  \colhead{$\mean{R_e}$} & \colhead{Max. $R_e$} & \colhead{Min. $R_e$}\\
		\colhead{} & \colhead{[mag]} & \colhead{[mag]} & \colhead{[mag]} & \colhead{[mag]} & \colhead{[kpc]} & \colhead{[kpc]} & \colhead{[kpc]}\\
		\colhead{(1)} & \colhead{(2)} & \colhead{(3)} & \colhead{(4)} & \colhead{(5)} & \colhead{(6)} & \colhead{(7)} & \colhead{(8)}\\
		%\\
	}
	\startdata
	J0300+0048 & 19.62 & 19.76 & 19.44 & 3.54 & 2.0 & 3.5 & 1.4 \\
	J1154+0300 & 20.75 & 21.38 & 20.09 & 3.45 & 5.2 & 9.3 & 3.1 \\
	J0819+4209 & 21.22 & 21.39 & 21.06 & 2.17 & $<$1.3 & -- & -- \\
	J1730+5850 & 20.12 & 20.26 & 19.98 & 2.99 & $<$1.6 & -- & -- \\
	\\
	\enddata
	\tablecomments{Our data do not strongly constrain the Sersic index, $n$. The average values of \mhost, and $R_e$ presented here are calculated based on three model fits per quasar, with $n=1$, $n=2.5$, and $n=4$, respectively; our simulations (Appendix \ref{sec:appendixB}) suggest that this averaging minimizes the uncertainty due to our ignorance of $n$. We also show the minimum and maximum values found for \mhost\,and $R_e$ given this range of $n$.} 
\end{deluxetable}
\clearpage
\begin{deluxetable}{lcccc}
	\tabletypesize{\small}
	\tablecaption{Potential neighbor galaxy counts\label{tab:neighbor}}
	\tablewidth{0pt}
	\tablehead{
		\colhead{Object} & \colhead{$N(R<25$ kpc)} & \colhead{$N(R<40$ kpc)} & \colhead{$N(R<80$ kpc)} & \colhead{$N(R<200$ kpc)} \\
		\colhead{(1)} & \colhead{(2)} & \colhead{(3)} & \colhead{(4)} & \colhead{(5)} \\
	}
	\startdata
	\multicolumn{5}{c}{\textbf{NICMOS/NIC2}}\\
	J0300+0048 & 1 & 2 & 4(+1F) & --- \\
	J1154+0300 & 1 & 1 & 5 & --- \\
	J0819+4209 & 1 & 1 & 3 & --- \\
	J1730+5850 & 0 & 1 & 3 & --- \\
	\multicolumn{5}{c}{\textbf{ACS/WFC}}\\
	J0300+0048 & 0 & 0 & (1F) & 8(+1F,+1A) \\
	J1154+0300 & 0 & 0 & 0 & 9 \\
	J0819+4209 & 1 & 1 & 5 & 25 \\
	J1730+5850 & 0 & 1 & 3 & 15 \\
	\enddata
	\tablecomments{Number $N$ of galaxies located at an angular separation corresponding to a transverse radius $R$ at the quasar redshift. We lack redshifts for these potential companion galaxies. Massive, unobscured starbursting galaxies situated at the quasar redshift are expected to be detectable in both bandpasses, whereas quiescent neighbor galaxies are not expected to be detected in the ACS/WFC images. For J0300+0048 we also see one foreground galaxy (denoted \emph{1F}), and one additional AGN (denoted \emph{1A}). The AGN is physically close to the FeloBAL, at $z=0.89$ \citep{Hall2003_0300caii}.}
\end{deluxetable}
\clearpage
\appendix

\section{Assembly and Testing of PSF Templates}\label{sec:appendix_psf}

Here, we describe the assembly of our empirical PSF templates, and the tests performed to determine the sensitivity of our study to false-positive host galaxy detections.

\subsection{Construction of PSF Templates}\label{sec:stackedpsf}

Dedicated PSF star observations were not obtained for this observing program. In agreement with \citet{Mechtley2012}, we find that the currently available analytical PSF models for NICMOS and ACS/WFC \citep[TinyTim,][]{Krist1995} lack the accuracy required for the study of high-redshift quasar host galaxies: the \ttim PSF, assembled from multiple \ttim\,models using the same dither pattern and processed using the same \md settings as our data, is generally too narrow. Instead, we use available \emph{HST} data to construct empirical PSF templates. The \emph{HST} PSF displays both temporal and position-dependent PSF variations \citep{Kim2008}. For the \emph{HST} WFC3 detector, \citet{Kim2008} find that PSF mismatch is minimized by observing the PSF star within around a month of the science observation, and within 100 pixels of the detector position of the object to be modeled. It is also advantageous to use a PSF template with high signal-to-noise (S/N) in the PSF wings, to minimize the additional noise introduced into the modeling. The ideal empirical PSF template is therefore a bright but unsaturated star, observed quasi-contemporaneously to the science observation, at the same detector position.

\paragraph*{PSF stars for ACS/WFC observations:}

Each ACS/WFC combined quasar image contains 7 to 10 stars suitable for use as PSF templates. Use of these stars removes the risk of PSF mismatch due to time evolution of the PSF. However, each of these stars suffer from some degree of mismatch with respect to the quasar PSF, due to position-dependent aberrations. To maximize the S/N in the PSF template, we construct stacked stellar PSF templates for the ACS imaging using these field stars; the stacking should also mitigate the field dependence of the PSF shape to some extent. We shift each sky-subtracted star to a common centroid using a linear sub-pixel interpolation kernel, and co-add the stars using uniform weights. 

\paragraph*{PSF stars for NICMOS observations:}

Due to the small field of view of the NIC2 detector, bright stars are seldom observed serendipitously; our data contain no suitable stars. We therefore turn to calibration star observations. For each quasar observation we construct a stacked PSF using all \emph{HST} MAST\footnote{The Mikulski Archive for Space Telescopes, online resource: https://archive.stsci.edu/hst/} archival observations of the G2-V star P330-E within a three-month window\footnote{Adhering to a one-month window, as recommended by \citet{Kim2008}, yields too few stars to provide a high-S/N stacked PSF.}, during which 2-3 datasets exist with the same instrumental setup as the quasar observations. While NICMOS did observe a few calibration stars other than P330-E within this time window, the observations were too short to adequately sample the PSF wings, and their inclusion did not improve the overall signal-to-noise of the PSF template. We use the same stacking method as used for the ACS/WFC. The S/N of each stacked PSF template exceeds that of the quasar, with the exception of J0300+0048(NIC2/F110W), for which the PSF has a significantly lower S/N (\S \ref{sec:results_nic}).

\subsection{Testing method for PSF templates}\label{sec:psftests}

To establish detection limits for our analysis, we perform PSF-only and PSF-plus-Sersic modeling of archival stellar observations, hereafter referred to as \emph{test stars}. The PSF-only modeling quantifies the amount and distribution of residual flux we can expect due to PSF mismatch, while the PSF-plus-Sersic modeling quantifies the maximum flux of a spurious 'host galaxy' component fitted to a true point source, for each instrument and filter combination. For the ACS/WFC data, we use the bright, uncrowded stars used to construct the stacked PSF template as test stars; we remove the test star from the stacked PSF prior to performing the fitting. For the NICMOS data, we use archival images of stars other than P330-E as test stars. We first turn to a sample of NICMOS test stars observed at a separation of no more than three months from the quasar observation; three are available for filter F110W, four for F160W. Unfortunately, these are all short-exposure images, with low S/N in the PSF wings. The quasar observations themselves are deep images with high S/N. To test the modeling of high-S/N point sources, we turn to an additional sample of NICMOS test stars, observed outside the 3-month window (hereafter, the \emph{high-S/N sample}). 

 For the purposes of these tests, we define a \emph{significant deviation} from the radial profiles of the test stars as a deviation with a significance of greater than $3\sigma$ over a contiguous interval of three pixels or greater. 

\subsection{Accuracy of PSF Templates, ACS/WFC}\label{sec:appendix_acs}

\begin{figure*}
		\includegraphics[trim={0 1cm 0.5cm 0.5cm},clip,scale=0.3]{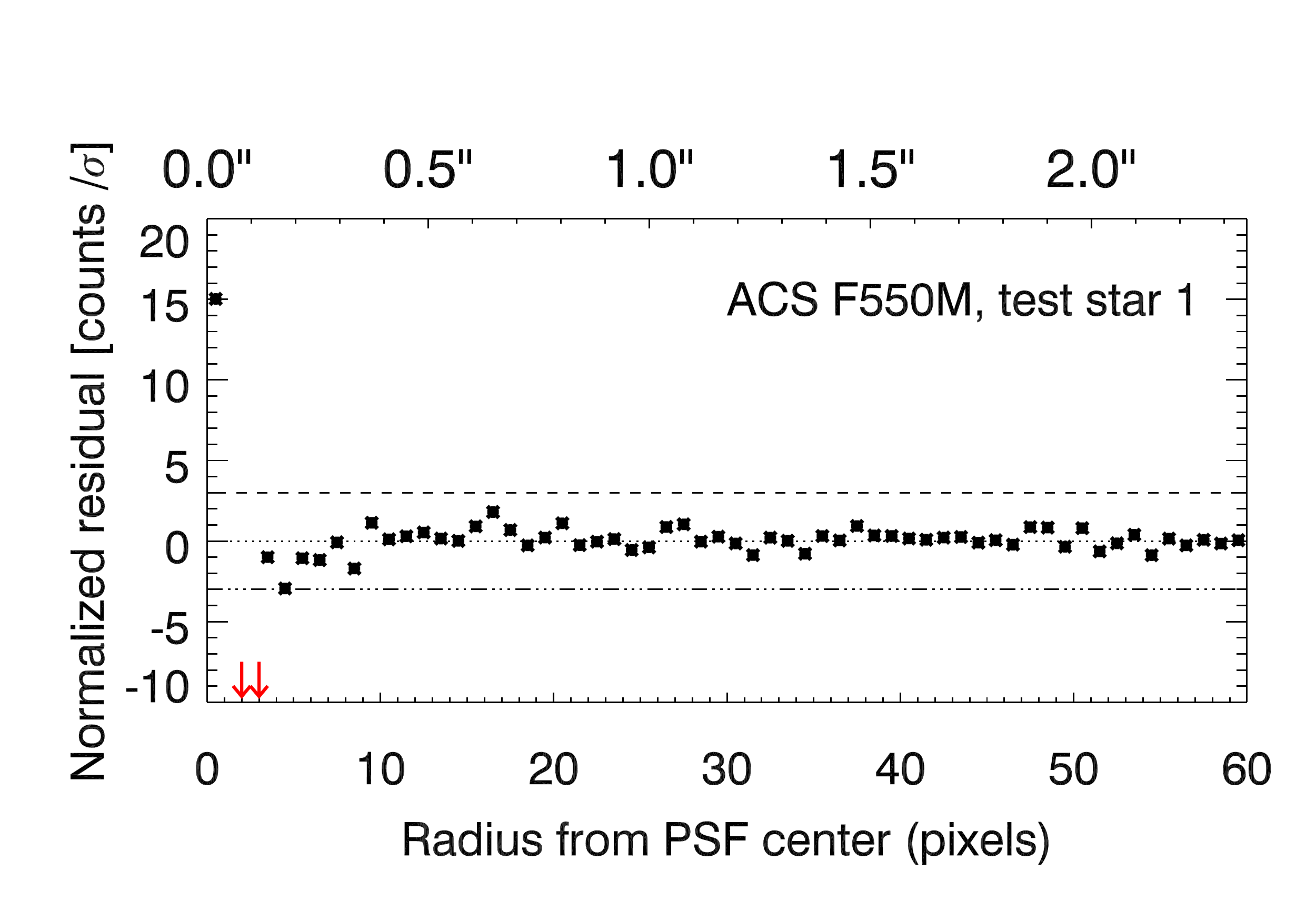}
		\includegraphics[trim={0 1cm 0.5cm 0.5cm},clip,scale=0.3]{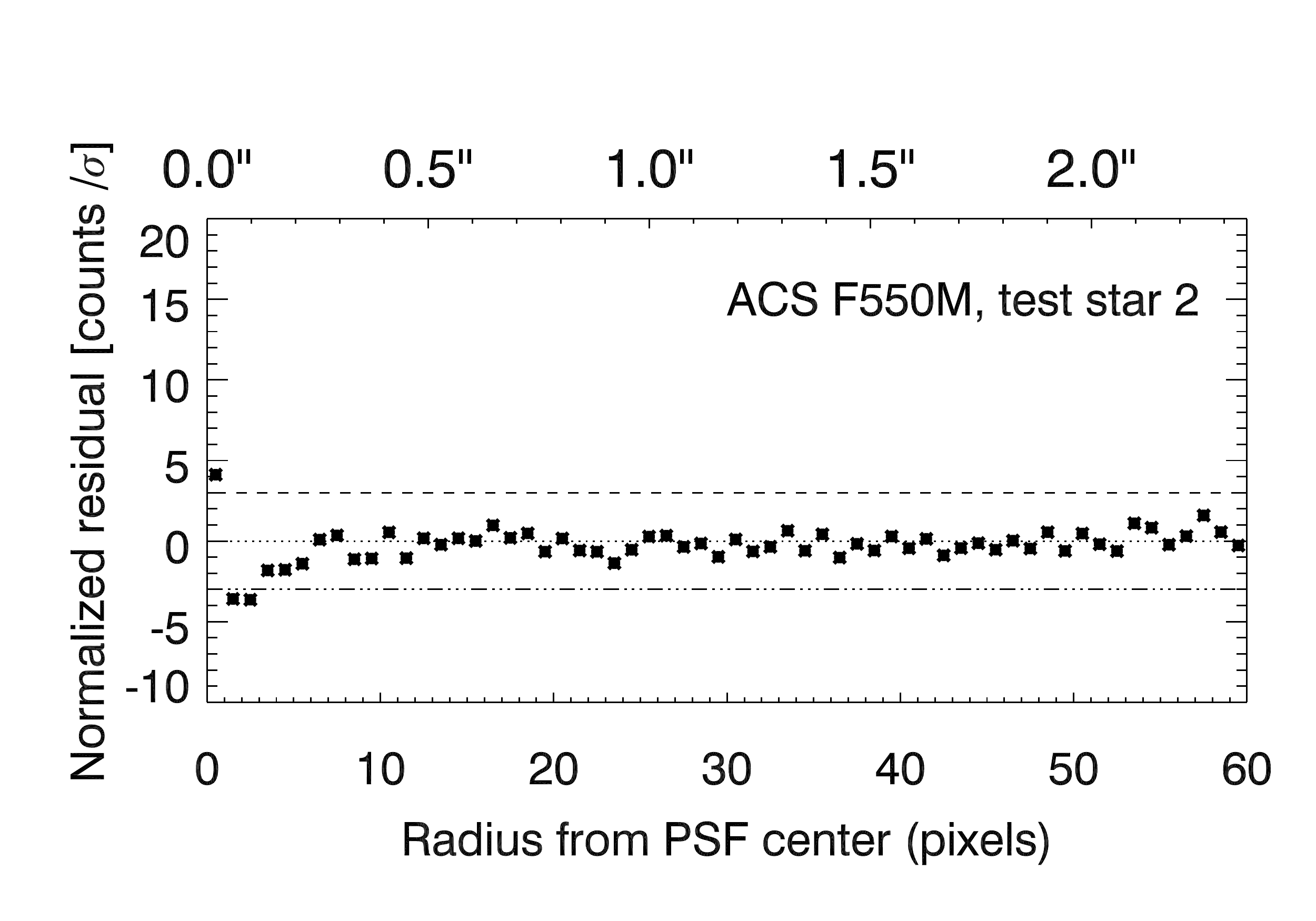}
		\includegraphics[trim={0 1cm 0.5cm 0.5cm},clip,scale=0.3]{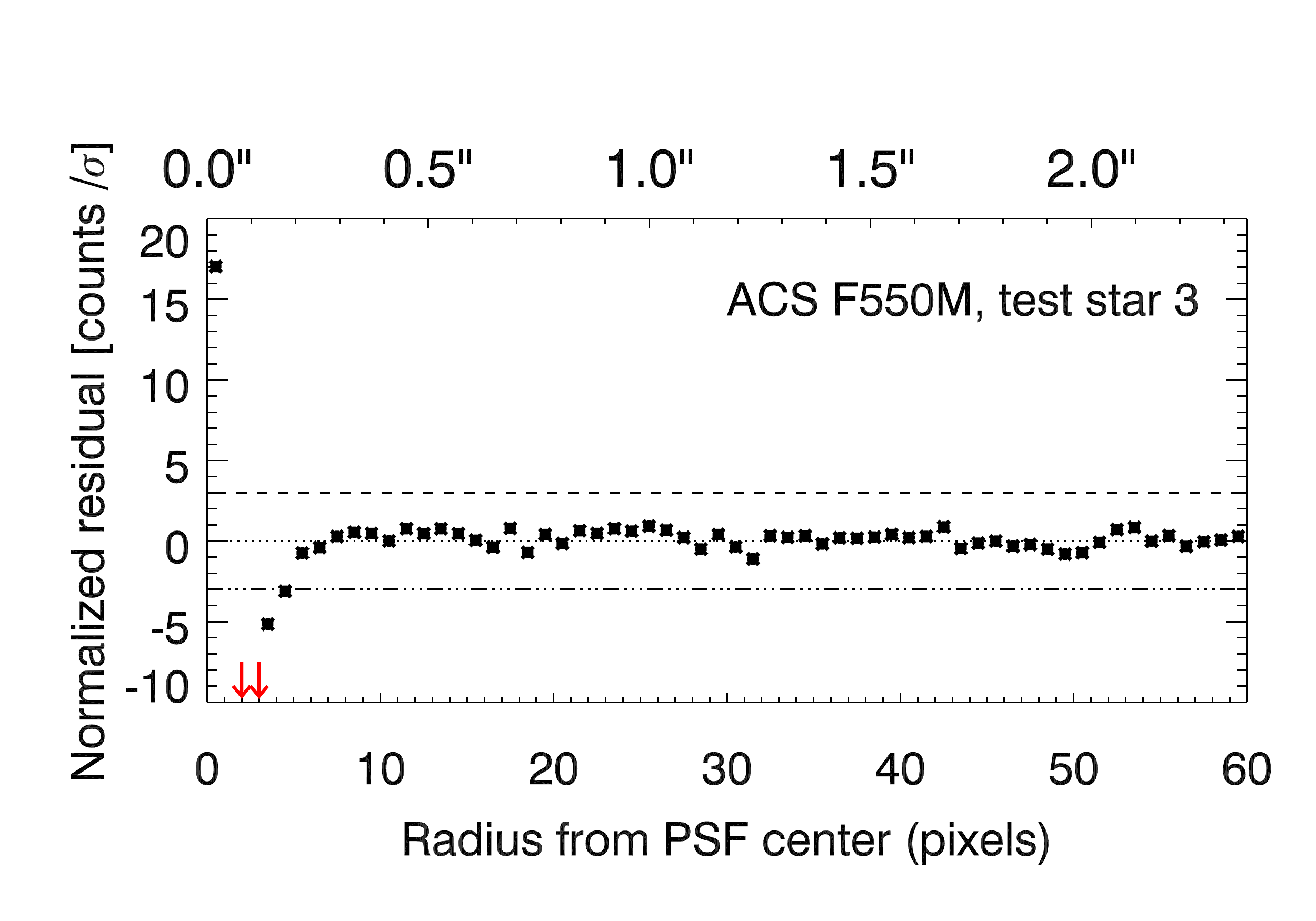}
		\includegraphics[trim={0 1cm 0.5cm 0.5cm},clip,scale=0.3]{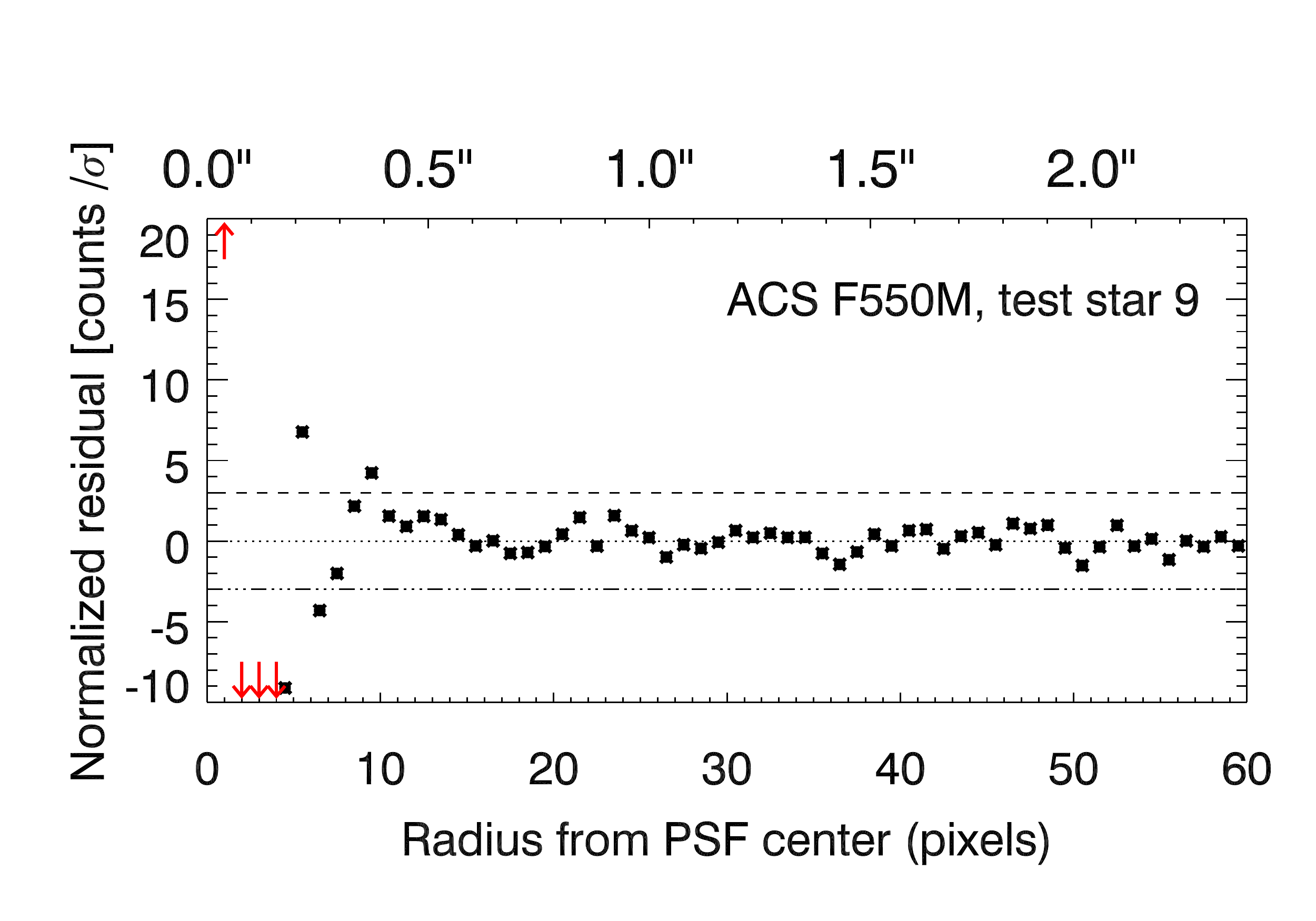} 

	\caption{Radial intensity plots for PSF template fits to test stars 1--3 and 9 in the ACS J1154+0300 field. In total we have 10 test stars for J1154+0300. Of these 10 stars, test star 9 shows the most significant deviation from the scaled PSF template. Black points: azimuthally averaged offset of the best-fit stacked stellar PSF from the test star PSF, expressed in units of the $1\sigma$ error on the data, as a function of radius. Dotted, dashed and dash-dot lines show the zero-point and positive, negative $3\sigma$ levels, respectively.}\label{fig:teststarfits_acs_1154}	
\end{figure*}

\paragraph*{PSF-only modeling:} Most of our PSF-only fits display significant deviations from the template radial profile in the PSF core. We therefore define an `inner radius' for each filter, inside of which we expect deviations from the PSF template even for a bona-fide point source. We find the following inner radii: 0.2'', 0.3'' and 0.4'' for filters ACS/F425W, ACS/F550M, and ACS/F625W, respectively. We use these inner radii to establish the detection criteria for PSF-only modeling (Appendix \ref{sec:appendix_detection}. We present radial profiles of four representative test stars in Figure \ref{fig:teststarfits_acs_1154}.

\paragraph*{PSF-plus-Sersic modeling: } For the PSF-plus-Sersic fits, the best-fit extended components are negligibly faint in most cases, as expected if the PSF template is accurate. However, for one of our test stars (of a sample of 34) our method finds a spurious extended component only 3.4 mag fainter than the PSF component, i.e., consistent with the contrast regime expected for bright quasars. This demonstrates that the PSF+Sersic fit method is in rare cases susceptible to false-positive detections. This motivates us to use the PSF-only modeling results as the primary test for extended emission in \S \ref{sec:fitting}.

\subsection{Accuracy of PSF Templates, NICMOS, Low S/N Sample}\label{sec:appendix_nicmos}

These test stars are observed within a time interval of three months with regards to the science observation. However, the observations are short, and therefore have low S/N in the PSF wings.

\paragraph*{PSF-only modeling:} PSF subtraction of the test stars leaves a characteristic `criss-cross' pattern in the 2D residuals (Figure \ref{fig:starfits_f110w}). The NICMOS PSF templates tend to oversubtract the test stars, leaving negative residuals, and are thus a conservative test for extended emission. However, one of our seven NICMOS test stars is problematic: WD 1657+343 shows residual flux in the PSF wings (Figure \ref{fig:starfits_f110w}, bottom row), suggesting the presence of an extended component. We suspect this white dwarf to be a member of an X-ray binary system given its \emph{Swift} XRT detection in hard X-rays \citep{ODwyer2003}, in which case the light from a faint, undetected companion, or any outflowing material due to their interaction, may explain the broadening of the surface brightness profile. The $\chi^2_\nu$ fit statistic does not favor the inclusion of higher-order background curvature for this short exposure.

\newcommand{\fourplotscale}{0.20}
\begin{figure*}
	\advance\leftskip-3cm
	\centering
	\includegraphics[scale=\fourplotscale]{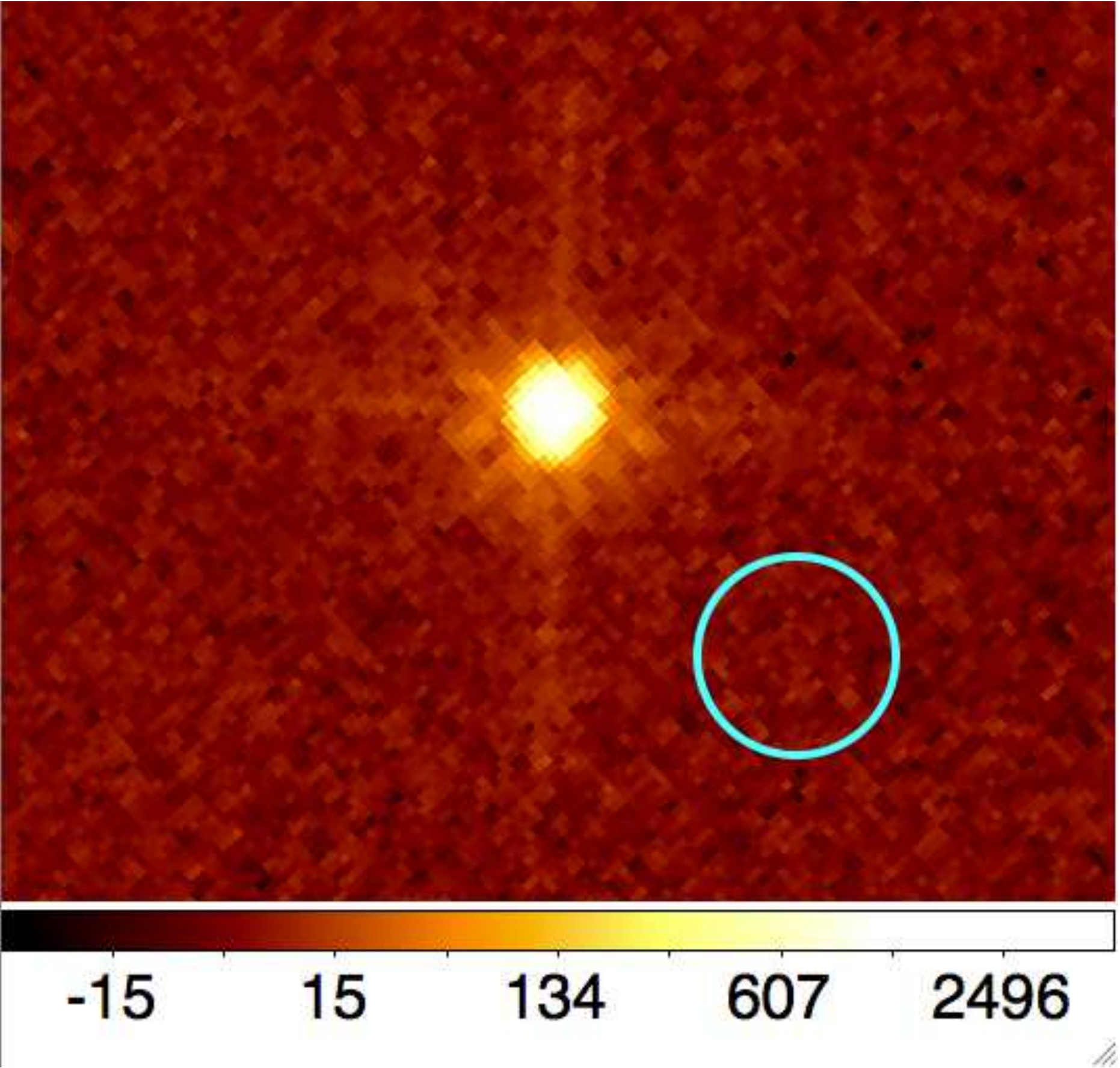}
	\includegraphics[scale=\fourplotscale]{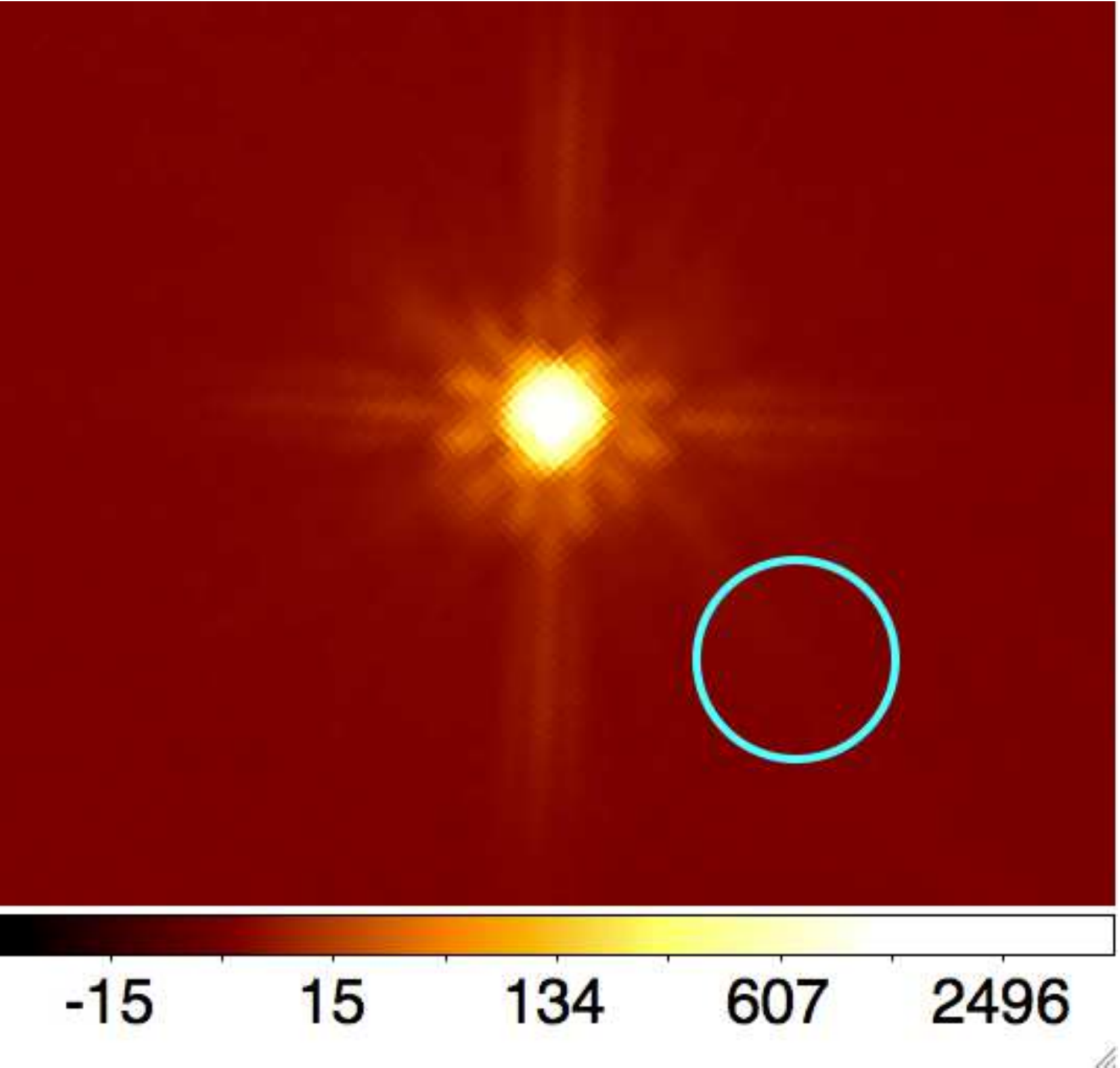} 
	\includegraphics[scale=\fourplotscale]{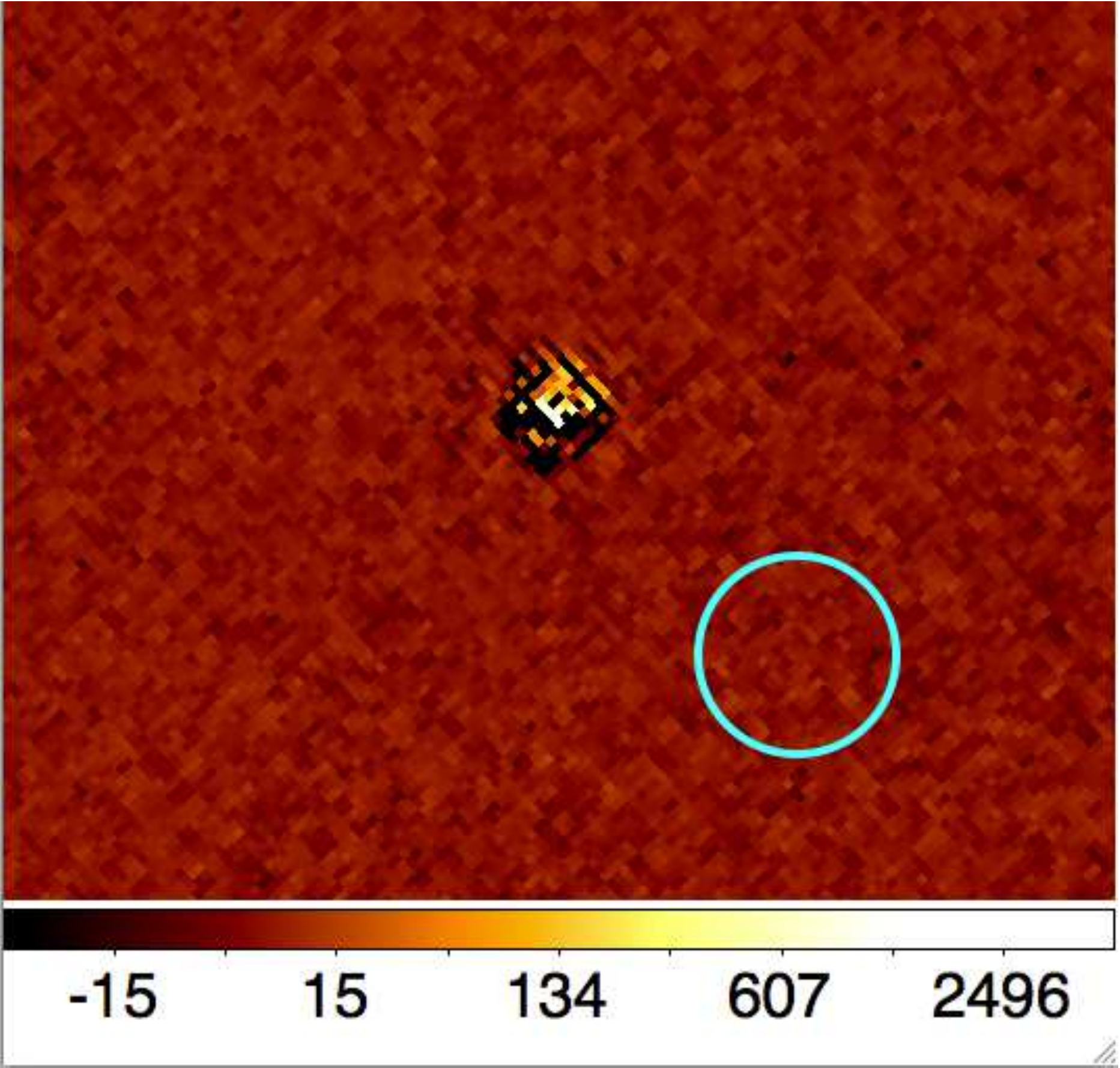}
	\includegraphics[scale=\fourplotscale]{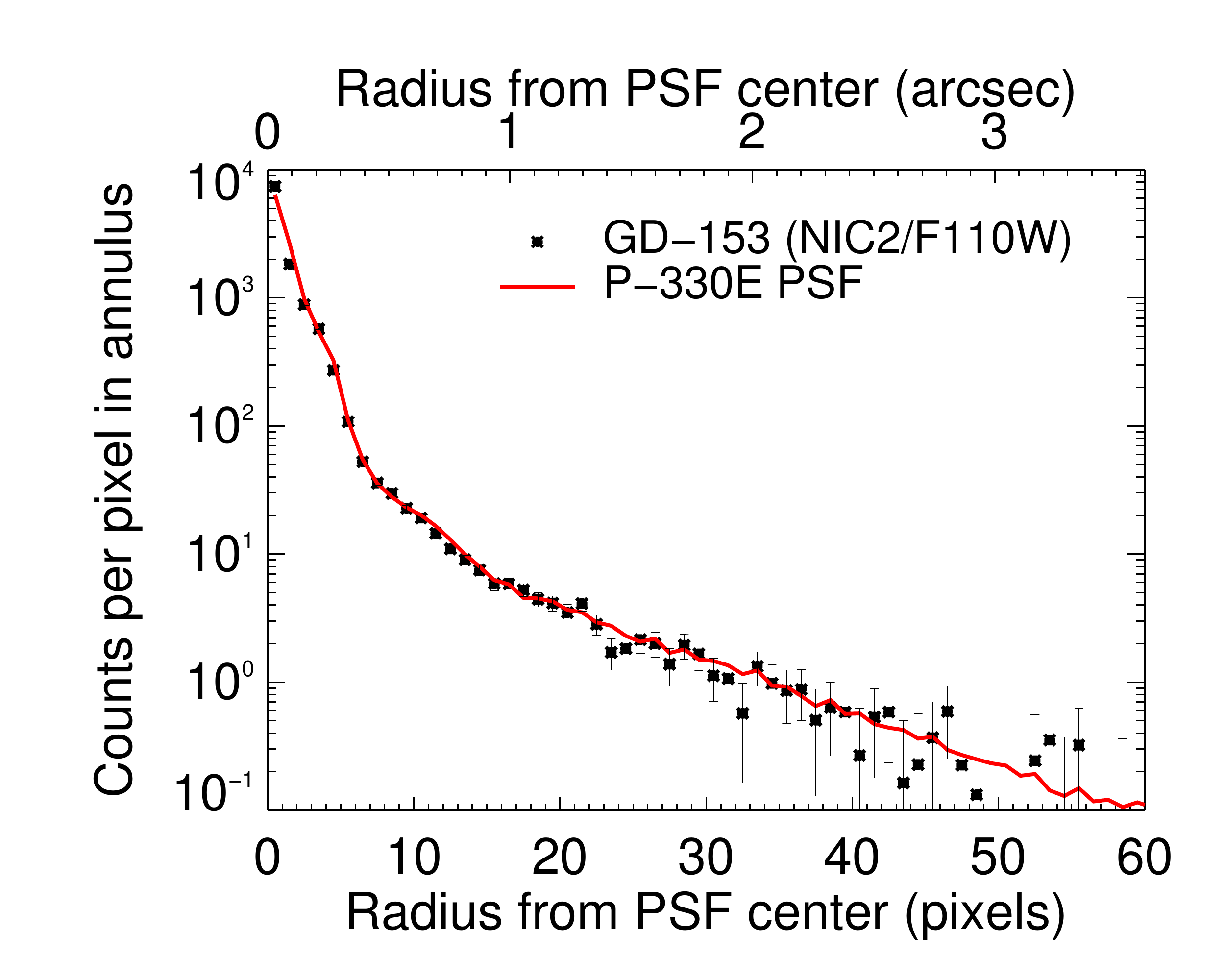}
	\includegraphics[scale=\fourplotscale]{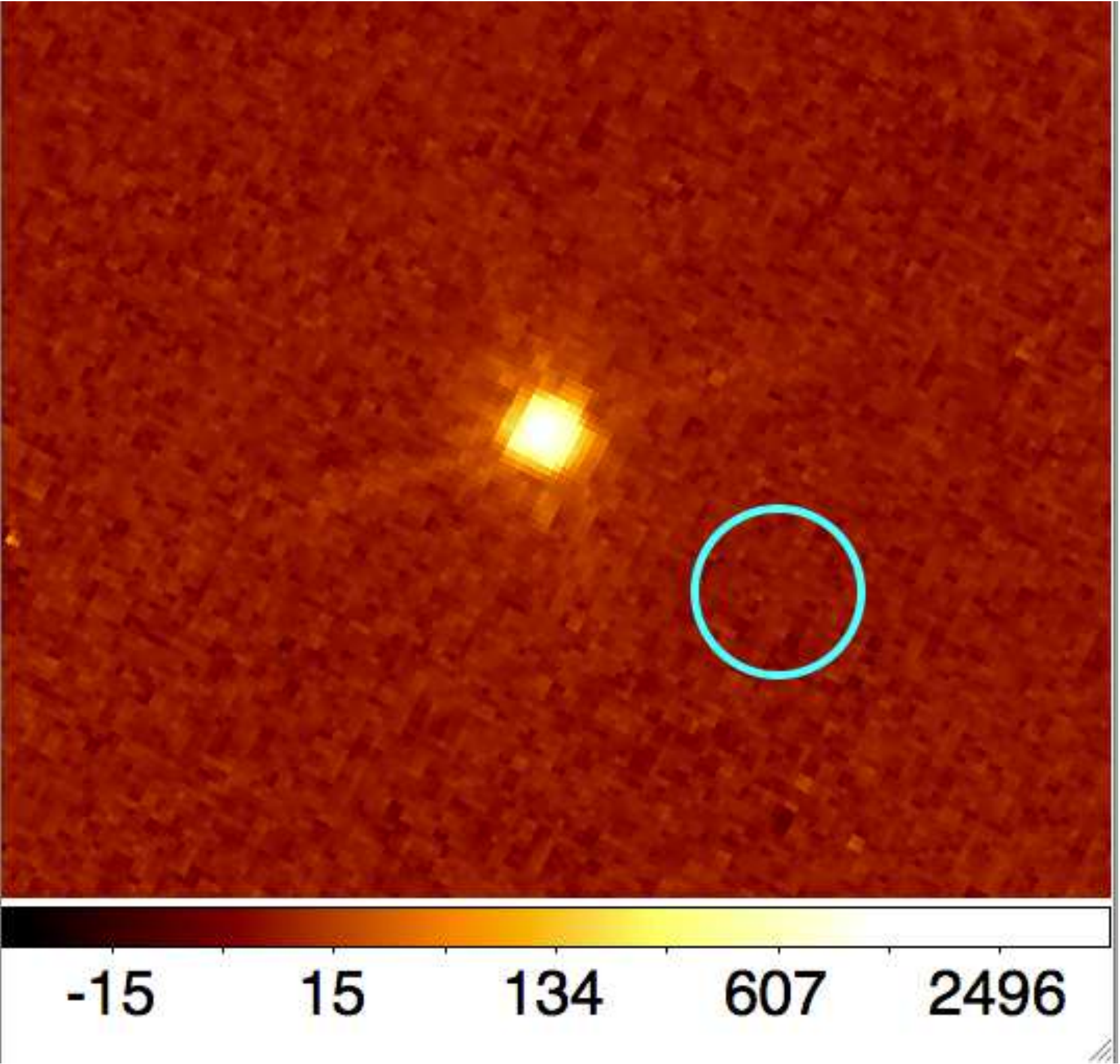}
	\includegraphics[scale=\fourplotscale]{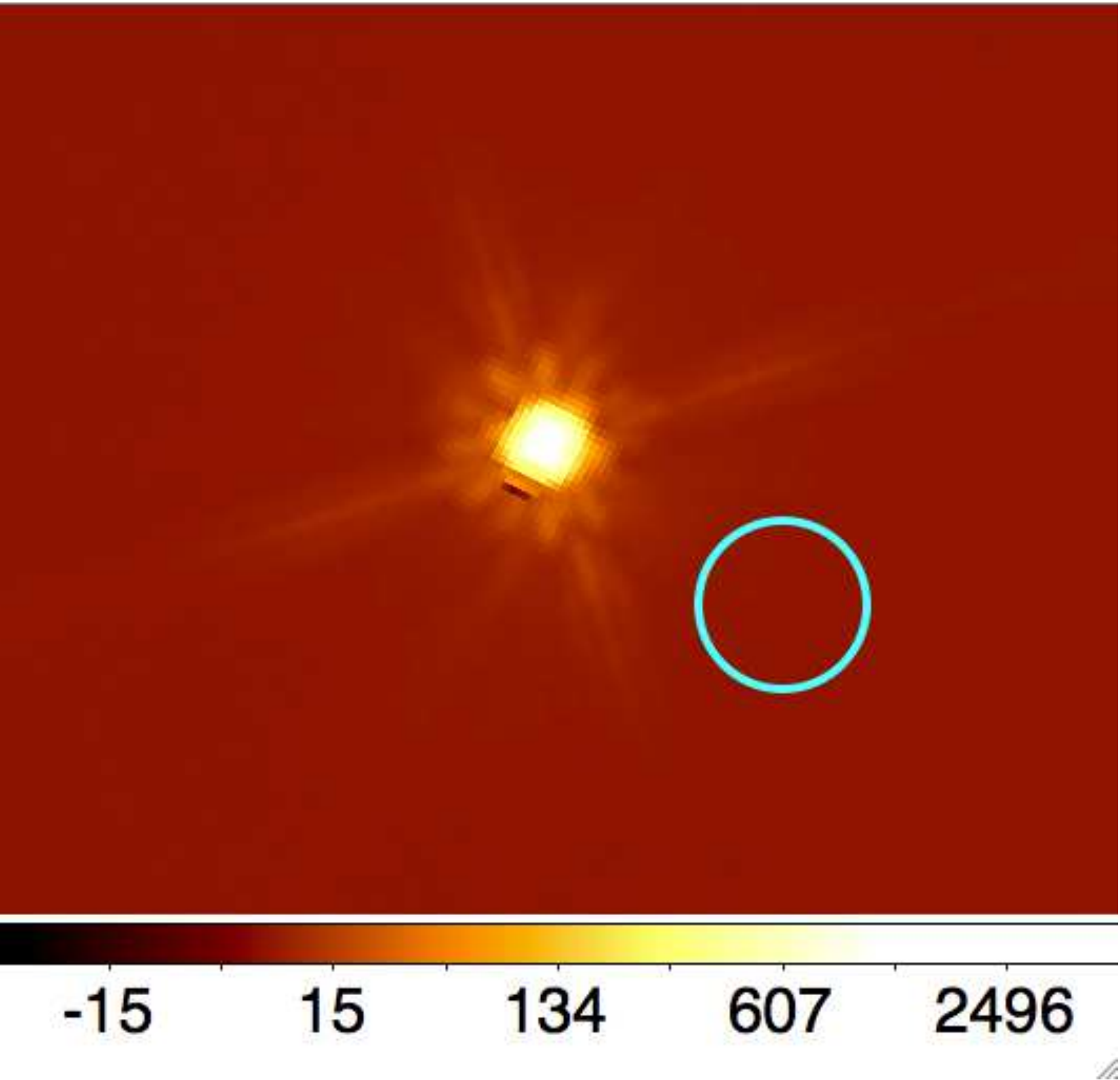} 
	\includegraphics[scale=\fourplotscale]{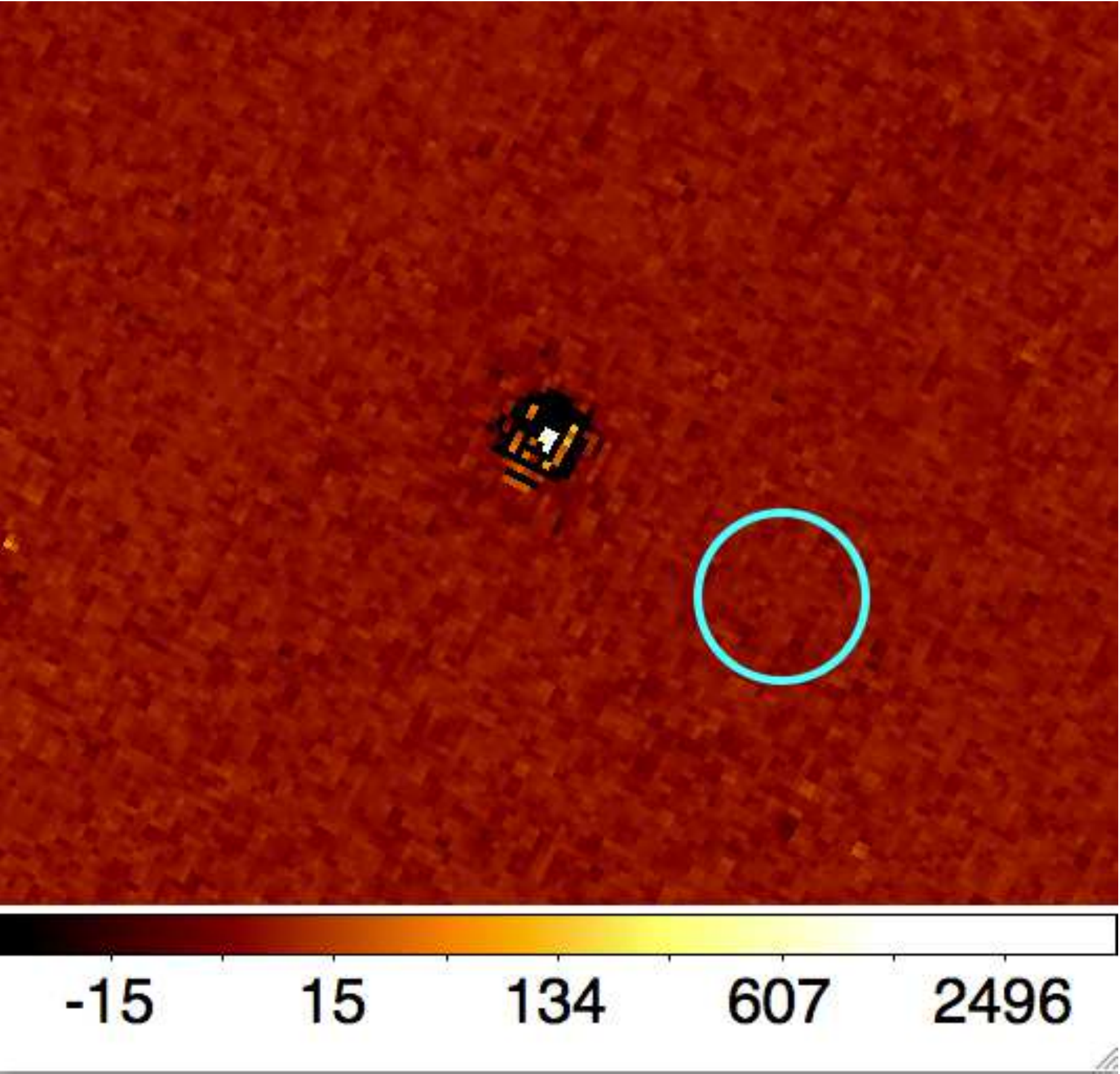}
	\includegraphics[scale=\fourplotscale]{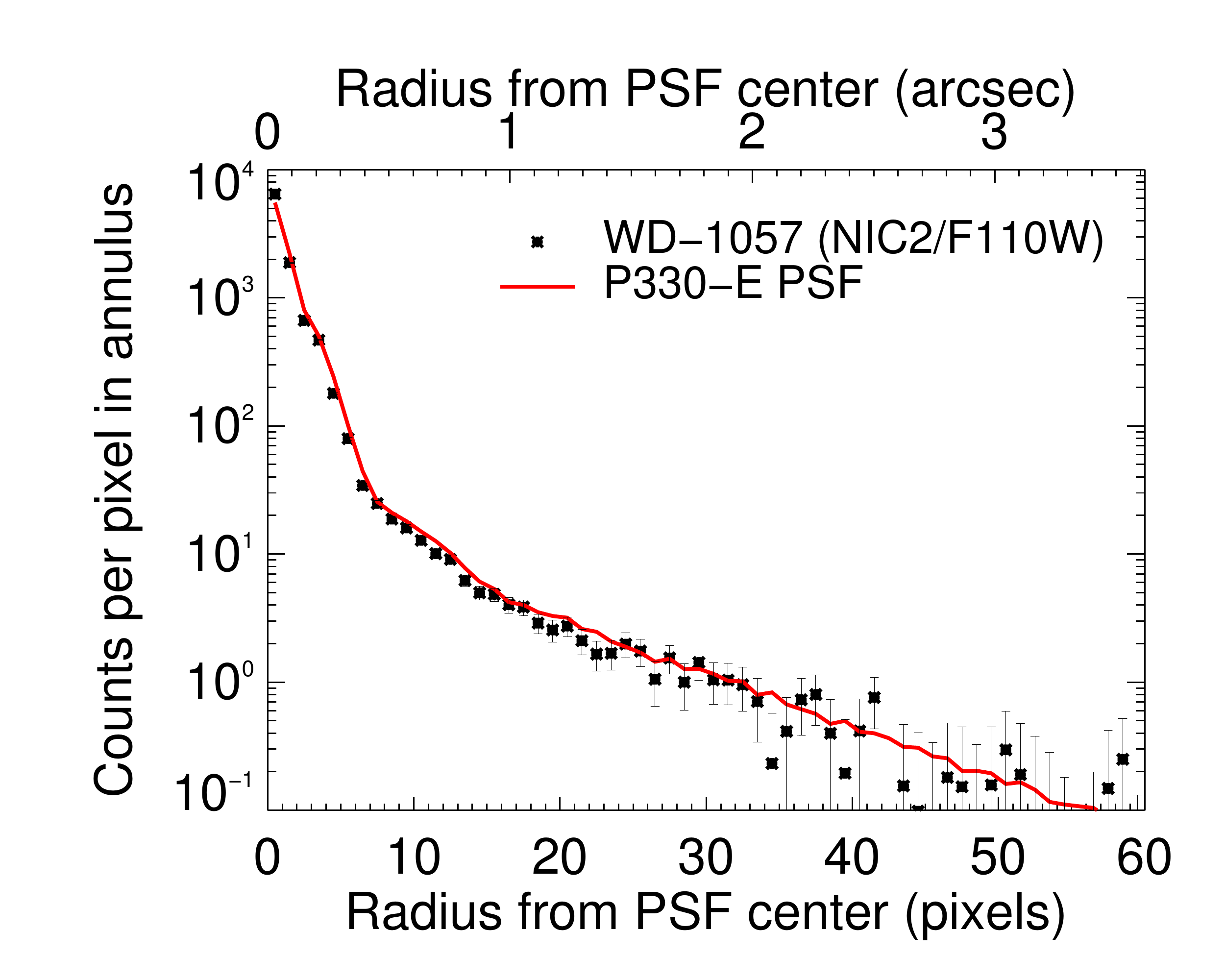}
	\includegraphics[scale=0.193]{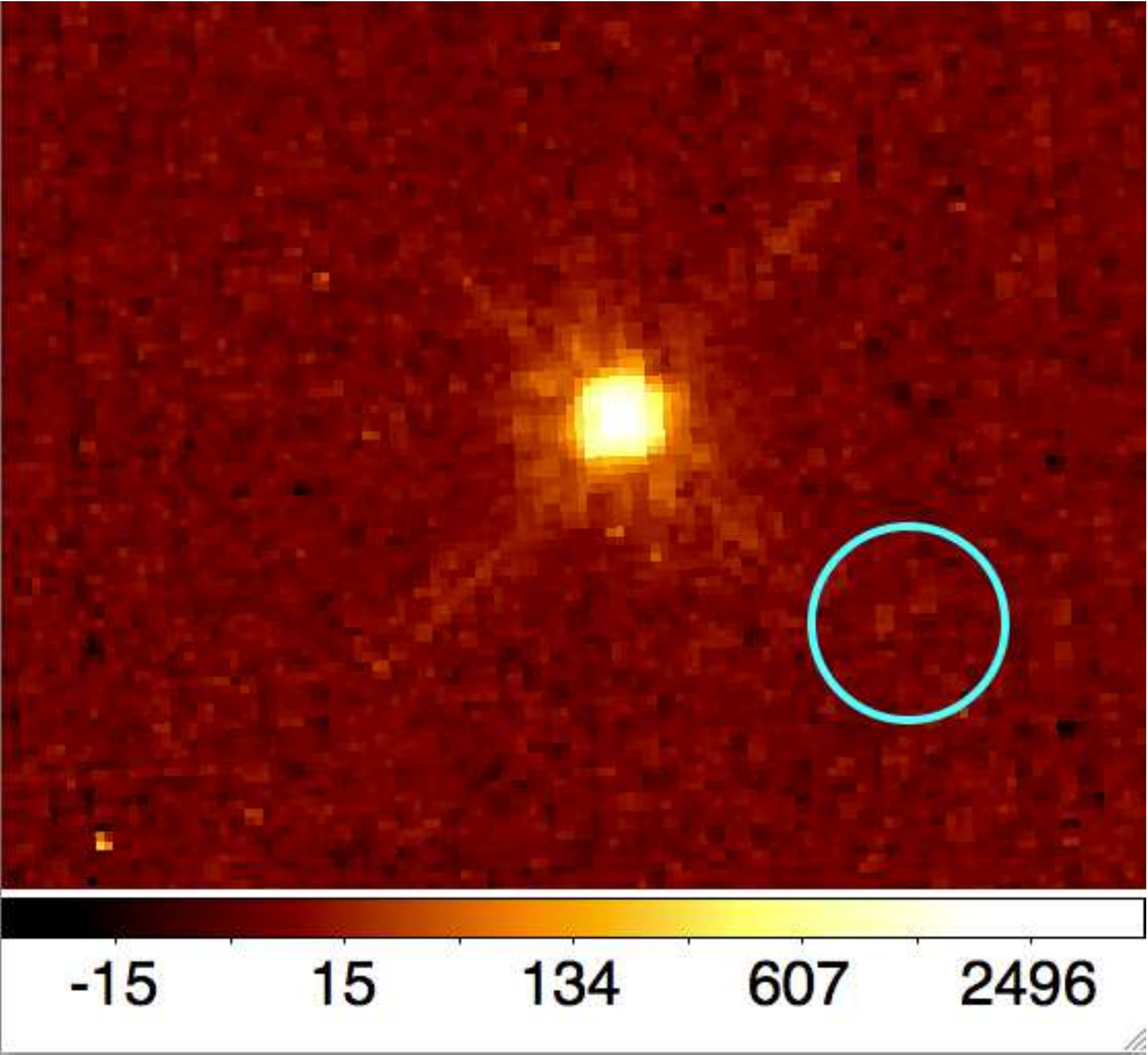}
	\includegraphics[scale=0.193]{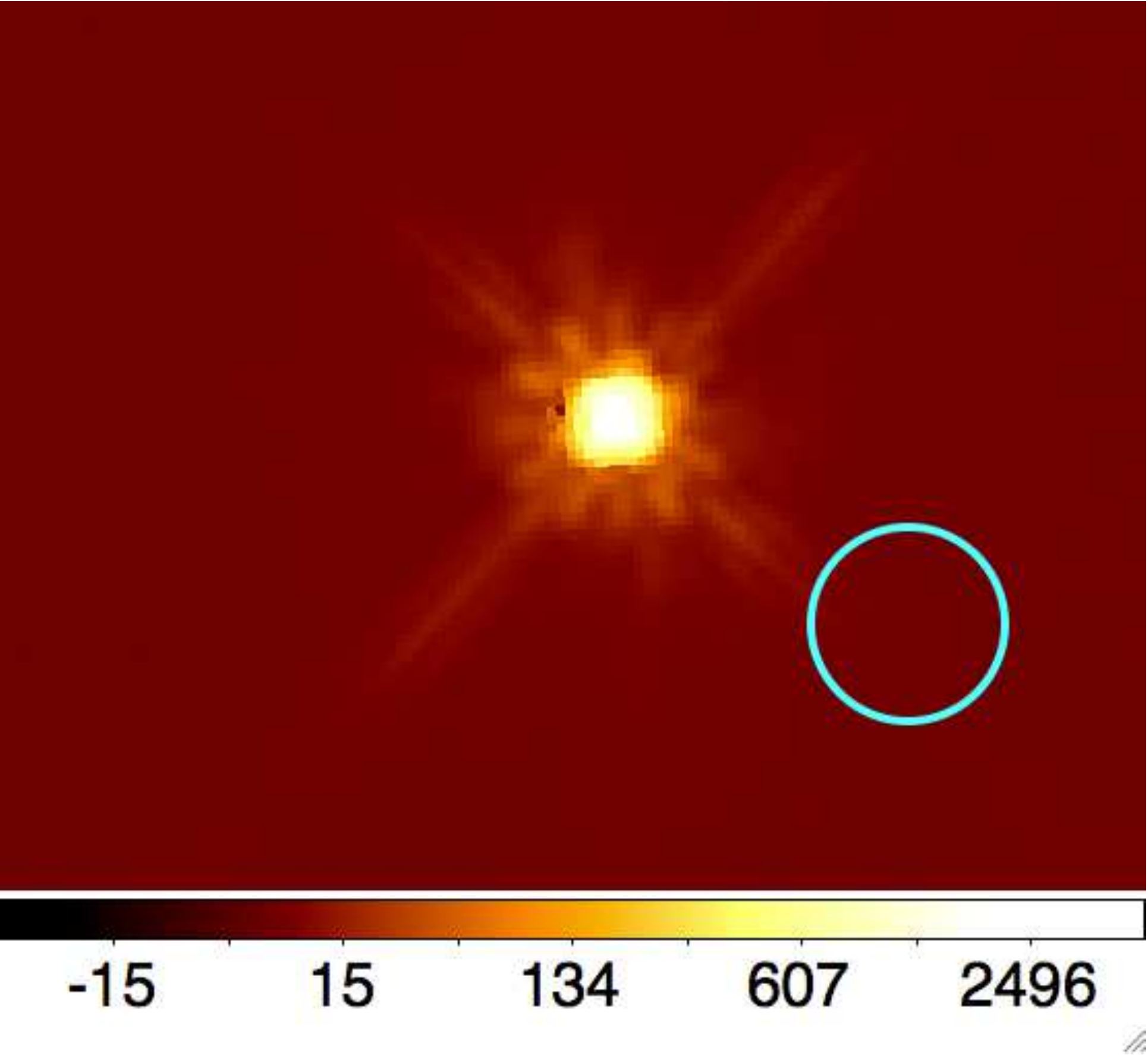} 
	\includegraphics[scale=0.193]{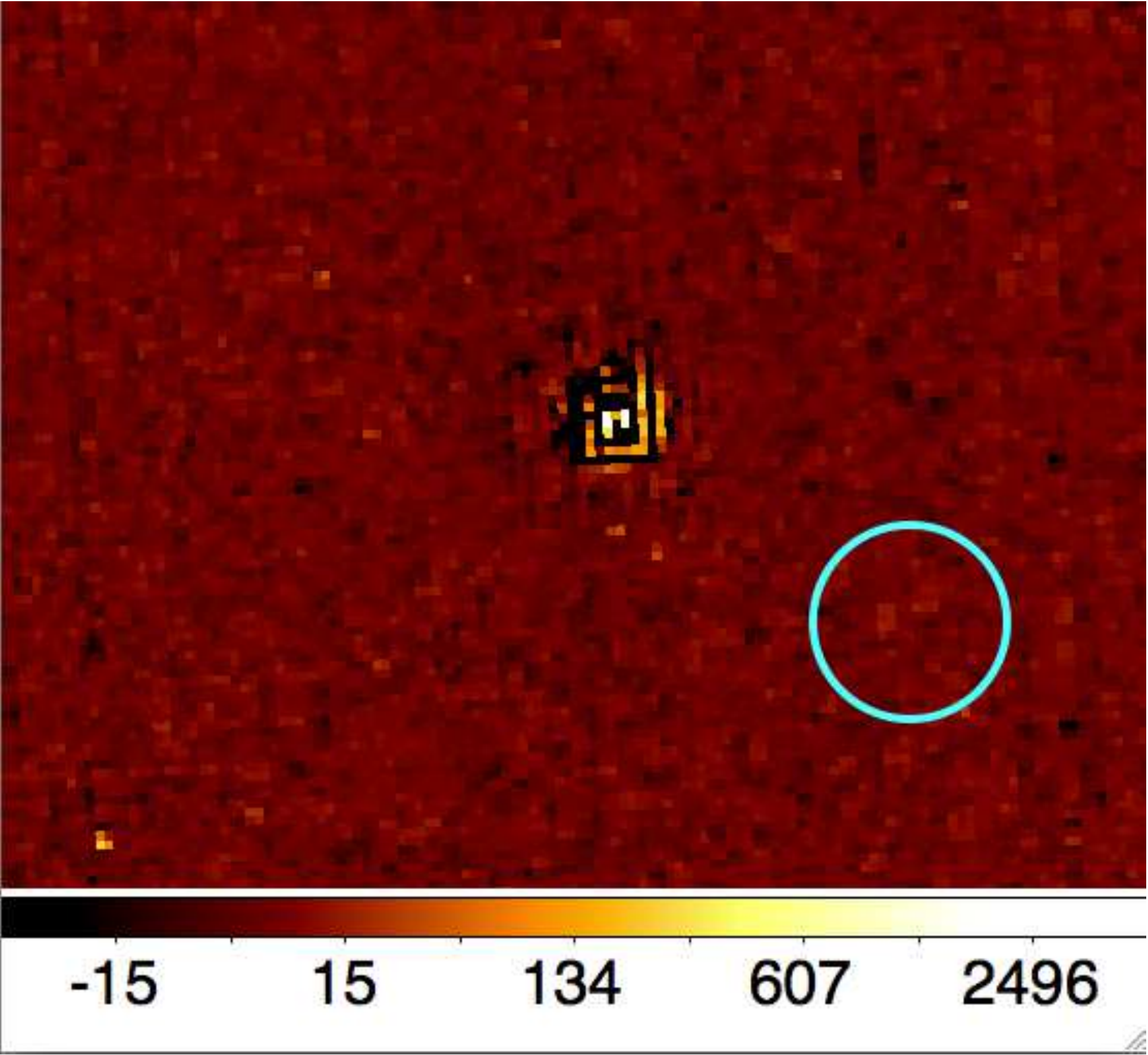}
	\includegraphics[scale=\fourplotscale]{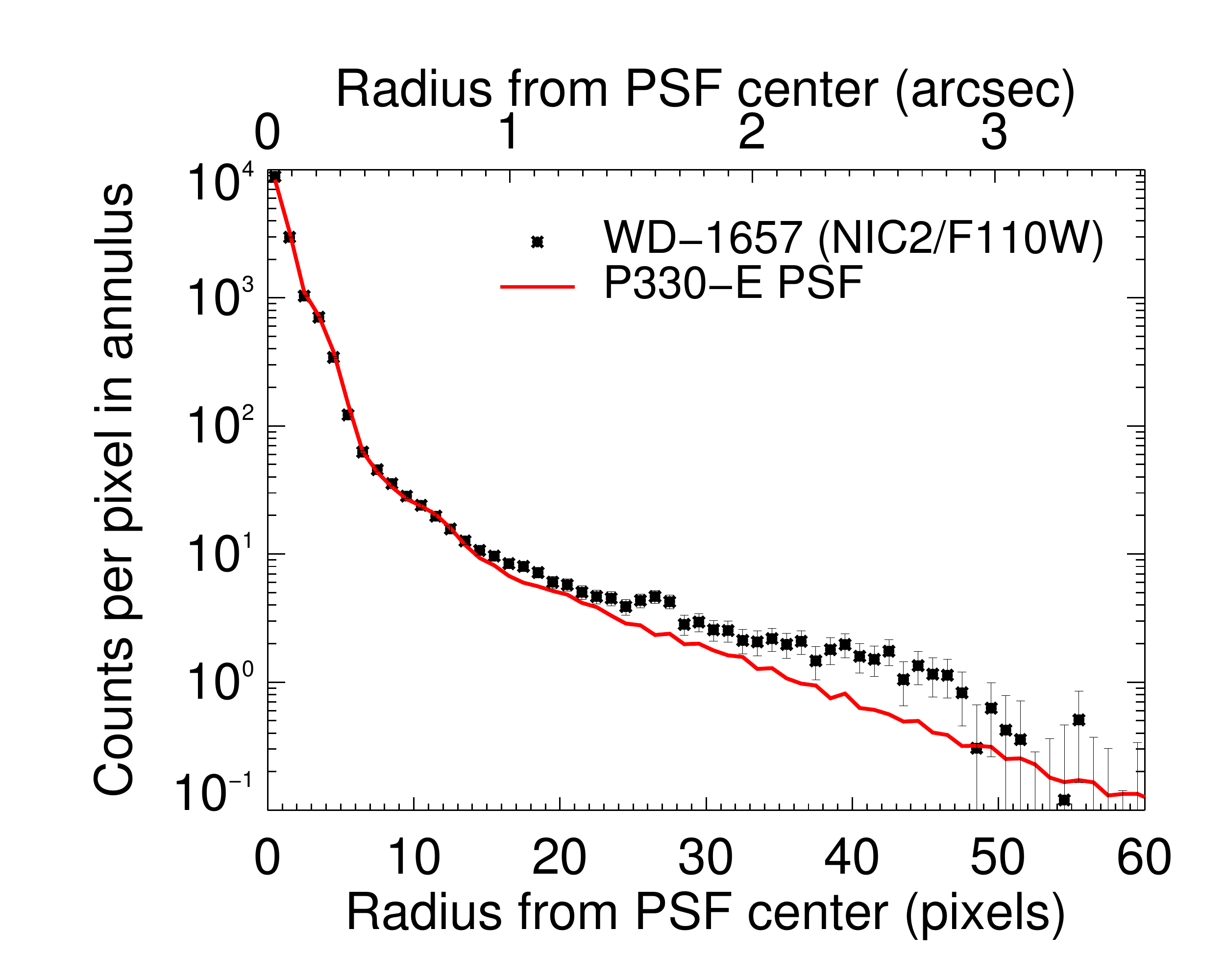}	  	   	  	      	
	\caption{Stacked P330-E stellar PSF fits to representative test stars observed with the NIC2/F110W filter. From top to bottom: GD-153, WD 1057, WD 1657. North is up and east is to the left. The cyan circle depicts a 1 arcsecond diameter. \emph{Left:} Original image. \emph{Center left:} model image consisting of a point source fitted to the star and a sky component. \emph{Center right:} residuals after subtracting the model. \emph{Right:} azimuthally-averaged intensity profile of test star (black points) and stacked stellar PSF template (red line) as a function of distance from the point-source centroid. WD-1657 (top right), which displays a significantly broadened PSF compared to the P330-E stacked PSF, is likely an X-ray binary (\S \ref{sec:appendix_nicmos}).}\label{fig:starfits_f110w}
\end{figure*}

\paragraph*{PSF-plus-Sersic modeling: } In multicomponent modeling we find that the PSF templates are robust against false positive detections: we measure nucleus-to-host contrasts of \contrast$\apprle$7 mag for test stars (including for the `problematic' star WD 1657+343 mentioned above).

\subsection{Results of PSF template tests for NICMOS, high-S/N test star sample}\label{sec:appendix_nicmos_highsnsample}

We retrieved an additional sample of 10 test stars that have high S/N in the PSF wings, but were observed at large time separations (of order 1-2 years) from the science observation\footnote{All these stars were observed later than the HST Servicing Mission 3B, during which the NICMOS detector was repaired, likely altering the PSF characteristics.}. These stars were generally only observed in two dither positions, forcing us to set the \md output pixel size to be identical to the input pixel size. All other steps in the data processing and modeling procedure are carried out identically to that of our quasar images. 

\paragraph*{PSF-only modeling: } We find that empirical PSF templates constructed from the high-S/N sample subtract the PSF wings more cleanly than for our low-S/N sample. The mismatch within the inner 1'' of the PSF centroid is more statistically significant here than for PSF-only fits to our low-S/N sample; this is unsurprising, as we expect PSF shape mismatch to scale linearly with total flux, unlike shot noise. 

\paragraph*{PSF-plus-Sersic modeling:} We perform multicomponent modeling of each star in our high-S/N sample, using one of the other high-S/N stars as a PSF template. Some of these fits yield a spurious `host galaxy' at brightness contrasts of $m_\mathrm{nuc}-m_\mathrm{host}\approx2$ mag. However, for each of these false-positive detections, the modeling converges on an `unresolved' host galaxy component, with the scale size at its lower limit $R_e=\mathrm{FWHM}_\mathrm{PSF}$. This motivates us to be suspicious of `unresolved' host galaxy components when modeling the quasar emission.

\paragraph*{Alternative FeLoBAL analysis using high-S/N PSF:} To test the robustness of our host galaxy detection for J1154+0300 to our choice of NICMOS PSF template, we repeat the modeling of \S \ref{sec:results_nic}, using stars from the high-S/N test star sample as PSF templates. In all cases, we find significant residual flux for J1154+0300. In most cases we find more residual extended flux using these PSF stars, compared to the original analysis using the stacked P330-E PSF (\S \ref{sec:results_nic}). This behavior is consistent with our expectation that the stacked PSF may be broadened somewhat due to centring issues. However, the difference in residual strength may also be due to long-timescale variation of the PSF shape: all stars in our high-S/N sample are observed at time separations of order 1-2 years before the observation of J1154+0300. Due to the risk of long-term PSF variation, we rely on the P330-E PSF template, with time separations of no more than three months, in our final analysis. We also repeat the PSF-plus-Sersic modeling using the high-S/N PSF templates, and find that while the host galaxy brightness is consistent with that found using P330-E to within the uncertainties presented in \S \ref{sec:results_nic}, the scale size $R_e$ converges at its lower limit for some of the high S/N PSF stars. This confirms that $R_e$ is not well-determined by our data, and is sensitive to PSF mismatch.
\section{Simulated quasar observations}\label{sec:sims}\label{sec:appendixB}

Here we describe simulated quasar observations used to quantify the uncertainties inherent to our image analysis (\S \ref{sec:fitting}). We aim to simulate as accurately as possible the appearance (in \emph{HST} observations) of quasars residing at the redshifts of our target FeLoBALs. We use these simulations to estimate upper limits on the host galaxy flux for non-detections, and to quantify our measurement uncertainties for detected host galaxies. 

One approach to the generation of simulated observations is to use suitably degraded images of real galaxies, with a superimposed point source corresponding to an AGN nucleus. We use this approach for our NICMOS data (Appendix~\ref{sec:nicmos_sims}). Another option is to use analytical galaxy models representing the quasar host, with a superimposed point source representing the nucleus. The use of analytical profiles runs the risk of underestimating parameter uncertainty, as real galaxies deviate from simple models. However, we do not expect the detailed morphology to be important when observing at the limit of marginal detection. We therefore use analytical profiles to simulate host galaxies in ACS data (Appendix~\ref{sec:acs_sims}), as none of our ACS observations yield host galaxy detections. In general, for either instrument, we find that the primary source of uncertainty is the addition of a bright point source.

\subsection{Simulated NICMOS observations using FERENGI}\label{sec:nicmos_sims}

\paragraph*{Sample of low-redshift galaxy templates:}\label{sec:ferengi_galaxy_sample}

We use artificially redshifted images of low-redshift non-AGN galaxies to represent quasar host galaxies for the NICMOS simulations. We require wide-field, multi-band imaging of template galaxies as input for the artificial redshifting procedure. We select six low-redshift galaxies, spanning a range of morphologies, imaged by the Sloan Digital Sky Survey (SDSS). We choose a morphologically diverse sample of large, luminous galaxies that we expect to be visible and resolved at $z\sim1$ in NICMOS imaging, and that have accurate distance measurements in the literature, allowing the determination of their absolute magnitude and physical size. The selected objects are as follows. \emph{1)} NGC 474, a face-on S0 galaxy which displays shells or tidal tails in the SDSS imaging; \emph{2)} PCG 6110, a massive E3 galaxy; \emph{3)} Arp 220, the lowest-redshift archetypical Ultraluminous Infrared Galaxy (ULIRG), displaying an irregular morphology with two distinct cores in the SDSS imaging; \emph{4)} NGC 5746, an edge-on SB(r)bc with a star formation rate of $1.2M_{\astrosun}$ yr$^{-1}$; \emph{5)} NGC 151, a face-on starbursting SB(r)bc galaxy. \emph{6)} NGC 6166, a cD galaxy.  

%\paragraph*{Single-component versus multicomponent galaxy modeling:}\label{sec:single_vs_multicomp}

%We use a single-component Sersic profile to model the quasar host galaxy in our analysis; the deviations of real galaxies from such a profile are a source of error. We therefore compare the brightness measured for our template galaxies using a Sersic fit to that measured using a more involved multicomponent model. Unsurprisingly, single Sersic components do not model the detailed surface brightness profile of well-resolved, nearby galaxies well. However, the difference in total brightness between a single-component fit and a multicomponent fit is less than 0.2 mag for an elliptical galaxy. This uncertainty is small compared to the other uncertainties described below.

\paragraph*{Artificial redshifting, cosmological dimming and rebinning of template galaxies:}\label{sec:ferengi_code}

We use the \textsc{Ferengi} software \citep{Barden2008} to generate images of the SDSS template galaxies as they would appear at high redshift as observed using the NIC2 instrument. First, the images are scaled to account for the change in angular size at the target output redshift. Next, a cosmological dimming corresponding to the output redshift is applied, and the image is rebinned to the pixel scale of the NIC2 instrument. SED template fitting is then applied on a per-pixel basis using the \textsc{IDL} routine \emph{kcorrect} \citep{Blanton2007}, so as to determine the K-corrected flux in the relevant NICMOS bandpass. The galaxy image is then scaled to the target exposure time and superimposed on a background image containing background signal and simulated readout noise at the level measured in our \emph{HST} observations. We generate an artificially redshifted image for each galaxy template at each of the target redshifts $z=0.89$, $z=1.46$ and $z=1.93$; all \textsc{Ferengi} output parameters (e.g., exposure time) are set to mimic the Fe\-LoBAL observations.\footnote{The redshift of and observing conditions for the last member of our Fe\-LoBAL sample, J1730+5850(NIC2/F160W), were similar to that of J0819+4209. Tests performed on simulated images for the J0819+4209 observation therefore apply to both objects.} Lastly, we add Poisson noise due to the host galaxy and the background component; the noise in the galaxy template itself is negligible due to the much smaller amount of pixels in the output images. We show the \textsc{Ferengi} output images, with no superimposed point source, in Figures \ref{fig:ferengi_galaxies} and \ref{fig:ferengi_galaxies_2}. To quantify the uncertainty due to our redshift range and instrumental setup, we first perform single-component modeling of these output images, prior to introducing a central point source. The measured absolute magnitudes are consistent with those of the input SDSS imaging to within 0.3 mag. The physical scale sizes of the single-component fits are too small by up to a factor 2. The Sersic index was recovered to within a value of $\pm1$ at all redshifts. It is difficult to determine the galaxy type via visual inspection, especially at $z=1.46$ and $z=1.93$ (Figures \ref{fig:ferengi_galaxies}, \ref{fig:ferengi_galaxies_2}). Faint features such as the tidal tails / shells, visible for NGC 474 in the SDSS imaging, are not detected at cosmological distances.

\newcommand{\smallscale}{0.21}

\begin{figure*}
	\advance\leftskip-3cm
	\centering
	\includegraphics[scale=0.5]{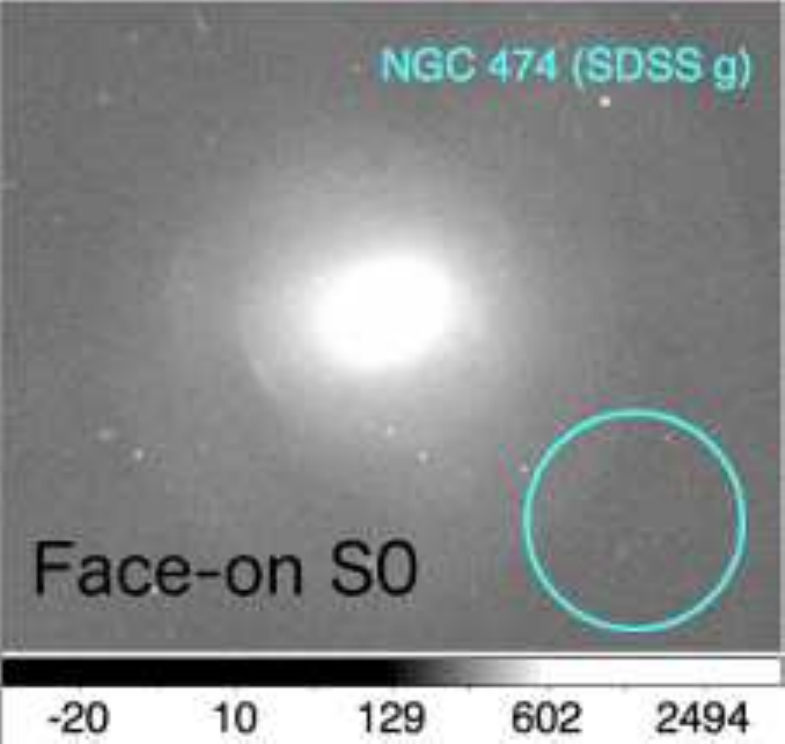}
	\includegraphics[scale=\smallscale]{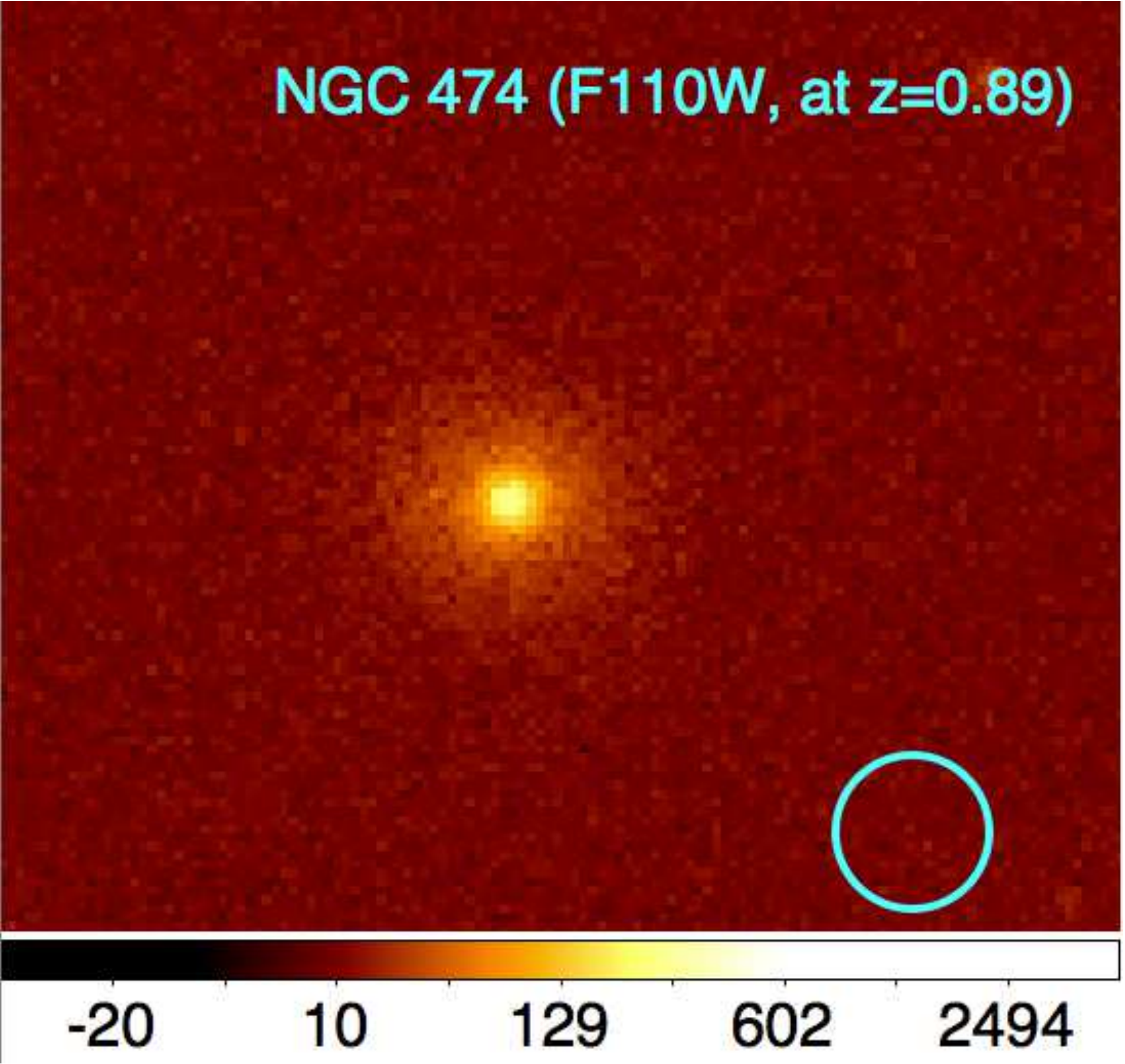}	 
	\includegraphics[scale=\smallscale]{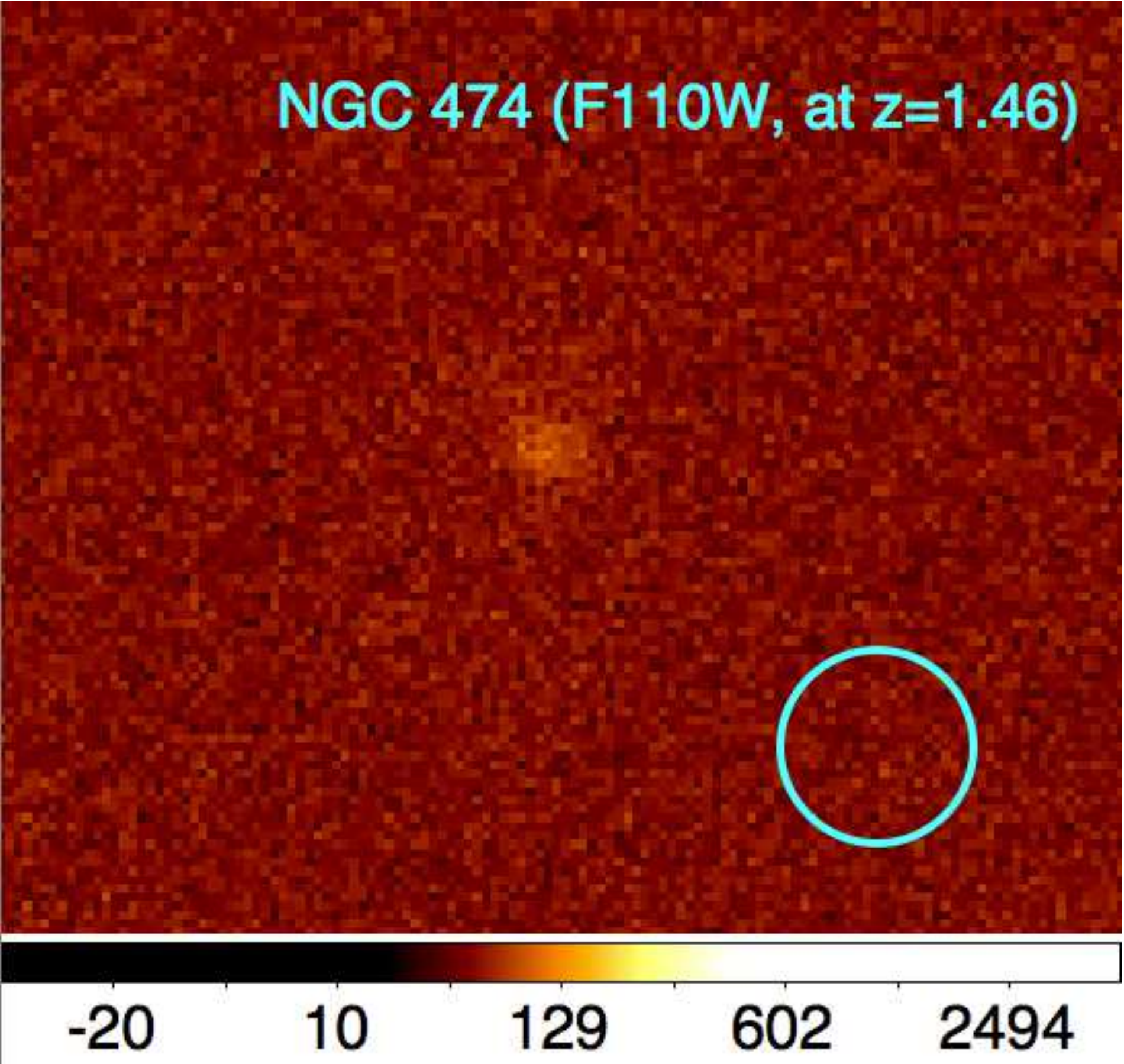}
	\includegraphics[scale=\smallscale]{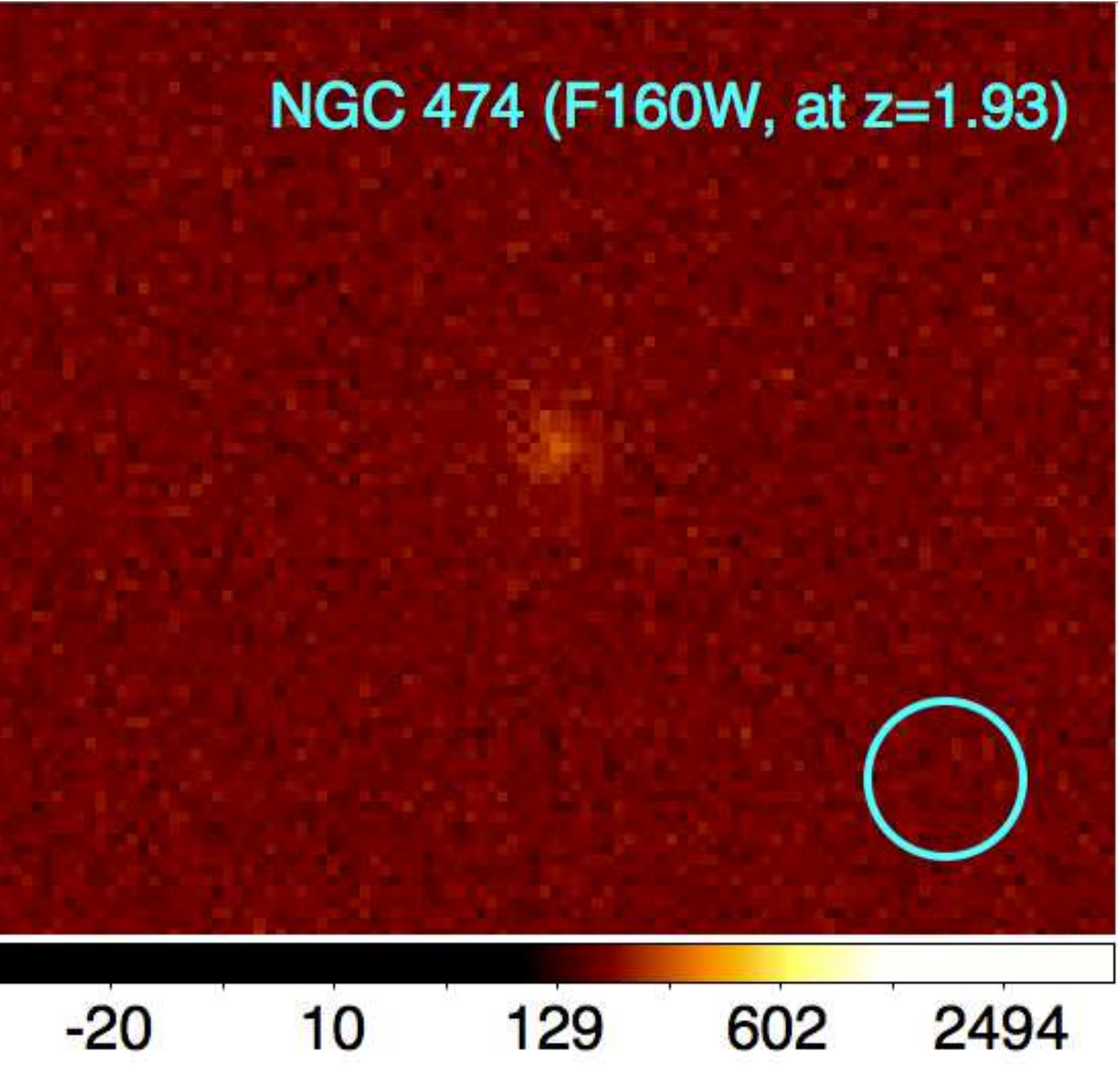}
	\includegraphics[scale=0.5]{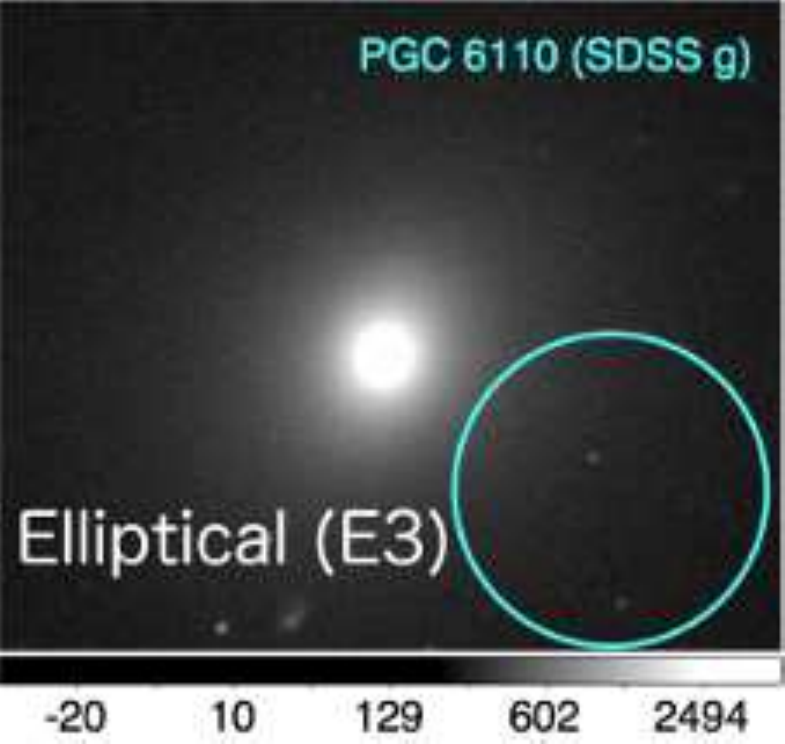}
	\includegraphics[scale=\smallscale]{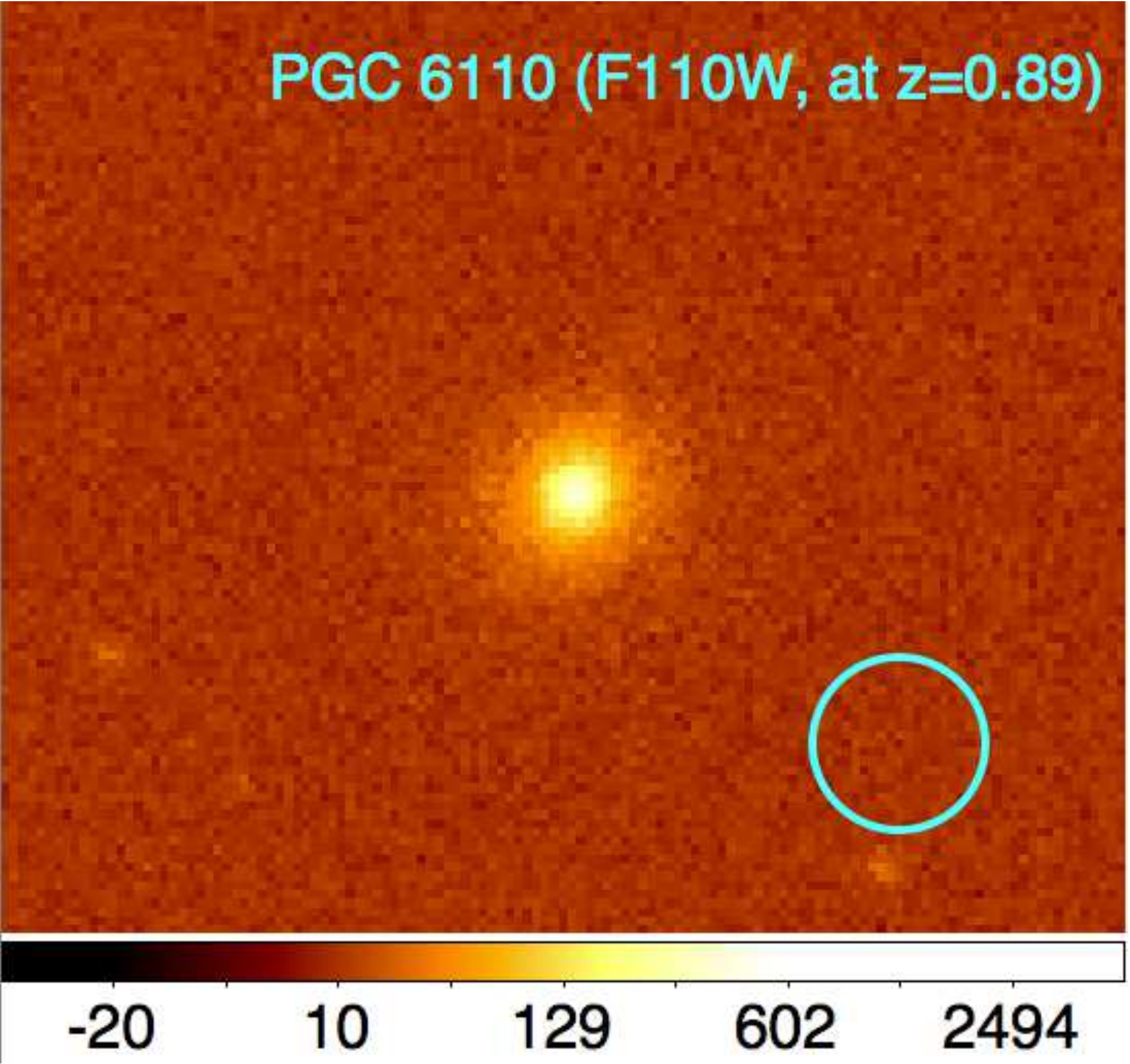}	 
	\includegraphics[scale=\smallscale]{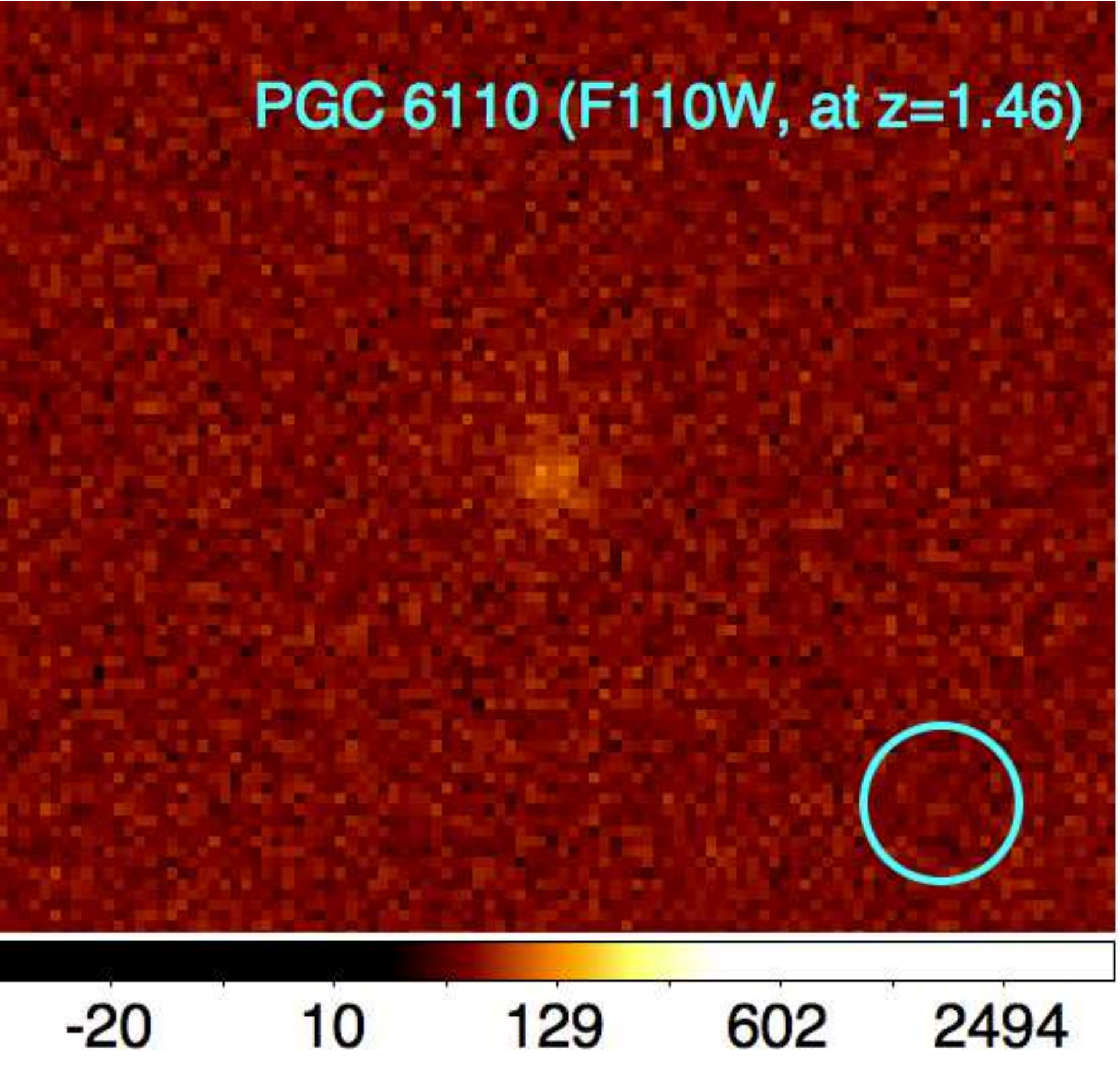}
	\includegraphics[scale=\smallscale]{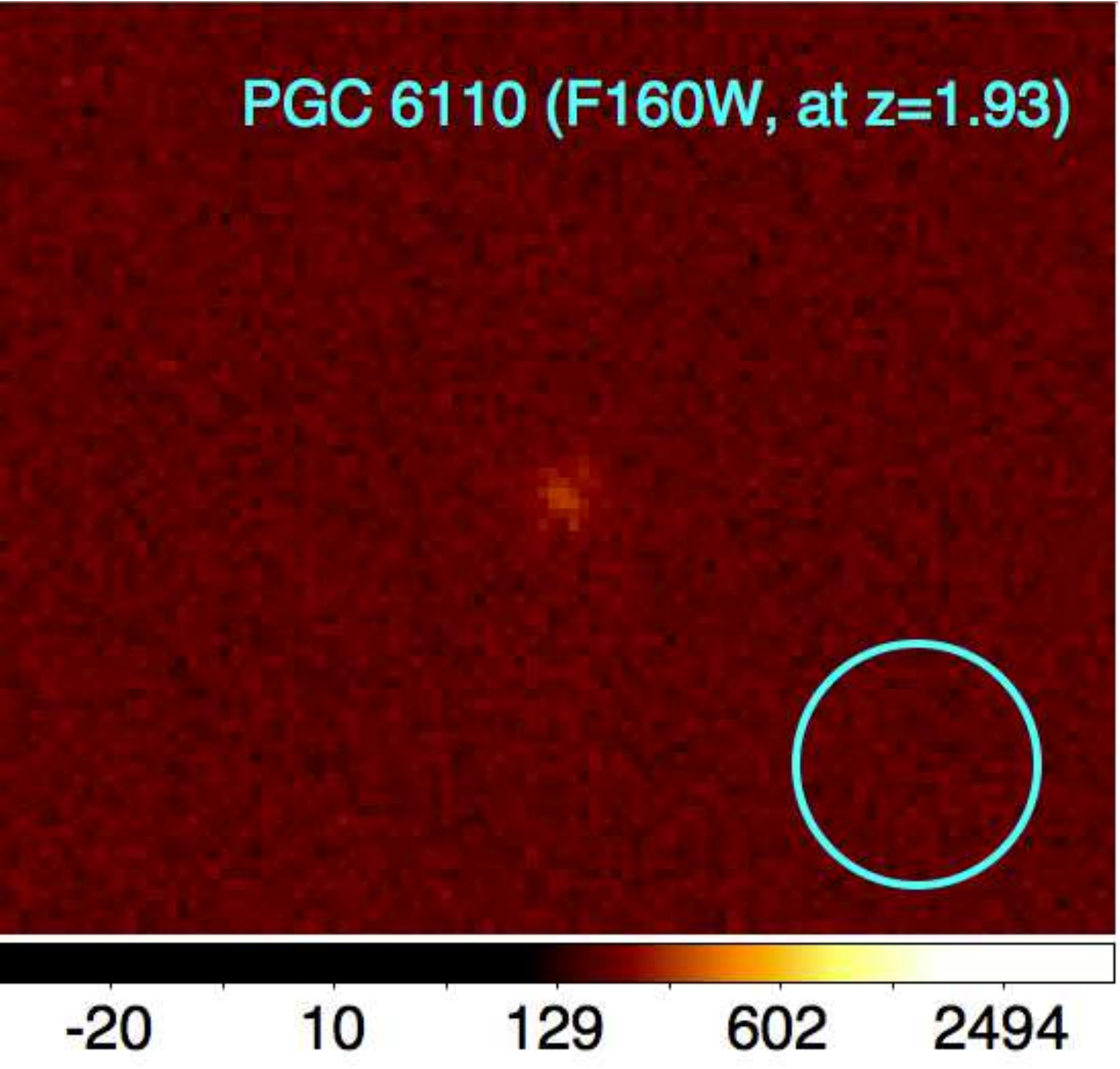}	
	\includegraphics[scale=0.5]{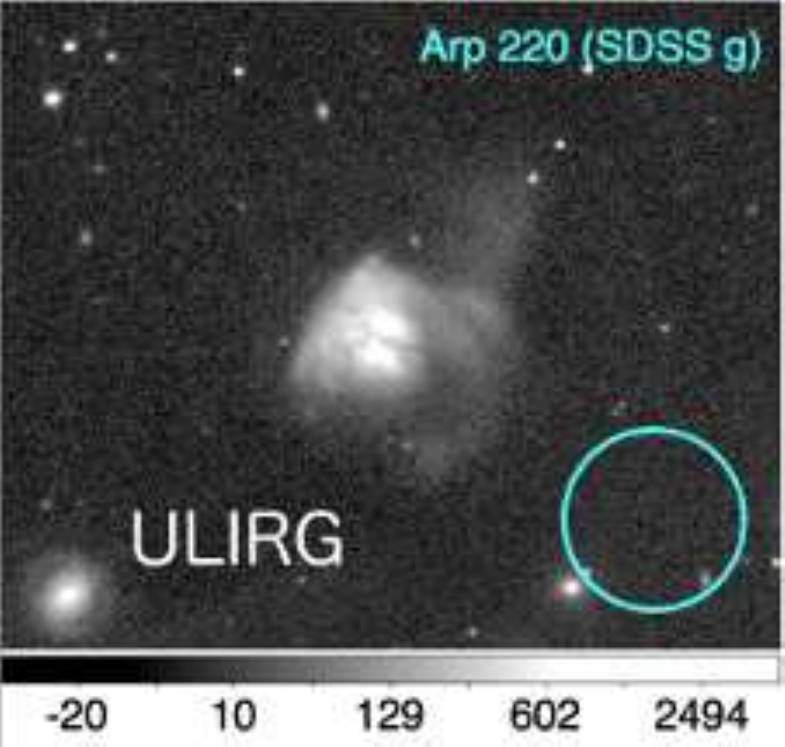}
	\includegraphics[scale=\smallscale]{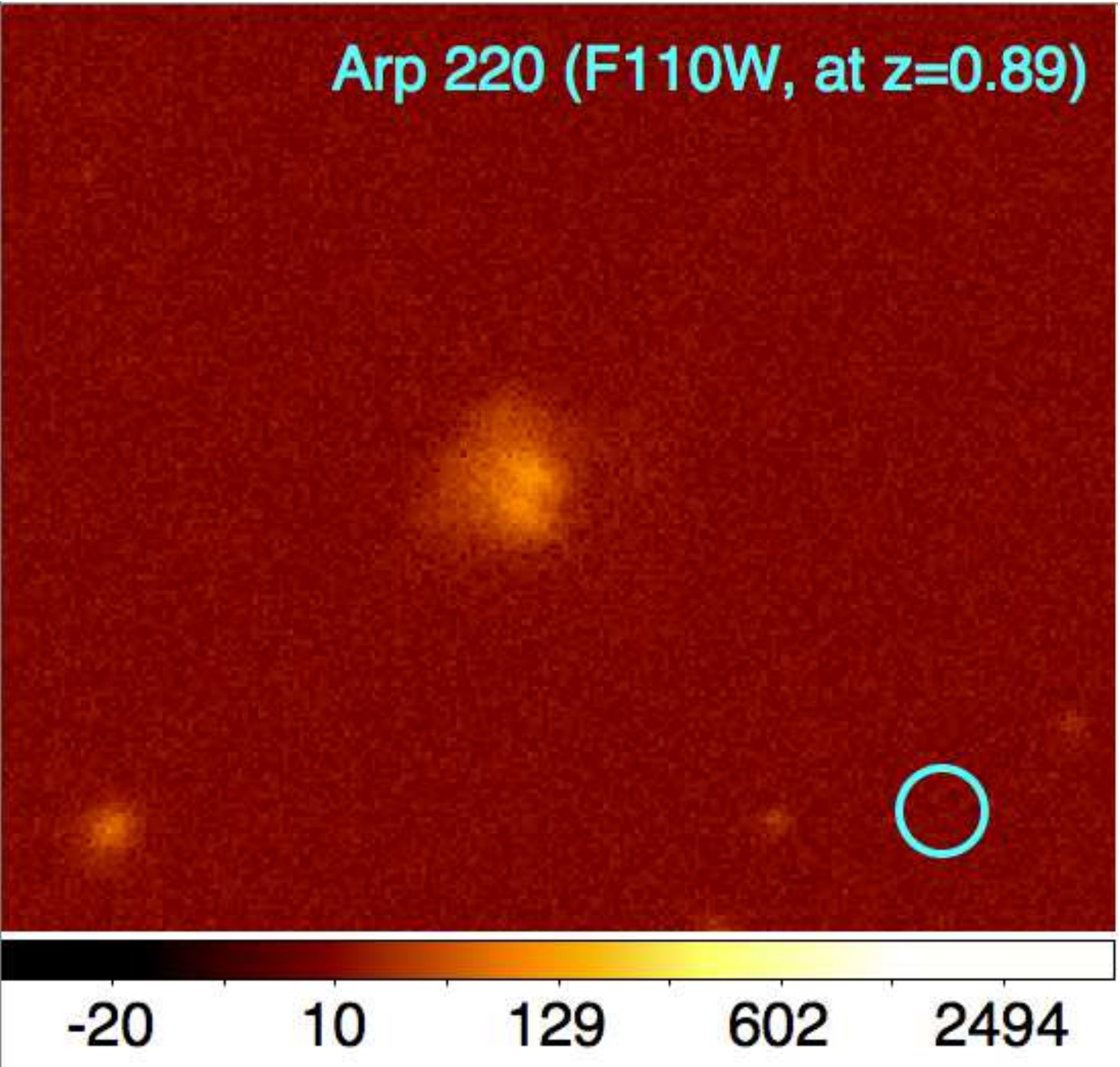}	 
	\includegraphics[scale=\smallscale]{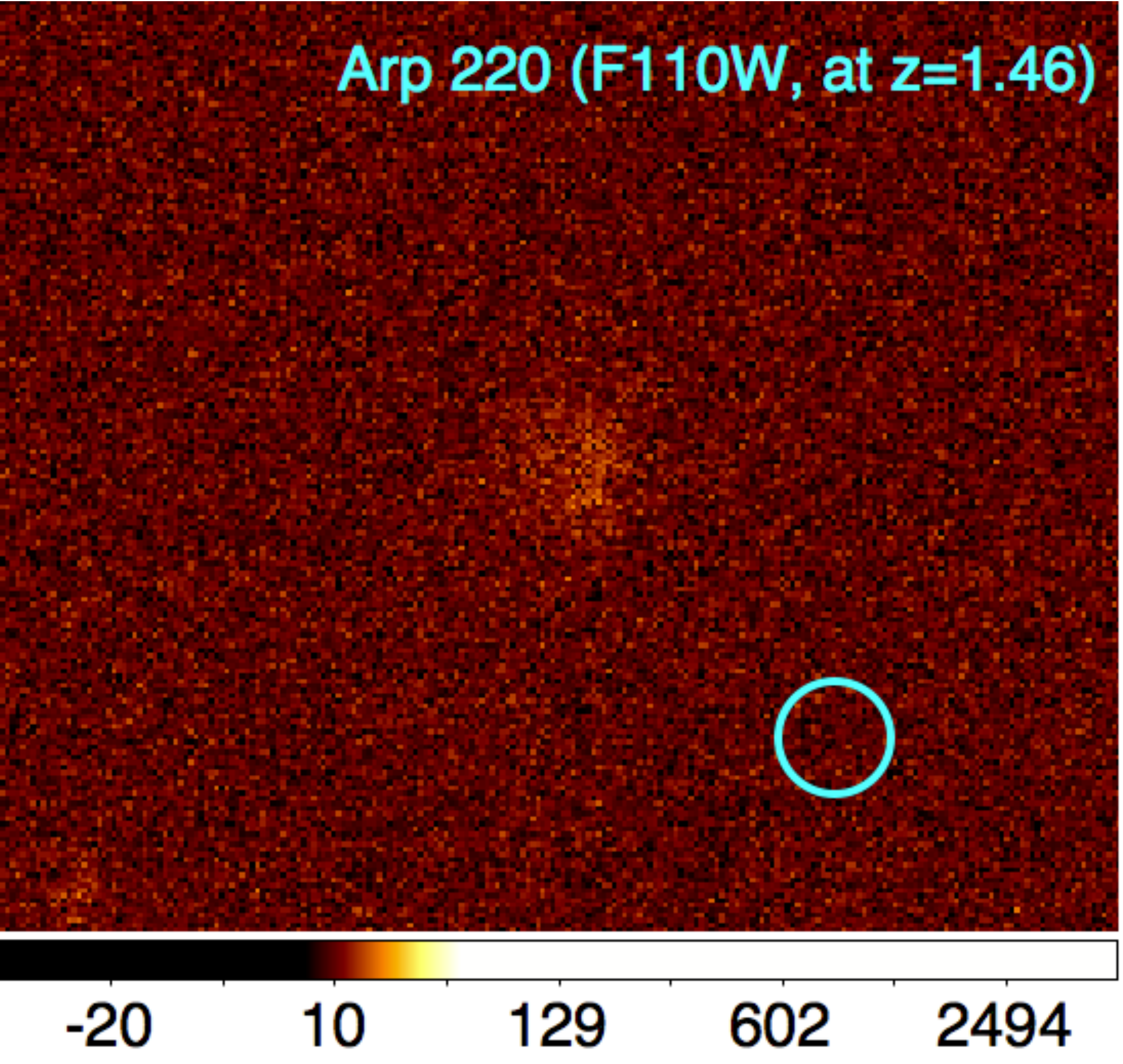}
	\includegraphics[scale=\smallscale]{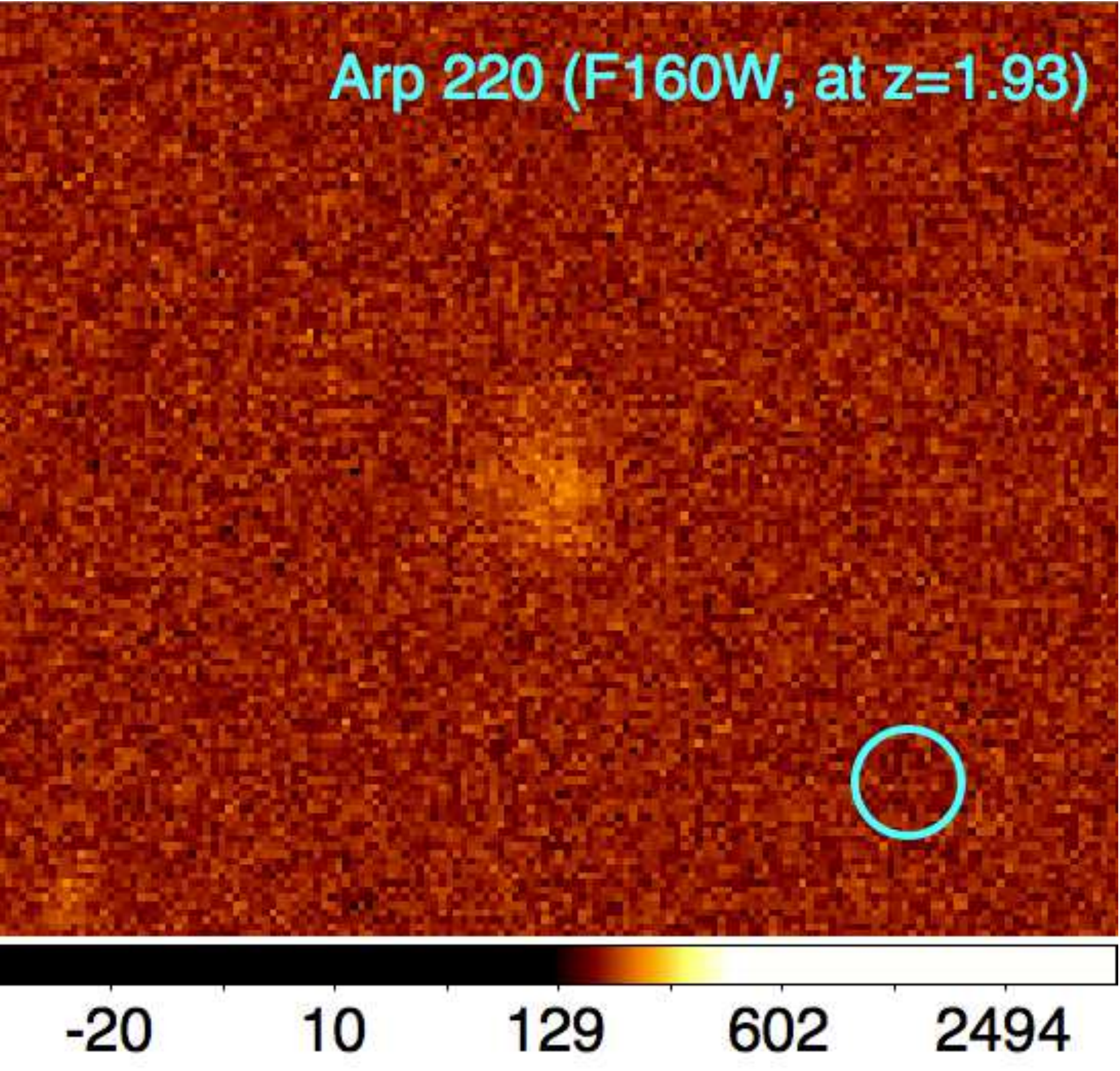}
	\includegraphics[scale=0.5]{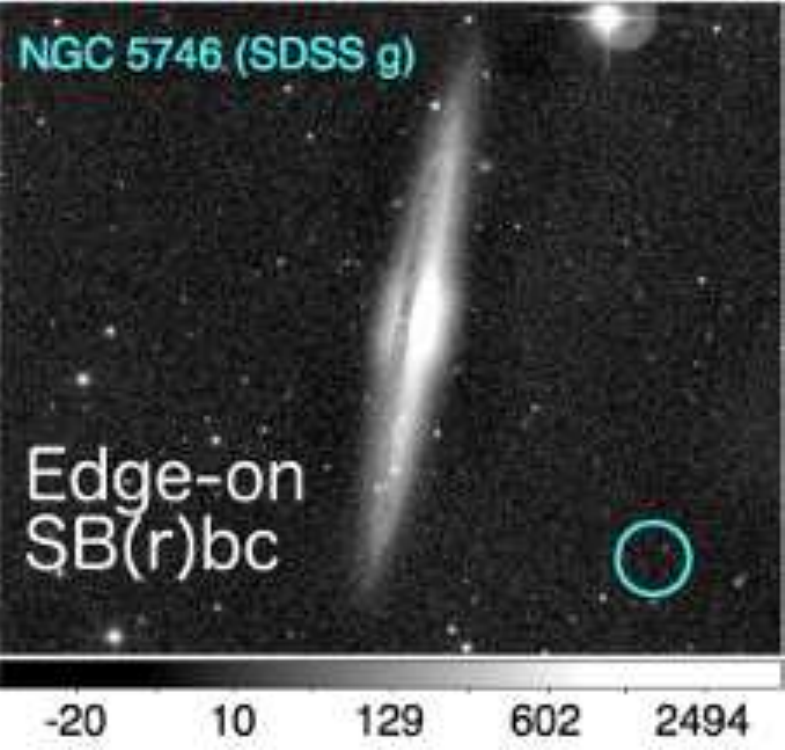}
	\includegraphics[scale=\smallscale]{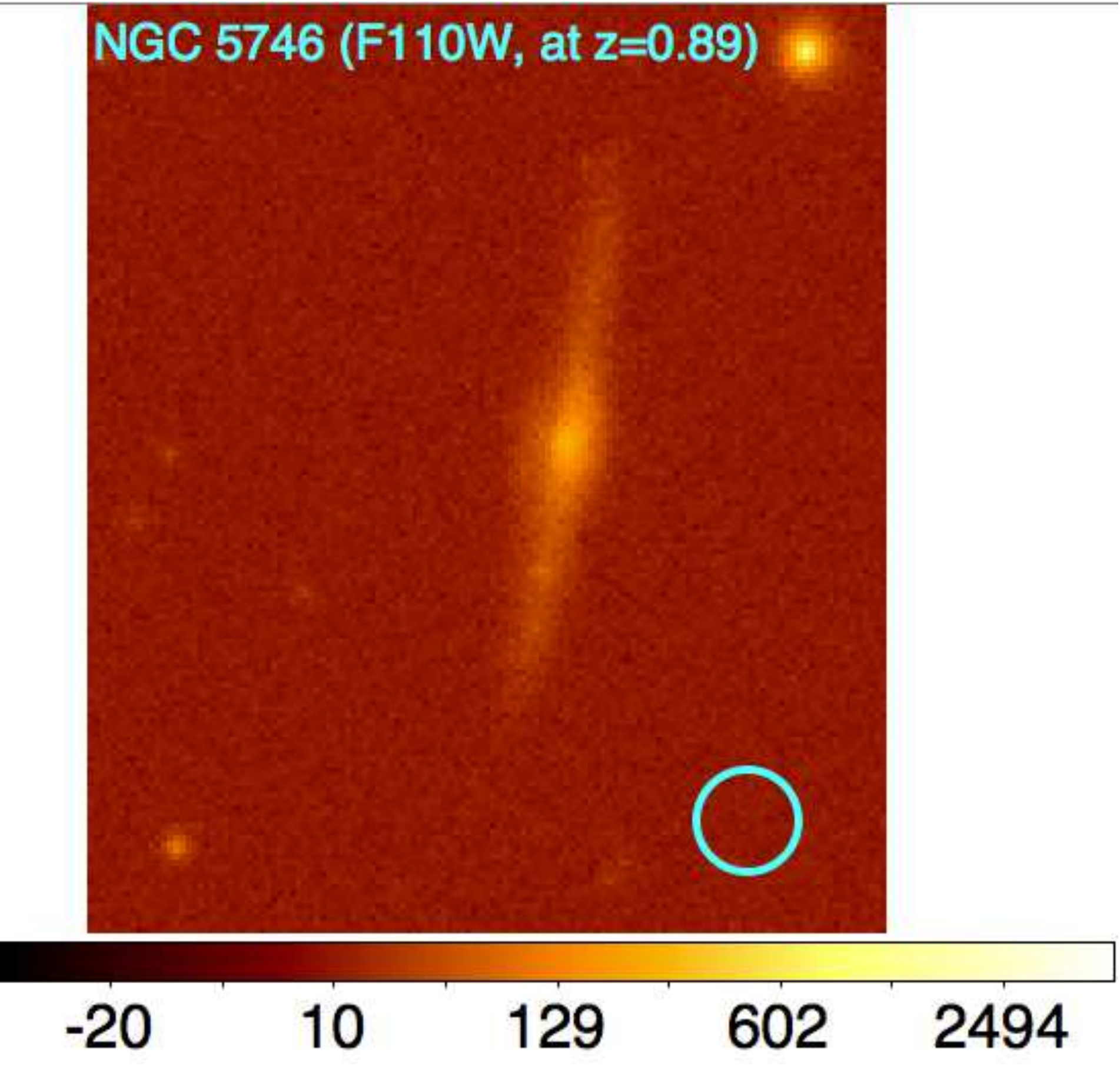}	 
	\includegraphics[scale=\smallscale]{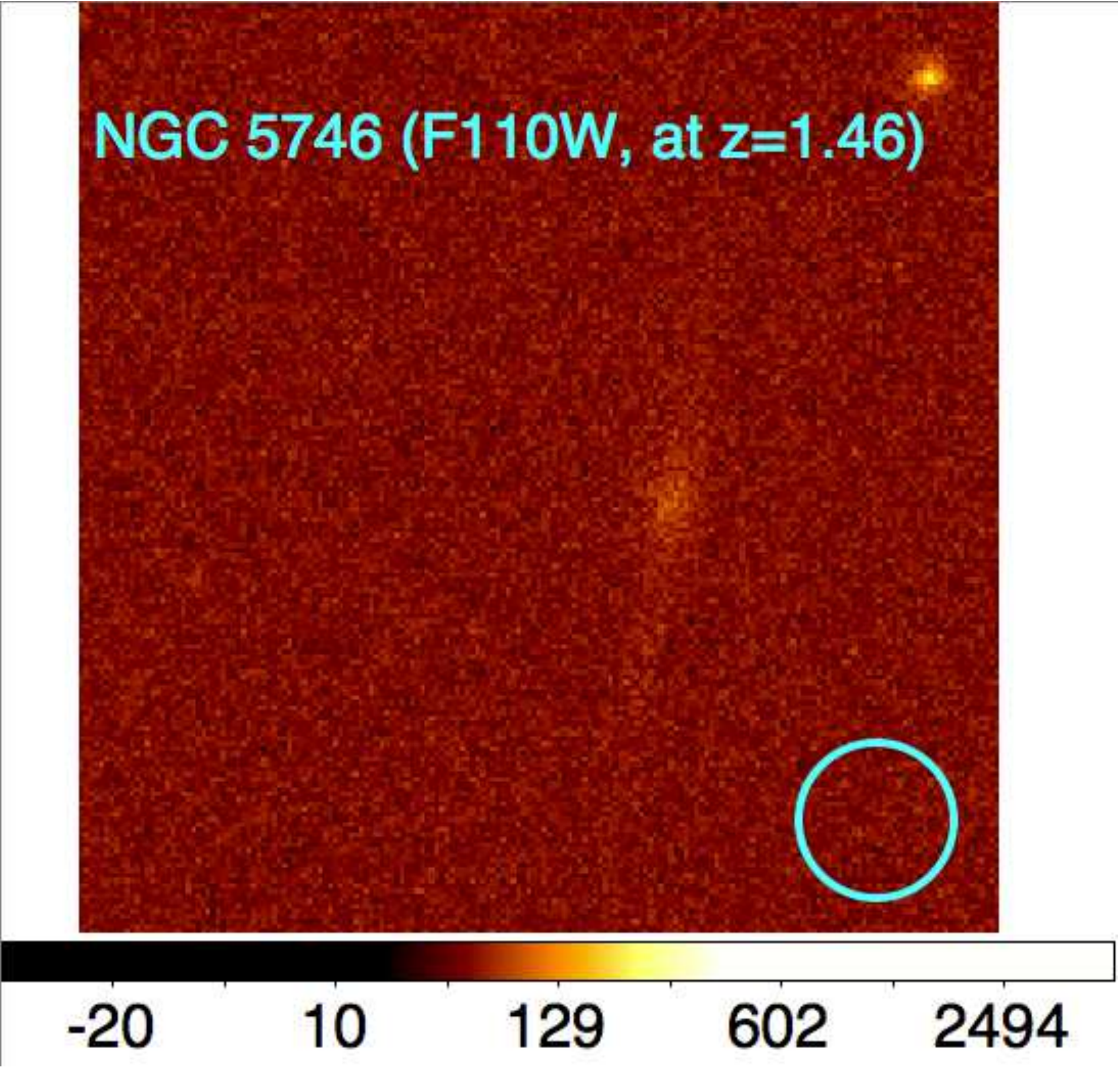}
	\includegraphics[scale=\smallscale]{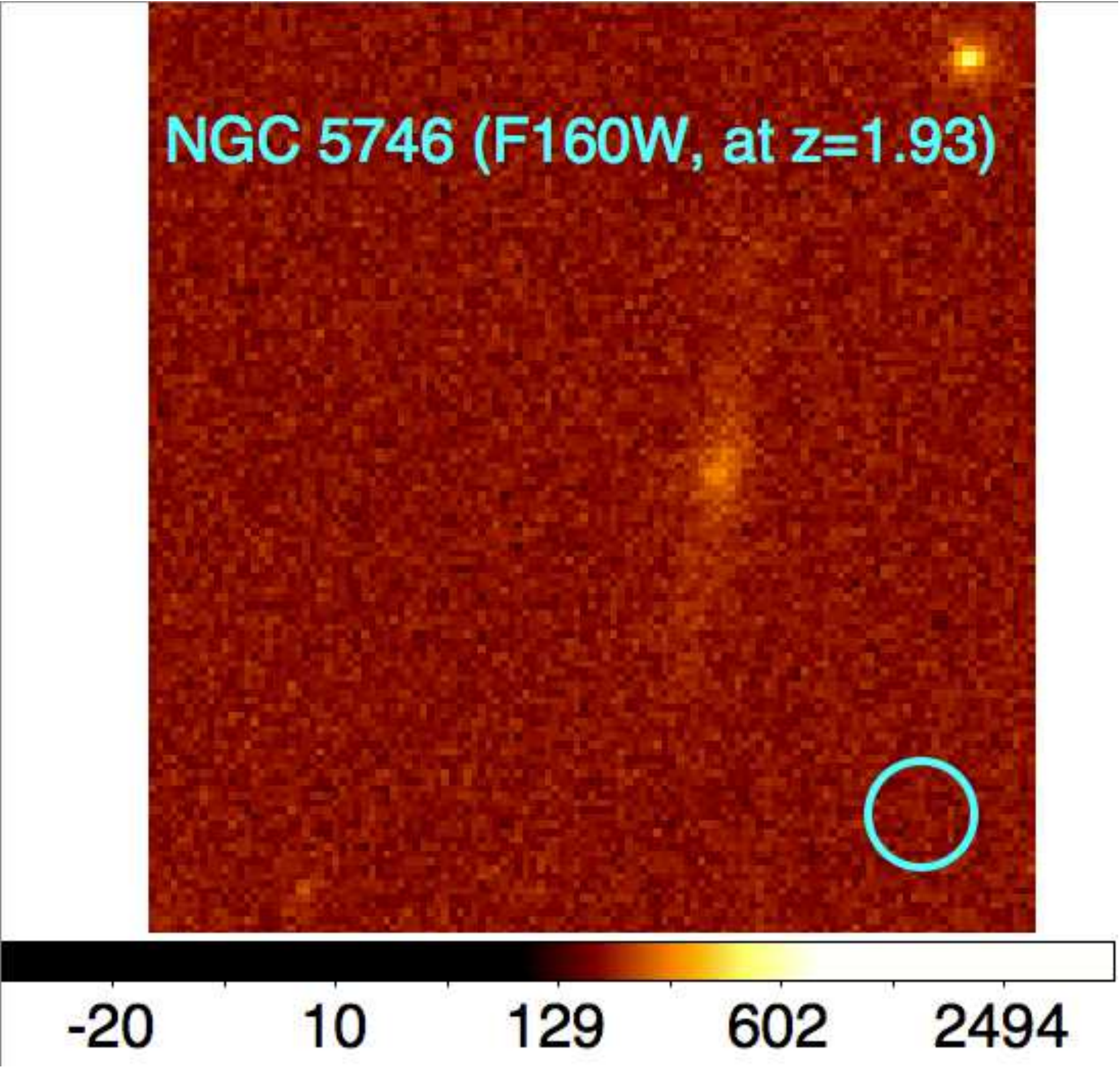}
	\includegraphics[scale=0.5]{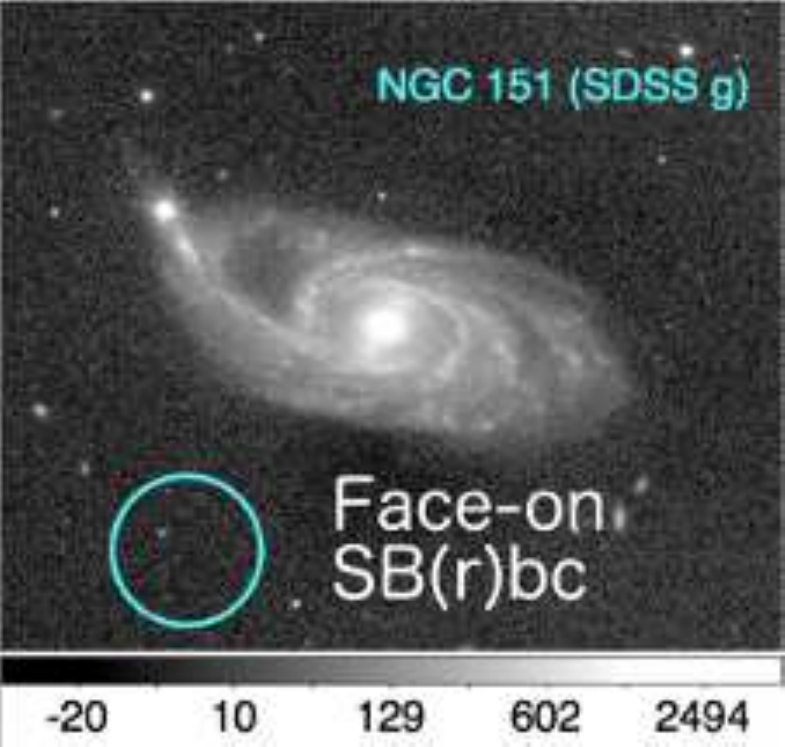}
	\includegraphics[scale=\smallscale]{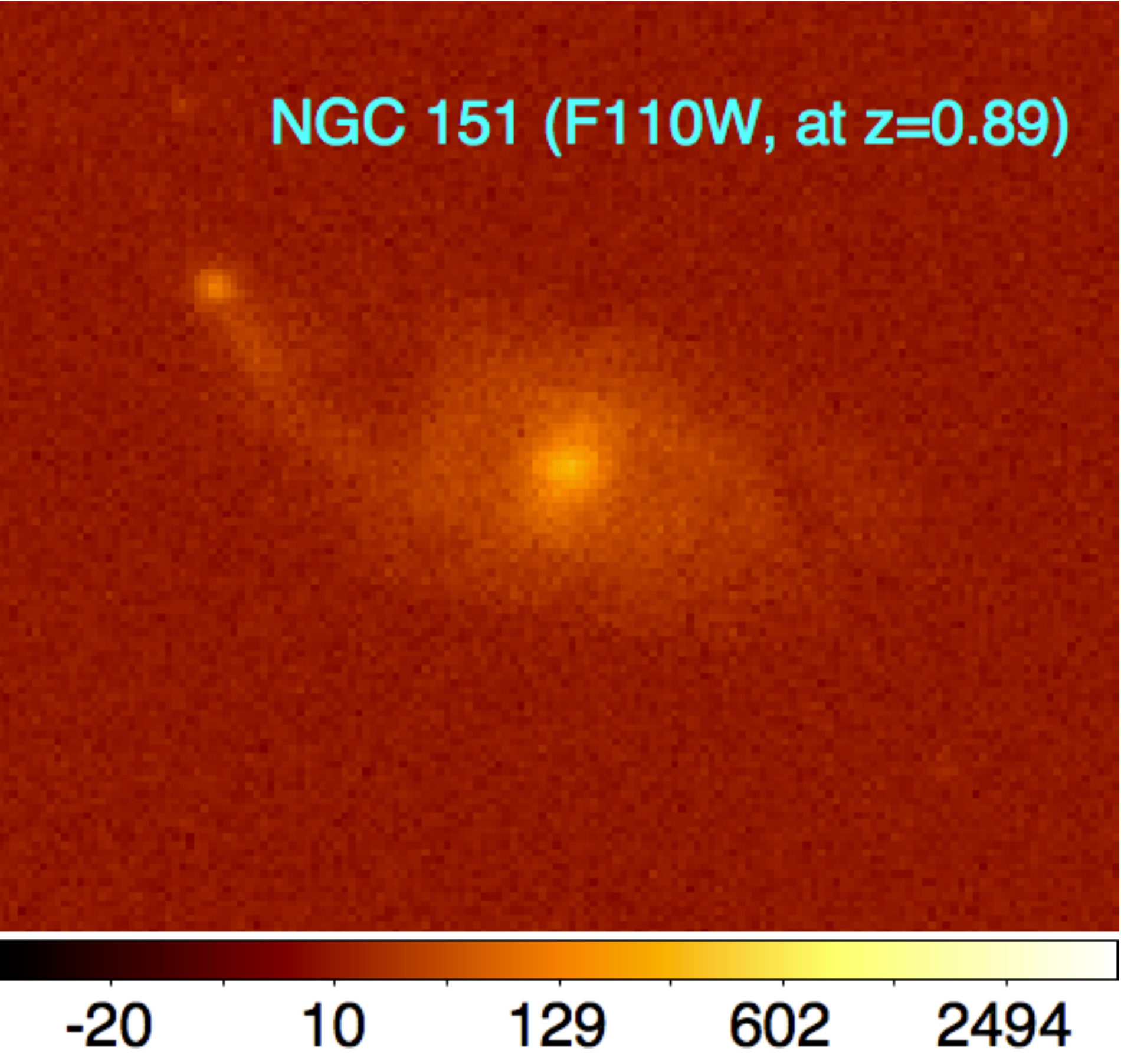}	 
	\includegraphics[scale=\smallscale]{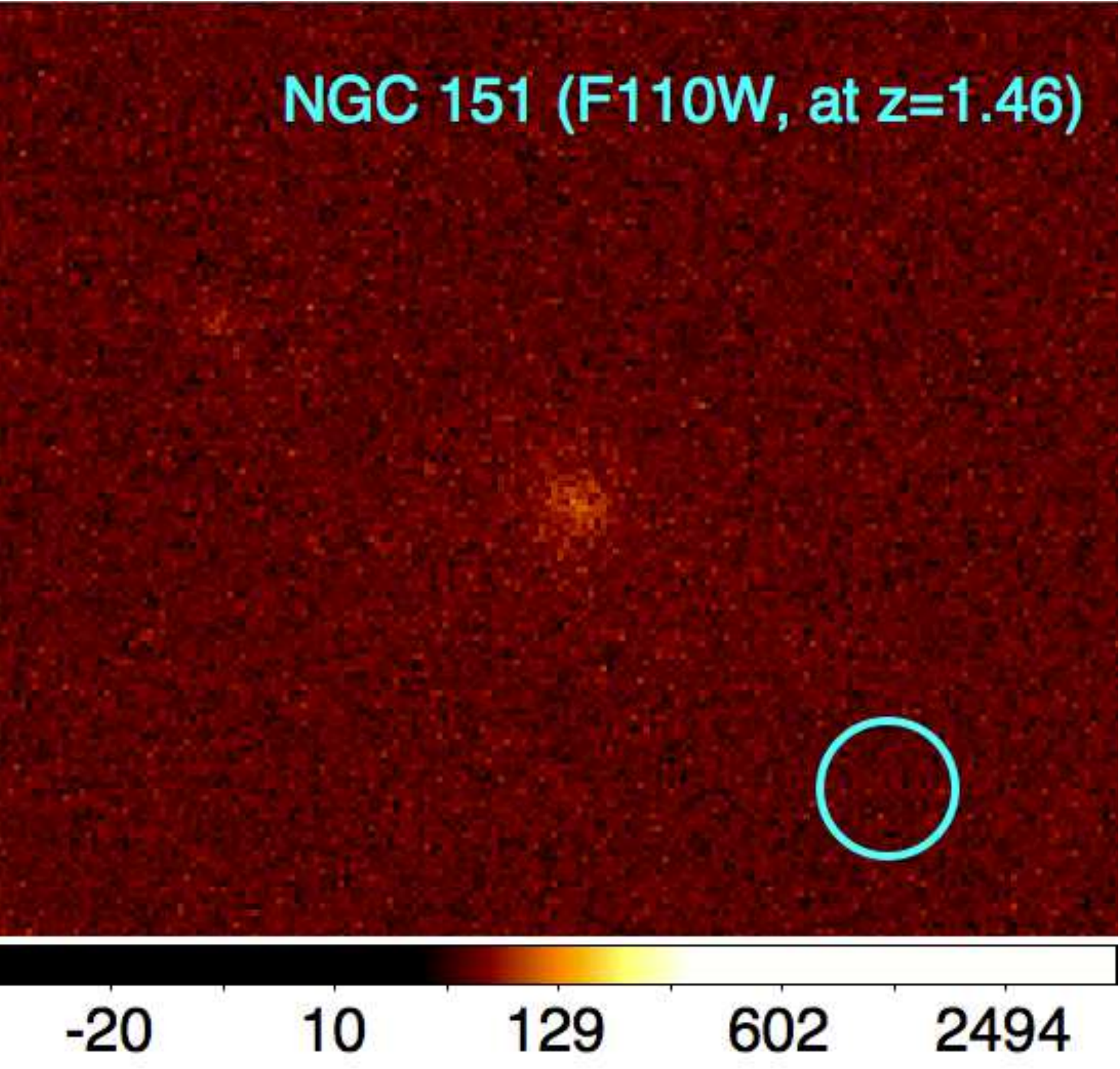}
	\includegraphics[scale=\smallscale]{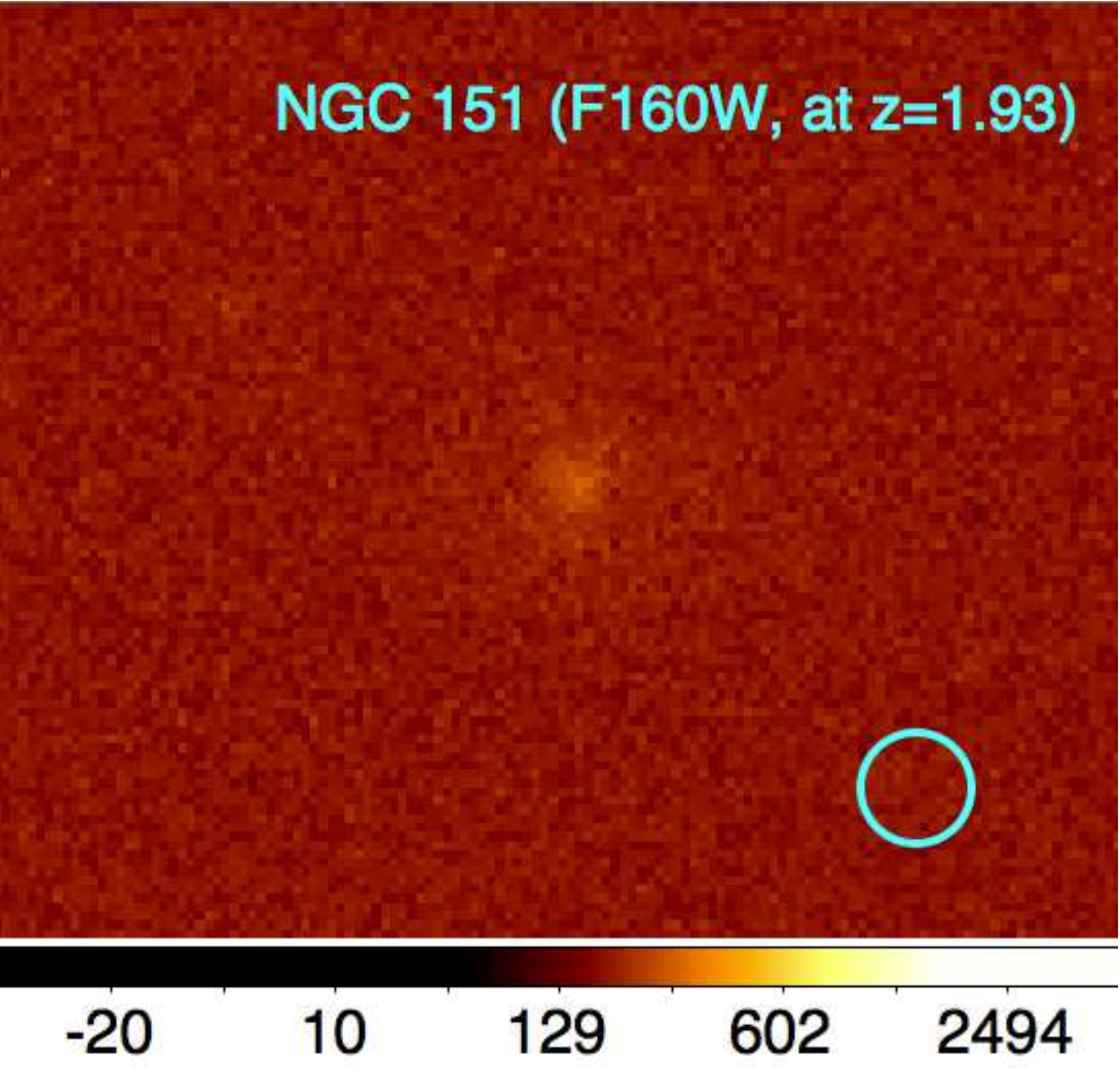}
	\caption{What would local galaxies look like if they were as distant as the Fe\-LoBAL quasars in the present sample? \textsc{Ferengi} images of five low-redshift galaxies as they would appear in NIC2 images at z=0.89, z=1.46 and z=1.93, given exposure times of 2688, 5141 and 8695 seconds, respectively. No luminosity evolution (Appendix \ref{sec:nicmos_sims}) is applied here. The leftmost column shows the input SDSS g-band images. Cyan circles represent a diameter of 1 arcminute in the SDSS image, 1 arcsecond in the \textsc{Ferengi} output images. The remaining template galaxy (NGC 6166) is shown in Figure \ref{fig:ferengi_galaxies_2}.}\label{fig:ferengi_galaxies}
\end{figure*}

\begin{figure*}
	\advance\leftskip-3cm
	\centering
	\includegraphics[scale=0.5]{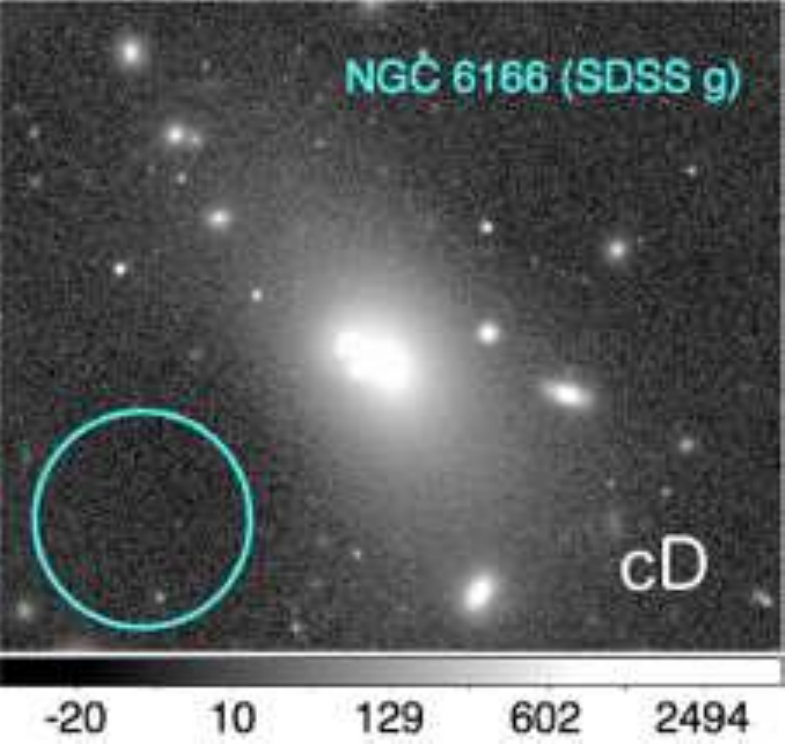}
	\includegraphics[scale=\smallscale]{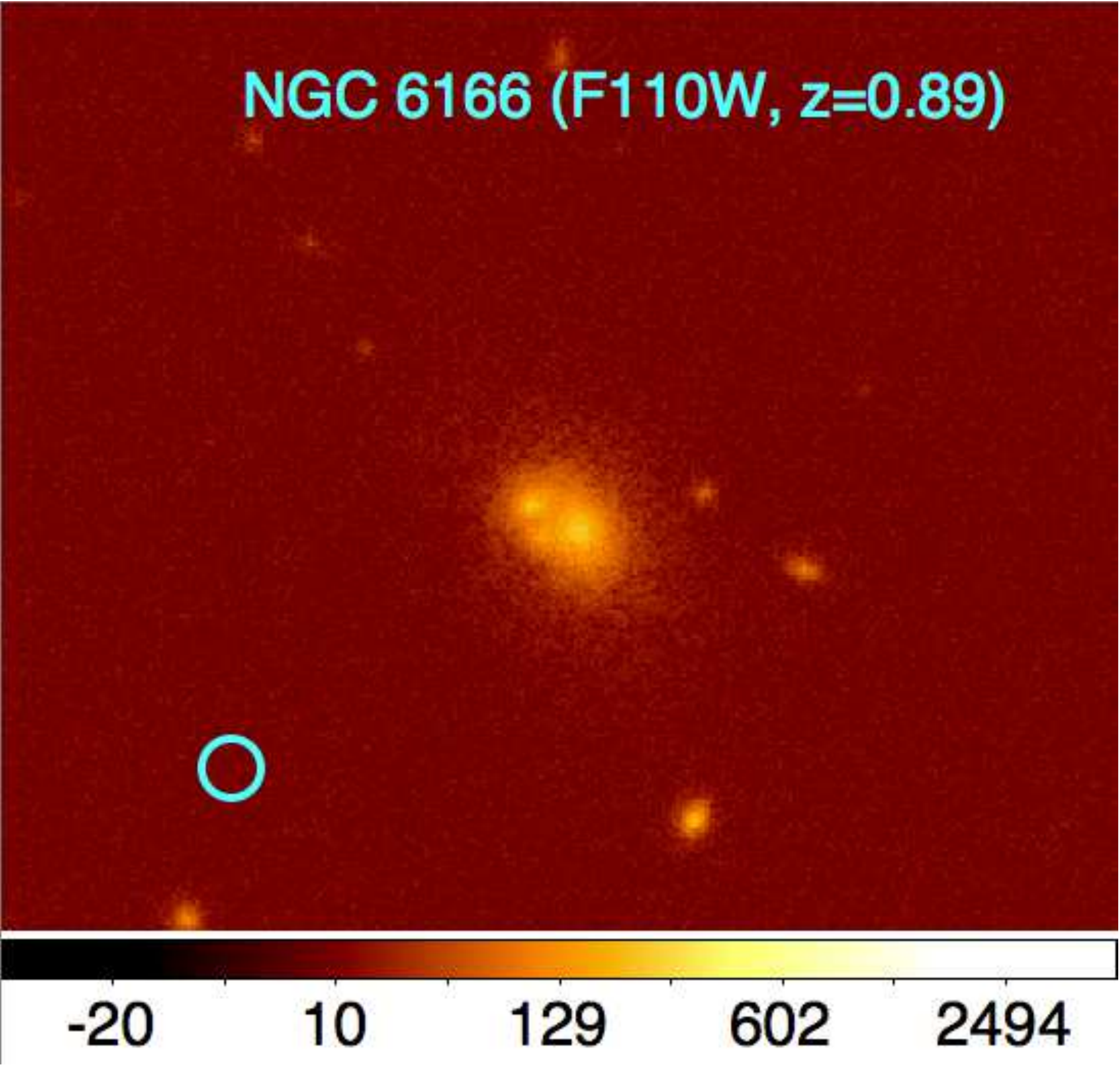}	 
	\includegraphics[scale=\smallscale]{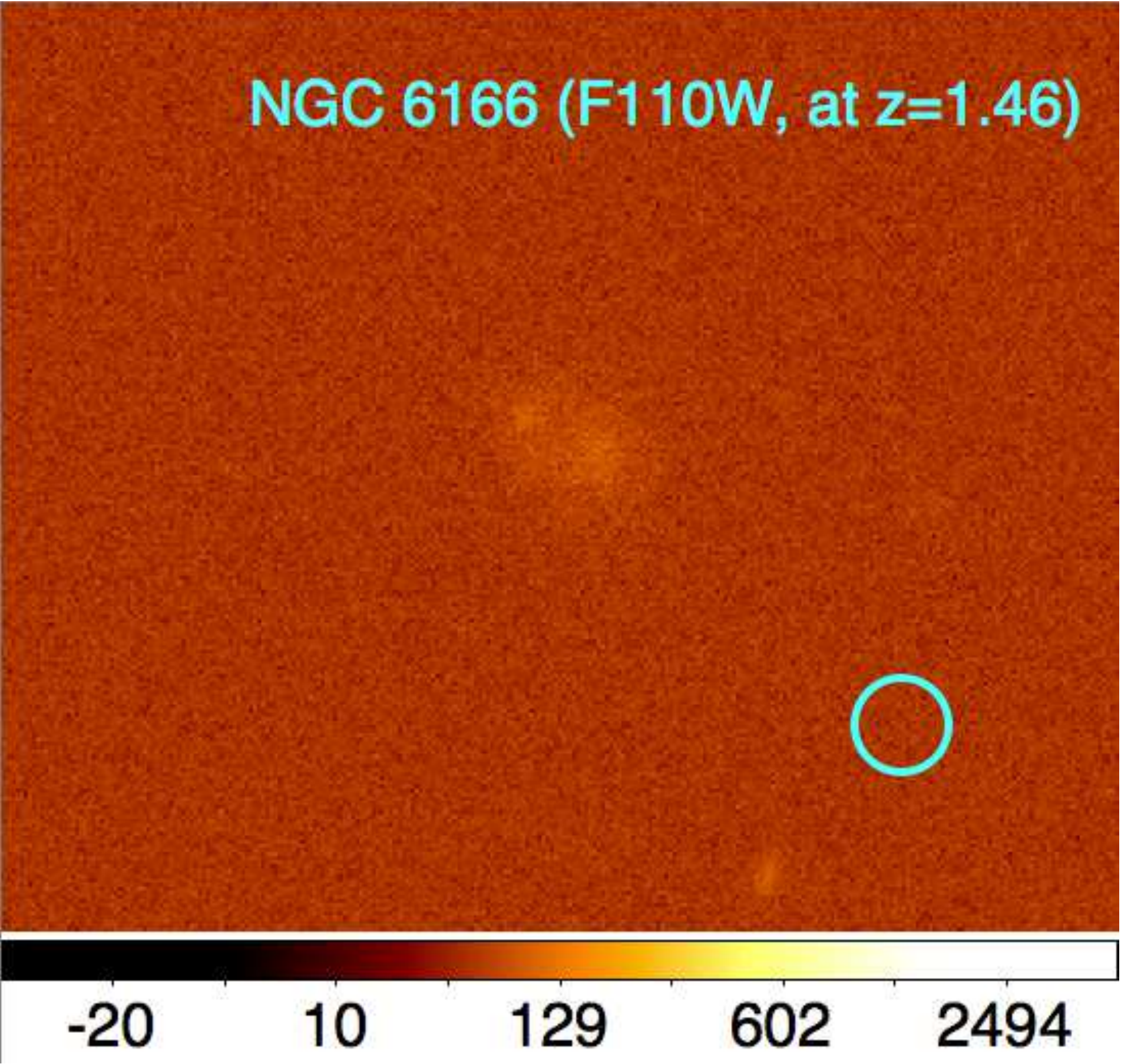}
	\includegraphics[scale=\smallscale]{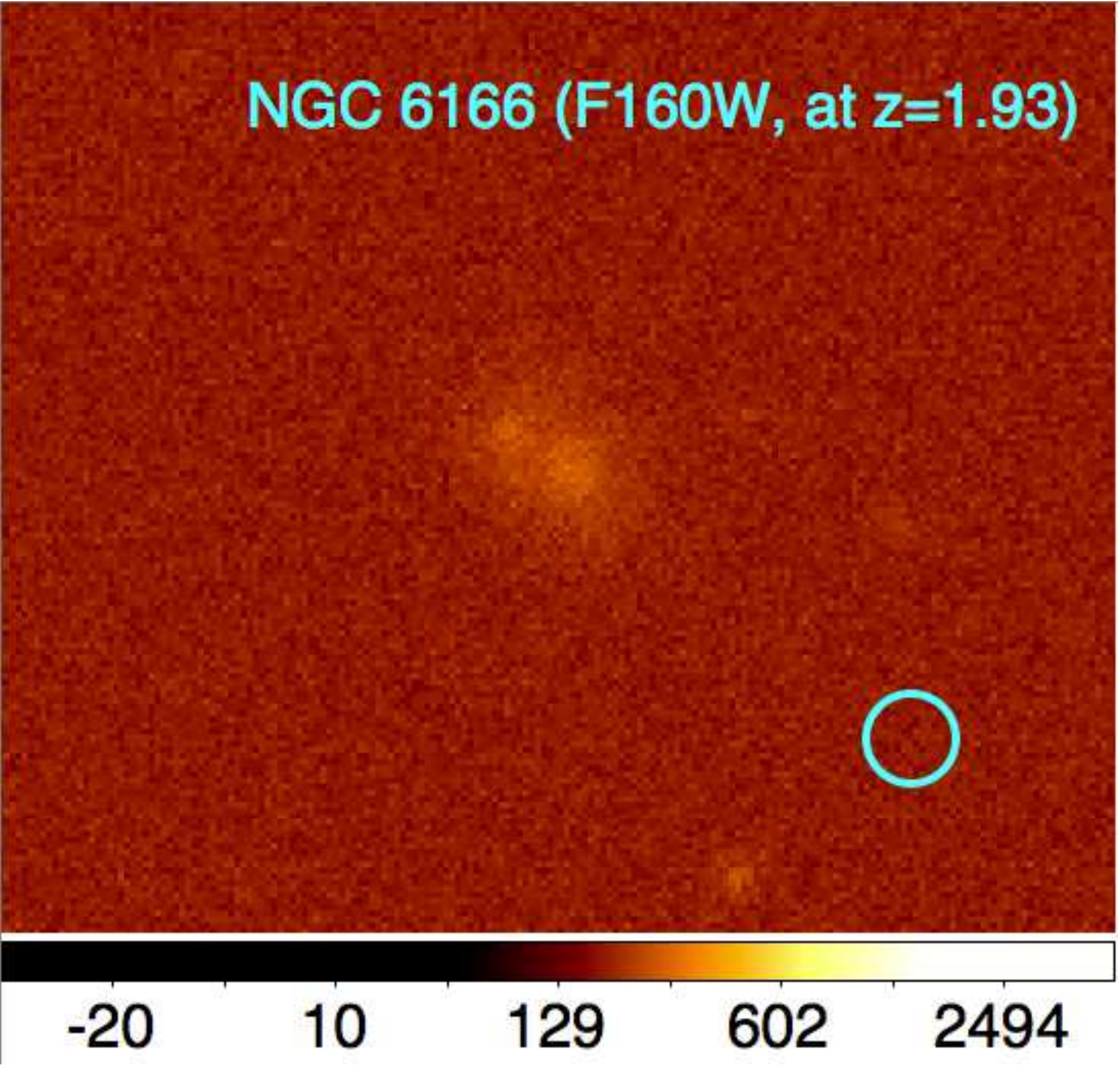}
	\caption{FERENGI output images for the final galaxy in our template sample. Panels arranged as in Figure \ref{fig:ferengi_galaxies}.}\label{fig:ferengi_galaxies_2}
\end{figure*}

\paragraph*{Accounting for luminosity evolution:}\label{sec:luminosity_evolution}\label{appendix:luminosity_evolution}

Galaxies are expected to dim with time throughout their post-starburst lifetime, due to the quiescent aging of the stellar populations \citep[e.g.,][]{Bruzual2003}. We find it necessary to include luminosity evolution for the simulated galaxies, as they otherwise become too faint at the source redshifts, especially compared to the host galaxy component we detect for J1154+0300. To explore the possibility of brightness evolution, we generate two sets of simulated NICMOS observations: one without brightness evolution, and another where the host galaxies become 1 mag brighter per unit redshift. We set this parameter so as to generate apparent brightnesses comparable to that which we measure for the host of J1154+0300 - it is, however, in broad agreement with \citet{Ramos2011}, who find that the Schechter characteristic galaxy luminosity decreases by 0.7 mag between $z = 1.8$ and $z = 0.3$ . Without luminosity evolution, the redshifted galaxy templates have apparent brightnesses as faint as 24th magnitude at $z=1.46$ (Table \ref{tab:psffits_ferengi_limits}); as we detect FeLoBAL hosts with apparent magnitudes of $\apprle$ 22 mag, it is appropriate to include luminosity evolution in our simulations.

\paragraph*{Tests of PSF-Host Decomposition:}\label{sec:ferengi_fitresults}

To quantify the uncertainties of our PSF-host decomposition method, we add a point-source component to the centroids of the galaxy templates in the \textsc{Ferengi} output images. We utilize one or two stacked observations of the NICMOS calibration star P330-E (Appendix \ref{sec:appendix_psf}) to provide the point source; the remaining P330-E observations in the appropriate filter are then used as a PSF template for the \textsc{Galfit} analysis. This provides an approximation of the temporal PSF variation relevant to our study. For each galaxy template, at each output redshift, we generate a range of images at host-to-nucleus brightness contrasts $0\le\contrast\le4$ mag, in increments of 0.25 mag. These images are analyzed identically to the real NICMOS observations, i.e., we firstly perform PSF-only modeling to determine whether a host galaxy is detected, and secondly perform PSF-plus-Sersic modeling to determine the host galaxy brightness and scale size.

For the PSF-only fits, Table \ref{tab:psffits_ferengi_limits} presents the limiting nucleus-to-host brightness contrasts beyond which we can no longer detect extended emission. The limiting brightness contrast depends on both the host galaxy type and on whether any luminosity evolution is applied; without luminosity evolution, the limiting contrast values are small (reaching 0.5 mag for the elliptical/S0 template at $z=1.93$). We note that the host galaxy detections presented in \S \ref{sec:fitting} are brighter than the no-evolution templates at $z>1$. Thus, while the no-evolution simulations indicate that we would not detect \emph{typical local-Universe galaxies} at $z>1$ (where they have \mhost$\apprge23$ mag), the with-evolution simulations confirm that we indeed are able to detect a $\sim$22nd magnitude host at $z=1.46$. Further supporting the FeLoBAL host detections, we note that any false-positive detections in NICMOS imaging are much fainter than \mhost$\sim22$ mag, according to our PSF tests (Appendix \ref{sec:appendix_nicmos}).

\begin{figure*}
	\advance\leftskip-5cm
	\centering
	\includegraphics[scale=0.40,trim=80 20 50 50]{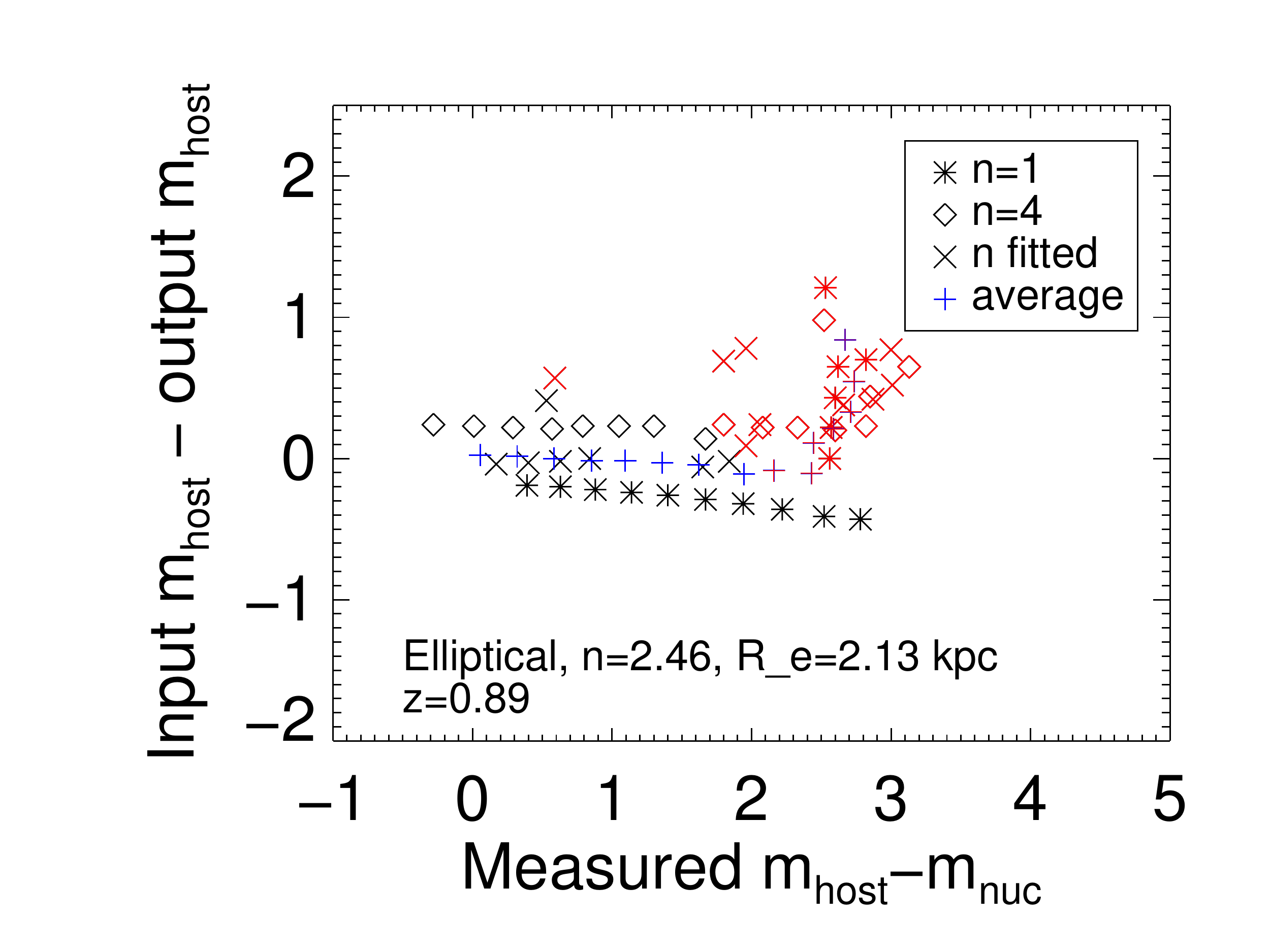}
	\includegraphics[scale=0.40,trim=50 20 50 50]{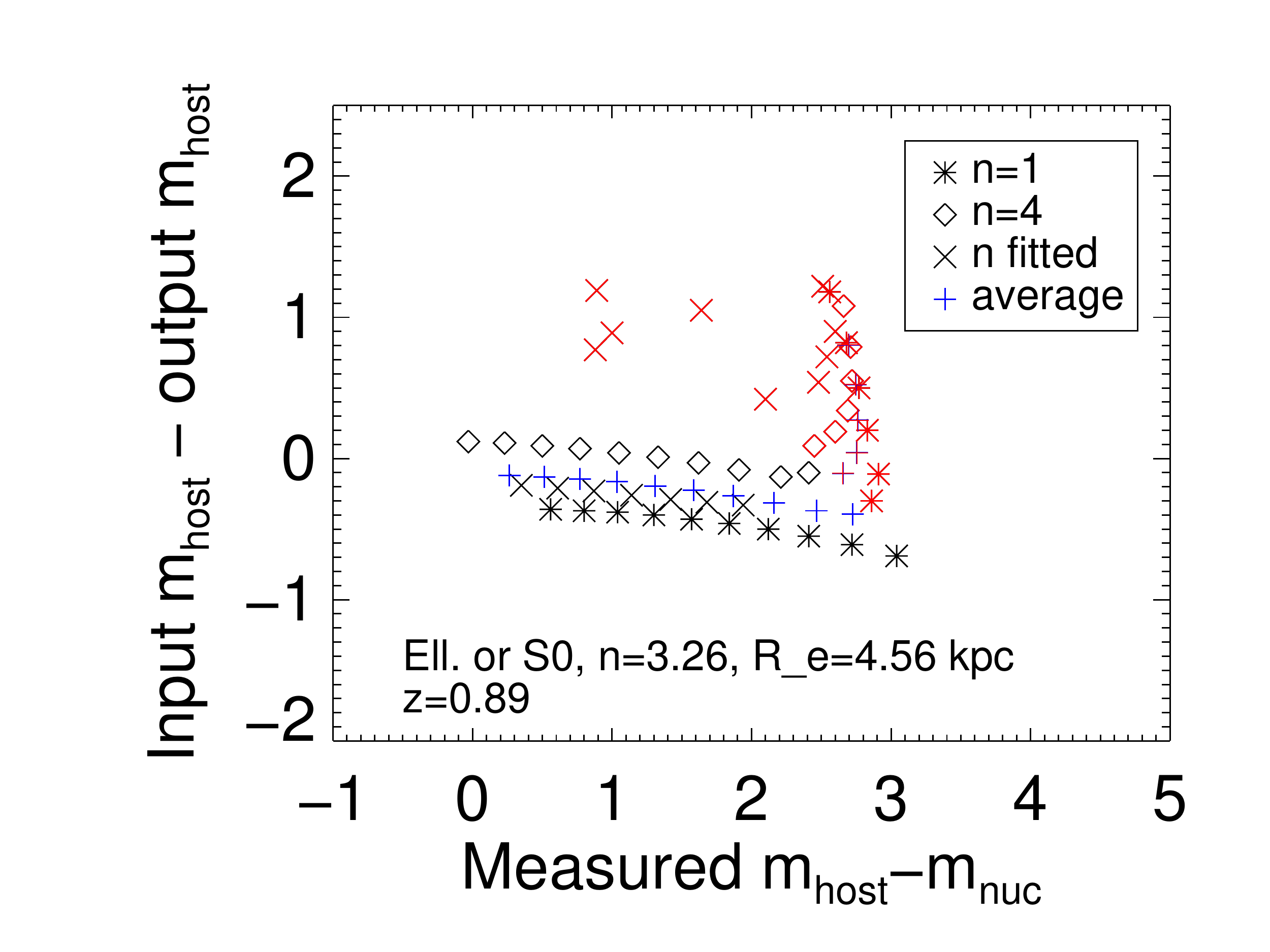}
	\includegraphics[scale=0.40,trim=80 20 50 50]{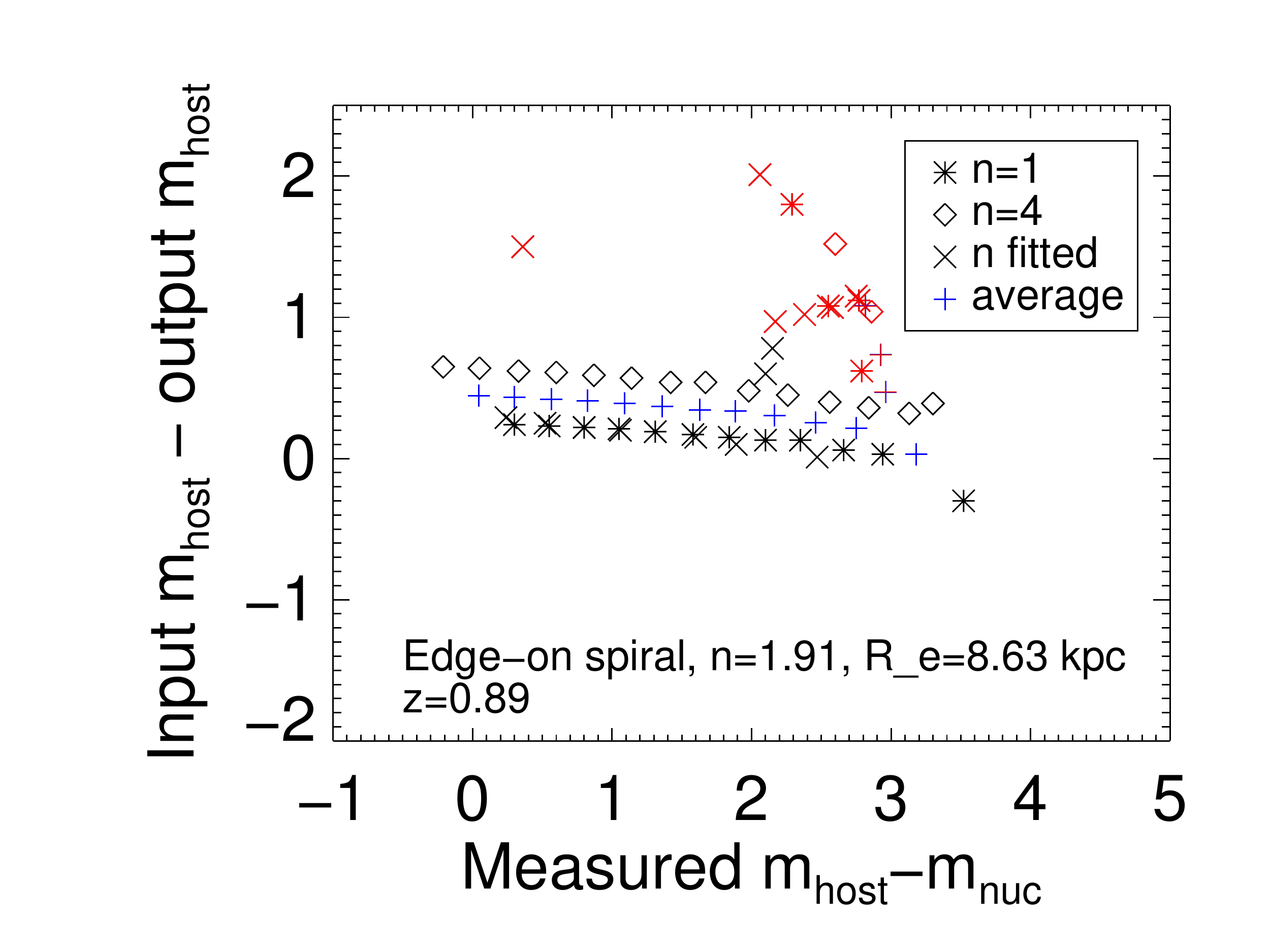}
	\includegraphics[scale=0.40,trim=50 20 50 50]{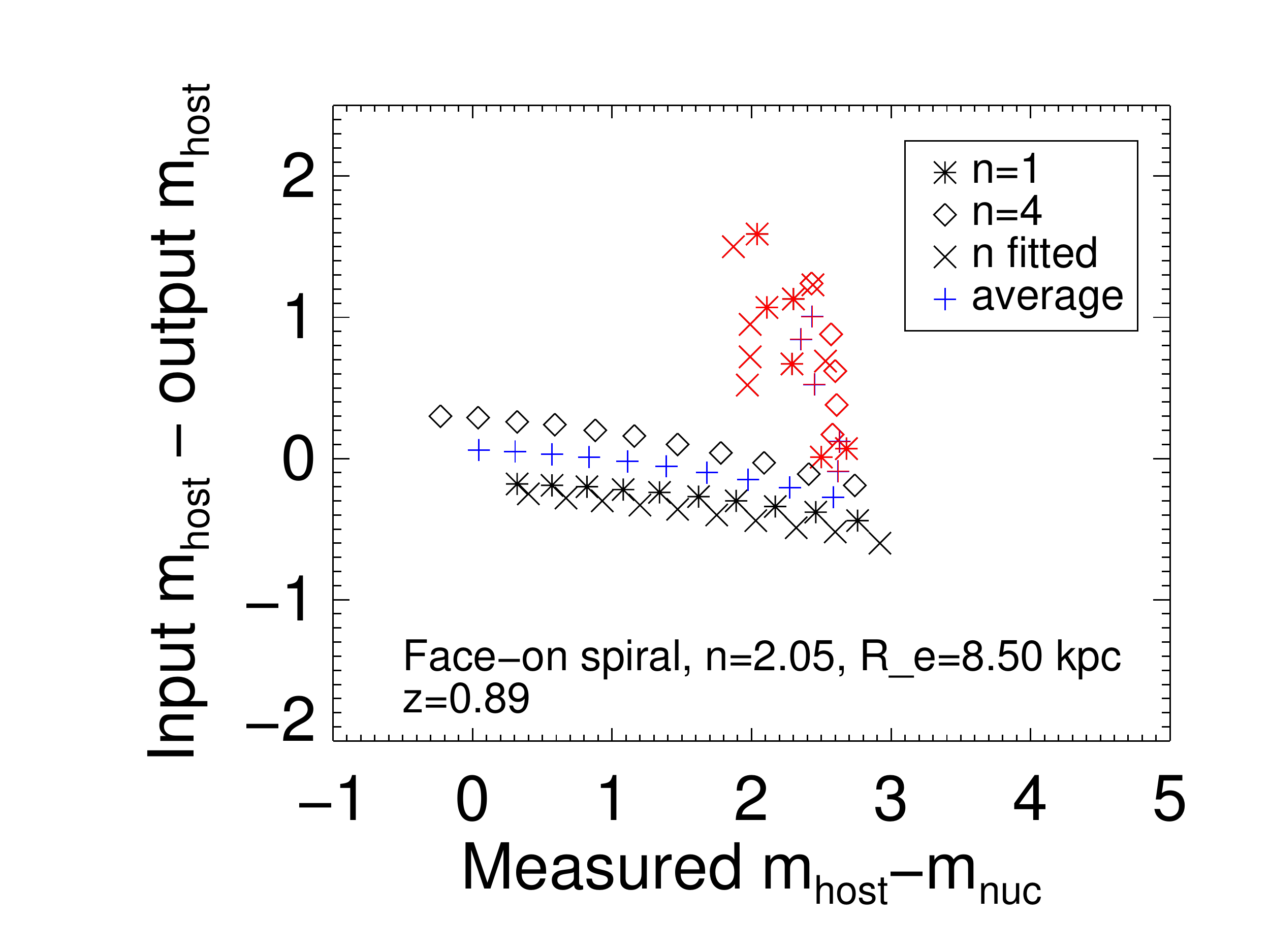}
	\includegraphics[scale=0.40,trim=80 20 50 50]{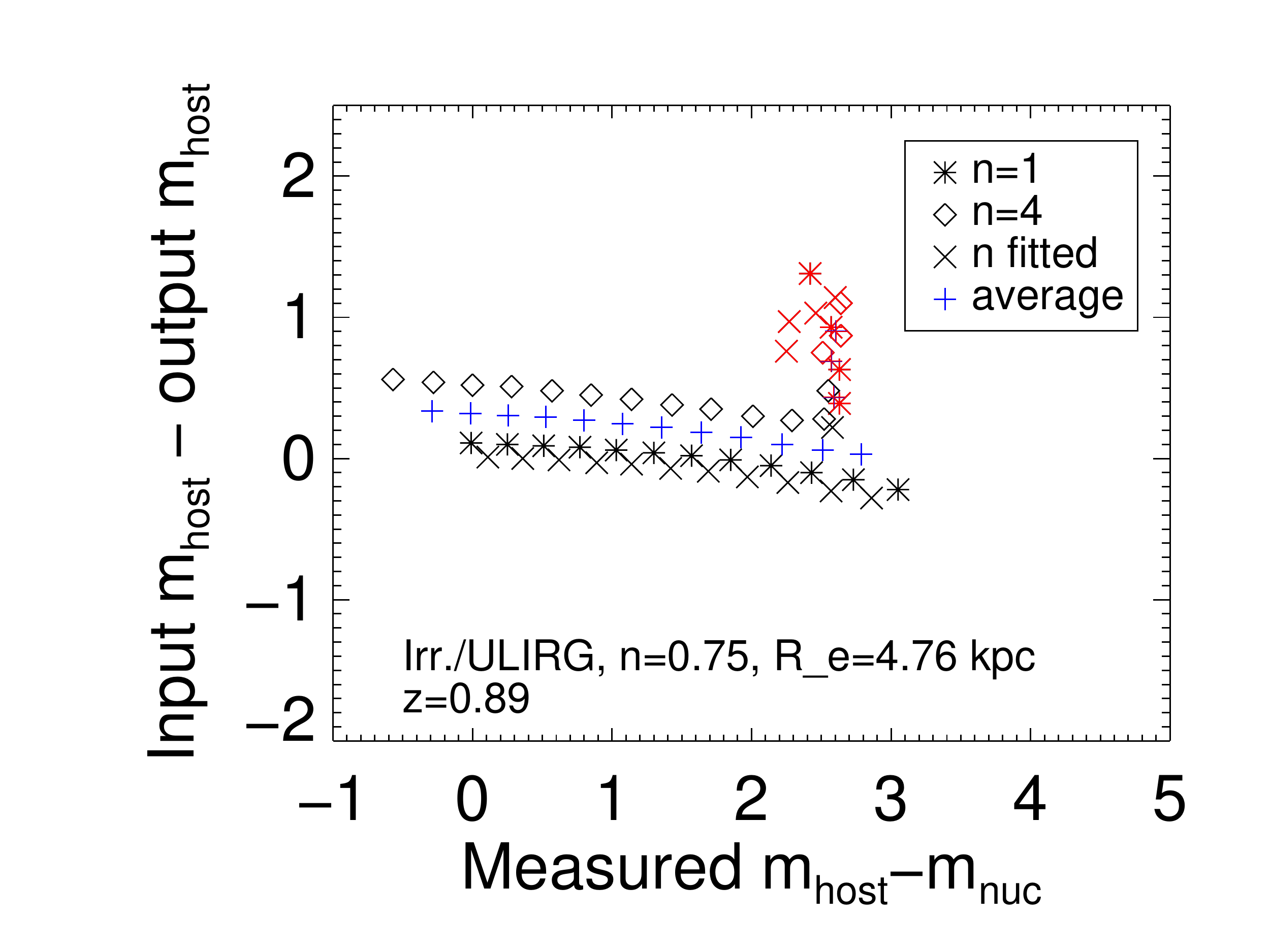}
	\includegraphics[scale=0.40,trim=50 20 50 50]{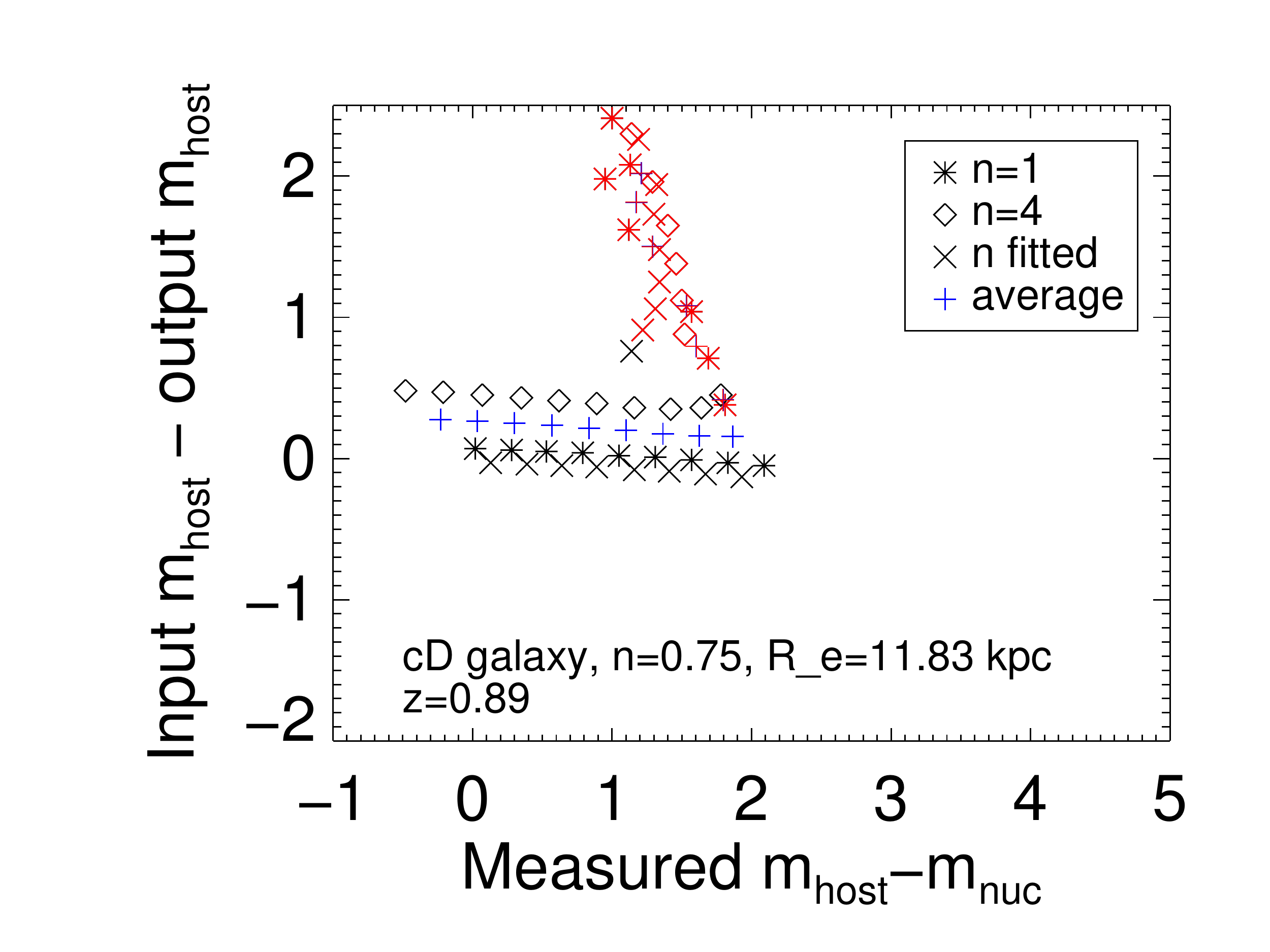}	
	\caption{Uncertainty in measured host galaxy magnitudes for simulated quasar images at $z=0.89$ as observed using NIC2, with $0\le\contrast\le4$ mag, in increments of 0.25 mag. The horizontal axis shows \contrast\, as measured using GALFIT. The vertical axis shows the difference between input and measured $m_\mathrm{host}$, i.e., positive values on the $y$ axis correspond to underestimated \mhost\,(thus, overestimated host brightness). Input \mhost\, is measured using a single Sersic component, prior to adding a superimposed point source to the image. Black crosses show models for which the Sersic index is fitted as a free parameter, while diamonds and asterisks denote constant Sersic indices; we note that the free-$n$ fits are often further away from the input value than either of the constant-$n$ models. Blue plus-symbols show the average (in log-space) $m_\mathrm{host}$ and average $m_\mathrm{nuc}$ between the $n=4$ and $n=1$ models. Models which converge with $R_e$ equal to the PSF FWHM have red symbols; these fits are typically unreliable, and tend to overestimate \mhost.}\label{fig:0300_nic2_sim_hostfits}
\end{figure*}

\begin{figure*}
	\advance\leftskip-5cm
	\centering
	\includegraphics[scale=0.40,trim=80 20 50 50]{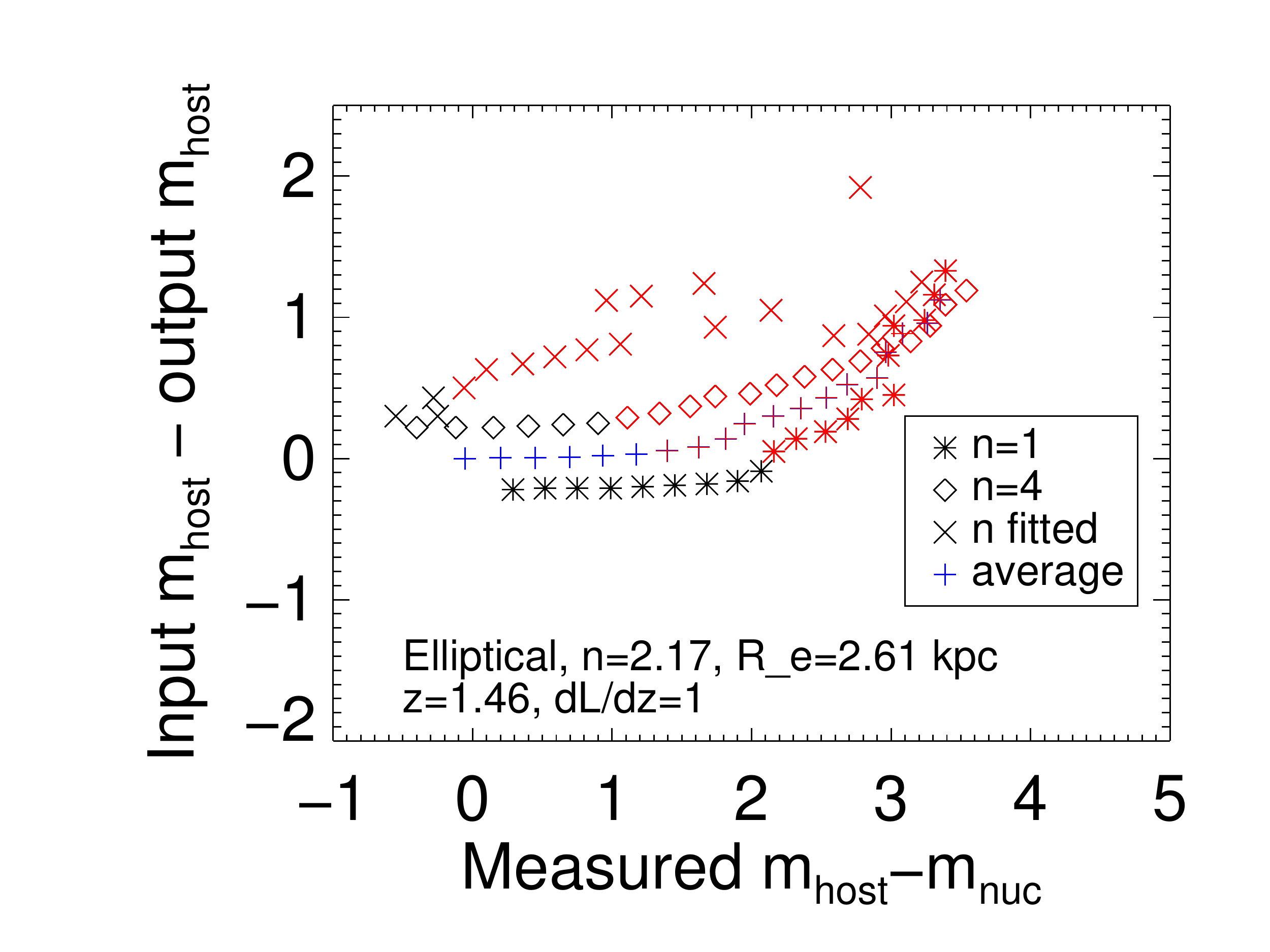}
	\includegraphics[scale=0.40,trim=50 20 50 50]{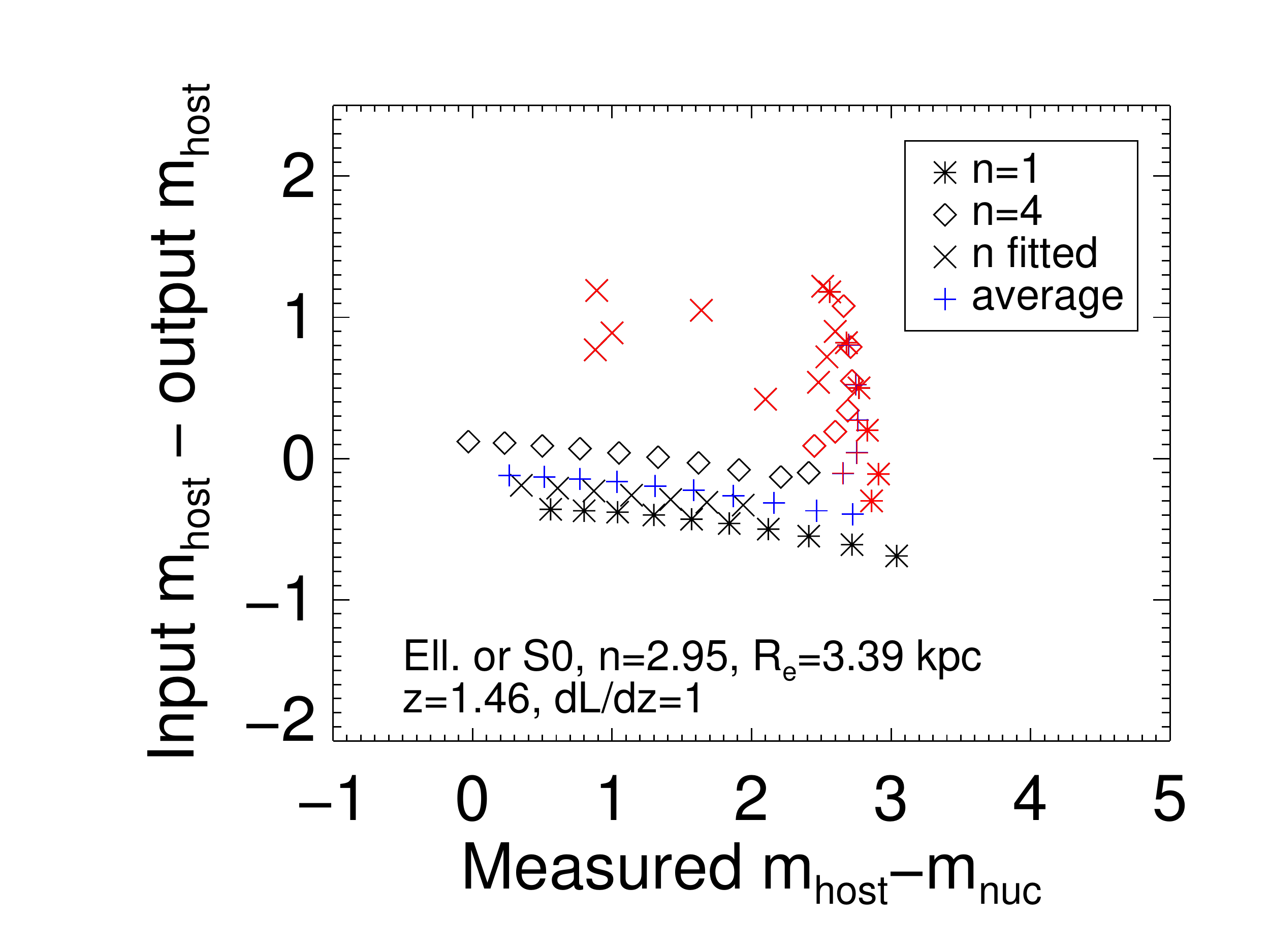}
	\includegraphics[scale=0.40,trim=80 20 50 50]{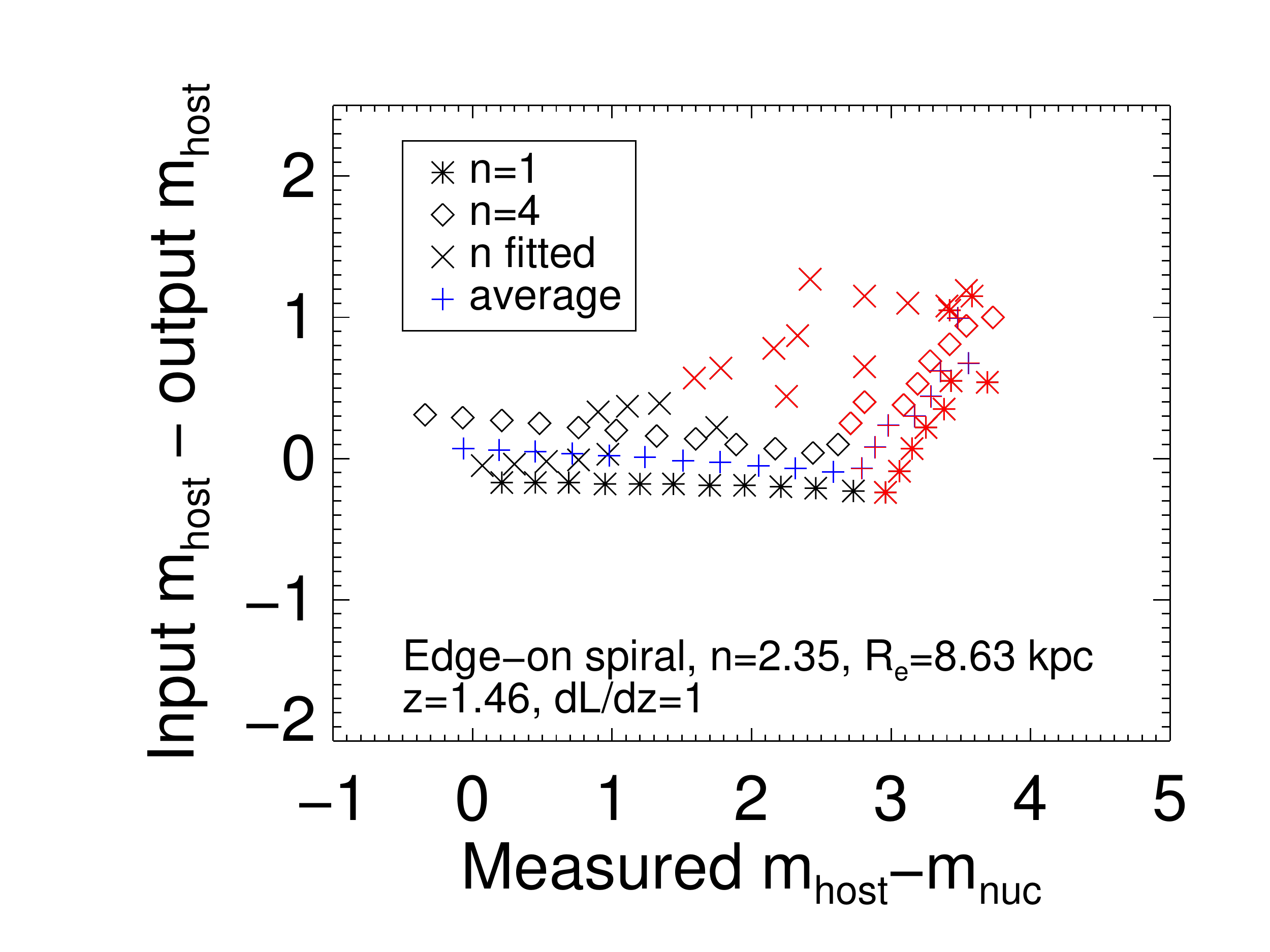}
	\includegraphics[scale=0.40,trim=50 20 50 50]{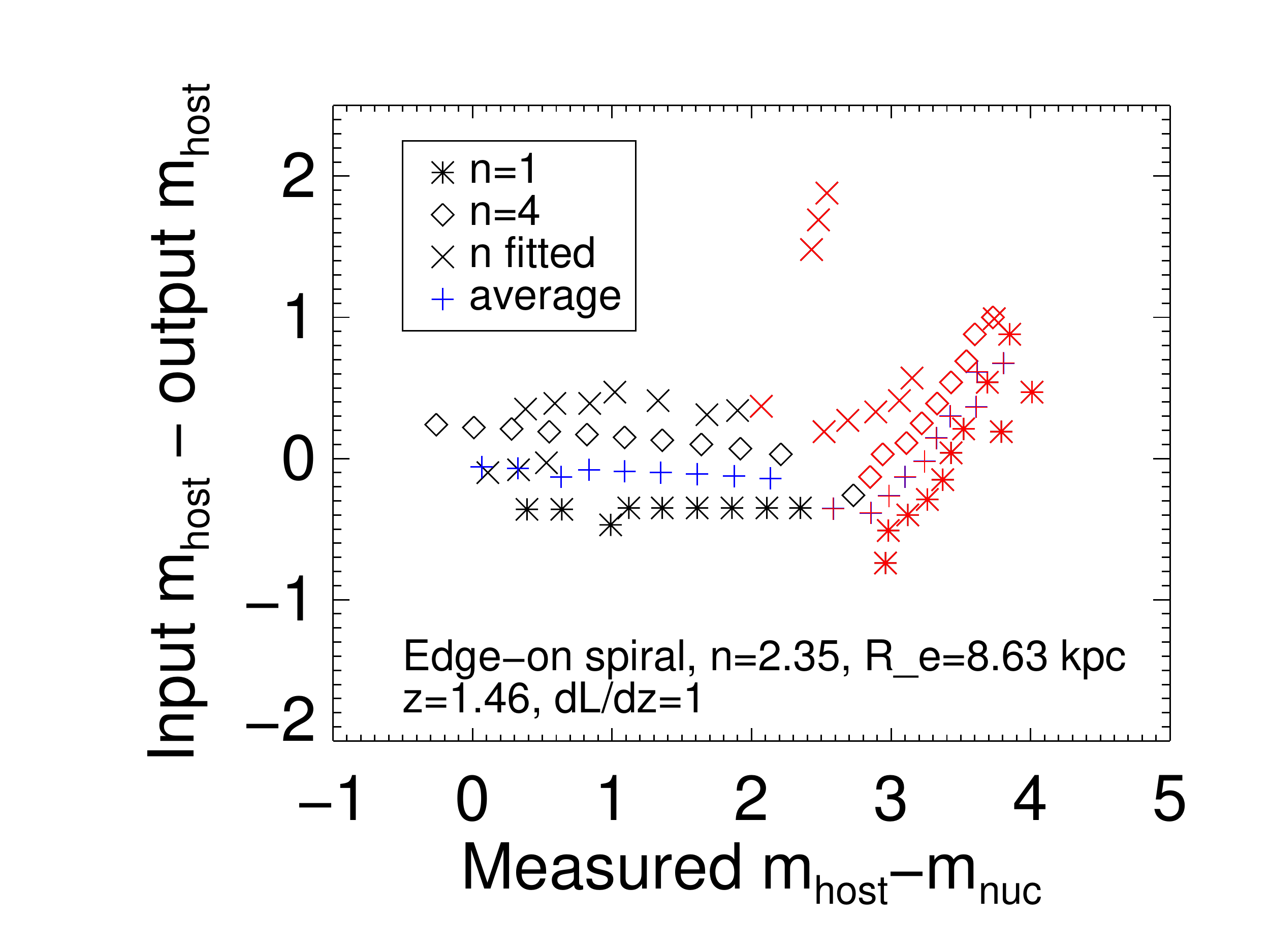}
	\includegraphics[scale=0.40,trim=80 20 50 50]{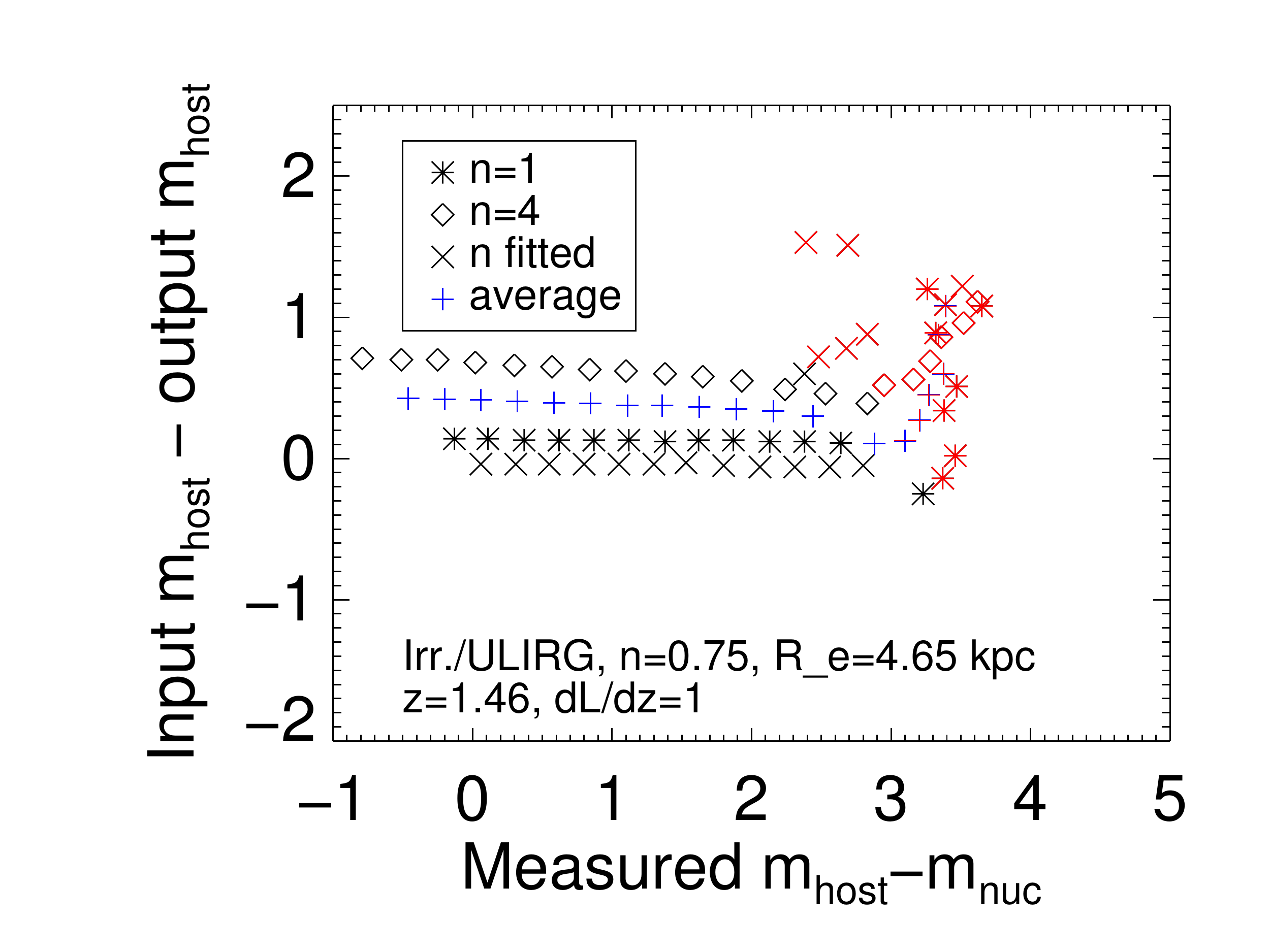}
	\includegraphics[scale=0.40,trim=50 20 50 50]{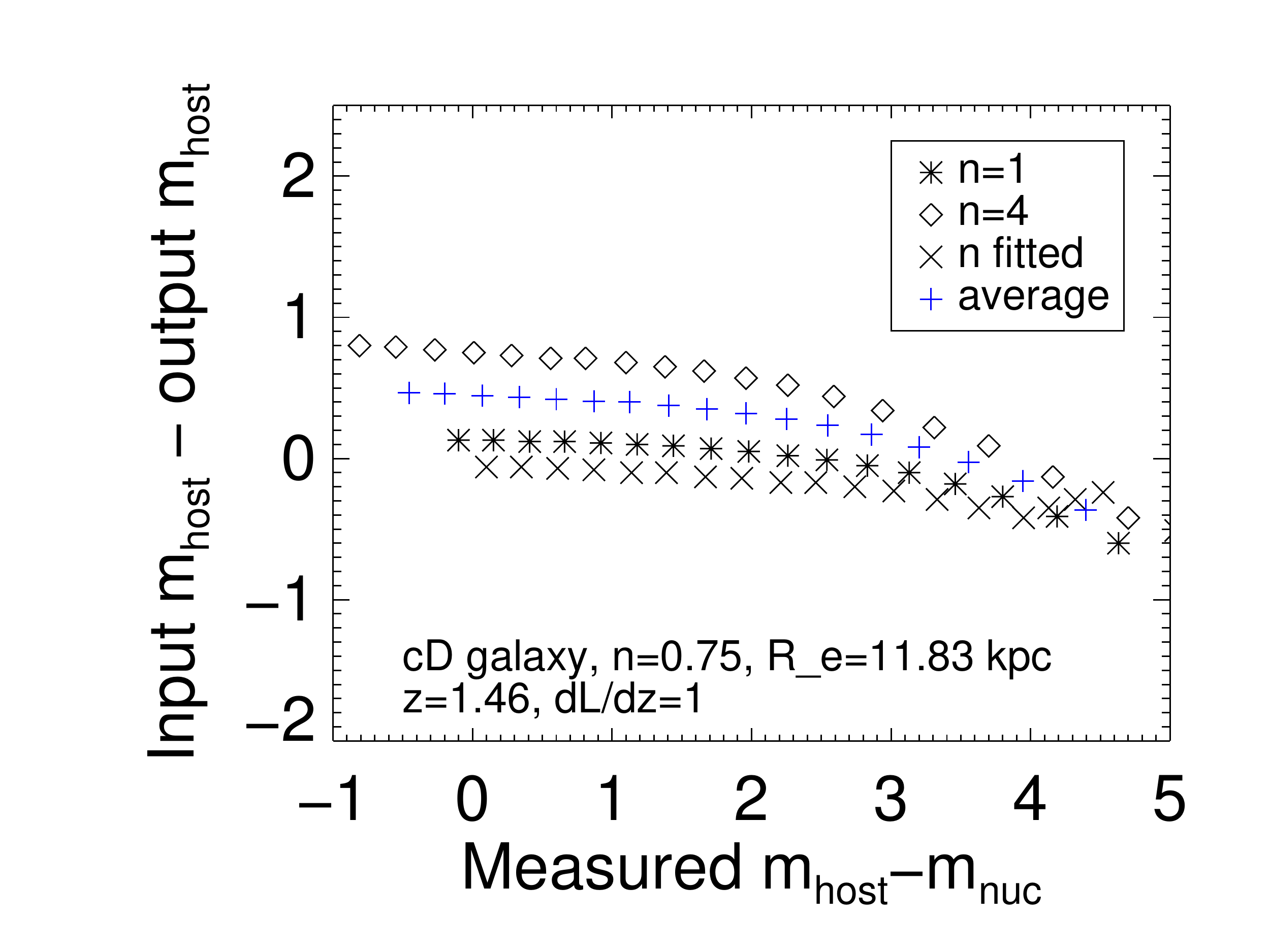}	
	\caption{Error in measured host galaxy magnitudes for simulated quasar images at $z=1.46$ as observed using NIC2, assuming 1 mag of luminosity evolution per unit redshift, such that the host galaxies in the simulated observations are artificially brightened compared to the local-Universe template galaxies, as described in \S \ref{sec:luminosity_evolution}. Symbols coded as for Figure \ref{fig:0300_nic2_sim_hostfits}.}\label{fig:simfits_z146_evo}
\end{figure*}

For the PSF-plus-Sersic fits, Figures \ref{fig:0300_nic2_sim_hostfits} and \ref{fig:simfits_z146_evo} show the error on the measured host galaxy brightness as a function of the measured $m_\mathrm{host}-m_\mathrm{nuc}$; points located along the horizontal line `input \mhost - output \mhost = 0' represent ideal decompositions. The main source of uncertainty for \mhost\,is our ignorance of the Sersic index. This parameter is not well-constrained (in terms of the $\chi^2_\nu$ fit statistic) in our FeLoBAL analysis; this is in general also true of the `simulated quasars' discussed here. Assuming no prior knowledge of the Sersic index, `randomly' choosing a model with a constant Sersic index ($n=1$ or $n=4$) can lead to an error on \mhost\,of $\sim1$ mag. On the other hand, models for which $n$ is fitted as a free parameter can diverge drastically from the correct \mhost. Taking the average (in log-flux space) of the brightnesses of the $n=1$ and $n=4$ models provides a more reliable measurement of \mhost\,compared to either a freely fitted Sersic index, or assuming a single value of $n$. We therefore use these averaged values in our FeLoBAL analysis, instead of relying on any one model.

Unresolved host galaxy components (for which the best-fit $R_e$ converges on the lower bound, $R_e=\mathrm{FWHM}_\mathrm{PSF}$) tend to underestimate \mhost\,(i.e., overestimate the host flux) by up to 2 mag; we show these unresolved fits using red symbols in Figures \ref{fig:0300_nic2_sim_hostfits} and \ref{fig:simfits_z146_evo}. \citet{Simmons2008} report similar findings in their simulated ACS imaging of AGN host galaxies at $z\apprle1$: when the host component $R_e$ becomes small, the measured \mhost\,tends to be inaccurate by up to 1.3 mag. Thus, $R_e>\mathrm{FWHM}_\mathrm{PSF}$ is an important diagnostic for our modeling methodology. In particular, the recovered host component for J0819+4209 is unresolved, suggesting that our decomposition may severely overestimate the host flux. We therefore assume an uncertainty of 1.3 mag towards fainter \mhost\,for this quasar. For the quasars with resolved host galaxy components (J0300+0048 and J1154+0300), our simulations indicate that the uncertainty on \mhost\,is of order 0.5 mag.

\paragraph*{Summary of NICMOS simulations:} Due to our ignorance of the host galaxy Sersic index, quoting an average measured $m_\mathrm{host}$ between $n=1$ and $n=4$ models reduces the uncertainty on the host galaxy brightness compared to using $m_\mathrm{host}$ measured for either model directly. Applying this averaging approach, our PSF-plus-Sersic modeling of the Fe\-LoBAL quasars measures the host galaxy magnitude with an accuracy of $\pm0.5$ mag or less for quasars for which the best-fit models are resolved. For unresolved host galaxies (i.e., those for which the fit converges at the limiting $R_e$), the uncertainty on the host galaxy brightness is much larger, up to 1.3 mag.

\subsection{Simulated ACS Observations}\label{sec:acs_sims}

\begin{figure*}
	\advance\leftskip-3cm
	\centering
	\includegraphics[scale=0.21]{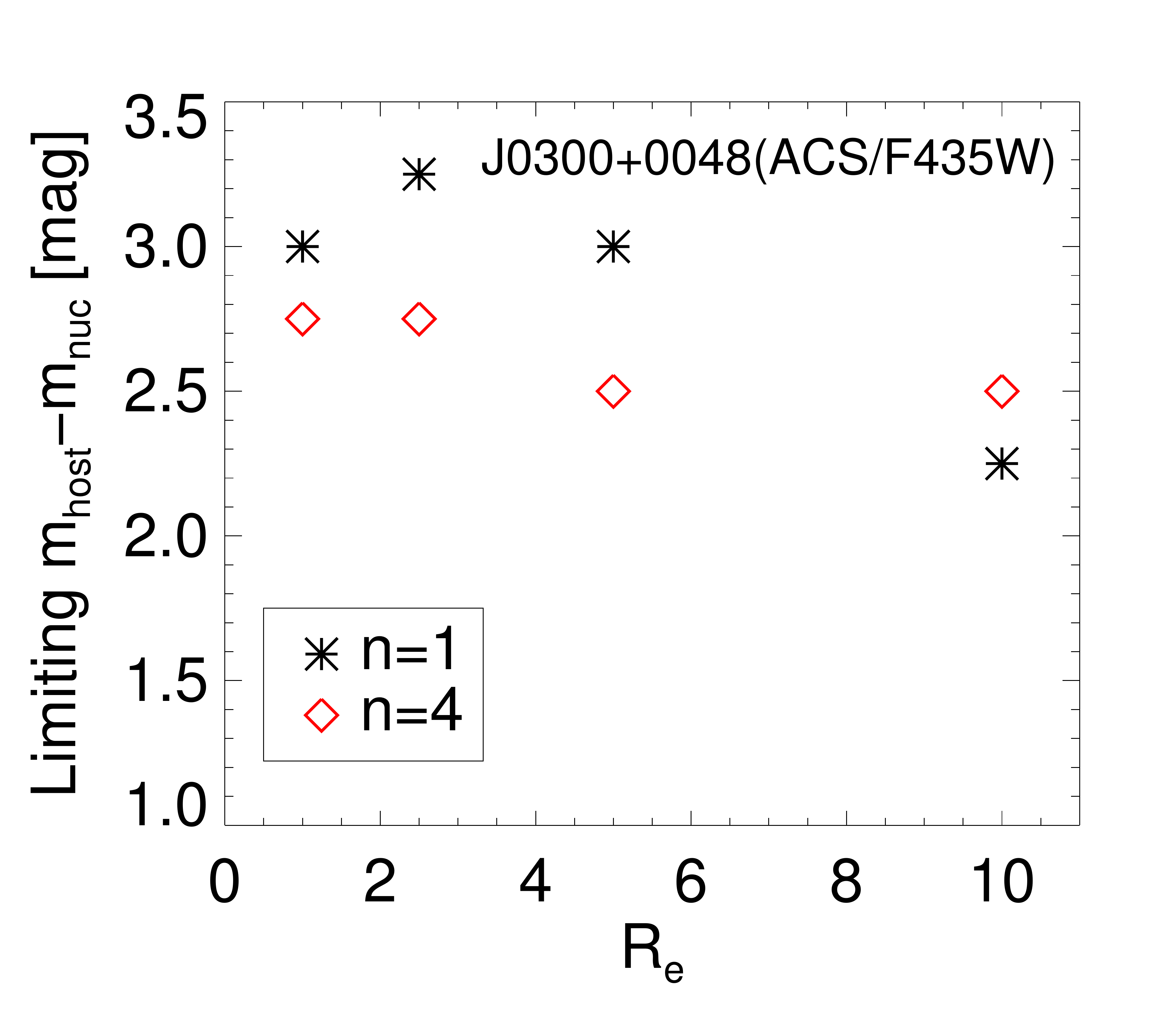}
	\includegraphics[scale=0.21]{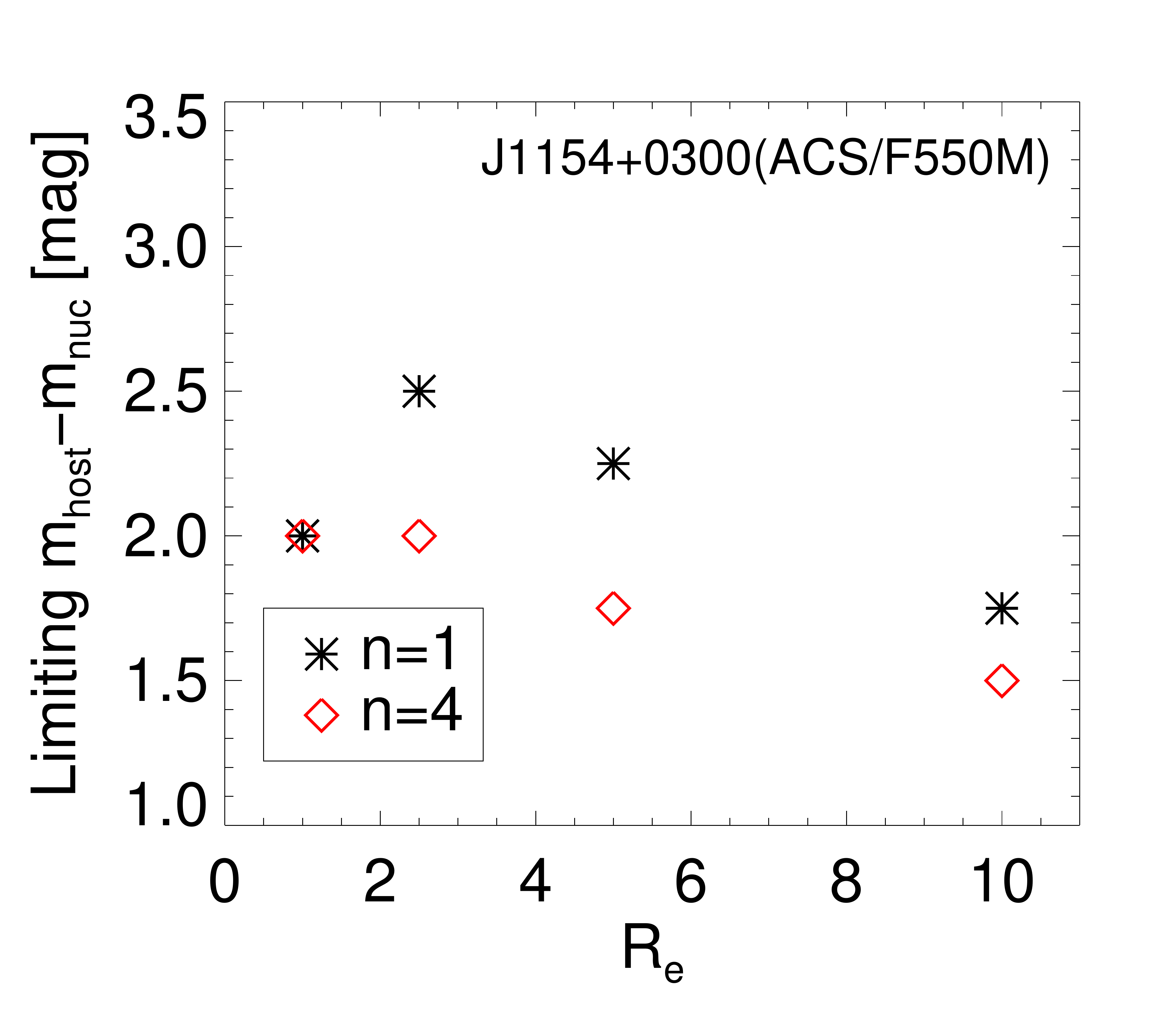}
	\includegraphics[scale=0.21]{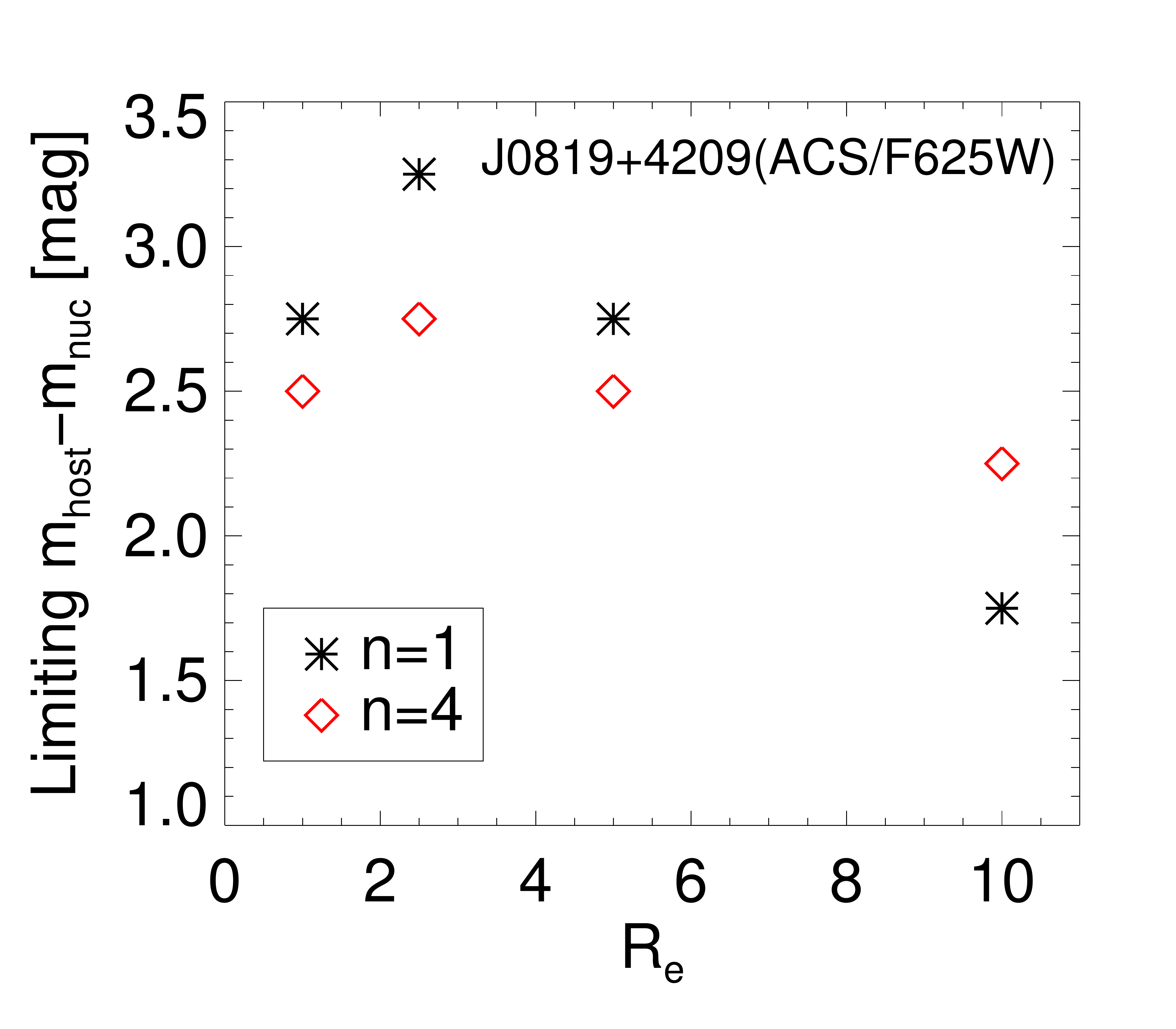}
	\includegraphics[scale=0.21]{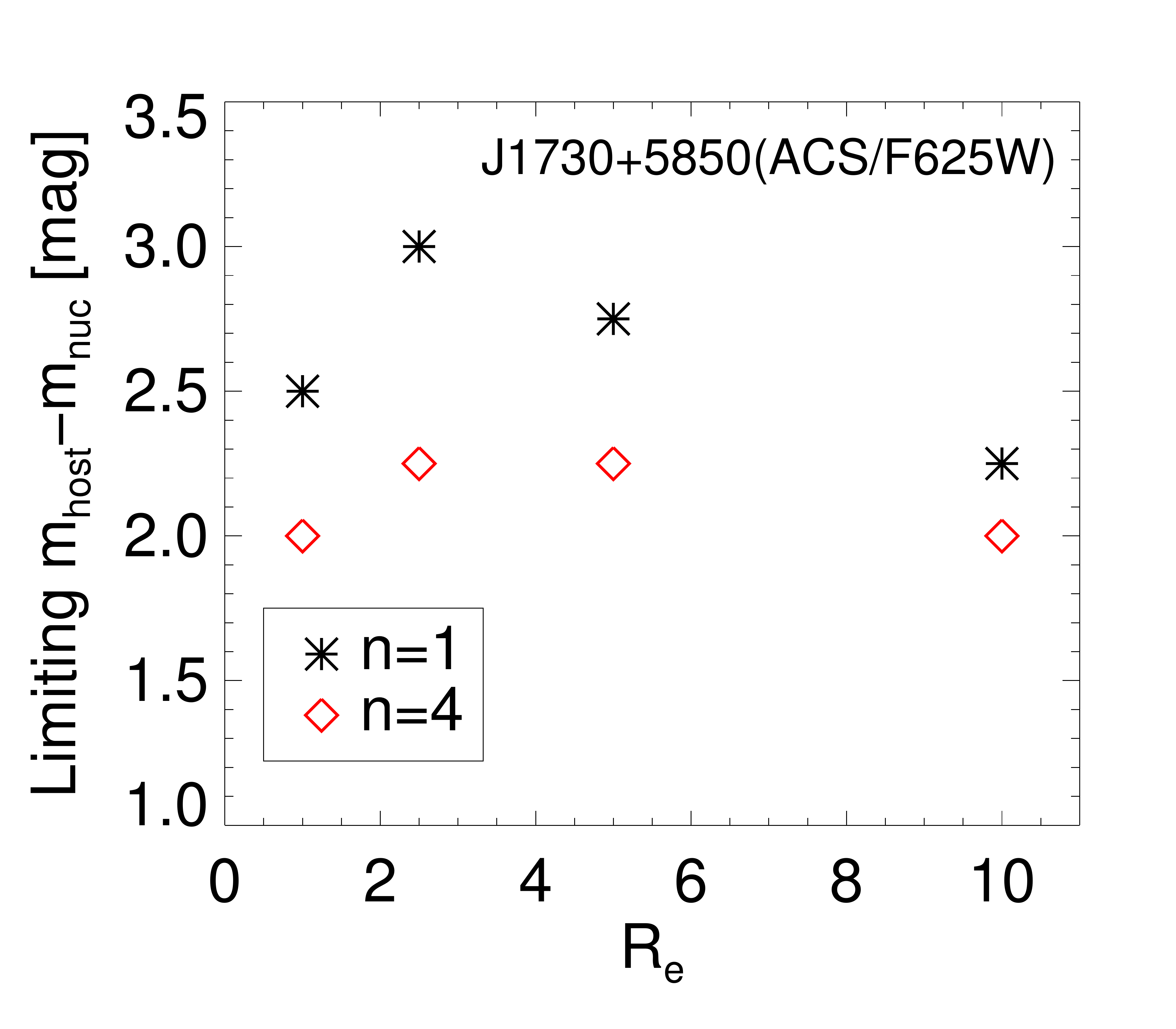}
	\caption{Detection limits for extended components in ACS observations of Fe\-LoBAL quasars, expressed in terms of the nucleus-to-host brightness contrast. These measurements allow us to place upper limits on the rest-frame UV host galaxy components (\S \ref{sec:fitting}). The upper limits depend both on the host galaxy properties (being weaker for very compact and very extended hosts), and on the observational situation (filter bandpass, exposure time, etc.). As we are ignorant of the host galaxy properties for a non-detection, it is appropriate to use the smallest limiting value of \contrast\,for the parameter range explored. E.g., for J0300+0048, the limiting contrast is \contrast=2.25 mag, as a $n=1$ galaxy with $R_e=10$ kpc becomes undetectable at this level.}\label{fig:acssims}
\end{figure*}

Using \textsc{Ferengi} to artificially redshift SDSS galaxies to $z\approx1$ as observed with ACS/WFC would involve an uncertain extrapolation of the SDSS imaging into the UV, an approach that \citet{Barden2008} advise against. Instead, we use analytical profiles to generate simulated observations for ACS. While these analytical-profile simulations do not account for the detailed morphological structure of real galaxies, we do not find this level of detail necessary for the present study, as we do not detect any of the FeLoBAL hosts in ACS imaging. Thus, we simply need to determine the values of \contrast\,at which the host galaxies are no longer detectable, so as to provide upper limits on the host galaxy magnitudes. We therefore produce a set of simulated images spanning nucleus-to-host contrasts $0\le\contrast\le4$ mag, scale sizes 0.5 kpc $\le R_e\le$ 10 kpc, and with Sersic indices of either $n=1$ or $n=4$. We prepare the simulated images as follows: we use the \textsc{Galfit} software to generate a Sersic profile representing a quasar host galaxy, and add Poisson noise due to this host component. We then add a point-source and a sky background component. A cutout image of a PSF star from the relevant ACS observation provides both the point-source and background components. We use the PSF star most similar in brightness to the Fe\-LoBAL quasar PSF. This procedure provides an image with a similar point-source brightness to the Fe\-LoBAL quasar, and with the same observing conditions (e.g., exposure time, background level and readout noise) as the real observation. We perform a PSF-only fit to each of the simulated observations, and classify them as detections or non-detections. Figure \ref{fig:acssims} shows the limiting nucleus-to-host brightness contrasts at which a host component of a given scale size and Sersic index is still detected, given the observing conditions of each ACS Fe\-LoBAL observation. As we are ignorant of the host galaxy properties for a non-detection, it is appropriate to use the smallest limiting value of \contrast\,for the parameter range explored. This yields the most conservative, i.e., brightest, upper limit on \mhost. We use these limiting values of \contrast\,to calculate the upper limits in Table \ref{tab:psffits_acs_limits}. In general, the more centrally-concentrated ($n=4$) hosts disappear at smaller limiting contrasts, as do the very  extended ($R_e=10$) hosts. The observations J1154+0300(ACS/F550M) and J1730+5850(ACS/F625W) require relatively bright host galaxies (\contrast$<1.5$ mag and $<2.0$ mag, respectively) if they are to provide a detection. This is due to these observations being less deep than the other ACS observations: in the case of J1154+0300(ACS/550M) this is due to the use of the narrower F550M filter, while J1730+5850(ACS/F625W) has a shorter total exposure time than originally granted, due to a technical failure during the observation.

\clearpage
\begin{deluxetable}{ccccccc}
	\tabletypesize{\footnotesize}
	\tablecaption{Host galaxy detection limits for NICMOS imaging\label{tab:psffits_ferengi_limits}}
	\tablewidth{0pt}
	\tablehead{
		\colhead{Host galaxy} & \colhead{$m_\mathrm{F110W}$\tablenotemark{b}} & \colhead{Lim. contrast\tablenotemark{c}} & \colhead{$m_\mathrm{F110W}$\tablenotemark{b}} & \colhead{Lim. contrast\tablenotemark{c}} & \colhead{$m_\mathrm{F160W}$\tablenotemark{b}} & \colhead{Lim. contrast\tablenotemark{c}} \\
		\colhead{template\tablenotemark{a}} & \colhead{z=0.89} & \colhead{z=0.89} & \colhead{z=1.46}  & \colhead{z=1.46} & \colhead{z=1.93} & \colhead{z=1.93} \\
		\colhead{(1)} & \colhead{(2)} & \colhead{(3)} & \colhead{(4)}  & \colhead{(5)} & \colhead{(6)} & \colhead{(7)} \\
	}
	\startdata
	\textbf{No luminosity} & & & \\
	\textbf{evolution:} & & & \\
	Ell./S0 & 21.35 & 3 mag & 23.62 & 1.5 mag & 23.54 & 0.5 mag\\
	Elliptical & 21.20 & 3 mag & 24.15 & 0.5 mag & 24.07 & 0.75 mag \\
	Edge-on spiral & 20.75 & 2.75 mag & 24.15 & 2 mag & 22.87 & 2.25 mag\\
	Face-on spiral & 20.03 & 3.5 mag & 23.37 & 3 mag & 22.31 & 3.5 mag \\
	Irr., ULIRG & 21.13 & 3.5 mag & 23.37 & 4 mag & 23.34 & 3 mag \\
	cD & 19.62 & $>$ 4.75 mag & 23.10 & $>$ 4.75 mag & 21.85 & 4.25 mag\\
	\textbf{$\mathbf{dL/dz=1}$} &  & & & \\
	Ell./S0 & 20.50 & 3.5 mag & 22.16 & 4 mag & 21.54 & 3.25 mag\\
	Elliptical & 20.88 & 3 mag & 22.60 & 3.5 mag & 22.03 & 3 mag\\
	Edge-on spiral & 19.88 & 3 mag & 21.69 & 4.5 mag & 20.97 & 3.25 mag \\
	Face-on spiral & 19.33 & $>$ 4.75 mag & 20.80 & $>$ 4.75 mag & 20.37 & 3.75 mag \\
	Irr./ULIRG & 20.25 & $>$ 4.75 mag & 21.93 & $>$ 4.75 mag & 21.38 & 4 mag \\
	cD & 18.71 & $>$ 4.75 mag & 20.58 & $>$ 4.75 mag & 19.88 & 4.25 mag\\
	\\
	\enddata
	\tablecomments{The first part of the table shows results for host components redshifted to the target redshift, applying cosmological dimming and assuming no luminosity evolution for the galaxies; these galaxies are rather faint as observed in NICMOS imaging. The second part of the table shows results for the same galaxies, assuming 1 mag of luminosity evolution per unit redshift. We use the latter results in our analysis (\S \ref{sec:fitting}), as the values of \mhost\,obtained assuming 1 mag of luminosity evolution more closely resemble those measured in our NICMOS images.}
	\tablenotetext{a}{The galaxy type of the host galaxy template. See Appendix \ref{sec:appendixB} for details of the specific galaxies used. }
	\tablenotetext{b}{The apparent brightness of the galaxy after artificially redshifting it to the target redshift, as imaged in the specified filter. This brightness was measured using a single Sersic component fit to the galaxy, without any superimposed point source component.}
	\tablenotetext{c} {The nucleus-to-host brightness contrast (\contrast) at which the host galaxy is only marginally detected upon fitting and subtracting a PSF component and examining the residuals.}
\end{deluxetable}

\clearpage
\section{Establishing Host Galaxy Detection Criteria and Upper Limits}\label{sec:appendix_detection}

Here, we discuss the host galaxy detection criteria applied in our analysis of the FeLoBAL imaging (\S \ref{sec:fitting}). First, we must consider the modeling of \emph{bona fide} point sources presented in Appendix \ref{sec:appendix_psf}. Due to PSF mismatch, the residual images of star-star subtractions display a characteristic `criss-cross' residual pattern (Figure \ref{fig:starfits_f110w}). When we add an extended component (Appendix \ref{sec:appendixB}), the PSF-only fit tends to oversubtract the center of the object, and leaves a broad 'fuzzy ring' of positive residuals. As the contrast between the nuclear and host galaxy apparent magnitudes, \contrast\,increases, the residual ring becomes fainter relative to the original flux. At a limiting value of \contrast, the residuals become dominated by PSF mismatch and PSF shot noise, and are thus indistinguishable from the residuals shown in Figure \ref{fig:starfits_f110w}. Based on these tests and simulations, we establish the following detection criteria for extended emission:

\emph{1)} At least three contiguous 1-pixel bins in our azimuthally averaged intensity plots show a positive residual at a significance $>3\sigma$ after PSF-only modeling. This must occur outside the inner 0.2''  of the radial profile for NICMOS and ACS/F425W, outside the inner 0.3'' for ACS/F550M, or outside the inner 0.4'' for ACS/F625W. These radial constraints are due to the severe mismatch in the PSF core, and a general tendency to oversubtract the inner regions when performing PSF-only fits in the presence of extended flux.

\emph{2)} The PSF wings are not significantly ($3\sigma$) oversubtracted in any azimuthally-averaged bin exterior to the radial constraints given in \emph{1)}. This requirement selects against spurious detections due to mismatch in the PSF curvature: true extended emission should be broader than a point source at all radii.

Only one star in our test sample (Appendix \ref{sec:appendix_psf}) yields a spurious detection according to these criteria. Namely, WD 1657+343, as imaged in NICMOS F110W. We suspect that this test star is an X-ray binary, which may therefore display real extended emission. As our simulations reveal a small risk ($\sim3$\%) of a false-positive host galaxy detection (at \contrast$\sim$3 mag) in PSF-plus-Sersic modeling (Appendix \ref{sec:appendixB}), we base these criteria exclusively on PSF-only modeling.

\paragraph*{Upper Limits for Non-Detections:}\label{sec:upper_limits}

For the observations where we do not detect extended flux, we estimate an upper limit on the brightness of an extended host galaxy, based on the limiting values of \contrast\,found in our simulations (Appendix \ref{sec:appendixB}). As we are ignorant of the morphology of the Fe\-LoBAL host galaxies, we establish detection limits for a range of galaxy types, and quote the most conservative (i.e., brightest) upper limit for a given observation.  Our limiting \contrast\,values are generally smaller for our ACS analysis compared to those obtained for NICMOS. There are two reasons for this. Firstly, we consider smaller values of $R_e$ (down to $R_e=0.5$ kpc) for the Sersic component in our rest-frame UV simulations, as the star formation may be concentrated towards the nucleus. Secondly, the quasars themselves are fainter in the ACS imaging, causing host galaxies at a given contrast level to be fainter in terms of flux. The asymmetry of the galaxy templates used in our NICMOS simulations may also aid detection; for the ACS analysis we used analytical, radially symmetric host galaxy profiles. We note that adopting larger (i.e., less conservative) limiting values of \contrast\,for the ACS analysis would only strengthen the conclusions drawn in \S \ref{sec:discussion_templates} with regards to the host galaxy spectral energy distribution.

\end{document}